\documentclass[twoside,10.9pt,nofootinbib]{article}
\usepackage{cite}
\usepackage{epsfig}
\usepackage{amsfonts}

\usepackage{titling}

\newcommand{\be}{\begin{equation}}
\newcommand{\ee}{\end{equation}}
\newcommand{\bea}{\begin{eqnarray}}
\newcommand{\eea}{\end{eqnarray}}

\usepackage{bm}
\usepackage{graphicx}
\usepackage{amsmath,amssymb}
\usepackage[utf8]{inputenc}

\usepackage[nottoc]{tocbibind}

\usepackage{lipsum}
\usepackage{needspace}
\usepackage{makecell}
\usepackage{multirow}
\usepackage[hang]{footmisc}

\usepackage[dvipsnames]{xcolor}
\usepackage[normalem]{ulem}
\usepackage{xspace}

\newcommand{\UNIT}[1]{\ensuremath{\,{\rm #1}}\xspace}
\newcommand{\bvec}[1]{\ensuremath{\boldsymbol{#1}}}
\newcommand{\refcite}[1]{Ref.~\cite{#1}}
\newcommand{\dd}{\mathrm{d}}
\newcommand{\MeV}{\UNIT{MeV}}
\newcommand{\GeV}{\UNIT{GeV}}
\newcommand{\AMeV}{\ensuremath{\,A{\rm MeV}}\xspace}
\newcommand{\AGeV}{\ensuremath{\,A{\rm GeV}}\xspace}
\newcommand{\MeVc}{\ensuremath{\,{\rm MeV}\!/c}\xspace}

\newcommand{\fm}{\UNIT{fm}}
\newcommand{\mb}{\UNIT{mb}}
\newcommand{\ba}{\begin{eqnarray}}
\newcommand{\ea}{\end{eqnarray}}

\DeclareRobustCommand{\sout}{\bgroup\markoverwith{\textcolor{red}{\rule[.5ex]{2pt}{1pt}}}\ULon}

\def\diff{\mathrm{d}}

\def\A{\mathrm{A}}
\def\B{\mathrm{B}}
\def\C{\mathrm{C}}
\def\D{\mathrm{D}}

\def\exp{\mathrm{exp}}

\def\vecr{\mathbf{r}}
\def\vecp{\mathbf{p}}
\def\Ntest{N_{\textrm{test}}}

\usepackage{comment}

\makeatletter
\def\l@subsubsection#1#2{}
\makeatother

\usepackage{tikz}
\usepackage{authblk}

\usepackage[colorlinks,citecolor=blue,urlcolor=blue,linkcolor=blue]{hyperref}
\usepackage{cleveref}
\providecommand{\keywords}[1]{\textbf{\textit{Keywords:}} #1}
\begin{document}

\title{ \vspace{1cm} Transport Model Comparison Studies of Intermediate-Energy Heavy-Ion Collisions}
\author{Hermann\ Wolter,$^{1,}\thanks{hermann.wolter@lmu.de}~$ Maria\ Colonna,$^{2,}\thanks{colonna@lns.infn.it}~$ Dan~Cozma,$^{3,}\thanks{dan.cozma@theory.nipne.ro}~$ Pawel\ Danielewicz,$^{4,5,}\thanks{danielewicz@frib.msu.edu}~$ Che~Ming~Ko,$^6$\thanks{ko@comp.tamu.edu} ~Rohit~Kumar,$^4$\thanks{kumarro4@msu.edu} Akira~Ono,$^{7,}\thanks{ono@nucl.phys.tohoku.ac.jp}~$ ~ManYee~Betty\ Tsang,$^{4,5,}\thanks{tsang@frib.msu.edu}~$ ~Jun~Xu,$^{8,9,}\thanks{xujun@zjlab.org.cn}~~$ Ying-Xun~Zhang,$^{10,11,}\thanks{zhyx@ciae.ac.cn}~$ ~Elena~Bratkovskaya,$^{12,13}$ Zhao-Qing~Feng,$^{14}$ Theodoros~Gaitanos,$^{15}$ Arnaud~Le~Fèvre,$^{12}$ Natsumi~Ikeno$^{16}$, Youngman~Kim,$^{17}$ Swagata~Mallik,$^{18}$ Paolo~Napolitani,$^{19}$ Dmytro~Oliinychenko,$^{20}$ Tatsuhiko~Ogawa,$^{21}$ Massimo~Papa,$^{2}$ Jun~Su,$^{22}$ Rui~Wang,$^{9,23}$ Yong-Jia~Wang,$^{24}$ Janus~Weil,$^{25}$  Feng-Shou~Zhang,$^{26,27}$ Guo-Qiang~Zhang,$^{9}$ Zhen~Zhang,$^{22}$ Joerg~Aichelin,$^{28}$ Wolfgang~Cassing,$^{25}$ Lie-Wen~Chen,$^{29}$ Hui-Gan~Cheng,$^{14}$ Hannah~Elfner,$^{12,13,20}$ K.~Gallmeister,$^{25}$ Christoph~Hartnack,$^{28}$ Shintaro~Hashimoto,$^{21}$ Sangyong~Jeon,$^{30}$ Kyungil~Kim,$^{17}$ Myungkuk~Kim,$^{31}$ Bao-An~Li,$^{32}$ Chang-Hwan Lee,$^{33}$ Qing-Feng Li,$^{24,34}$ Zhu-Xia~Li,$^{10}$ Ulrich~Mosel,$^{25}$ Yasushi~Nara,$^{35}$ Koji~Niita,$^{36}$ Akira~Ohnishi,$^{37}$ Tatsuhiko~Sato,$^{21}$ Taesoo~Song,$^{12}$ Agnieszka~Sorensen,$^{38,39}$ Ning~Wang,$^{11,40}$ Wen-Jie Xie$^{41}$\newline
{(\large TMEP collaboration)}\\

$^1$Faculty of Physics, University of Munich, D-85748 Garching, Germany, \\
$^2$INFN-LNS, Laboratori Nazionali del Sud, 95123 Catania, Italy,\\
$^3$IFIN-HH, 077125 M\v{a}gurele-Bucharest, Romania,\\ 
$^4$Facility for Rare Isotope Beams, Michigan State University, East Lansing, Michigan 48824, USA \\
$^5$Department of Physics and Astronomy, Michigan State University, East Lansing, Michigan 48824, USA\\
$^6$Cyclotron Institute and Department of Physics and Astronomy, Texas A$\&$M University, \\College Station, Texas 77843, USA \\
$^7$Department of Physics, Tohoku University, Sendai 980-8578, Japan \\
$^{8}$Shanghai Advanced Research Institute, Chinese Academy of Sciences, Shanghai 201210, China \\
$^{9}$Shanghai Institute of Applied Physics, Chinese Academy of Sciences, Shanghai 201800, China \\
$^{10}$Department of Nuclear Physics, China Institute of Atomic Energy, Beijing 102413, P.R. China \\
$^{11}$Guangxi Key Laboratory Breeding Base of Nuclear Physics and Technology, Guilin 541004, China\\
$^{12}$GSI Helmholtzzentrum f\"{u}r Schwerionenforschung, Planckstr. 1, 64291 Darmstadt, Germany \\
$^{13}$Institute for Theoretical Physics, Goethe University, Max-von-Laue-Strasse 1, 60438 Frankfurt am Main, Germany \\
$^{14}$School of Physics and Optoelectronics, South China University of Technology, Guangzhou 510640, China \\
$^{15}$Department of Physics, Aristotle University of Thessaloniki, GR-54124 Thessaloniki, Greece\\
$^{16}$Department of Life and Environmental Agricultural Sciences, Tottori University, Tottori 680-8551, Japan\\
$^{17}$Rare Isotope Science Project, Institute for Basic Science, Daejeon 34000, Korea\\
$^{18}$Physics Group, Variable Energy Cyclotron Centre, 1/AF Bidhan Nagar, Kolkata 700064, India\\
$^{19}$Université Paris-Saclay, CNRS/IN2P3, IJCLab, 91405 Orsay, France\\
$^{20}$Frankfurt Institute for Advanced Studies, Ruth-Moufang-Strasse 1, 60438 Frankfurt am Main, Germany\\
$^{21}$Division of Environment and Radiation Sciences, Nuclear Science and Engineering Directorate, Japan Atomic Energy
Agency, Shirakata-Shirane, Tokai, Ibaraki 319-1195, Japan\\
$^{22}$Sino-French Institute of Nuclear Engineering $\&$ Technology, Sun Yat-sen University, Zhuhai 519082, China\\
$^{23}$Key Laboratory of Nuclear Physics and Ion-beam Application~(MOE), Institute of Modern Physics, Fudan University, Shanghai 200433, China\\
$^{24}$School of Science, Huzhou University, Huzhou 313000, China\\
$^{25}$Institut f\"ur Theoretische Physik, Universit\"at Giessen, Giessen, Germany\\
$^{26}$Key Laboratory of Beam Technology of Ministry of Education, College of Nuclear Science and Technology, Beijing Normal University, Beijing 100875, China\\
$^{27}$Institute of Radiation Technology, Beijing Academy of Science and Technology, 100875 Beijing, China,\\
$^{28}$SUBATECH, UMR 6457, IMT Atlantique, IN2P3/CNRS Universit\'e de Nantes, 4 rue Alfred Kastler, 44307 Nantes, France \\
$^{29}$School of Physics and Astronomy, Shanghai Key Laboratory for Particle Physics and Cosmology, and Key Laboratory for Particle Astrophysics and  Cosmology (MOE), Shanghai Jiao Tong University, Shanghai 200240, China\\
$^{30}$Department of Physics, McGill University, Montreal, Quebec, Canada H3A2T8\\
$^{31}$Department of Physics, Ulsan National Institute of Science and Technology, Ulsan 44919, Korea\\
$^{32}$Department of Physics and Astronomy, Texas A$\&$M University-Commerce, Commerce, TX 75429-3011, USA\\
$^{33}$Department of Physics, Pusan National University, Busan 46241, Korea\\
$^{34}$Institute of Modern Physics, Chinese Academy of Sciences, Lanzhou 730000, People's Republic of China\\
$^{35}$Akita International University, Akita 010-1292, Japan\\
$^{36}$Research Organization for Information Science and Technology, Shirakata-shirane, Tokai, Ibaraki 319-1188, Japan\\
$^{37}$Yukawa Institute for Theoretical Physics, Kyoto University, Kyoto 606-8502, Japan\\
$^{38}$Department of Physics and Astronomy, University of California, Los Angeles, CA 90095, USA\\
$^{39}$Lawrence Berkeley National Laboratory, 1 Cyclotron Road, Berkeley, California 94720, USA\\
$^{40}$Department of Physics and Technology, Guangxi Normal University, Guilin 541004, China\\
$^{41}$Department  of  Physics,  Yuncheng  University,  Yuncheng  044000,  China\\
}


\maketitle
\begin{abstract} 
Transport models are the main method to obtain physics information on the nuclear equation of state and
in-medium properties of particles from low to relativistic-energy heavy-ion collisions. The Transport
Model Evaluation Project (TMEP) has been pursued to test the robustness of transport model predictions
in reaching consistent conclusions from the same type of physical model. To this end, calculations
under controlled conditions of physical input and set-up were performed with various participating codes.
These included both calculations of nuclear matter in a box with periodic boundary conditions, which
test separately selected ingredients of a transport code, and more realistic calculations of heavy-ion collisions.
Over the years, six studies have been performed within this project. In this intermediate review, we
summarize and discuss the present status of the project. We also provide condensed descriptions of the
26 participating codes, which contributed to some part of the project. These include the major codes in
use today. After a compact description of the underlying transport approaches, we review the main
results of the studies completed so far. They show, that in box calculations the
differences between the codes can be well understood and a convergence of the results can be reached.
These studies also highlight the systematic differences between the two families of transport codes,
known under the names of Boltzmann-Uehling-Uhlenbeck (BUU) and Quantum Molecular Dynamics (QMD) type
codes. 
However, when the codes were compared in full heavy-ion collisions using different physical models, as recently for pion production, they still yielded substantially different results. This calls for further comparisons of heavy-ion collisions with controlled models and of box comparisons of important ingredients, like momentum-dependent fields, which are currently underway.  
Our evaluation studies often indicate improved strategies in performing transport simulations and thus
can provide guidance to code developers.
Results of transport simulations of heavy-ion collisions from a given code will have more significance
if the code can be validated against 
benchmark calculations such as the ones summarized in this
review.\\ \\
\keywords heavy-ion collisions, intermediate energy, transport theory, transport codes, nuclear equation-of-state
\end{abstract}
%




\tableofcontents 

\newpage

\section{Introduction}

One of the great challenges in nuclear physics today is the determination of the nuclear matter equation of state (EoS), i.e., the behavior of the energy or pressure of nuclear matter under the variation of density, temperature and neutron-proton asymmetry. The last aspect, namely the nuclear symmetry energy, is of particular importance in understanding the structure of heavy nuclei and complex astrophysical objects.
Over the years, information about the EoS has been obtained from astrophysical observations~\cite{lattimer2012nuclear} as well as from nuclear physics experiments~\cite{Horowitz:2014bja}, particularly those involving  heavy-ion collisions, which can achieve densities above the saturation density $\rho_0$ in the laboratory. In astrophysics, a number of observations of neutron stars with masses near or above two solar masses have set important limits on the pressure of neutron-rich matter at high density~\cite{Cromartie2020,Fonseca21,Zhang_2021}. The observation of a neutron star merger event has provided limits on the deformability of neutron stars~\cite{LIGO18}, and recent simultaneous determinations of masses and radii from milli-second pulsars begin to set limits on the neutron-star mass-radius correlation~\cite{raaijmakers2019nicer,Miller:2021qha,Riley:2021pdl}. 
While astrophysical observations can provide information on the global behavior of the EoS under stellar conditions, nuclear physics observations and their interpretations allow to investigate the EoS in considerable detail, like its density dependence from very low to supra-saturation densities, its temperature and asymmetry dependence, and its composition under these various conditions.  

For densities below $\rho_0$, rather strict limits on the nuclear symmetry energy have come from nuclear structure and reaction measurements and their interpretation~\cite{Li:2008gp,Tsang:2012se,Horowitz:2014bja,Colonna:2020euy,lynch2021decoding}.
 Precision measurements of masses and isobaric analog states~\cite{danielewicz2017symmetry} are mainly sensitive to the symmetry energy  at about
 (2/3)$\rho_0$.
Measurements of the neutron skin thickness in heavy nuclei show a dependence on the slope $L$ of the symmetry energy at this density~\cite{Xu:2020fdc}. The recent model-independent measurement of the neutron skin of the $^{208}$Pb nucleus with parity-violating electron scattering (PREX2)~\cite{PREXII}, and very recently also of the $^{48}$Ca nucleus (CREX)~\cite{PREX:2021uwt}, has provided additional constraints here. The analysis of the dipole polarizability of nuclei~\cite{Zhang:2015ava} appears sensitive to the symmetry energy $S$ below 0.5$\rho_0$~\cite{lynch2021decoding}.{}
Both the symmetry energy and its slope obtained below $\rho_0$ have been extrapolated to saturation density $\rho_0$.  However, this procedure is not without model dependence~\cite{lynch2021decoding}.

While nuclear structure experiments are generally limited to investigations below saturation density, heavy-ion collisions are an important way to obtain information on the EoS in the laboratory away from saturation. By selecting different collision systems with different isospins, collision energies,  and impact geometries, the phase diagram in the hadronic domain (and similarly for collisions in the partonic regime at ultra-relativistic energies) can be and has been explored intensively. For densities below saturation, tight constraints have been obtained from isospin diffusion~\cite{tsang04,Che05,SunTsang2010isodiff,tsang2009constraints} and fractionation~\cite{iso_frac}, and from the isotopic content of produced nucleons or light clusters~\cite{Filippo2012,morfouace2019constraining,lynch2021decoding}. At present, one of the major challenges is to obtain constraints on the density dependence of the symmetry energy above saturation density. Measurements of isospin-dependent momentum distributions (stopping and flow)~\cite{LeFevre:2015paj,Russotto-Asyeos,Cozma:2013sja} and of light cluster and particle yields~\cite{FOPI-pion,FOPI:2010xrt} have been proposed as promising ways to make progress towards this goal. A recent high precision experiment of collisions of different isotopes of Sn at 270 MeV/nucleon has already provided important  data~\cite{SpRIT:2020blg,estee2021probing} that is sensitive to the symmetry energy at around 1.5$\rho_0$\cite{Russotto-Asyeos,Cheng:2016pso}, and more data will be forthcoming.

Heavy-ion collisions create short-lived and dynamic states of nuclear matter that are out of equilibrium over large fractions of the evolution time. The interpretation of such events represents a great challenge to reaction theory. Non-equilibrium methods have to be applied to interpret the outcome of the collision and to extract information on the EoS of nuclear matter at equilibrium, and on the in-medium properties and cross sections of hadrons. To do so, semi-classical transport theory has been used for many years with considerable success, see  topical issue on "Challenges in nuclear dynamics and thermodynamics"\cite{Gulminelli:2006zc} or  Refs.~\cite{xu2019transport,Colonna:2020euy}.  

Transport theories are formulated in the full one-body phase space and are highly non-linear. They consist of a mean-field evolution of the phase-space distribution (Vlasov equation) and a collision term, usually of two-body nature, to describe the dissipation (Boltzmann collision term). Fluctuations are also of great importance, and they require the inclusion of stochastic elements into the treatment. There are basically two families of transport approaches for heavy-ion collisions that are dominated by hadronic degrees of freedom, namely the Boltzmann-Vlasov type (usually called the Boltzmann-Uehling-Uhlenbeck (BUU) method) and the molecular dynamics type (usually called the Quantum Molecular Dynamics (QMD) method). A short review of the physical approach used in these two families will be given in the next section. Because of their complexity, the relevant equations  are not solved directly in these models, but rather by numerical methods, in particular by simulations with test particle or molecular dynamics approaches.
Ideally, conclusions from comparing transport calculations with given experimental data should not depend critically on the implementation of the simulation. However, any implementation involves choices of strategies and approximations, which are not directly enforced by the underlying equations. Therefore, the question arises on how tightly the physical modeling of the collision is connected to the result and thus to the inferences from an experiment, or what role is played by the particular implementation of the simulation. In fact, recent investigations based on the same experimental data came to
different conclusions.  E.g., in interpreting the ratios of produced
oppositely charged pions different analyses extracted very different density dependences of the nuclear symmetry energy without changes of other ingredients of the physical models, see Ref.~\cite{Song:2015hua} and references therein. Another example is the double ratio of neutron over proton pre-equilibrium emissions~\cite{Coupland:2014gya,Kong:2015rla}.

In this situation, the idea of a Transport Model Evaluation Project (TMEP), i.e., of code comparisons under controlled conditions or of benchmark calculations, has arisen already some time ago. It was started in 2005 with a study of $K$ and $\pi$ meson production in the energy regime of 1 AGeV~\cite{ECT05}, and then with workshops in 2009 and 2014 moved to the intermediate energy regime of 100 to 400 AMeV, where most of the measurements on the longitudinal and transverse momentum distributions were performed. In the latter case, a comparison of Au+Au heavy-ion collisions showed differences in the observables between 13 and 30\%, depending on the energy~\cite{Xu2016}. However, in a full simulation of a heavy-ion collision, it is difficult to tie differences in the results to specific aspects of the simulation, as different effects influence each other. Thus, to better understand these discrepancies, simulations were made of infinite nuclear matter in a box with periodic boundary conditions. This method allows to investigate the different ingredients separately and compare them to exact limits.  Several investigations of this kind have already been completed: a study of the mean-field propagation in a Vlasov calculation~\cite{Colonna2021}, a study of the treatment of the collision term in a cascade calculation~\cite{Zhang2017}, and a study of pion production also in the cascade mode~\cite{ono2019}. Further investigations in a box are in progress of pion production in the presence of momentum-dependent mean fields~\cite{cozma2021}. In parallel, comparative studies are underway of pion production in relation to experiments of Sn+Sn collisions at 270 MeV/nucleon~\cite{SpRIT:2020blg,estee2021probing,xu2021}. The completed studies and their results are reviewed in Sec.~\ref{sec:review}. It should be noted that controlled comparisons of complex simulations have also been undertaken in other fields of physics, from atomic traps, through ultra-relativistic heavy ion collisions, to core collapse supernova calculations, and they have always been very fruitful for the respective fields~\cite{Xu:2004mz,lepers2010,2018JPhG45,just2018}.

Transport models are used in many fields of physics, in other sciences, and in applications. Here we mainly discuss transport models for nuclear systems with the aim to answer questions about the nuclear EoS and consequences for astrophysical systems. But there are numerous research areas where nuclear transport models can be important outside of these immediate questions. In many fields of basic research, nuclear targets are used in energetic collisions with electrons, neutrinos, possibly dark matter particles.  However, the energetic particles often disrupt the target or modify it in a way that cannot be treated perturbatively or statistically.
It is then important to know as precisely as possible the final state of the nuclear target to calibrate the energy of the incoming particle.
In many applications a detailed understanding of nuclear collisions is important: Examples are the design and performance of radiation shielding, the conditions of space travel, the design and simulation of detector responses, the investigation of the effectiveness of transmutation, and  therapy with particle beams. Transport models are also very good event generators for training deep-learning neural networks. 
Thus in many fields, quantitatively reliable transport models for nuclear systems are of great importance, and the TMEP initiative can contribute towards this goal. 

In the TMEP studies, the physics input and collision set-up were usually controlled, but the internal strategies of the simulations were left as in the normal use for interpreting experimental data from heavy-ion collisions.  Due to limitations in space, we gave some information on these strategies in Refs.~\cite{Xu2016,Colonna2021,Zhang2017,ono2019}.  However, this only gives limited information on the ideas of the code developers. Even though detailed  descriptions of most codes have been published, they
are scattered in the literature, and thus it is not easy to have rapid access to this information. One objective of the present article is to provide compact descriptions, written by the code authors, of all codes that have participated in at least one of the TMEP comparisons, and these are listed in Table~\ref{transport_models}.
We give the code names and the code correspondents, who usually are also the authors of the code descriptions, the energy range for which the code is intended,  and the treatment of the effects of relativity, i.e., of non-relativistic or relativistic kinematics and/or of a covariant treatment of the forces. Here we limited the upper energy range to 200 A GeV, but some codes can also be used for LHC energies up to 10 TeV, as discussed in the code descriptions. Note also that some codes can be run in different modes, e.g., GiBUU with relativistic kinematics or with covariant forces, IBUU and ImQMD in the regular mode or with the lattice-Hamiltonian method.

\begin{table}[h!]
\caption{List of transport models that participated in the TMEP code comparisons discussed in this paper. The columns give the information on the name of the code, the main correspondents of the code, the energy range intended for the code, the treatment of effects of relativity (see Sec.~\ref{sec:2.1}), and the comparisons in which the code participated. The different comparisons are listed in the last column in the table by a numbers n, which refer to the subsections~\ref{sec:review}.n, where they are described in detail:  n=1 for Au+Au collisions around 1 AGeV, n=2 for Au+Au collision at 100 and 400 AMeV, n=3 for box-Vlasov, n=4 for box-cascade with only nucleons, n=5 for box-cascade with pion and $\Delta$ resonance production, and n=6 for the prediction of pion ratios for Sn+Sn collisions.\label{transport_models}}
\begin{center}
\resizebox{\textwidth}{!}{%
\begin{tabular}{ccccc}
\hline\hline\noalign{\smallskip}
BUU Type & Code Correspondents & Energy Range~[A GeV] & Relativity & Comparisons\\\\
\noalign{\smallskip}\hline\hline\noalign{\smallskip}
BLOB & P. Napolitani, M. Colonna & 0.01-0.5 &non-rel&2\\
BUU-VM & S. Mallik & 0.02-1 &rel&3,4,5\\
DJBUU & Y. Kim, S. Jeon, M. Kim, C.-H. Lee, K. Kim &0.05-2 &cov &3\\ 
GiBUU & J. Weil, T. Gaitanos, K. Gallmeister, U. Mosel &0.05-40&rel/cov&1,2,3,4\\
IBL & W.J. Xie, F.S. Zhang & 0.05-2 &rel&2\\
IBUU & J. Xu, L.W. Chen, B.A. Li &0.05-2&rel&2,3,4,5\\
LBUU(LHV) & R. Wang, Z. Zhang, L.-W. Chen & 0.01-1.5 & rel & 3\\
pBUU & P. Danielewicz & 0.01-12 &rel&1,2,3,4,5,6\\
PHSD & E. Bratkovskaya, W. Cassing & 0.1-200 & rel/cov& 1,6\\
RBUU & T. Gaitanos & 0.05-2&cov&1,2\\
RVUU & Z. Zhang, C.M. Ko, T. Song  &0.05-2&cov&1,2,3,4,5\\
SMASH & D. Oliinychenko, H. Elfner, A. Sorensen & 0.5-200 &cov&3,4,5,6\\
SMF & M. Colonna, P. Napolitani &0.01-0.5&non-rel&2,3,4\\
{$\chi$}BUU & Z. Zhang, C.M. Ko &0.01-0.5&non-rel&6\\
\noalign{\smallskip}\hline\hline\noalign{\smallskip}
QMD Type & Code Corespondents & Energy Range~[A GeV] & Relativity &  Comparisons\\\\
\noalign{\smallskip}\hline\hline\noalign{\smallskip}
AMD & A. Ono &0.01-0.3&non-rel&2\\
AMD+JAM & N. Ikeno, A. Ono &0.01-0.3&non-rel+rel&6\\
BQMD/IQMD & A. Le Fèvre, J. Aichelin, C. Hartnack, R. Kumar &0.05-2&rel&1,2,6\\
CoMD & M. Papa &0.01-0.3&non-rel&2,4\\
ImQMD & Y.X. Zhang, N. Wang, Z.X. Li &0.02-0.4&rel&2,3,4\\
IQMD-BNU & J. Su, F.S. Zhang &0.05-2&rel&2,3,4,5,6\\
IQMD-SINAP & G.Q. Zhang &0.05-2&rel&2\\
JAM & A. Ono, N. Ikeno, Y. Nara, A. Ohnishi&1-158&rel&4,5\\
JQMD 2.0 & T. Ogawa, K. Niita, S. Hashimoto, T. Sato &0.01-3&rel&4,5\\
LQMD(IQMD-IMP) & Z.Q. Feng, H.G. Cheng &0.01-10&rel&2,3,4,5\\
TuQMD/dcQMD & D. Cozma&0.1-2&rel&1,2,3,4,5,6\\
UrQMD & Y. J. Wang, Q. F. Li, Y. X. Zhang &0.05-200
&rel&1,2,3,4,6\\
\noalign{\smallskip}\hline\hline
\end{tabular}
}
\end{center}
\end{table}

The last column in Table~\ref{transport_models} indicates the  completed comparisons in which the code participated. 
It is seen that essentially all codes that have been applied to heavy-ion collisions in the hadronic regime in recent years have participated in part of these studies, which thus give a good representation of the current activity in this field. The description of the codes collected here presents a unique overview of all the transport codes in use in this field today. Some codes joined in the later investigations, while others have dropped out partly due to lack of time by the participants to perform the required calculations. It should be noted that in the course of the comparisons, some codes introduced modifications, either because the result was very different from the general behavior, or because the codes deviated from exact limits in the box calculations. We see this as the positive effect of the project and a good strategy to evaluate any code. 

However, we did not attempt to develop a universal code for transport calculations in this energy regime, which did not appear to us a realistic and even desirable goal. 
Lessons learned from the box comparisons have been extensively discussed during the TMEP Collaboration meetings and some of
the codes have been improved accordingly.
This will become more evident when the codes are compared again for full heavy ion collisions with controlled input, a study of which is presently underway.  
We think that it would be desirable to give version numbers to the codes documenting their development, to indicate which features and modifications have been included, as is done already by some codes in the project. In the long run it may also be desirable to have a repository for transport codes in this energy regime, once they are sufficiently stable in their development. Presently we think that the consistency with benchmark calculations is the more practical way to assess the performance of a code.

We also emphasize that this study does not attempt to determine the correct physical models to describe an observable, as, e.g., the above-mentioned pion-ratio. For this the physical models in our comparisons are mostly not realistic enough anyway. Rather we want to  ascertain that simulations with the same physical model reach similar conclusions for an observable, or, if they do not, that we can identify the origins in the specific assumptions of a model or a class of models. In a certain sense such differences constitute a kind of systematic theoretical error in the interpretation of heavy-ion collisions by transport models, and in this project we aim to quantify and hopefully reduce this error.

This review is organized as follows. In Sec.~\ref{sec:transport_approaches}, we give a brief characterization of the two main families of transport approaches, the BUU- and QMD-type codes, in order to establish a unified framework and terminology. In Sec.~\ref{sec:review}, we summarize the main results of the comparisons completed so far and discuss the ongoing and planned studies as well as open problems. In Secs.~\ref{sec:buu} and \ref{sec:qmd}, we then collect the code descriptions provided by the corresponding authors for their BUU-type or QMD-type codes. The paper closes with some concluding remarks.

\section{Transport approaches}
\label{sec:transport_approaches}

In this section, we briefly characterize the two main approaches used for transport simulations. It is not intended as a comprehensive theoretical discussion of the derivation and validity of transport theories, which can be found in various reviews \cite{Bertsch:1988ik,Bass:1998ca,Malfliet90,Danielewicz:1982kk,bonasera1994,xu2019transport}. Rather, it should serve as a guide of the main characteristics, the methods of implementation, and the physical model ingredients of the transport approaches, and it establishes a terminology and sets a framework for the description of the codes. This section follows generally the corresponding sections of Refs.~\cite{Xu2016,Colonna2021}.

The primary method to describe the dynamics of nuclear collisions from Fermi to relativistic energies is semi-classical transport theories, such as the Nordheim approach, in which the Vlasov
equation for the one-body phase-space distribution, $f(\vec{r},\vec{p}; t)$, is extended with a Pauli-blocked Boltzmann collision term \cite{Bertsch:1988ik,bonasera1994}, which accounts for the average effect of the two-body residual interaction. The resulting transport equation, often called Boltzmann-Uehling-Uhlenbeck (BUU) equation, contains two main ingredients: the self-consistent mean-field potential and the two-body scattering cross sections. In order to introduce fluctuations and further (many-body) correlations in the treatment of the reaction dynamics, two main avenues have been taken (see Refs.~\cite{xu2019transport,Colonna:2020euy,Colonna2021,ono19_PPNP}). One is the class of molecular
dynamics (MD) models \cite{firstIQMD,Aic91,ono99,fel90,col98,ono92,ono1992a,Papa01,zha06}, while the other is represented by stochastic extensions of mean-field approaches of the Boltzmann-Langevin type~\cite{Ayik88,AbeAyik96,ran90,rep,Napolitani2013,Napolitani2015}. 

\subsection{{BUU-like models}} \label{sec:2.1}

In BUU-like approaches, the time evolution of the one-body phase-space distribution function of particle species $a$, $f_a(\vec{r},\vec{p};t)$, follows the equation
\begin{equation}
\Big(\frac{\partial}{\partial t}+ \vec{\nabla}_p\epsilon \cdot \vec{\nabla}_r-\vec{\nabla}_r \epsilon\cdot \vec{\nabla}_p\Big) f_a(\vec{r},\vec{p};t)=I_{\text{coll}} [f_a(\vec{r},\vec{p};t)] \, ,
\label{eq:BUU}
\end{equation}
where $\epsilon[f]$ is the single-particle energy, which is generally momentum-dependent and can usually be derived from a density functional, and $I_{\text{coll}}$ is the two-body collision integral due to the two-body scattering
$p+p_b\rightarrow p^\prime+p_b^\prime$,
\be
I_{\text{coll}}[f_a]=\sum_b\frac{g_b}{(2\pi\hbar)^3}\int d^3p_b\, d\Omega'\, v_{ab}\, \frac{d\sigma_{ab}^{\rm med}}{d\Omega'} [(1-f_a) (1-f_b) f_a^\prime f_b^\prime  - f_a f_b (1-f_a^\prime) (1-f_b^\prime)].
\ee
The distribution functions in Eq.(2) are all taken at the same position $\vec{r}$ and time $t$, and the momenta  $\vec{p}\,'$ and $\vec{p_b}'$ are determined by energy-momentum conservation and the scattering angle $\Omega'$.
The summation $b$ in the simplest case is over neutrons and protons and $g_b$ is the spin degeneracy, but it may be extended to include other particle species with evolution equations of their own phase-space densities of the type of Eq.\eqref{eq:BUU}. In the above, $d\sigma_{ab}^{\rm med}/d\Omega$ are the in-medium nucleon-nucleon elastic differential scattering cross sections, or, for the case of other particle species, the corresponding inelastic cross sections. The relative velocity factor $v_{ab}$ is given by the difference in velocities, non-relativistically and relativistically for collinear velocities, as  $v_{ab}=|\vec{v}-\vec{v_b}|$, and otherwise as
\be
v_{ab}=\frac{[(p \cdot p_b)^2-M_a^2M_b^2]^{1/2}}{p^0 p_b^0} ,
\label{eq:v-ab}
\ee
where  $M_a$ and $M_b$ are the bare masses, and the numerator on the r.h.s. involves a product of the 4-vectors $p=(p^0\equiv{\epsilon},\vec{p})$ and $p_b=(p_b^0,\vec{p_b})$. The particular form of Eq.~(\ref{eq:v-ab}) emphasizes the covariance of the relativistic Boltzmann equation, as $p_b^0$ in the denominator can be combined with the momentum integration to yield the invariant momentum measure $d^3p_b/p_b^0$. Moreover, both sides of the Boltzmann equation can be multiplied by $p^0$ to produce a covariant derivative in the first term on the l.h.s. The phase-space distribution, averaged over spin, is also a Lorentz scalar.
Actually, most codes described in this review use some kind of relativistic formulation.

To characterize the different formulations of the BUU equation, we introduce for each considered particle species (index $a$ omitted in the following for simplicity) the kinetic momentum ${p^*}^\mu = p^\mu - V^\mu$ and the energy $E^*\equiv {p^*}^0 = (\vec{p^*}^2 + {m^*}^2)^{1/2}$. Here $V^\mu$ represents the vector field and $m^*$ is the Dirac effective mass, given by $m^* = M - \Phi$, where $\Phi$ is the scalar field and $M$ denotes the bare mass of the particle.  The vector field depends on the baryon four-current $j^{\mu}(\vec{r};t)=(\rho,\vec{j})$, which is given self-consistently by 
\begin{equation}
j^{\mu} = g\int \frac{d^3p^*}{(2\pi\hbar)^3} ~ \frac{{p^*}^\mu}{E^*}~ f(\vec{r},\vec{p^*};t) ,
\label{eq:current}
\end{equation}
where $g$ is the degeneracy factor of the considered species ($4$ in the case of nucleons in symmetric nuclear matter). The scalar field $\Phi(\rho_S)$ depends on the scalar density $\rho_S(\vec{r};t)$ defined as
\begin{equation}
{\rho_S} = g\int \frac{d^3p^*}{(2\pi\hbar)^3} ~ \frac{{m^*}}{E^*}~ f(\vec{r},\vec{p^*};t).
\label{eq:scalar}
\end{equation}
In this case, the single-particle energy in Eq.(\ref{eq:BUU}) simply reads as $\epsilon = p^0 = E^* + V^0$. The specific dependence of the fields on the densities is detailed below. 
 
The different transport codes can be assigned to three main categories: 

\noindent (a) {\it Non-relativistic codes} (labeled as ``non-rel'' in Table~\ref{transport_models}). 

These codes can be described in the above general scheme by considering only the vector field and neglecting the spatial component of the baryon four-current ($\vec{j}$ = 0). Thus, the energy $E^*$ becomes $E$ with $E = \sqrt{\vec{p}^2 + M^2}$. Moreover, the non-relativistic limit is taken for $E$. The single-particle energy can then be written as $\epsilon = \frac{\vec{p}^2}{2M} + U(\rho) + M $, where $U(\rho)$ is the mean-field potential. The latter is usually introduced phenomenologically, and very often a Skyrme-like form is employed, i.e., $U(\rho) = a(\rho/\rho_0) + b(\rho/\rho_0)^\sigma$ with $\rho_0$ being the saturation density and the non-linear term taking into account the effect of many-body forces. In many applications, but not in the comparisons discussed here, a phenomenological momentum-dependence is also included in the potential $U(\rho,p)$.  

\noindent (b) {\it Codes with relativistic kinematics} (labeled as ``rel'' in Table~\ref{transport_models}).

The same ingredients as in the ``non-rel'' case, but the kinematics is treated relativistically. Hence, the single-particle energy is expressed as $\epsilon = E + U(\rho,p)$.

\noindent (c) {\it Covariant codes} (labeled as ``cov'' in Table~\ref{transport_models}).

We place into this category all codes that employ scalar and/or vector fields including the spacial component.  Many codes in this category follow the general scheme of the Walecka or Relativistic Mean-Field (RMF) model. Denoting by $m_\sigma$ and $m_\omega$ the masses of the isoscalar $\sigma$ (scalar) and $\omega$ (vector) mesons, respectively, and by $g_\sigma$ and $g_\omega$ their respective coupling constants, the following relations hold for the scalar and vector fields:  
\begin{equation}
\Phi = \frac{g_\sigma^2}{m_\sigma^2}\rho_S
;~~~ V^\mu = \frac{g_\omega^2}{m_\omega^2}j^\mu.
\end{equation} 
When one wants to take into account the nuclear symmetry energy, the isovector mesons $\rho$ 
(vector) and sometimes $\delta$ (scalar) are also included in a similar way.  For a more realistic 
description of symmetric and asymmetric nuclear matter{,} this model is usually extended, 
either by adding terms non-linear in the scalar field $\Phi$ in the relation between $\rho_S$ and $\Phi$ and/or non-linear 
coupling terms between mesons (non-linear RMF models), or by assuming that the coupling 
constants $g_\sigma$ and $g_\omega$, etc., are functions of the density. In these density-dependent coupling models, the functional form of the density dependence is either determined phenomenologically, e.g., by fitting nuclear masses~\cite{Typel:1999yq}, or by parametrizing results from nuclear-matter models, like the Brueckner Hartree-Fock or the chiral perturbation theory~\cite{Hofmann:2000vz,Zhang:2018ool}. Also, other assumptions on the scalar and vector fields exist, as detailed in the corresponding code descriptions.


Fluctuations of the one-particle density, which account for the effect of neglected many-body correlations, can be introduced by adding to the r.h.s. of Eq.(\ref{eq:BUU}) a stochastic term, representing the fluctuating part of the collision integral \cite{Ayik88,ayi90,AbeAyik96,ran90}. This leads to the Boltzmann-Langevin (BL) equation, in close analogy with the Langevin equation for the Brownian motion. 


The integro-differential non-linear BUU equation  is solved numerically. To this end, the continuous distribution function $f_a$ is represented in terms of a sum of finite elements, called test particles (TP)~\cite{Wong82}, as
\begin{equation}
f_a(\vec{r},\vec{p};t)=\frac{1}{g_a N_{TP}}\bigg(\frac{2\pi}{\hbar}\biggr)^3 \, \sum_{i=1}^{N_a N_\text{TP}} G(\vec{r}-\vec{R}_i(t)) \, \tilde{G}(\vec{p}-\vec{P}_i(t)) \, ,
\label{eq:fTP}
\end{equation}
where $N_a$ is the number of particles of type $a$ (in the case of only nucleons $N_a \equiv N \,\text {or}\, Z$ for the total number of neutron or protons), $N_{TP}$
 the number of TP per particle (often the same for each species $a$ but could also be different), $\vec{R}_i$ and $\vec{P}_i$  are the time-dependent coordinates and momenta of the TPs, and $G$ and $\tilde{G}$ are the profile functions in coordinate and momentum space, respectively, with a unit normalization. 
In particular, when $\delta$ functions are  adopted for the profile functions, inserting this ansatz into the left-hand side of Eq.~\eqref{eq:BUU} results in Hamiltonian equations of motion for the TP propagation,
\begin{equation}
\frac{d\vec{R}_i}{dt}=\vec{\nabla}_{{P}_i} \epsilon \hspace*{2em} \text{and} \hspace*{2em} \frac{d\vec{P}_i}{dt}=-\vec{\nabla}_{R_i} \epsilon \, .
\label{eq:prop}
\end{equation}
Solving the test particle equations of motion, Eq.~(\ref{eq:prop}), requires the calculation of the local single-particle energy, which depends on bulk quantities, e.g., the local density $\rho(\vec{r},t)$, which can be evaluated from Eq.\eqref{eq:current}. 

The phase-space distribution represented by Eq.~\eqref{eq:fTP} fluctuates strongly when $\delta$-functions are used for the profile functions. In this case, one uses a large number of TPs and/or profile functions of finite size especially for the spacial part of the TPs, e.g. Gaussians or triangular shapes, to smooth the distributions in coordinate space. A systematic procedure was introduced by Lenk and Pandharipande~\cite{Lenk89} by using the lattice-Hamiltonian method. Here, the coordinate space is divided into cubic cells (typically of volume $\Delta l^3$ =  1 fm$^3$) and the spatial density is evaluated at the cell site coordinates $\vec{r}_\alpha$ as $\rho_\alpha = \rho(\vec{r}_\alpha)$.  The potential part of the total Hamiltonian of the system is then given by $H_{\rm pot}=\Delta l^3\sum_{\alpha}e_{pot}(\rho_\alpha)$, where $e_{\rm pot}$ denotes the potential part of the energy density. We note that these quantities 
 depend on the TP coordinates, $\vec{R}_i(t)$, according to Eq. \eqref{eq:fTP}. 
The resulting canonical equations-of-motion from this Hamiltonian are 
\begin{equation}
\frac{d\vec{R}_i}{dt}=\vec{\nabla}_{{P}_i} \epsilon \hspace*{2em} \text{and} \hspace*{2em}
 \frac{d\vec{P}_{i}}{dt} = -\Delta l^3\sum_{\alpha}\frac{de_{\rm pot}}{d\rho_\alpha} \vec{\nabla}_{{R}_i}G_\alpha = -\Delta l^3\sum_{\alpha}\epsilon_{\rm pot}(\rho_\alpha)\vec{\nabla}_{{R}_i}G(\vec{r}_\alpha-\vec{R}_i(t)),
\end{equation} 
where the sum is over all cells that contribute in a relevant way. It was shown in Ref.~\cite{Lenk89} that equations-of-motion in this form strictly conserve the total energy. The Lattice-Hamiltonian method has been used for BUU codes of the type ``non-rel'' and ``rel'' and also for QMD codes.

\subsection{{QMD-like models}}

In quantum molecular dynamics (QMD) models, the many-body state is represented by a simple product wave function of single-particle states with or without anti-symmetrization \cite{ono99,Aic91}, usually assumed to have a fixed Gaussian shape. Although this ansatz corresponds to an independent-particle approximation, the use of localized wave packets induces classical many-body correlations both in the mean-field propagation and in the collision integral, where the latter is treated by the same stochastic methods as in BUU (see Sec.~\ref{sec:collision_term}). This way of introducing many-body correlations produces possible trajectory branchings, and has been proven to be particularly efficient for the description of fragmentation events, where nucleons are well localized inside separate fragments in the final state \cite{Aic91}. The time evolution of nuclear dynamics is formulated in terms of the changes in nucleon coordinates and momenta, i.e., the centroids of the wave packets, in a similar way as in classical molecular dynamics. They move under the influence of the mean-field potential, which is usually consistently accounted for by density functionals, but may also be formulated with two- or many-body interactions. This approach can be viewed as derived from the time-dependent Hartree method with a product trial wave function of single-particle states of Gaussian form,
\begin{align}
\label{eq:QMDwf}
&\Psi(\vec{r}_1,\dots, \vec{r}_A; t) =  \prod_{i=1}^A \phi_i(\vec{r}_i;t), \\
&\phi_i(\vec{r}_i;t) = \frac{1}{[2\pi (\Delta x)^2\big]^{\frac{3}{4}}}\exp\bigg[-\frac{[\vec{r}_i-\vec{R}_i(t)]^2}{4(\Delta x)^2}\biggr]\,\exp\bigg[(i/\hbar)\vec{P}_i(t)\cdot{[\vec{r}_i-\vec{R}_i(t)]}\biggr],\nonumber
\end{align}
with the centroid positions $\vec{R}_i(t)$ and momenta $\vec{P}_i (t)$ treated as variational parameters. However, the widths $\Delta x$ are fixed in order for the wave function to be able to describe finite distance structures, as observed in the fragmentation of colliding nuclei. 
{The 1-body Wigner function for the wave function of Eq.~\eqref{eq:QMDwf} is
\begin{align}\label{eq:wignerfxn}
&f(\vec{r},\vec{p})=\sum_{i=1}^A f_i(\vec{r},\vec{p}),\, \text{with}\\\nonumber
&f_i(\vec{r},\vec{p})=\bigg(\frac{\hbar}{\Delta x \Delta p}\biggr)^3 \, \exp\bigg[-\frac{(\vec{r}-\vec{R}_{i}(t))^2}{2\Delta x^2}-\frac{(\vec{p}-\vec{P}_{i}(t))^2}{2\Delta p^2}\biggr].
\end{align}
The prefactor in Eq.~\eqref{eq:wignerfxn} reduces to the value 8 due to the relation  $\Delta x\Delta p=\hbar/2$ from the ansatz of Eq.~\eqref{eq:QMDwf}.  However, the values of  $\Delta x$ and $\Delta p$ are often used independently of each other in the codes.
Eq.~\eqref{eq:wignerfxn} shows that the Wigner function is normalized to be dimensionless, and is explicitly positive-definite for this ansatz. Although the Wigner function can then be interpreted as a probability density, it may be larger than unity,
which then requires a special treatment for the Pauli blocking in the collision term (see Sec.~\ref{sec:collision_term}). 
The QMD ansatz with Gaussian wave packets yields the following equations of motion derived from the time-dependent variational principle
\begin{equation}
\frac{d\vec{R}_i}{dt}=\vec{\nabla}_{{P}_i} \big< H \bigr> \hspace*{2em} \text{and} \hspace*{2em}
\frac{d\vec{P}_i}{dt}=-\vec{\nabla}_{{R}_i} \big< H \bigr> ,
\end{equation} 
which are of the same form as those obtained for the TPs in BUU, Eq.~\eqref{eq:prop}, written in terms of the centroid positions $\vec{R}_i(t)$ and $\vec{P}_i (t)$ of the wave packets and by replacing the single-particle energy $\epsilon$ with the expectation value of the many-body Hamiltonian $\langle H \rangle$. The approach has been extended to include anti-symmetrization of the wave function in the Antisymmetrized Molecular Dynamics (AMD) method~\cite{ONOb,ONOc}, which results in more complicated equations of motion, but of similar structure.

\subsection{Fluctuations} \label{sec:fluctuations}

The main difference between the BUU and QMD models lies in the amount of fluctuations and correlations in the representation of the phase-space distribution, which affects the evolution both in the mean-field propagation~\cite{Colonna2021} and the collision term~\cite{Zhang2017}. In the standard BUU approach, the phase-space distribution function is seen as a one-body quantity, which is a smooth function of coordinate and momentum, and can be approximated increasingly better by increasing the number of TPs in the representation. In the limit of $N_{\text{TP}}\rightarrow \infty$, the BUU equation is solved exactly. In this limit, the solution is deterministic and does not contain fluctuations, also if the collision term is taken into account~\cite{Colonna2021,Zhang2017}.  However, as mentioned above, suitable stochastic extensions can be formulated, if fluctuations are considered to be important. Of course, numerical fluctuations are present in practical BUU calculations with a finite number of TPs~\cite{BONASERA1990169}.

In QMD, nucleon correlations arise from the representation in terms of a finite number of wave packets of finite width, leading to enhanced fluctuations of the one-body density relative to BUU. These fluctuations are strongly driven by the stochastic evaluation of the collision term, since complete nucleons are scattered in QMD, rather than small segments of phase space as in BUU. Thus, in the philosophy of QMD, one goes beyond the mean-field approach and includes correlations and fluctuations in the QMD ansatz of the wave function. However, these fluctuations can lead to a loss of the fermionic character of the system more rapidly than in BUU, as studied for the mean-field propagation in Ref.~\cite{Colonna2021} and for the collision term in Ref.~\cite{Zhang2017}. The fluctuations in QMD-type codes are regulated and smoothed by varying the parameter $\Delta x$, the width of the wave packet, see Eq.~\eqref{eq:QMDwf}. 
However, the parameter $\Delta x$  has also been related to the range of the nuclear interaction, and as such can only be varied within limits.
QMD can be seen as an event generator, where the time evolution of different events is solved independently, and therefore the effect of the fluctuations is not suppressed even in the limit of averaging over an infinite number of events.

\subsection{The collision term} \label{sec:collision_term}
 
In BUU, the collision term is commonly simulated by performing stochastic TP collisions. In QMD, the same procedure is used for the nucleons, and can formally be derived by putting $N_{TP}=1$. This procedure involves two steps: first, to determine if two test particles collide in a given time step, and second, to check whether the final state of the collision is allowed by the Pauli principle. The strategies for both of these steps are discussed in considerable detail in the box cascade comparison in Ref.~\cite{Zhang2017} and in the code descriptions; thus only some brief remarks are made here. The condition for a collision is usually determined by a geometric criterion, often called the Bertsch criterion, as it was first fully formulated in the review paper by Bertsch and Das Gupta~\cite{Bertsch:1988ik}. In this method, two TPs collide in a given time step if they reach a distance of closest approach given by the TP cross section $\sigma^\prime =\sigma^{\rm med}/N_{TP}$ within that step. The choice of the final momenta of the TPs is stochastic, with the condition that the total energy and momentum are conserved. In Ref.~\cite{Zhang2017}, it was found to be important to eliminate consecutive collisions of the same TPs, {and sometimes one even has to consider physical or unphysical higher-order correlations induced by the geometrical prescription.} Alternatively, statistical criteria for the collision probability have been used, based essentially on the ratio between the mean free path, which depends on the density and the cross section, and the relative distance traveled in time step $\Delta t$ by a pair of TPs~\cite{Zhang2017}.  

For each such ``attempted" collision, the Pauli blocking is checked by calculating the phase-space occupation for the final states $f'_a$ and $f'_b$. Here, an average over the phase-space cells of the final states has to be taken to obtain reasonably smooth results. The Pauli blocking probability is calculated in most cases as $1-(1-f'_a)(1-f'_b)$. As mentioned above, the QMD ansatz may result in an over-occupation of a final cell. Most codes in this case disallow the collision (i.e., force the occupation to have the value 1), others constrain the distribution function to remain smaller than one (CoMD, see the corresponding code description).  In principle, with Fermi statistics implemented at the beginning of the reaction and the Pauli principle enforced in the collision term, the fermionic nature of the system should be preserved in the evolution. However, in Ref.~\cite{Zhang2017}, it was seen that the phase-space occupation for the final states is subject to fluctuations, which may allow collisions that should have been forbidden. This destroys the Fermi distribution of an isolated system more or less quickly, depending on the number of TPs.  It has been shown~\cite{AbeAyik96} that the coarse-graining, which is implicit in the representation of the phase-space distribution by finite elements, acts as a dissipation, which eventually evolves the distribution to a classical Maxwell-Boltzmann distribution.

For BUU codes, the method of solving the collision term with a geometric criterion, as explained above, is called the full-ensemble method. It is numerically expensive since it scales like $(A^2\cdot N_{TP}^2)$. By considering that the range of the nuclear interaction is finite, one can limit the coordinate space region where one has to check for collision partners, such that the computation time scales like $(A\cdot N_{TP}^2)$. In most calculations, the parallel-ensemble method is used, where the total number of test particles is randomly divided into $N_{TP}$ ensembles of $A$ particles each~\cite{Welke:1989col}. Collisions are then allowed to occur only within each ensemble with the full cross section $\sigma^{\rm med}$, while the Pauli blocking and the mean fields are calculated by averaging over the test particles from all ensembles. This method then scales like $(A^2\cdot N_{TP})$, which often is faster, since the number of TPs is large. It was checked that, in typical cases, this procedure gives results similar to the full ensemble method.
A numerically even less expensive method is related to the lattice-Hamiltonian method~\cite{Lenk89} and works with statistical concepts in sub-ensembles. It was first used in Ref.~\cite{Danielewicz:1991dh} and was then motivated more rigorously in Ref.~\cite{lang93}, where it was called the local-ensemble method, and where it was shown to scale as $(A \cdot N_{TP})$. It does not employ any geometric interpretation, and is, therefore, free of problems with time-ordering of collisions and Lorentz covariance.

At collision energies above particle production thresholds, new particles appear and particle production observables, e.g., the yields of pions, are considered as promising probes for the investigation of the EoS, but at the same time this increases the complexity of the simulations. Additional distribution functions appear for each particle with their corresponding transport equations, which are coupled to those of nucleons by the collision term with inelastic two-body reactions or by the decay or formation of unstable particles. This requires additional physical input, such as the mean fields of these particles and the in-medium inelastic cross sections or decay amplitudes, which often are not known from experiment and have to be modeled. Additionally, when considering elastic, inelastic, and decay  processes simultaneously, the sequence of treating these processes in the collision term becomes relevant. This was studied in the comparison of pion production in a periodic box~\cite{ono2019}.  

When considering particle production near the threshold, the newly produced particles will be very rare. To obtain reasonable statistics for the evaluation of observables, one has to simulate a very large number of events, which is numerically costly. This difficulty can be circumvented in two ways. One can often treat the production perturbatively, such that the new particles do not influence the evolution of the nucleonic distribution function. An exact method is the partition method~\cite{KoZhang21,PRC_JAM9_01c}, which amounts to using a much larger test-particle number for the rare species and by gauging the inelastic cross sections accordingly. 

In an inelastic two-body collision, the energy-momentum conservation
determines the threshold for particle production. Since the collision happens in the medium, the potential energies of all involved particles have to be taken into account, which shifts the thresholds away from their values in the vacuum. These threshold shifts depend on the choices for the mean fields of the produced particles, which are often not known, e.g., in the case of $\Delta$-resonances, and have to be specified in models, which may differ between the codes. The situation becomes even more complicated with momentum-dependent mean fields. Various schemes have been proposed to ensure the energy-momentum conservation on the local or global level~\cite{Cozma:2014yna}, but we will not describe details here.

In principle, the mean fields or self-energies in quantum transport theories are complex. The imaginary part corresponds to a finite width of the spectral function of the particle due to an intrinsic width and/or the collisional broadening. This is treated explicitly in off-shell transport theories~\cite{leupold:2000off,leupold:2000ma}. These have been implemented in the codes GiBUU~\cite{Buss:2011mx} and PHSD~\cite{Linnyk:2015rco}, but we will not go into detail here. However, the finite width of a particle may become important in particle production, when the energy of the collision is near or even below the two-particle threshold. This has often been the case for pion production at intermediate energies, which proceeds via $\Delta$-resonance excitation, which has a width of approximately 120 MeV. Rather than performing off-shell transport calculations, this effect is often treated by assuming a mass distribution for the unstable particle and choosing the mass stochastically from this distribution under the constraint of energy-momentum conservation. In such an assumption, the detailed-balance condition is modified from the on-shell process~\cite{Danielewicz:1991dh}. Returning to on-shell conditions at large times is no problem here, since by then all the unstable particles have decayed.

Finally, we would like to mention the treatment of fragment and cluster production. The production of larger fragments can be seen as triggered by fluctuations when the system is inside the spinodal region, which are amplified to realistic fragments by the mean field. These can be identified at later times in the simplest case by coalescence methods, but also by more refined approaches, which can identify fragments already at an earlier stage in the collision~\cite{lefevre2019}. However, the production of light clusters up to He isotopes, is not well described in transport models by this mechanism, because the structure of the clusters is due to true quantum correlations with discrete bound states, which are absent in semi-classical transport descriptions, and which may additionally be modified by medium effects. At the same time, light clusters are copious in the final state of an intermediate-energy heavy-on collision and could provide important observables to determine the EoS, particularly the symmetry energy{~\cite{Chen:2003qj,Chen:2003ava,Zhang:2014sva}}. A few codes treat the production of light clusters dynamically~\cite{ono19_PPNP,Danielewicz:1991dh,dan92}. Briefly stated, they either modify the phase-space of the nucleons forming a cluster according to the cluster wave function (AMD code), or treat light clusters as separate particle species (pBUU code). These have their own distribution functions and transport equations, and are coupled to the nucleons by in-medium production vertices, which have to involve a third particle to ensure energy-momentum conservation.

\newpage

\section{Review of Results of Comparisons of Transport Simulations under Controlled Conditions} \label{sec:review}

Up to now, there have been 6 published papers within the code evaluation project~\cite{ECT05,Xu2016,Zhang2017,ono2019,Colonna2021,SpRIT:2020blg}. These publications go into considerable detail to analyze the similarities and differences of the results of the codes under controlled conditions. Thus, it may be useful to summarize the most important findings of these investigations in brief reviews including the most significant figures. This is the main goal of this section. These summaries follow a logical order. We start with two comparisons of full heavy-ion collisions at relativistic and intermediate energies. Realizing that there are rather large differences, which are not easy to disentangle for a full collision, we turn to the simpler set-up of calculations in a box with periodic boundary conditions (``box-calculations'') to study the different aspects of transport simulations in a very controlled way. We investigated, in sequence, the mean-field propagation with a simple density functional, the collision term with elastic collisions only, and the collision term including inelastic collisions leading to pion production. In view of upcoming detailed experimental data of pion production in Sn+Sn collisions at 270 AMeV, we then asked for a prediction of pion yields by the various codes without prior knowledge of the data. This showed that significant differences among the codes still remained,
which are thought to be due to differences of the physical models, whose implementations are presently studied in further comparative calculations. These and further extensions of transport models are discussed in the last two sections of this chapter.

\begin{figure}[hb]
\centering
\vspace*{-1.10cm}
\includegraphics[width=0.8\textwidth,angle=-00]{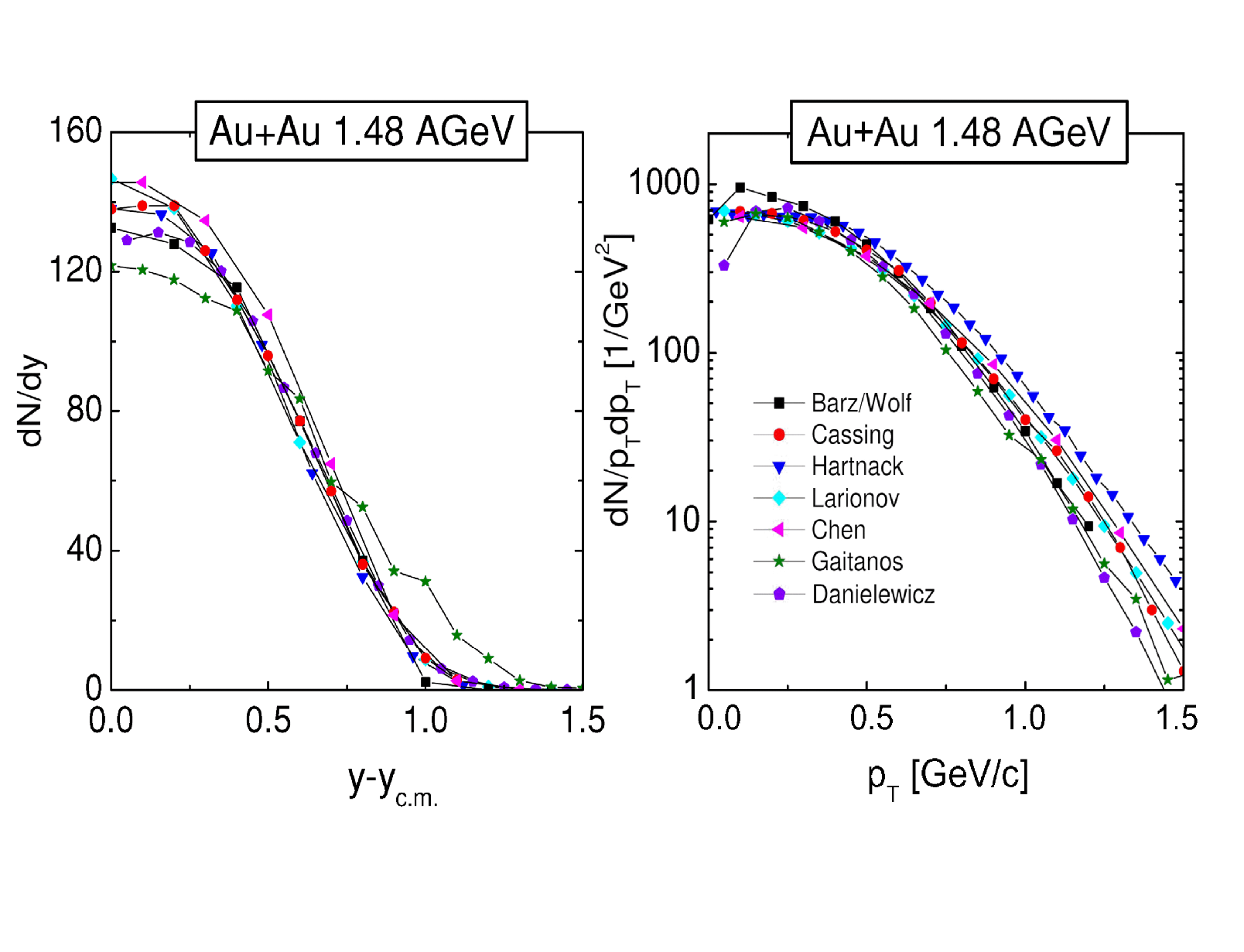}
\vspace*{-2.cm}
\caption{Longitudinal (left) and transverse (right) momentum distributions of protons from Au+Au collisions at $E_{lab}$ = 1.48 AGeV and b = 1 fm. The lines with symbols are the results of the participating transport codes, identified in the legend by the names of the authors (for more information on these codes, see Ref.~\cite{ECT05}). Figure adapted from Ref.~\cite{ECT05}.}
\label{prot1GeV}
\end{figure}

\begin{figure}[tb]
\centering
\vspace*{0mm}
\includegraphics[width=0.8\textwidth,angle=-0]{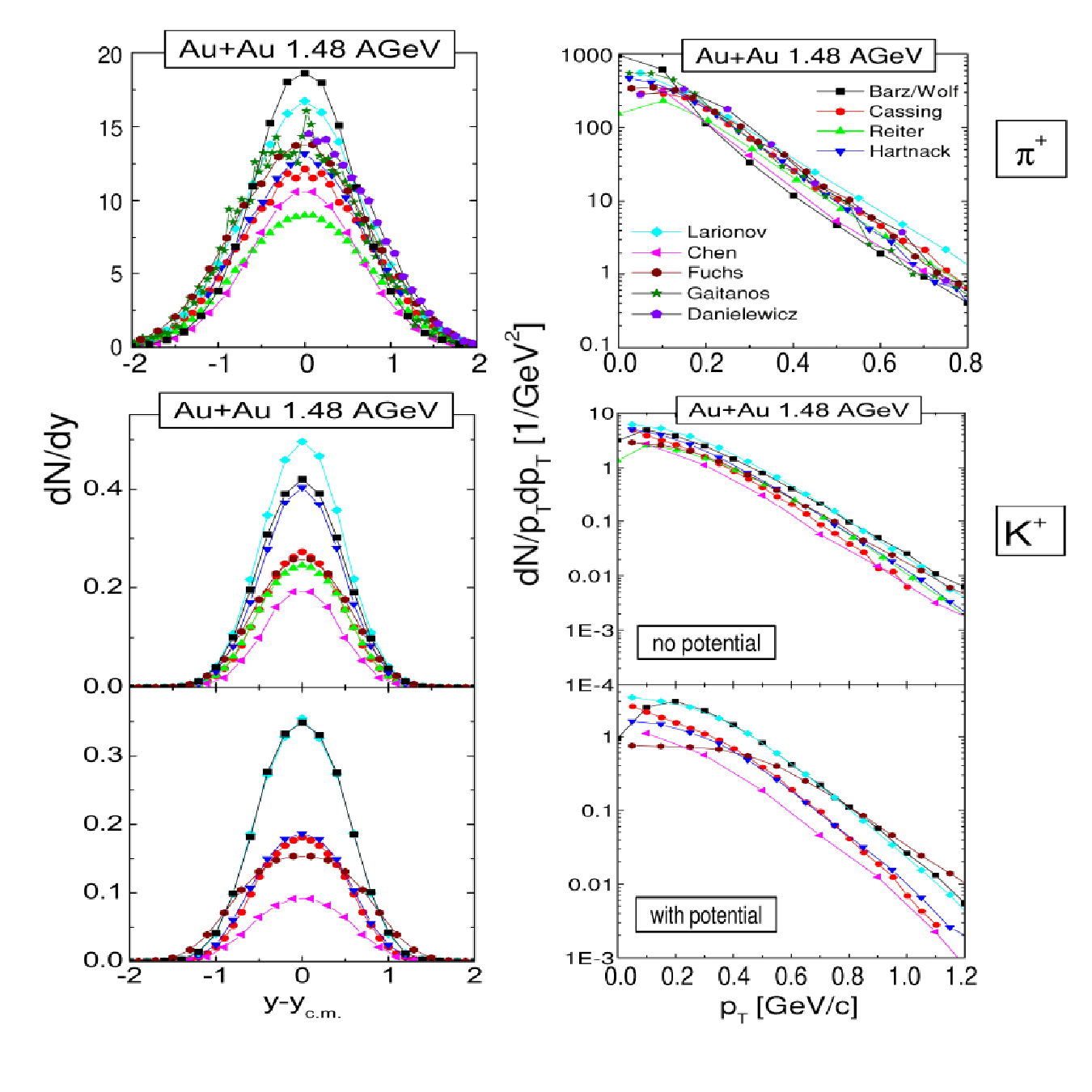}
\vspace*{-10mm}
\caption{Longitudinal (left) and transverse (right) momentum distributions for $\pi^+$ (upper row) and $K^+$ (lower two rows) for the same Au+Au reaction as in Fig.~\ref{prot1GeV}. The kaon spectra are shown without and with the use of a kaon potential. The lines with symbols are the results from the different participating transport codes as in Fig.~\ref{prot1GeV}. Figure adapted from Ref.~\cite{ECT05}.}
\label{piK1GeV}
\end{figure}

\subsection{Heavy-Ion Collisions at Energies Around 1 AGeV: Pion and Kaon Production}
\label{HIC-1GeV}

An early comparison of several cascade and transport codes with experimental data for La+La collisions at $E_{lab}$ = 800A MeV~\cite{Aichelin:1989jm} studied coalescence-invariant inclusive proton cross sections and found generally good agreement among the codes, but systematic deviations from the data, which were later found to be incorrect due to this study. A more systematic comparison of 8 transport codes for heavy-ion collisions was undertaken during a workshop held in 2004 at the ECT* in Trento, with the results published in Ref.~\cite{ECT05}. In this study, reactions of Au+Au at $E_{lab}$ = 1 and 1.48 AGeV and of Ni+Ni at 1.92 AGeV were compared. 
The main emphasis was on the production and momentum distributions of $\pi$ and $K$ mesons, since at the time the $K^+$ production was of particular interest for the determination of the EoS of symmetric nuclear matter. The physical input into the codes, such as the parametrization of cross sections, potentials, and decay width, was left as in the normal use of the codes. Thus, the emphasis was put more on the physics of meson production than on the convergence of transport simulations. The longitudinal and transverse momentum distributions of protons are shown in Fig.~\ref{prot1GeV} for Au+Au collisions at 1.48 AGeV for the different codes participating in the comparison at that time. The results of these codes agree with each other fairly well, which is probably due to the fact that at this energy the mean field is not very important. It should be noted that the codes in this study are in some cases early versions of the codes described in sections 4 and 5 in this paper. There are the following correspondences between the codes in Fig.~\ref{prot1GeV} and codes described here
(see also Table~\ref{transport_models}):
Cassing-PHSD, Reiter-UrQMD, Hartnack-IQMD, Larionov-GiBUU, Chen-RVUU, Fuchs-TuQMD, Gaitanos-RBUU, and Danielewicz-pBUU. Some of the code correspondents of this comparison have changed in the meantime. The code denoted 'Barz/Wolf' is no longer active and a description is not included here.

Fig.~\ref{piK1GeV} shows the spectra of positive pions (upper row) and kaons (lower two rows) with and without a kaon potential for the same Au + Au collision as in Fig.~\ref{prot1GeV}. Large differences are seen between the codes. These were partly due to the use of different models for the production and properties of the $\Delta$ resonances and the cross sections for strangeness production. 
This was taken up again in a review paper by Ch. Fuchs on kaon production at intermediate energies, where it was shown that the discrepancies in the $K^+$ observables resulted mainly from the use of different models of the kaon interactions, and that a much better convergence can be achieved between codes by using similar models, also between codes of BUU and QMD type~\cite{Fuchs:2005zg}. 
In any case the differences were less evident when considering ratios of observables. At this time, a much-discussed observable was the ratio of $K^+$ production in Au+Au and C+C collisions, which is expected to be sensitive to the iso-scalar compressibility~\cite{Aichelin:1985rbt} in such a way that it is higher in the Au+Au collision due to the higher compression reached, while the ratio is not much affected by the absolute $K^+$ yields. It was seen that in spite of the different absolute values, the ratio was rather consistent between the different codes~\cite{ECT05}, and thus the evidence for a soft symmetric EoS with momentum dependence was judged to be robust~\cite{Fuchs:2000kp}.

\subsection{Heavy-Ion Collisions at 100 and 400 A MeV} \label{HIC_100AMEV}

In a further workshop in Trento in 2009, the beam energy of Au+Au collisions was lowered to 100 and 400 A MeV with controlled physical input, and substantial differences in the results of the codes were observed. A similar set up was taken up again during a workshop at Shanghai Jiao Tong University in 2014, with a considerably enlarged participation of transport codes, especially codes developed in the 1990's in China. This formally started the Transport Model Evaluation Project (TMEP). 

The physical input in this project was simple and only roughly realistic, i.e., a Skyrme-like mean field or a non-linear relativistic mean-field (RMF)  in ($\sigma\omega$)-parametrization and a constant NN cross section. Conditions were specified for the initialization (a Woods-Saxon density profile in coordinate space and a local Fermi sphere in momentum space), the impact parameters, and the number of runs (events). The codes were run in different modes: cascade (only collisions), Vlasov (only mean field) and full. For each code, the comparison monitored the evolution of the density distributions, the rate, energy, and final-state blocking of the collisions, and certain observables, namely the rapidity distributions and transverse flows in the final state. The details and results of this comparison were published in Ref.~\cite{Xu2016}.

\begin{figure}[h!]
\centering
\hspace*{-6mm}
\includegraphics[width=0.44\textwidth,angle=-00]{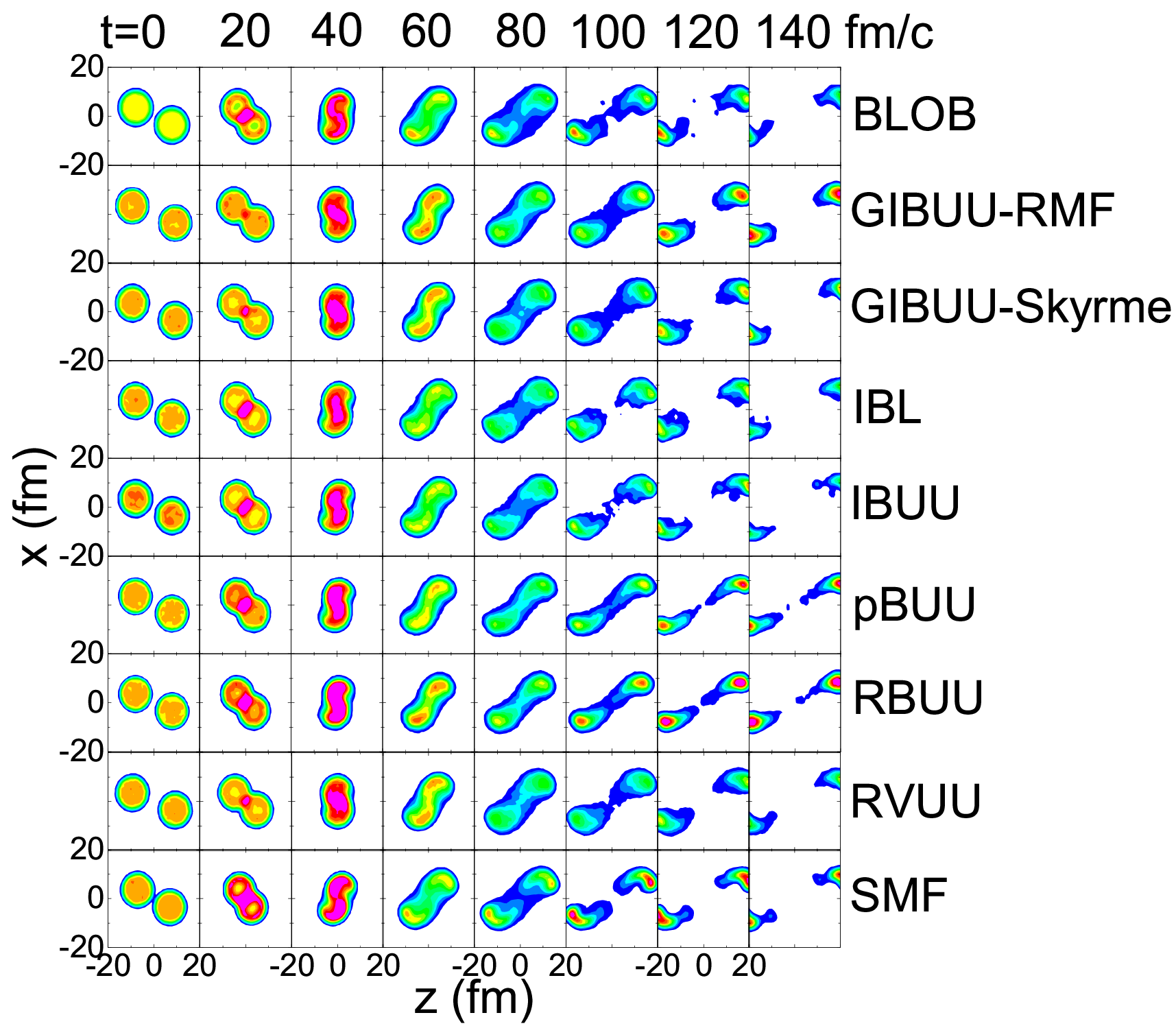}
\hspace*{-8mm}
\includegraphics[width=0.58\textwidth,angle=-00]{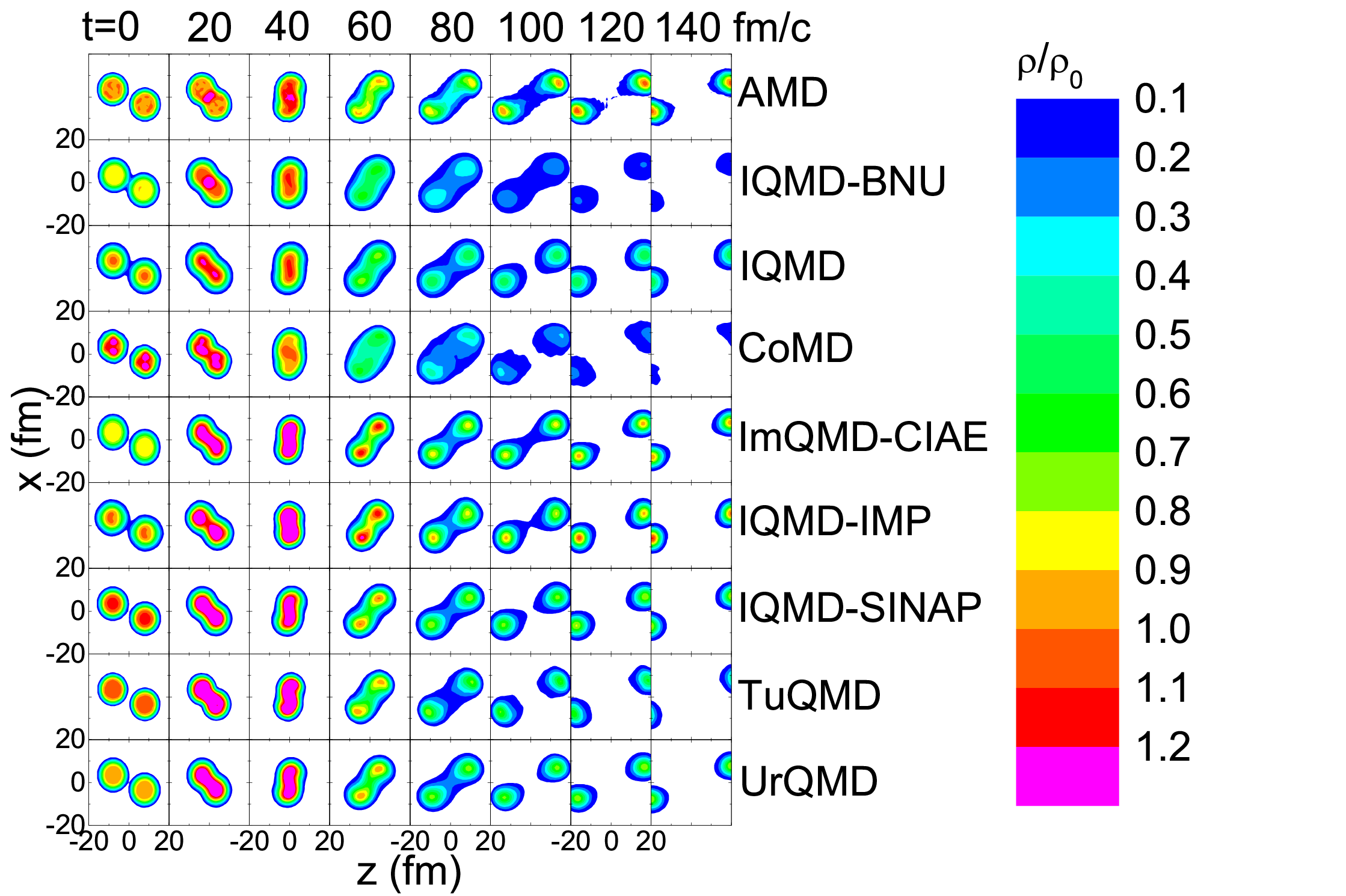}
\caption{Average density contours in steps of 20 fm/$c$ in Au+Au collisions at impact parameter b = 7 fm and beam energy 100 A MeV from BUU-type (left) and QMD-type (right) models in the full mode. Figure adapted from Ref.~\cite{Xu2016}.}

\label{evolution}
\end{figure}

An impression of the collision dynamics is given in Fig.~\ref{evolution}, which displays the time evolution, in steps of 20 fm/$c$, of the density contour plot averaged over the runs of the comparison for  BUU (left panel) and QMD (right panel) codes at the beam energy of 100 AMeV. For BUU models, the averaged density contours are similar to those of a single run, while for QMD models they strongly fluctuate from event to event. The general progression of a heavy-ion collision is exhibited in all models: the merging and maximum compression until about 40 fm/$c$, the development of a sideward flow from about 60 to 80 fm/$c$, and the formation and subsequent breaking of a neck at about 100 fm/$c$. From then onward, one observes the formation and evolution of the projectile- and target-like residues, which are clearly highly excited and develop their own dynamics. In this case, one can consider 140 fm/$c$ as the freeze-out time, after which the de-excitation of the primary fragments can be calculated with a statistical code for a comparison to experiment. The figure shows differences between the codes already at the initial time, but also significant differences are seen in the subsequent evolution.

\begin{figure}[tb]
\centering
\vspace{-0mm}
\includegraphics[width=0.7\textwidth,angle=00]{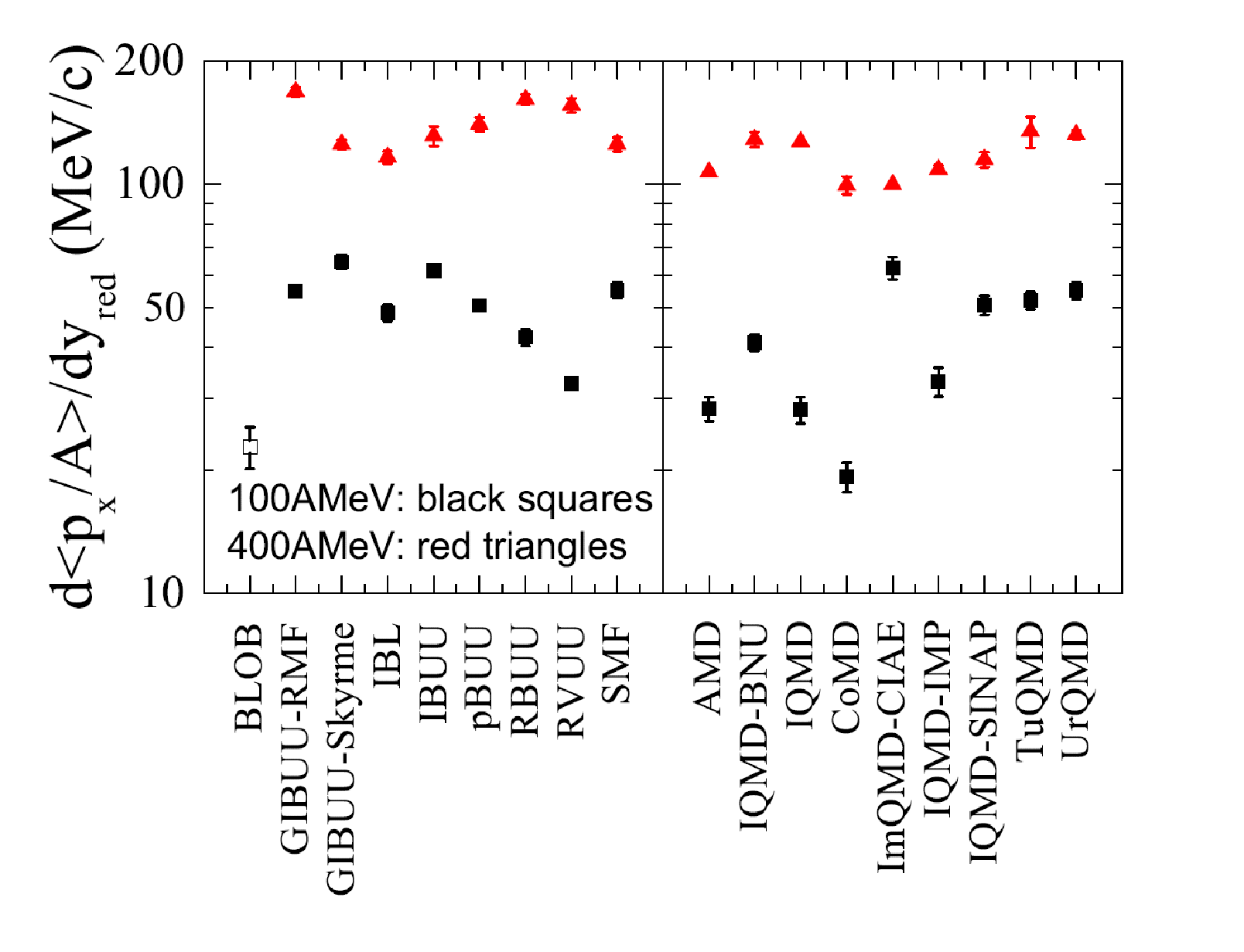}
\vspace{-6mm}
\caption{Flow values for 8 BUU-type (left) and 9 QMD-type (right) models in Au+Au collisions at the impact parameter $b = 7$ fm and the beam energy of 100 A MeV (black squares) and 400 A MeV (red triangles).
The sideward flow is defined as the average momentum per nucleon in $x$-direction with respect to the reduced rapidity $y_{\rm red}=y/y_{\rm beam}$, and the flow is the slope of this curve at zero rapidity. The error bars are the fitting uncertainties, and where they are not seen they are smaller than the symbols. Figure adapted from Ref.~\cite{Xu2016}. (For the open BLOB point, see text and footnote.)
}
\label{slope}
\end{figure}

A first observation was that the initially prepared nuclei were not necessarily stable in a free propagation. The prescribed density profiles were not always well reproduced by the codes, particularly in QMD. Moreover, they are not necessarily the ground states for the given mean-field interaction, and thus often are not stable, but instead oscillated or settled into another configuration. This happens on a time scale of the order of the collision time of the nuclei, and can lead to differences in the configuration during the collision, which influence the subsequent evolution. As a solution for this problem, it was proposed to initialize the collision with a configuration consistent with the interaction used in the propagation, e.g., with a Thomas-Fermi ground state.

Furthermore, considerable differences in the energy distributions of the attempted (before Pauli blocking) collision rates were seen. The biggest differences occur for collision energies below the Fermi surface where most of the $NN$ collisions happen, while the convergence is better at higher energies (where the blocking is less important). The distributions of nucleon rapidities and transverse momenta show differences, especially at the collision energy of 100 A MeV. This can be traced to the fact that at 100 A MeV, one is near the balance energy, where the attractive mean field and the repulsive collisions nearly balance each other, making this a particularly sensitive region. For a quantitative measure of the differences, the flow value was used, i.e., the slope of the $p_x$ sideward momentum curve at zero rapidity. The standard deviation from the average result was about 30\% at 100 and 13\% at 400 A MeV as shown in Fig.~\ref{slope}.  BUU and QMD codes gave similar values, but also show some systematic difference. Subsequently, this was found to be due to a systematically lower value of the forces  in QMD codes, resulting from a different evaluation of the many-body component of the density functional (see the next subsection). We also note that the results in this figure represent the status of the codes at the time of the publication of Ref.~\cite{Xu2016}, and that several codes have since been modified, in particular in response to the results of the code comparisons\footnote{\samepage In particular, the low value of the code BLOB was found to be due to an early version of the code and very low statistics. A more recent value from an improved version is given in P.~Napolitani, {\it et al.}, J. Phys.: Conf.~Ser. 1014, 012008 (2018)}.
The comparison of nucleonic observables in heavy-ion collisions, such as the flow shown here, is taken up again in the ongoing study of Sn+Sn collisions at 270 AMeV~\cite{xu2021}, where an improved convergence of the results will be seen.

It is not easy to trace these differences to particular features of the simulations, but there are indications that the initializations and the treatment of the collision term are the main reasons. In full heavy-ion collisions, different effects interact with each other, e.g., different collision rates lead to different degrees of violence for the collision and thus different densities may be probed in the  codes. It is also difficult to determine which strategies are more advantageous purely from a comparison between the codes, since the average result of the codes obviously is not necessarily indicative of the exact result.

Progress can be made with calculations in a box with periodic boundary conditions, which effectively simulate infinite nuclear matter of a given density and temperature. In a box, identical initial conditions are easily constructed. By performing Vlasov and Cascade calculations, one can test the mean-field propagation and the collision term separately. Furthermore, there are exact analytical or numerically obtained values for many quantities, e.g., the collision rates, to which the results of the codes may be compared rather than having to compare the codes among each other. 
In this way much more definite statements could be made about the performance of the codes, and, in fact, modifications were implemented in many cases.

\subsection{ Mean-Field Dynamics in a Box}
The transport equation of Eq.~(\ref{eq:BUU}) has two aspects: the Vlasov dynamics on the l.h.s., which describes the evolution of the phase-space distribution function under the action of a self-consistent mean field, and the collision term on the r.h.s., which represents the two-body dissipation. In this comparison, we tested the mean-field dynamics in more detail by solving the Vlasov equation in a periodic box~\cite{Colonna2021}. The emphasis of this study was to investigate the difference between the BUU and QMD approaches with respect to the different amounts of fluctuations and the method of calculating the forces. The same simple force parametrizations were used as in the study of the previous subsection. The box was initialized with a standing density wave in $z$-direction oscillating around nuclear matter at saturation density and zero temperature. This corresponds to the zero-sound  propagation, which can be solved for in Landau theory, and can be compared to the calculations in the limit of small amplitudes. A numerically exact solution is further provided by solving a partial differential equation for the deformed Fermi-surface (DFS) $p_{surf}(z,p,\theta_p,t)$, where $p$ and $\theta_p$ are the values of the momentum and its angle with respect to the $z$-axis. These analytical and exact solutions were compared to the results of 9 BUU-type and 5 QMD-type codes.  

\begin{figure}[tb]
\centering
\vspace{-0.3cm}
\includegraphics[width=0.7\textwidth]{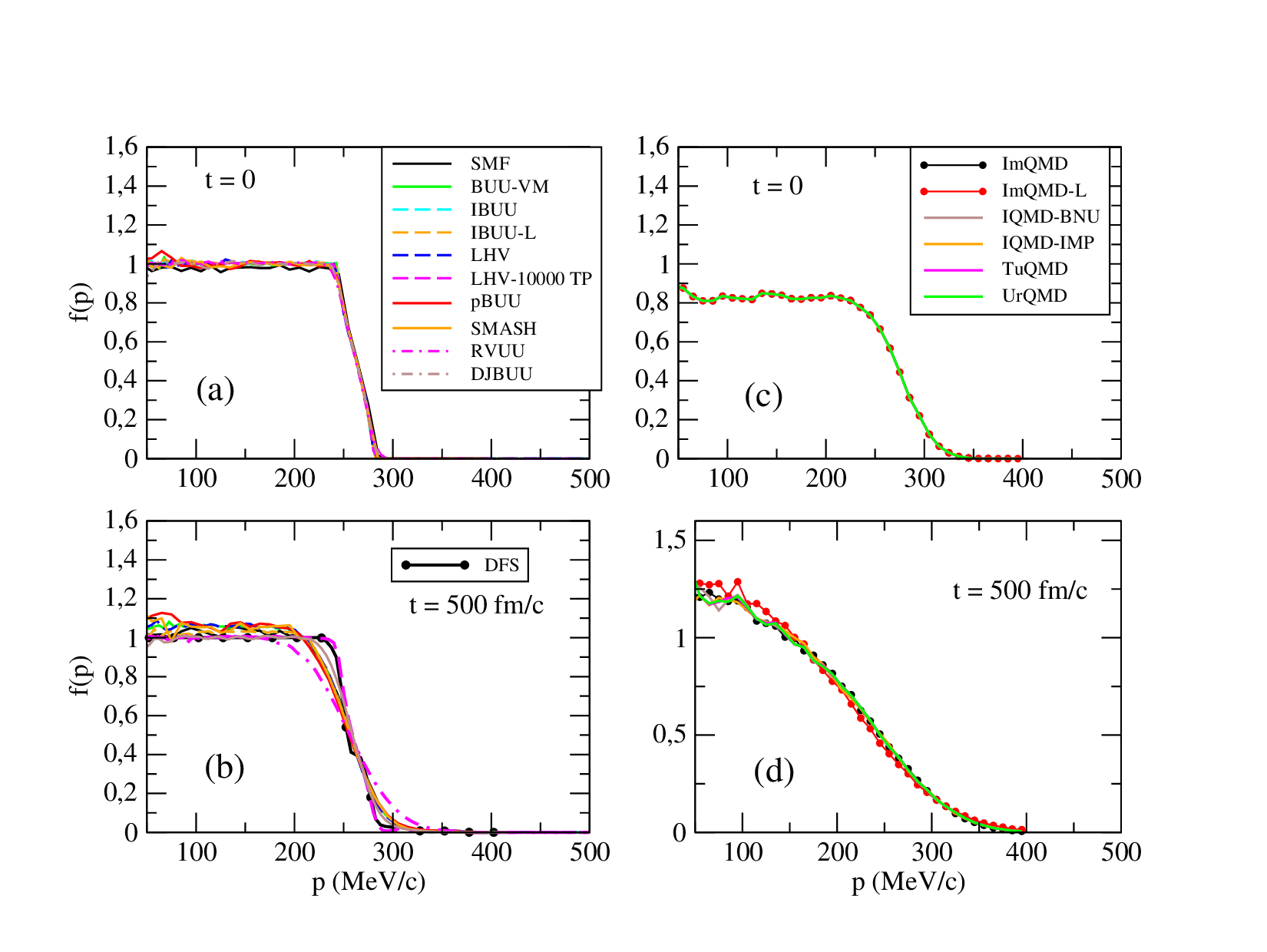}
\vspace{-10mm}
\caption{Momentum distributions as obtained in the different codes at initial (top panels) and final (bottom panels) times: BUU-like in the left panels and QMD-like in the right panels. The QMD codes have used an identical initialization of the nucleon positions and momenta. In the lower left panel, the numerical result of the exact Deformed Fermi Surface model (DFS) is shown at the final time. Figure adapted from Ref.~\cite{Colonna2021}.}
\label{momdis}
\end{figure}

\begin{figure}[t!]
\centering
\vspace{-10mm}
\includegraphics[width=0.7\textwidth]{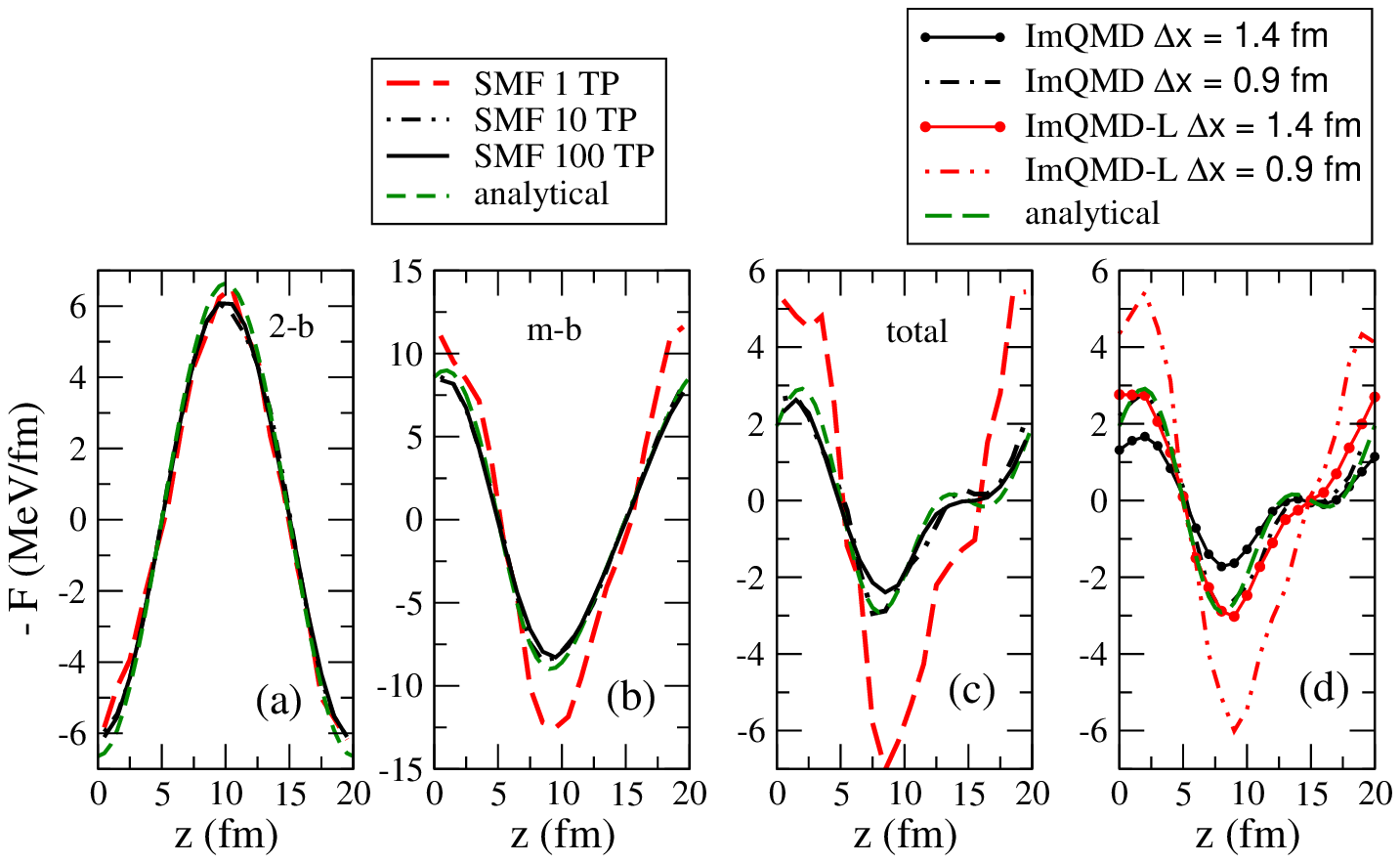}
\vspace{-6mm}
\caption{Gradient of the mean-field potential. Panels (a)-(c) correspond to SMF calculations, with several options for the TP number, from two-body (2-b), many-body (m-b) and total contributions, respectively, with the legend given by the left panel on top of the figure. Panel (d) corresponds to ImQMD and ImQMD-L calculations using several options for the Gaussian width, with the legend given by the right panel on top of the figure. The analytical results are given by dashed green lines. Figure taken from Ref.~\cite{Colonna2021}.}
\label{force}
\end{figure}

In a Vlasov calculation, the initial momentum distribution should be approximately preserved in the evolution. Fig.~\ref{momdis} presents the momentum distributions at the initial (top) and final  (bottom) times for BUU (left) and QMD (right) codes. In many cases, the curves overlap strongly, so that the individual results are difficult to distinguish, but here we want to discuss the general behavior.  The BUU-type codes rather well preserve the Fermi-distribution of the initial condition. By increasing the number of test particles to 10000 for the LHV code (LHV-10000 TP), the solution is close to that of the DFS calculation. For the QMD approach, fluctuations are not suppressed even in the limit of many events. The consequence of this is seen in the lower right panel of Fig.~5 for the final QMD distributions (here the initial distributions are taken to be identical), where the larger fluctuations relative to BUU act as a stronger dissipation and drive the distributions more rapidly towards the classical Maxwell-Boltzmann distribution for the temperature corresponding to the initial excitation.

The force in the Vlasov equation, i.e., the gradient of the mean-field potential, drives the evolution of the wave. It is plotted as a function of $z$ in Fig.~\ref{force} to show its dependence on the precision of the representation of the phase-space and on the type of transport approach. In the left 3 panels, the force is shown for the 2-body term (panel~(a)), the effective many-body term (panel~(b)), and for the total force (panel~(c)) at the initial time for a typical BUU-code (SMF) for different TP numbers, together with the analytical result in each case. With the typical value of 100 TP/nucleon, the simulation is very close to the analytical result, while for the extreme case of 1 TP/nucleon (equivalent to QMD) the increased fluctuations lead to  stronger gradients. In panel~(d), the gradients are shown for a typical regular QMD code (ImQMD) for different values of the Gaussian width $\Delta x$ and also for the case of the lattice-Hamiltonian formulation of the code (ImQMD-L). The width $\Delta x$ in QMD influences the magnitude of the fluctuations, and stronger gradients are seen for smaller values. The regular method leads to systematically smaller gradients due to an approximation usually employed when evaluating the many-body term.  This, however, can be avoided in the lattice-Hamiltonian method. With a properly chosen $\Delta x$, it is also possible to come close to the analytical result for the force.
Additionally, the mean-field dynamics is strongly influenced by the amount of fluctuations, which act as a source of damping to the density oscillations.

\begin{figure}[tb]
\centering
\includegraphics[width=0.7\textwidth]{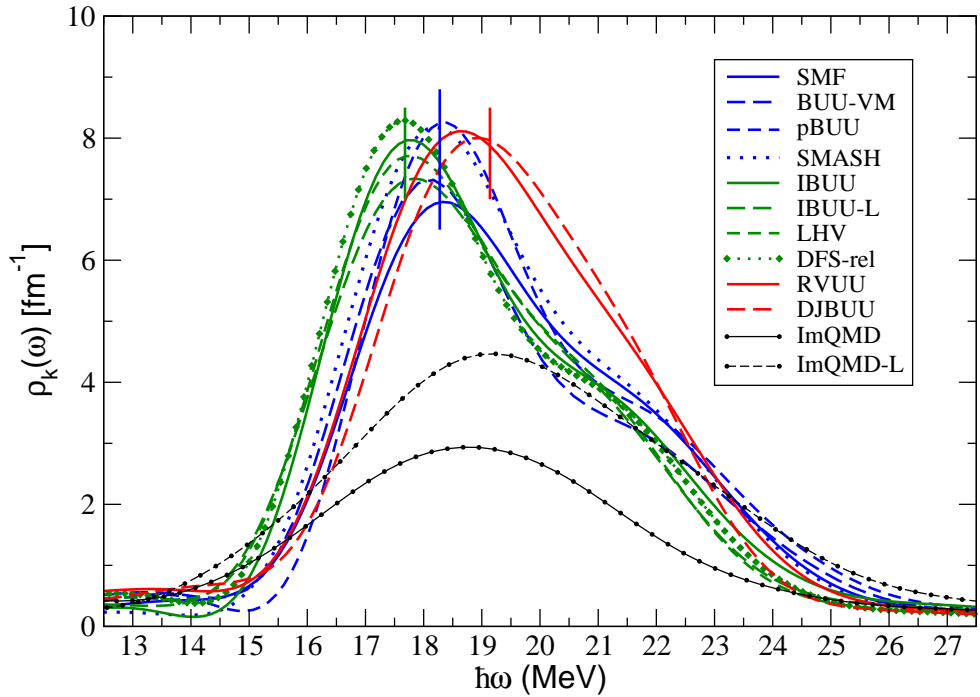}
\vspace{-2mm}
\caption{Response function $\rho_k(\omega)$, i.e., the Fourier transform with respect to space and time of the averaged density distribution, from nine BUU-like and two QMD-like calculations. 
The results for the BUU codes are grouped according to their treatment of the dynamics (and are distinguished by the color of the lines): non-relativistic kinematics (blue), relativistic (green), covariant (red).
The vertical lines indicate the analytical zero-sound energies for the different code types. See text and Ref.~\cite{Colonna2021}, from which this figure is adapted.}
\label{sound}
\end{figure}

The behavior of the different codes can be compactly shown by the response function, i.e., the Fourier transform of the oscillating density with respect to space and time (see, e.g.,  Eq.(19) of Ref.~\cite{Colonna2021}),  which is shown as a function of energy $E = \hbar\omega$ in Fig.~\ref{sound} for all BUU codes and for one QMD code (ImQMD) (due to the same initialization, all QMD codes give identical results in the regular method). The BUU codes have slightly different forces, depending on the treatment of relativity, which is distinguished by the color of the lines (see Table~\ref{transport_models}): non-relativistic  kinematics (SMF, BUU-VM, and, effectively, pBUU and SMASH, blue lines), relativistic kinematics (IBUU, IBUU-L, LHV, DFS, green lines), covariant treatment (DJBUU, RVUU, red lines). The expectations for the energy  of the zero-sound modes from Landau theory are shown as vertical lines of the corresponding color. The maxima of the different BUU codes generally agree well with these values (the greater deviation of RVUU is understood~\cite{Colonna2021}).  The exact numerical result (DFS, relativistic kinematics) is well reproduced by the codes with the same kinematic treatment. The same also holds for the width of the response function, which is due to the non-linearity of the Vlasov equation for the chosen density functional with a many-body term and the rather large initial amplitude of the density perturbation. The shoulder seen on the high energy side is due to the excitation of the next harmonic. The QMD code (black line) in the regular mode exhibits a larger width due to the larger damping effects from increased fluctuations relative to BUU,  and a shift of the maximum due to the approximation used in calculating the gradients of the potential from many-body forces. In the lattice-Hamiltonian version of the code (ImQMD-L), the maximum of the response function is close to the energy of the corresponding relativistic zero-sound solution.

\subsection{Collision Integral in a Box}

In this comparison, we investigated the second main ingredient of a transport code: the treatment of the collision term, which describes the dissipation in a collision, and is the important step beyond a quantum or semi-classical mean-field propagation. The different aspects of the calculation of the collision integral are discussed in some detail in Sec.~\ref{sec:collision_term}. The collision integral can be tested exclusively in a cascade calculation by turning off the mean field. The comparison of box calculations in the cascade mode was carried out with 7 BUU-type and 8 QMD-type codes participating in this study~\cite{Zhang2017}. Here, we treated the simplest cascade calculation by including only elastic NN collisions. The system was initialized at saturation density and for temperatures of 0 and 5 MeV.

\begin{figure}[h]
\centering
\vspace{-0mm}
\includegraphics[width=0.7\textwidth]{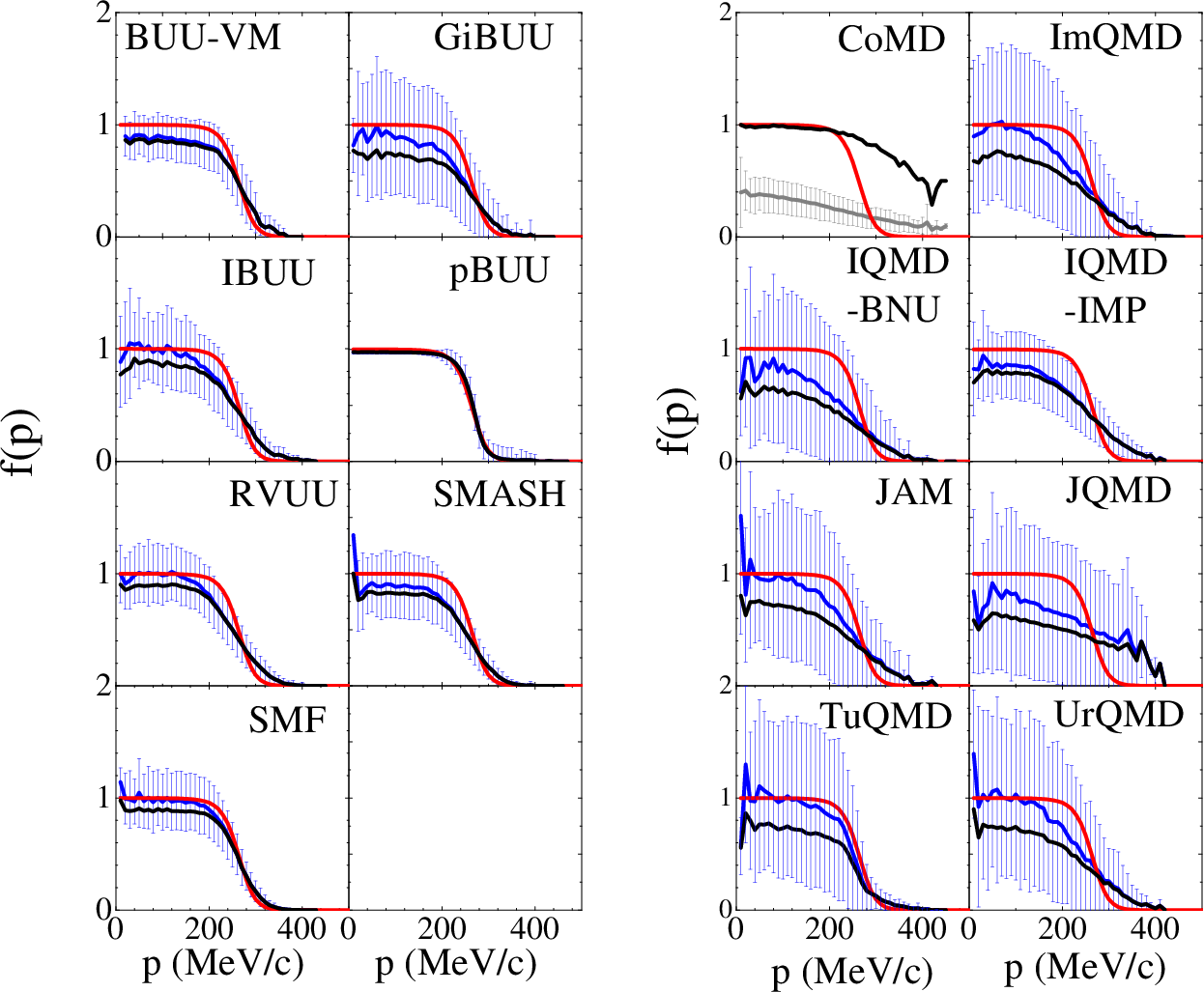} 
\vspace{-1mm}
\caption{Distribution of occupation probabilities (blue) in the first time step of the simulation for the T = 5 MeV initialization with the mean and variance shown by the blue curve and the blue error bars. Left panels show results for BUU-type codes and right panels for QMD-type codes. The average effective blocking probabilities are shown as the black curve (see text). The Fermi-Dirac distribution with T = 5 MeV used for initialization is represented with the solid red line. The gray line and error bars for CoMD are discussed in the description of the code in Sec.~\ref{comd}. Figure taken from Ref.~\cite{Zhang2017}.}
\label{pauli}
\end{figure}
\begin{figure}[tb]
\centering
\includegraphics[width=0.7\textwidth]{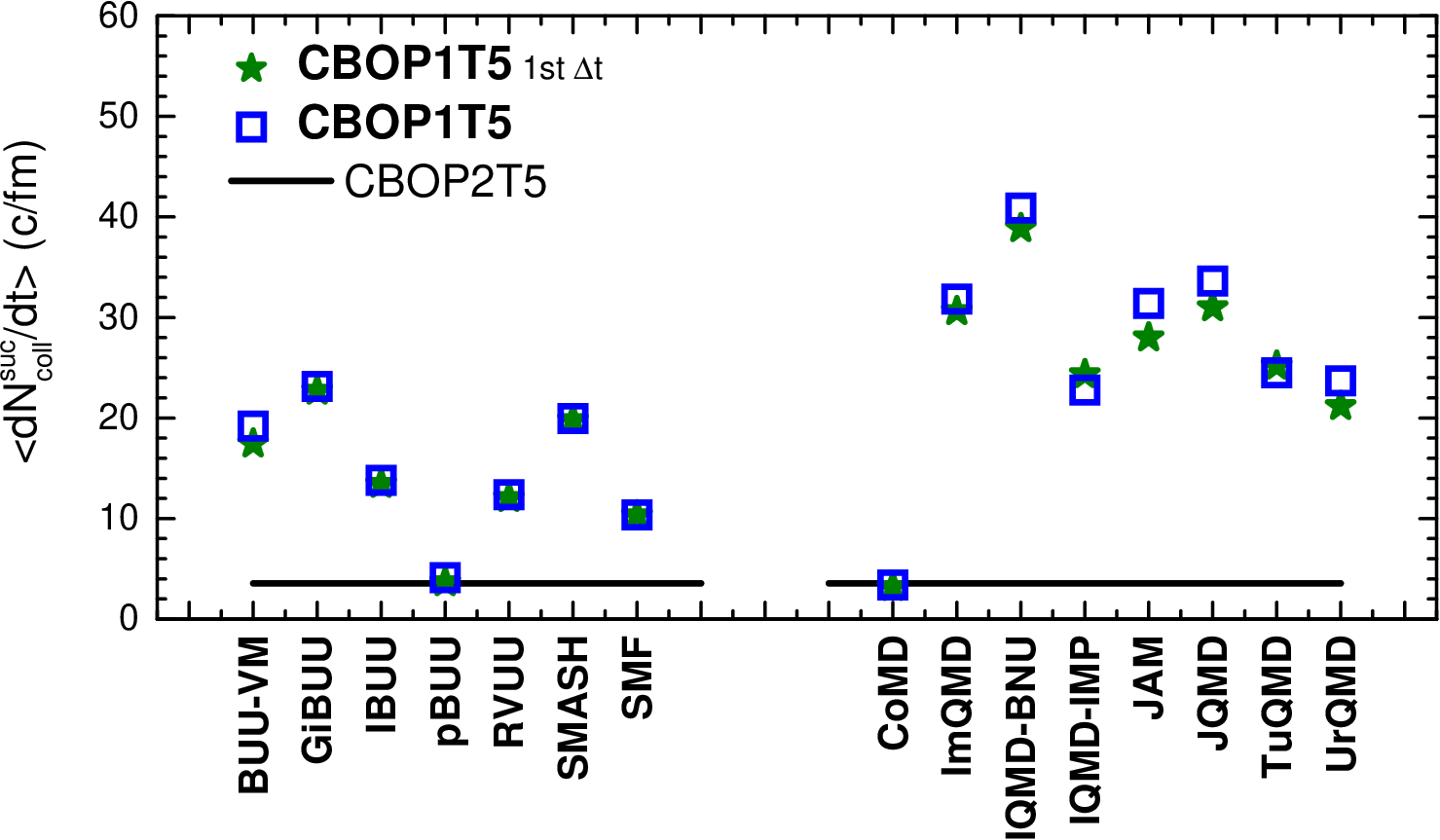}
\caption{Successful collision rates from the different models in simulations with Pauli blocking for T = 5 MeV initializations (called CBOP1T5 mode in the figure). The square symbols show the results averaged over the time interval 60 - 140 fm/$c$, while stars give the successful collision rates for the first time step. The black line represents the reference value calculated with the basic cascade code for a fixed Fermi-Dirac blocker (CBOP2T5 mode). Figure adapted from Ref.~\cite{Zhang2017}.}
\label{collision}
\end{figure}
As discussed above, the treatment of the collision integral involves two steps: first, making the decision whether two given (test) particles collide, and second, determining the Pauli-blocking of the final state of the two particles after the collision. In cascade calculations without blocking, one tests the first ingredient. This case corresponds to a classical ideal gas, where the collision rate can be calculated easily from kinetic theory both for non-relativistic and relativistic kinematics. Generally, all codes agreed with this limit within 1\%. To achieve this, however, it was necessary to explicitly eliminate repeated collisions between the same particles in the same or subsequent time steps; the latter was not guaranteed in the original Bertsch prescription~\cite{Bertsch:1988ik}. Repeated collisions represent a correlation between collisions that are not considered in kinetic theory. Very small remaining differences to kinetic theory are seen and are probably caused by higher-order correlations, which still remain in a simulation. The Bertsch prescription, also in the modified form, relies on distance and collision time criteria, which can be quite involved with relativistic kinematics, and details are given in tabular form in Ref.~\cite{Zhang2017} and in the code descriptions in sections 4 and 5. Rather than using the intuitive geometric criterion, statistical strategies are used locally, by assuring that the density dependence of the mean free path is obeyed (see Sec.~\ref{sec:collision_term}). These methods lead to slightly better agreement with the exact limits for the ideal gas and are actually easier to implement.

The second step of determining the Pauli blocking of the collision showed considerably larger differences among the codes. This step requires to calculate the phase-space occupation probabilities  of the two particles in their final states after the collision, which has to be done by averaging in a certain neighborhood of the final-state phase-space. In a box, this average should be given by the initialized Fermi-Dirac (FD) distribution and should be stable in time. The final-state occupation probabilities accumulated from all collisions in the first time step are shown in Fig.~\ref{pauli}.  Because of the discretized representation of phase space in a simulation (different in BUU and QMD), there are considerable fluctuations of these occupation probabilities. In BUU, these depend on the number of test particles per nucleon, and in the limit of the TP number going to infinity, there are no fluctuations and one should obtain an exact solution of the BUU equation. In the simulations, by varying the number of test particles, one can see that, in principle, the limit can be reached. With QMD as an event generator, the fluctuations are not suppressed in the limit of infinitely many events. The amount of fluctuations is regulated by the averaging procedure, generally by the width of the wave packet in coordinate and momentum space. With the parameters usually chosen, fluctuations are considerably larger than in BUU. 

Due to fluctuations, the occupations can be larger (even larger than unity), but also smaller than the initialized FD distribution for the given temperature. In the latter case, collisions that should not occur in the Fermi system take place, and this leads to collision rates that are higher than the exact or numerically obtained reference values. If the occupation is larger than unity, most codes block the collision completely, and thus effectively set it to 1. In Fig.~\ref{pauli}, the average of the occupation probabilities determined in this way is given by the black line. It is seen that it is below the prescribed FD distribution inside the Fermi-sphere, but at the same time reaches much beyond it. This is particularly seen for the QMD codes as a result of the large fluctuations in the representation of the phase space. As a consequence, the initialized Fermi-Dirac distribution is not stable, but changes to a classical Maxwell-Boltzmann distribution on a time scale of 10-100 fm/$c$, depending on the code. The resulting  collision rates for the case of T = 5 MeV are shown in Fig.~\ref{collision} (for the BUU codes on the left and the QMD codes on the right) for the first time step and for a chosen time interval later in the collision (the difference is small). These results are compared to the result of a basic cascade code, which enforces the FD occupation probabilities at each time step, and thus can serve as a reference representing the result of kinetic theory. It is seen that the collision rates in essentially all models tend to deviate significantly from the reference values, and considerably more so for the QMD codes. This strong influence of fluctuations on the collision rates was seen for the first time in this comparison.  
The reference line in Fig. 9 is the result of the kinetic theory without any fluctuation. Thus it is not the result that is desirable in a transport simulation, since fluctuations are physical. The deviation from the kinetic result corresponds to different physical models, how to introduce fluctuations, and is thus connected to the question mentioned in Sec.~\ref{sec:fluctuations} about the proper treatment of fluctuations in transport simulations.

\subsection{Collision Integral with Pions and $\Delta$ Resonances in a Box} \label{pion_box}

In this comparison, we tested pion production in transport simulations~\cite{ono2019}. This is of particular importance since the charged pion ratio, $\pi^-/\pi^+$, is expected to be a good probe of the symmetry energy. For densities above saturation, it is one of the few probes accessible in the laboratory. At intermediate energies, it reflects the $n/p$ ratio, which is directly controlled by the symmetry energy, at the position where the $\Delta$'s and pions are produced. However, as discussed in the introduction, the analysis of this pion ratio for Au+Au collisions at intermediate energies~\cite{FOPI-pion} by several transport codes has yielded widely different conclusions on the stiffness of the symmetry energy. Thus, it is of great interest to investigate the origin of these discrepancies. This is done in this box comparison and in the dedicated study of a full heavy-ion collision discussed in the next subsection.

\begin{figure}[tb]
\centering
\vspace*{-10mm}
\includegraphics[width=1.0\textwidth]{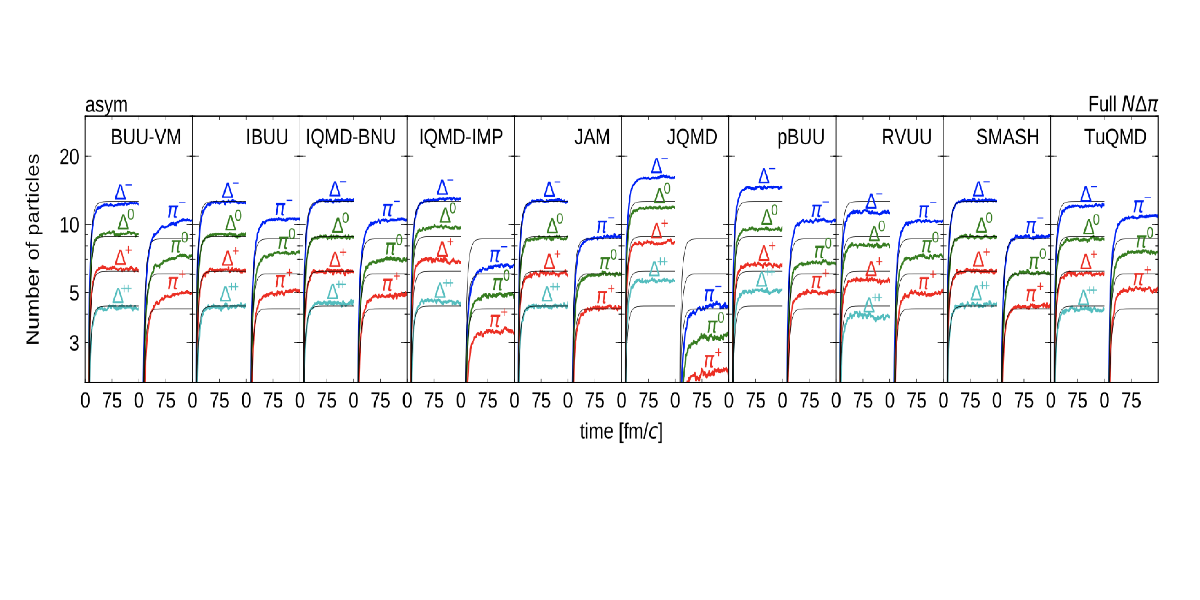}
\vspace*{-30mm}
\caption{Time evolution of the numbers of the different charge states of $\Delta$'s and $\pi$'s in an asymmetric system ($\delta = 0.2$) for the full $N\Delta\pi$ dynamics, i.e., including both $NN\leftrightarrow N\Delta$ and $\Delta\leftrightarrow N\pi$ processes. In the panel for each code, the evolution of the $\Delta$'s is shown on the left and that of the $\pi$'s on the right. Solutions of the rate equation are represented by thin black lines. Figure taken from Ref.~\cite{ono2019}.}
\label{pindelta}
\end{figure}

To set up the problem in a simple yet meaningful way, we initialized nuclear matter in a box at saturation density with $N/Z$ = 1 and 1.5 (asymmetry $\delta=(N-Z)/A$ = 0 and 0.2) and $T = 60$ MeV as a typical temperature in heavy-ion collisions at intermediate energy. We turned off the mean field, which is not crucial for this comparison of pion production, and also the Pauli blocking to avoid the effects of fluctuations seen in the comparison discussed in the previous section. Thus we performed a cascade calculation in a box with pions and $\Delta$s. The production of pions was assumed to proceed solely via the  $\Delta$ resonance, which is achieved by implementing inelastic collisions $NN\leftrightarrow N\Delta$ with energy- and mass-dependent cross sections, a width of the  $\Delta$, and the decay $\Delta\leftrightarrow N\pi$. The results were compared with two exact limits: the ideal $N \Delta \pi$ Boltzmann gas at equilibrium for the long-time behavior, and the solution of rate equations for the time evolution, where thermal (but not chemical) equilibrium is assumed at all times. We monitored the multiplicities of pions and $\Delta$s, the reaction and decay rates, ratios of isospin states of pions and $\Delta$s, and the isospin conservation. Ten codes of BUU and QMD types participated, and here there should be no systematic differences between them, since the mean field and the Pauli blocking are turned off.

The evolution of the multiplicities of the different charge states of pions and $\Delta$s as a function of time is shown in Fig.~\ref{pindelta}. For each code, the time evolution of the $\Delta$s is shown on the left side of the panel and that of the pions on the right. The thin black lines represent solutions of the rate equations. One can observe large differences among the codes. A similar figure (not shown here) for the case with the $\Delta$ decay turned off, i.e., where only the $NN\leftrightarrow N\Delta$ process was considered, showed much smaller differences~\cite{ono2019}. Thus, the discrepancies have to do with the simultaneous occurrence of two inelastic processes. On the other hand, the spacings between the curves for the different charge states in Fig.~\ref{pindelta} is rather similar among the codes and relative to the rate equation. From this observation, we expect that ratios of multiplicities could be more consistent, as will be seen below.

\begin{figure}
\centering
\begin{tikzpicture}
\tikzstyle{nuc}=[very thick, solid];
\tikzstyle{delta}=[very thick, double];
\tikzstyle{interact}=[thick, dotted];
\tikzstyle{extra}=[solid];
\tikzstyle{cor}=[color=black, draw, rounded corners];
\draw[nuc](0,0) -- node[above]{$N_i$} ++(1,0) coordinate(Ni1)
-- node[above]{$N_i$} ++(1,0) coordinate(Ni'1)
-- node[above]{$N_i$} ++(2,0) coordinate(Ni');
\draw[nuc](0,-1) -- node[below]{$N_j$} ++(1,0) coordinate(Nj1);
\draw[nuc](Nj1) -- node[below]{$N_j$} ++(3,0) coordinate(Nj');
\draw[extra] (Ni'1) ++(-0.5,1) node[right]{$X$} -- ++(0.5,-1) -- ++(0.5,1);
\draw[interact] (Ni1) -- (Nj1);
\draw[nuc](Ni') -- node[above]{$N_i'$} ++(1,0);
\draw[delta](Nj') -- node[below]{$\Delta_j'$} ++(1,0);
\draw[interact](Ni') -- (Nj');
\end{tikzpicture}
\caption{\label{fig:correlation} Higher-order correlation induced between $N_i$ and $N_j$ after the $N_iN_j$ elastic collision and the scattering of $N_i$ (or $N_j$) by another particle $X$. This correlation enhances the possibility of the second $N_iN_j$ collision leading to $N_iN_j\to N_i'\Delta_j'$.
}
\end{figure}
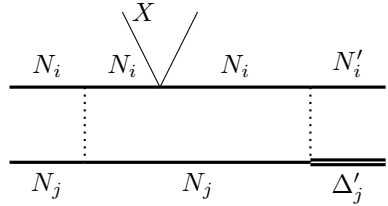

A detailed analysis revealed two main sources for the differences in the multiplicities. The simulation usually proceeds in time steps $\Delta t$, during which all particles are first propagated (here in straight lines) and then the different collision and decay processes are tested and performed, usually with the geometrical Bertsch prescription as explained above. Since the multiplicities are monitored at the end of each time step, they can depend on the sequence by which this is done. If the collisions are performed first and then the decays, the $\Delta$ multiplicities will be low, resulting in high $\pi$ multiplicities. The opposite sequence would result in high $\Delta$ and low $\pi$ multiplicities. Of course, this effect is smaller if some intermediate prescription is used, but this is code-dependent. The effect will also be smaller for a smaller time step. However, there is a second effect due to correlations between collisions that can also influence the multiplicities. As discussed above, the simplest correlation, namely the repeated collision between the same particles, has been eliminated. But higher-order correlations, which are not present in the Boltzmann equation, exist and cannot easily be eliminated. One such correlation is illustrated in Fig.~\ref{fig:correlation} for the process $N_iN_j\rightarrow N_i^{\prime}\Delta_j^{\prime}$, in which $N_i$ scatters with another particle $X$ before it scatters again with $N_j$. This process is not eliminated and enhances the probability for $\Delta$ production, since the nucleons $N_i$ and $N_j$ are still close in coordinate space. This effect will be larger if the time step $\Delta t$ is smaller. These two effects on the multiplicities from  sequence of collisions and higher-order correlations thus depend in opposite ways on the time step, so that they counteract each other and cannot altogether be eliminated. 

\begin{figure}[tb]
\centering
\vspace{-0mm}
\includegraphics[width=1.0\textwidth]{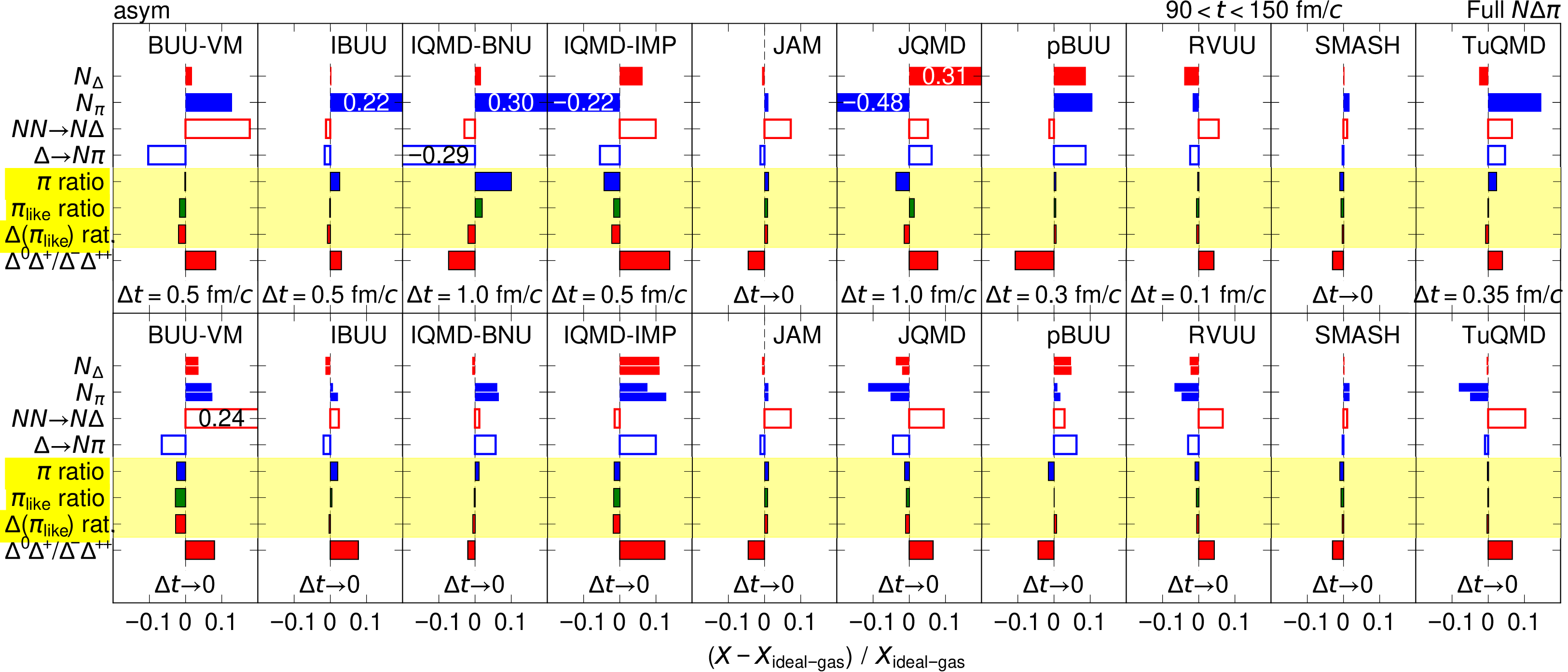}
\vspace{-4mm}
\caption{\label{fig:dtsummary} Relative deviation of the results of different codes from the value obtained in the ideal Boltzmann gas mixture for multiplicities of $\Delta$s and pions, collision and decay rates, and different isotopic ratios, averaged over $90<t<150$ fm/$c$ in the asymmetric system ($\delta=0.2$). Of particular interest are the particle ratios, highlighted in yellow, and among these especially the $\pi_{like}$ ratio (see text), which corresponds to the observable $\pi^-/\pi^+$ ratio in heavy-ion collisions. The upper panel shows the case for a time-step $\Delta t$ as used in calculations of the codes published in the past.  The lower panel shows the limiting case of $\Delta t\to 0$ obtained by linear extrapolation (JAM and SMASH do not rely on time steps).  In the lower panel, the two bars for the multiplicities show the result of linear extrapolations of $X$ (upper bar) and $1/X$ (lower bar), where $X\equiv N_\Delta$ and $N_\pi$, respectively. Figure adapted from Ref.~\cite{ono2019}.
}
\end{figure}

A possible solution here is to do the calculations for two time steps and extrapolate in some way to $\Delta t\rightarrow 0$. The values of several quantities using this procedure are shown in Fig.~\ref{fig:dtsummary}: multiplicities of 
$\Delta$ and $\pi$, reaction rates $NN\rightarrow N\Delta$ and $\Delta\rightarrow N\pi$, several particle yield ratios, and a quantity testing the isospin symmetry, all averaged over late times of the collision. The upper part gives the results of the codes for the usual size of the time step used by the code, while the lower part shows the (linear) extrapolation to $\Delta t\rightarrow 0$ (the codes JAM and SMASH are time-step-free, see Secs.~\ref{sec:JAM} and \ref{sec:SMASH}). Shown is the relative deviation of the various quantities from the values as obtained from the ideal Boltzmann gas. One can see in the upper part the large deviations in the multiplicities for the finite time step, as already seen in Fig.~\ref{pindelta}. As expected from that figure, the deviations are smaller for the multiplicity ratios, which are highlighted by the yellow bands. Generally, the deviations are significantly reduced after the extrapolation  $\Delta t\rightarrow 0$ in the lower part of the figure.
\footnote{Time-step-free methods, in which the collisions are performed in the order of their occurrence by a forward-projection of the trajectories of the particles, are particularly effective in the case of cascade calculations with straight trajectories. In the presence of mean-field potentials they are less advantageous and rather time-consuming, and anyway their use requires a substantial modification of a code.}

The deviations are particularly small for the yield ratios. Of particular interest is the $\pi_{like}$ ratio defined as~\cite{Li:2002qx} 
\begin{equation}
R(\pi_{like}) = \frac{\pi^- +\Delta^- +\frac{1}{3} \Delta^0}{\pi^+ +\Delta^{++} +\frac{1}{3} \Delta^+}.
\end{equation}
It corresponds to the $\pi^-/\pi^+$ ratio after all $\Delta$s have decayed, and thus reflects the $\pi^-/\pi^+$ ratio that is measured in a heavy-ion collision.
\footnote{The $\Delta(\pi_{like})$ ratio in Fig. 12 is the $\pi^-/\pi^+$ratio, which would result, if only the decay of the $\Delta$-resonances was considered 
$R(\Delta(\pi_{like}))=(\Delta^-+\frac{1}{3}\Delta^0)/(\Delta^{++}+\frac{1}{3}\Delta^+).$}
It is seen that the agreement of this ratio between the codes and with the reference value is of the order of a few percent and is thus generally very good. Therefore, this ratio appears as a robust observable for the determination of the symmetry energy. 
The fact, that the codes differed very much in their conclusions in the past must be due to different physical models in the description of  heavy-ion collision not tested in this cascade box-calculation,
such as momentum-dependent mean-field interactions, energy conservation in inelastic processes, or different assumptions for the inelastic cross sections and for pion and $\Delta$ potentials, but it may also be that codes have been improved in the course of the code comparisons. The question of the consistency of different codes in predicting pion ratios in heavy-ion collisions is taken up in the next subsection.

\subsection{Symmetry energy investigation in pion production from Sn+Sn systems}

In this comparison, we returned to the investigation of pion production close to the threshold in realistic heavy-ion collisions~\cite{SpRIT:2020blg}. In contrast to the study of Sec.~\ref{HIC-1GeV}, the beam energy is much lower and the mean field plays a much more significant role. We consider simultaneously inelastic collisions with particle production and the mean field, which is of particular interest for obtaining information on the symmetry energy at densities above saturation. From the study of pion production in a box (Sec.~\ref{pion_box}), we know that the pion ratios in simple conditions are consistent among codes. It is now of interest to see how this is reflected in a heavy-ion collision. Finally, we take advantage of a new experiment designed to measure the multiplicities of negatively and positively charged pions with high accuracy for central collisions of $^{132}$Sn+$^{124}$Sn, $^{112}$Sn+$^{124}$Sn, and $^{108}$Sn+$^{112}$Sn at $E/A=270$ MeV with the S$\pi$RIT Time Projection Chamber. The uncertainties of the individual pion multiplicities are measured to 4\%, and those of the charged pion multiplicity ratios to 2\%.

\begin{figure}[tb]
\centering
\vspace{-0mm}
\includegraphics[width=0.7\textwidth, angle=0.00]{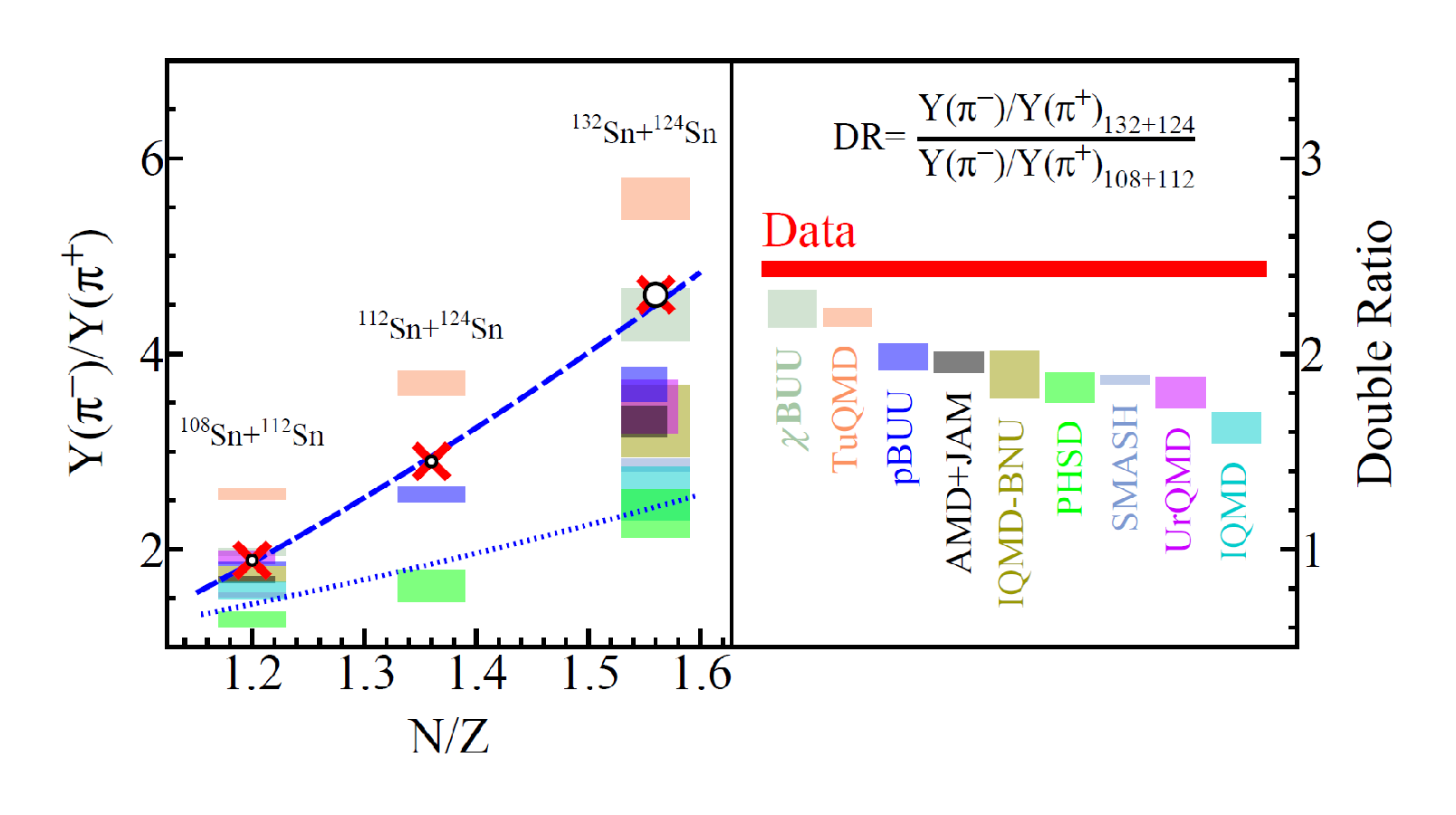}
\vspace{-6mm}
\caption{(Left panel) Charged pion yield ratios as a function of $N/Z$. The experimental data are shown as crosses with the circles representing the experimental errors. The results of the calculations are represented by colored boxes for the different codes identified by their color in the right panel. The upper and lower boundaries of the boxes give the result for the soft and stiff symmetry energy choices for each code, i.e., the height of the boxes is representative for the sensitivity to the stiffness of the symmetry energy. The dashed blue line is a power-law fit with the function $(N/Z)^{3.6}$, while the dotted blue line represents $(N/Z)^2$ of the system. (Right panel) Double pion yield ratios for $^{132}$Sn +$^{124}$Sn and $^{108}$Sn +$^{112}$Sn. The data and the uncertainty are given by the red horizontal bar, while the results of the transport models are shown by the colored boxes, in a similar way as in the left panel. Figure adapted from Ref.~\cite{SpRIT:2020blg}.}
\label{pion}
\end{figure}

In view of previous experience with the interpretation of the FOPI pion data~\cite{FOPI-pion}, the comparison was designed to preclude the adjustment of the input parameters by the code authors to better reproduce the data. Rather, the authors were asked to provide predictions prior to the knowledge of the data. Thus, there were no requirements on the physical model input, but the authors should use the best settings for their codes as used in previous analyses. This included choices of the momentum dependence of the mean fields, models for the inelastic cross sections, and choices of the symmetry energy density functional. In most cases, these were adjusted to fit the Au+Au data of Ref.~\cite{FOPI-pion}, and are close to the information given in the code descriptions in the next sections. The authors were asked, however, to provide predictions for two behaviors of the symmetry energy, a soft and a stiff one of their choice, which had values of the symmetry energy slope $L$ around 55 and 150 MeV. Thus, this was not a study ``under controlled conditions" as the previous ones, but rather a test of the present state of the predictive power of simulations of heavy-ion collisions with respect to the symmetry energy. It also establishes a benchmark for future comparisons. Seven well-known codes participated in this study, and two additional ones (IQMD and PHSD) joined later and are also included here. 

Figure~\ref{pion} shows the predictions of the codes for the single pion yield ratio $\pi^-/\pi^+$ (left panel), and for the double ratio for the two reactions with the highest and lowest neutron content (right panel), together with the data of the S$\pi$RIT experiment.  The single pion ratios are plotted against the $N/Z$ ratio of the collision system (only three codes provided results for the intermediate system, which was not asked for in the comparison). The color assignment for the codes is given in the right panel. The results of the codes are represented as boxes, where the upper and lower boundaries are the results of the two choices of the symmetry energy. In most cases, the soft symmetry energy gives higher ratios, but the opposite is observed for two codes (TuQMD, SMASH). In the left panel, the dotted line shows a quadratic dependence of the single pion ratio on the neutron to proton ratio, $(N/Z)^2$, which would be expected from a $\Delta$-resonance model for pion production, while the dashed line is a fit to the data, which yields a dependence of  $(N/Z)^{3.6}$. A thermal model for the pion production yields a power-law dependence where the exponent is proportional to $E_{sym}/T$, thus the stronger dependence seen in the experiment clearly indicates a sensitivity to the symmetry energy.

The calculations follow the trend of the experiment only qualitatively, but the differences among the codes are large and are of the same order as the differences from the experiment\footnote{\samepage The PHSD code has
been used primarily for collisions studies at GeV energies and was thus not very well adjusted to this energy regime. This may explain the rather large deviations for the single pion ratios.}. Also, the
sensitivity of the calculations to the symmetry energy (height of the boxes) is rather small compared to the differences among the codes and from the experiment. A similar picture is seen for the results of the
double ratios in the right panel. The calculated values are somewhat more consistent between the codes, but the difference from the experiment is still large. Clearly, the predictions of the calculations by the
different codes differ too much to allow the extraction of reliable constraints on the symmetry energy from the data, based solely on the pion multiplicities and their ratios. 
The large differences seen here may also explain the contradictory conclusions about the density dependence of the symmetry energy obtained previously from the pion data in the Au+Au system mentioned in the introduction~\cite{FOPI-pion}.

In view of the results of the box-pion comparison discussed in Sec.~\ref{pion_box}, these differences are probably not due to the treatment of the inelastic processes. Their parametrization may be different here, but this will not be the main reason for the discrepancies, because they are mostly based on fits to the same experimental data. An important difference, not tested before, is the momentum dependence of the isoscalar and isovector mean fields, whose treatment can be rather different in the codes. In connection with inelastic processes, this also implies questions of threshold effects for pion production at subthreshold energies and of energy conservation. Additionally, the algorithms to calculate the Coulomb potential, which significantly affects the pion ratio, is different in the different codes. Generally, different evolutions of the reaction in the density-temperature plane will affect pion production. Therefore, also nucleonic observables, like collective flow, should be monitored at the same time. A follow-up study of the same systems with more controlled input and analyzed output is in progress~\cite{xu2021}.

During this comparison, we realized that the total pion multiplicities may not be the best observables to test the symmetry energy at high density.
The multiplicities are dominated by low-energy pions, which undergo multiple collisions throughout the evolution of the system and are thus not good
probes specifically of the high-density phase of the collision. The low-energy part of the spectra depends strongly on assumptions on the $\Delta$ and pion potentials, and it is usually not well described by transport codes, and the results differ widely among each other. As already pointed out in previous studies~\cite{Li:2015hfa,SpRIT:2016nrj,Cheng:2016pso}, the high-energy part of the pion spectra and the spectral ratios should be a better probe of the symmetry energy. Recently, a study has been published (outside of the TMEP project) where a single code (TuQMD/dcQMD)~\cite{Cozma:2021tfu} was applied not only to the pion multiplicities, but also to the transverse momentum spectra of the pions~\cite{estee2021probing}. This calculation included essentially all needed ingredients for calculating pions reliably, which are partly not included in other codes. Besides conserving total energy, it has a more reliable treatment of the Coulomb potential and also included the pion potential (see description in Sec.~\ref{tuqmd}). It is found that in the space of the stiffness of the symmetry energy (slope $L$) and the isovector momentum-dependence (neutron-proton effective mass splitting $\delta m^*_{np}=m^*_n-m^*_p$), it is possible to obtain constraints from the comparison with the experimental spectra.
Meanwhile a study appeared of the same reaction~\cite{Yong:2021nwn}, which obtains results for the single and double pion total multiplicity ratios much closer to the experimental values. The main difference relative to the calculations reported here seems to be the use of a high-momentum tail in the initial distribution attributed to short-range correlations (see Sec.~\ref{sec:discussion}).  A further study appeared~\cite{Wei:2021}, where the total yield ratios are reproduced by modifying the momentum-dependence of the symmetry potential. These studies were not part of the comparison reported here, which was a prediction before the experimental result was known, but were published after Ref.~\cite{SpRIT:2020blg} had appeared.
These latest results also underscore the conclusions drawn above that total pion yield multiplicities and their ratios alone are unreliable in extracting the symmetry energy information.

\subsection{Summary of the Results of the TMEP Project }
\label{sec:summary}

The TMEP Collaboration was formed in 2014 to address the conflicting conclusions about the symmetry energy obtained from various transport models from studying the negatively and positively charged pions in central Au+Au collisions at 1.5 AGeV measured by the FOPI Collaboration~\cite{FOPI-pion}.  
The goal of the Collaboration is to identify reasons for such conflicting conclusions and try to improve the predictive power of analyses of heavy-ion collisions by the transport approach.  Although one would generally expect transport models to give similar results, as they are supposed to solve the same kinetic equations for intermediate-energy heavy-ion collisions, comparison studies from the TMEP Collaboration have shown that this is often not the case.
The reasons for this are not easy to uniquely identify in a simulation of a heavy-ion collision, because differences in the treatment lead to different evolutions of a reaction in density and excitation energy, and thus test different regions of the EoS. A better insight is obtained by studies in a box with periodic boundary conditions, which approximates infinite nuclear matter, where the different effects can be studied separately. Additionally, exact limits are often available, and thus one is not confined only to comparing codes with each other. In these studies, the various aspects of a transport model, such as mean-field propagation, collision term, and treatment of particle production and decay, have been investigated. Not only could one identify reasons for different results, but also understand many details of the transport approach.
Through the five papers that have been published by the TMEP Collaboration up to now, these differences have been mostly understood, and improved methods of treatments have been pointed out.  

An important result has been to identify clearly the differences between the BUU- and QMD-like approaches to transport theory. The main difference is the amount of fluctuations in the representation of the phase-space distribution. This difference is motivated by physics. In BUU, the objective is the evolution of the smooth one-body distribution function under the action of a mean field and the dissipation due to nucleon scattering, which does not naturally include fluctuations. To include these, one has to extend the approach to a Boltzmann-Langevin method, which is done consistently only by the codes SMF and BLOB~\cite{Colonna1998,Napolitani2017}. In QMD, the ansatz in the form of molecular dynamics goes beyond the mean field by including classical correlations and fluctuations. 
These are gauged by the parameter $\Delta x$, the width of the nucleon wave packets, which is only roughly constrained by the range of the nuclear interaction, but strongly influences the outcome of a collision.
The comparison has highlighted the consequences of these different amounts of fluctuations in the two approaches in a new way. In particular, they strongly influence the Pauli blocking probabilities, which in turn impact particle production. Fluctuations also act effectively as a dissipation and thus influence the damping of the motion, as was seen in differences in the width of the response functions studied in Sec.~\ref{HIC_100AMEV}. The calculation of effective many-body forces in the two approaches can also be different, thus affecting the frequency of the collective modes, which can be solved by the lattice-Hamiltonian method, as recently implemented in some QMD codes. 

For pion production, the box calculations have identified sources of differences in simulations including simultaneously inelastic processes and particle decay. Here, the sequence in treating the different processes and higher-order correlations between collision processes, which are not present in kinetic theory, lead to differences in particle production depending on calculational parameters such as the size of the time step. However, in the box calculations it was found that these differences are less important for the pion ratios. 

This did not lead to a similar consistency of the pion ratios in a realistic heavy-ion collision. This indicates the importance of additional effects in a realistic description of pion production. Using momentum-dependent forces leads to threshold shifts  in $\Delta$ and $\pi$ production and decay in the $NN\leftrightarrow N\Delta$ and $\Delta\leftrightarrow N\pi$ reactions, due to the difference in the total mean-field potentials in the initial and final states. These effects affect both the pion yield and the ratio of negatively to positively charged pions.
Except for the TuQMD/dcQMD model~\cite{Cozma:2016qej}, the RVUU model~\cite{Song:2015hua}, and the $\chi$BUU model~\cite{Zhang:2018ool}, none of other transport models in these studies include these threshold effects.
Additionally, the total energy of the colliding system may be violated in $NN$, $N\Delta$, and $\Delta\Delta$ scatterings.
Although a method has been developed in Ref.~\cite{Cozma:2014yna} to conserve the total energy in these scatterings, it has only been implemented in the dcQMD model. 
Medium effects on pions have not been included in most transport codes, which can be taken into account either through modified pion dispersion relations~\cite{Zhang:2017mps} or by pion optical potentials~\cite{Cozma:2016qej}. There is much information on these from the study of pion scattering from nuclei~\cite{Itahashi:2000pi-nucl} and from pionic atoms~\cite{Geissel:pionic-atoms}. 
Two studies are underway within the TMEP collaboration to study these effects, the pion production in a semi-realistic heavy-ion collision with controlled input but without momentum-dependence of the forces~\cite{xu2021} and medium effects, and box calculations to study specifically the threshold, the global energy conservation, and medium effects in transport models~\cite{cozma2021}. 

An important influence on the momentum spectra of charged pions is due to details of the Coulomb potential. Coulomb forces are included essentially in all transport codes on protons, charged nucleon resonances, and charged pions. They are often calculated using point particles, which, to avoid singularities, involves the introduction of a cutoff distance, below which the Coulomb force is assumed not to change. A more accurate treatment of the Coulomb potential is to find the electric and magnetic fields acting on a charged particle from the solutions of the Poisson equations with source terms given by the local charge and current densities, which for the electric fields is implemented in some codes. The effects will be clarified in the ongoing controlled heavy-ion collision study~\cite{xu2021}. 

For observables that involve rare particles, like pions in medium-energy nuclear collisions, it is inefficient to solve the BUU or QMD codes by treating these particles on the same footing as the nucleons. An efficient way to enhance the statistics of these particles is to treat them by the partition method~\cite{KoZhang21,Zhang:1998tj}, i.e., each of these particles is replaced by $N$ test particles in QMD, while in BUU the number of test particles per particle is enhanced for these species by a factor $N$. The scattering cross section of a test particle is then reduced by a factor of $N$. For example, the number of pions in Sn+Sn collisions at 270 A MeV is about one per event, which is a factor of about 200 smaller than the nucleon number, so one can choose $N=200$ to obtain the same statistics for pions and nucleons. 

\subsection{Discussion and Outlook} \label{sec:discussion}

In this subsection, we mention some open problems of transport approaches to heavy-ion collisions  with a view on improvements and future developments of transport codes, where, however, an exhaustive discussion is beyond the scope of this review. We also include some comments on the importance in applications of quantitative predictions of transport models for collisions of particles or nuclei on nuclei.

Since the abundance of light clusters up to $He$ nuclei produced in medium-energy heavy ion collisions is large~\cite{FOPI-pion}, their inclusion is needed in transport models to properly describe the collision dynamics.  Adding a cluster-finding algorithm at the end of a heavy-ion collision has allowed QMD-like codes to address light cluster production in medium energy heavy-ion collisions, see Sec.~\ref{sec:BQMD}. A more realistic approach is to include light clusters as dynamic degrees of freedom, as pioneered in the study of Ref.~\cite{Danielewicz:1991dh,dan92} by separate transport equations for the different clusters coupled through the collision terms via production and destruction cross sections. E.g., the production cross section of deuterons from the reaction $NNN\to Nd$ can be obtained from the empirical deuteron dissociation cross section by a nucleon, $Nd\to NNN$ . Since light cluster production involves many-nucleon scattering, the traditional geometric method for treating nucleon-nucleon scattering~\cite{Bertsch:1988ik} becomes not applicable. A stochastic method based on the transition probability~\cite{Xu:2004mz} needs to be used.

An issue of great current interest concerns how to effectively and correctly incorporate nucleon-nucleon short-range correlations (SRCs), within transport theory, which have been clearly identified in nucleon knock-out reactions~\cite{Hen:2016kwk}. In microscopic many-body calculations, it is seen that these introduce a high-momentum tail (HMT) into the momentum distribution, and that the kinetic symmetry energy is reduced relative to the Fermi gas kinetic energy, basically because SRCs are more effective in symmetric relative to asymmetric matter~\cite{Carbone:2013cpa,Rios:2013zqa}. These effects can be important in the interpretation of heavy-ion collisions to determine the EoS and particularly the symmetry energy, e.g., in the production of particles near thresholds. Phenomenological approaches have been proposed to study the above two SRC effects. One is to modify the phase-space distributions of nucleons in the incident nuclei by adding by hand a high-momentum tail ~\cite{Hen15} to study initial-state effects from first-chance collisions~\cite{ Guo:2021zcs,Zhang:2016vcc,Yong:2021nwn }. A second is to reduce the kinetic symmetry energy by a phenomenological factor, with the remaining kinetic correlation energy added to the potential symmetry energy~\cite{ PLi:2014vua}, to study the energetic effects of SRCs during the collision. In both approaches strong effects have been observed, but both methods are ad-hoc and not consistent in themselves and between each other.

A consistent inclusion of SRCs in transport seems to require an off-shell transport approach, i.e., the inclusion of dynamical spectral functions for all particles. This has been studied in extensions of transport theories by the groups of Mosel at al.~\cite{Effenberger:1999uv} and Cassing et al.~\cite{Cassing:1999mh}, with some differences in detail, and is implemented in the codes GiBUU  and PHSD (see sections~\ref{sec:GiBUU} and \ref{phsd}). In Ref.~\cite{Lehr:2000ua} it is demonstrated that in this approach the momentum distribution automatically develops a high momentum tail. The importance for pion production, however, is not large in these investigations and somewhat different in the two approaches, which has, however, not been systematically investigated. It has also been suggested that three- or many-body collisions may be a way of treating SRCs in transport approaches. Three-body collisions for pion production processes, like $NNN \rightarrow NN\Delta$, have been investigated by Bertsch, et al.~\cite{Bertsch:1995ig}, where it is found that the SRCs between two of the incident nucleons give a noticeable contribution. A different approach, based on a mean-free-path approximation to the collision integral, has been proposed by Bonasera, et al.~\cite{Bonasera:1992kjm}, where rather large effects are observed even on bulk observables. The incorporation of $n$-body collisions in transport equations in a schematic cluster approximation was also studied by Batko, et al.~\cite{ Batko:1991xd}, where effects were found to be rather small. None of these methods have so far been widely exploited in the description of heavy-ion reactions. The effects of SRCs on particle production close to threshold and the approximations used in treating them thus clearly deserve further study.

Besides using transport models to extract the properties of nuclear matter at various density, temperature, and isospin asymmetry, transport models can also be used
to relate the predictions from the nuclear many-body theory based on realistic nucleon-nucleon interactions to observables measured in medium-energy nuclear collisions. Some time ago, a connection was made between relativistic Dirac-Brueckner-Hartree-Fock calculations of symmetric and asymmetric nuclear matter at finite temperature and their use in heavy-ion collisions~\cite{gaitanos99,Fuchs:2005yn,Plohl:2006hy}, where also non-equilibrium effects were included in an approximate way.  Another step in this direction has recently been taken by the $\chi$BUU model~\cite{Zhang:2018ool} based on the momentum-dependent nuclear mean-field potentials from chiral effective theory.

Finally, we would like to expand somewhat on the remarks in the Introduction about the importance of reliable transport models in other fields of physics and applications. 
Examples are long-baseline neutrino experiments, which aim to extract neutrino mixing parameters, CP violating phases and the neutrino mass ordering\cite{Diwan2016lonbas}. For this extraction, the incoming neutrino energy is needed. Due to the generation of the beam, its energy is, however, known only with very large uncertainties,  and it has to be reconstructed from the observation of the final state. This is often modeled by simple Monte-Carlo cascade approaches, and reliable transport descriptions could add significantly to the success of these studies. This is of direct concern for the US experiment Deep Underground Neutrino Experiment (DUNE), and also for the ongoing experiments NuMI Off-axis $\nu_e$ Appearance (NOvA) and Tokai to Kamioka (T2K)~\cite{mosel2016lonbas,mosel2017nupion}. Also, the detection of dark matter particles faces similar problems.
Another example is the semi-inclusive electron scattering at Jefferson Lab (JLAB), where reactions such as (e,e'p) on nuclear targets need a firm basis to describe the background. These experiments aim for solving fundamental questions, such as about color transparency~\cite{HallC:2020ijh}, short-range correlations~\cite{Wright:2021dal}, and hadronization inside a nuclear medium~\cite{CLAS:2021jhm}. Experiments in the fixed target mode at the planned Electron-Ion Collider (EIC) will also need a good description of final state interactions.

In many applications of nuclear research and technology, the quantitative prediction of the effects and products of the bombardment of material with energetic particles is of great importance. For these applications, special programs have been developed, e.g., the program PHITS (Particle and Heavy Ion Transport code System) as a general-purpose package to simulate the transport of many types of particles (nucleons, heavy-ions, electrons, photons) for a wide range of eV to TeV energies~\cite{phits2013,sato2015}. As models of nuclear reactions, this package incorporates the cascade model JAM~\cite{JAM} and the transport code JQMD2.0~\cite{JQMD2.0}, see Secs.~\ref{sec:JAM} and \ref{sec:JQMD2.0}, together with the evaporation code GEM~\cite{GEM}, but also nuclear data libraries like JENDL4.0~\cite{JENDL}. Examples of applications of this codes are the design of accelerator facilities, like the spallation source and transmutation facility at Japan Proton Accelerator Research Complex (J-PARC)~\cite{Harada2005} or the fragment pre-separator at Facility for Rare Isotope Beams (FRIB). Dose calculations are essential in particle therapy and have been performed with PHITS and other codes~\cite{Sato2009dose}, and transport models are needed for the calculation of the secondary dose from fragments of the first collision~\cite{Yonai2012dose}. For radiation therapy, the codes SMF and BLOB have also been coupled with the GEANT4 general purpose Monte-Carlo toolkit~\cite{Ciardiello:2020dtr,mancini2019geant}. Dose calculations are also mandatory for the feasibility of long-duration space travel. An example is the MATROSHKA experiment in and around the International Space Station (ISS)~\cite{sihver2008dose}. The dose exposition due to solar flares on travel and communications on earth is of great relevance~\cite{kataoka2018flare}. Transport models are also good event generators and have applications, e.g., to design or simulate new detectors \cite{barney2021,tsang2021applying}, or to provide training and validating events in machine learning projects \cite{lif2021}. The design of the Multi-Purpose Detector (MPD) at the Nuclotron-based Ion Collider fAcility (NICA) in Dubna was studied using the PHQMD code~\cite{kolesnikova2020}, a variant of the PHSD code described in Sec.~\ref{phsd}. For all these and many other applications, the availability of  reliable nuclear transport models is very important.

\newpage
\section{Boltzmann-Uehling-Uhlenbeck(BUU)-like codes} \label{sec:buu}
In this section, we collect the descriptions of the codes of BUU type, and in the next those of QMD type.
The genesis of these descriptions is the same for both families: The code authors were asked to write a brief 2-3 page description of the code, addressing the history of the code, the initialization, the standard choice of forces, the treatment of the collision term with the Pauli blocking, the in-medium cross sections, and to give the main references. These descriptions were cast into a uniform appearance but otherwise not much edited. 
Note, in particular, that the notation in these prescriptions has not been standardized, but should be, of course, consistent and self-explanatory for each description. Under each code heading, we list the names of the main authors who provided the descriptions. 

\subsection{The Boltzmann-Langevin One-Body (BLOB) Code}
\vskip 0.1in
P. Napolitani, M. Colonna
\vskip 0.1in
\subsubsection{Boltzmann-Langevin dynamics}

A full implementation of the Boltzmann-Langevin (BL) equation in a transport model for heavy-ion collisions
is necessary to handle large-amplitude phase-space fluctuations, and to describe processes characterized
by instabilities. 
This requirement applies particularly to the Fermi energy domain, where a variety of exit channels is accessible when progressing from the same entrance channel. In this regime, as opposed to more conventional treatments, the implementation of the BL equation can describe the competition between energetically favored mechanisms, such as multi-fragmentation, fusion, binary splits and neck fragmentation in dissipative collisions.
At higher incident energy, the inclusion of large-amplitude fluctuations, less relevant for the entrance-channel
kinematics, becomes important for the description of the bulk behavior of the hot sources formed in the system.

The BLOB~\cite{Napolitani2013,Napolitani2017} model, as well as the SMF model~\cite{Colonna1998} from which it evolved (see Sec.~\ref{sec:SMF}), aims at solving the BL equation for the distribution function $f$,
\begin{equation}
	\partial_t\,f - \left\{H[f],f\right\} = {\bar{I}[f]}+{\delta I[f]} \;.
\label{eq1}
\end{equation}
The left-hand side gives the Vlasov evolution for $f$ in its own self-consistent mean field, and
the right-hand side introduces the unknown N-body correlations through the residual interaction.
The latter consists of the average Boltzmann hard two-body collision integral
$\bar{I}[f]$ and the fluctuating term $\delta I[f]$, both written in terms of the one-body distribution function.

\subsubsection{Heritage and specificity of the BLOB code}

While the BLOB model inherits the mean-field description from the SMF model, different
approaches are involved in treating the collision integral $\bar{I}[f]$ and the fluctuation term $\delta I[f]$ (i.e., the
right hand side of Eq.~\ref{eq1}). The SMF method consists of projecting the fluctuations on suitable subspaces (in this case the configuration space).
The BLOB approach solves, on the other hand, the BL equation in full phase space.
In both SMF and BLOB, the fluctuating term ${\delta I[f]}$ acts on the dynamical trajectories during the whole
evolution, while in some studies aiming at introducing small-amplitude fluctuations, it can be reduced to a
stochastic definition of initial states. As will be discussed in the following, the numerical implementation
of the BLOB approach assures that the residual term $\bar{I}[f]+{\delta I[f]}$ can affect a more extended portion of the
phase space in each single scattering event.

The BLOB model differs substantially from an earlier similar strategy, used in the
approach by Bauer, Bertsch, and Das Gupta~\cite{Bauer1987}, because it constrains the fluctuating term ${\delta I[f]}$ to act on phase-space
volumes with the correct occupancy variance. Such constraint ensures that the Pauli blocking is not violated,
and it imposes special attention to the metric of the phase space (see discussion in Ref.~\cite{Chapelle1992}).
Calculations in a periodic box for unstable nuclear matter, in one dimension~\cite{Rizzo2008} and in three
dimensions~\cite{Napolitani2017,Napolitani_Messina}, have shown that the BLOB approach describes the growth rate of the
corresponding (spinodal) unstable modes consistent with the form of the mean-field potential, as
ruled by the dispersion relation~\cite{colonna_new}.

The BLOB model for heavy-ion collisions is constructed based on this efficient description of the dispersion relation.

%
%
\begin{figure*}[t]\begin{center}
	\includegraphics[angle=0, width=0.5\columnwidth]{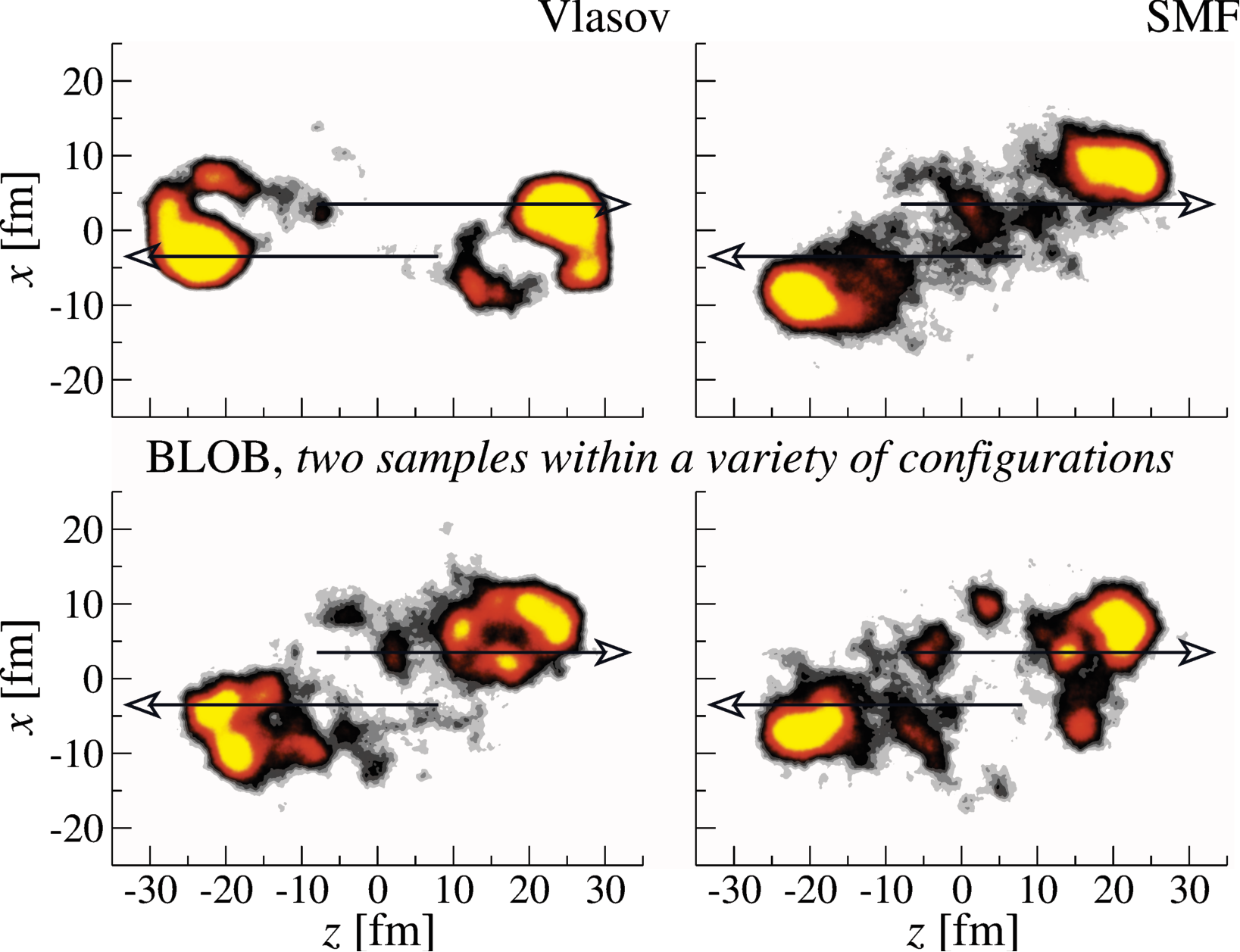}
\end{center}\caption
{
Simulation of the $^{197}$Au$+^{197}$Au collision at $100\,A$MeV for an impact parameter of 7 fm, at the time $t=140$ fm/c, by three approaches which share the same initialization: Vlasov (top, left), SMF (top, right) and two different exit-channel configurations obtained with BLOB (bottom).
The arrows indicate the direction of the target and projectile; their origin indicate the centers of target and
projectile at the initial time $t=0$ in the simulation.
}
\label{fig1}
\end{figure*}

\subsubsection{Solution of the Boltzmann-Langevin equation in full phase space}

The solution of the BL equation in full phase space is obtained by replacing the conventional Uehling-Uhlenbeck
average collision integral by a similar form where one given binary collision does not act only on the two test particles `a' and `b', but
rather it involves extended phase-space agglomerates of test particles of equal isospin A$={a_1,a_2,\dots}$,
and B$={b_1,b_2,\dots}$
to simulate wave packets:
\begin{equation}
	{\bar{I}[f]}+{\delta I[f]}
	= g\int\frac{\diff^3\vecp_b}{h^3}\,
	\int \diff\Omega\;\;
	W({\scriptstyle\A\B\leftrightarrow\C\D})\;
	F({\scriptstyle\A\B\rightarrow\C\D})\;,
\label{eq2}
\end{equation}
where $W$ is the transition rate, given in terms of the relative velocity between the two colliding agglomerates and the differential nucleon-nucleon cross section,
\begin{equation}
	W({\scriptstyle\A\B\leftrightarrow\C\D}) = |v_\A\!-\!v_\B| \frac{\diff\sigma}{\diff\Omega}\;,
\label{eq3}
\end{equation}
and $F$ contains the products of occupancies and vacancies of initial and final states calculated for the
test-particle agglomerates
$	F({\scriptstyle\A\B\rightarrow\C\D}) =
	\Big[(1\!\!-\!\!{f}_\A)(1\!\!-\!\!{f}_\B) f_\C f_\D - f_\A f_\B (1\!\!-\!\!{f}_\C)(1\!\!-\!\!{f}_\D)\Big].$
At each interval of time, by scanning all phase space in search of collisions, and by redefining all
test-particle agglomerates accordingly in phase-space cells of volume $h^3$, nucleon-nucleon correlations
are introduced. Since $\Ntest$ test particles are involved in one collision, and since those test particles
could be sorted again in new agglomerates to attempt new collisions in the same interval of time when
the collision is not successful, the nucleon-nucleon cross section contained in the transition rate $W$ is divided by $\Ntest$: $\sigma = \sigma_{NN} / \Ntest$. Special attention is paid to the metric in defining the test-particle agglomeration, i.e., the agglomerates are searched by requiring that they are the most compact configuration in the phase space that violates neither Pauli blocking in the initial and in the
final states nor the energy conservation in the scattering. The localization in the momentum space makes the collisions
more effective in agitating the phase space, and the localization in coordinate space is needed to keep
hydrodynamic effects like the flow dynamics.

The correlations produced through this approach are then exploited within a stochastic procedure, which compares the effective collision probability $W\times F$ with a random number. As a consequence,
fluctuations develop spontaneously in the phase-space cells  of volume $h^3$ with the correct fluctuation amplitude,
determined by a variance, which at equilibrium is equal to $f(1-f)$. A precise shape-modulation
technique~\cite{Napolitani2012} is applied to ensure that the occupancy distribution does not exceed unity
in any phase-space location in the final states. This leads to the correct Fermi statistics for the distribution
function $f$ in terms of its mean value and variance.

\subsubsection{Fluctuations and collective effects at intermediate energy}

In nuclear reactions, from the deep-inelastic regime to Fermi energy, the inclusion of fluctuations through the BLOB approach gives a more accurate description of the variety of mechanisms
that are related to a given entrance channel~\cite{Napolitani_ECHIC2014}. In central collisions, the variety of mechanisms, which compete with fusion span a larger energy interval in the BLOB approach compared to the SMF approach; in particular, the BLOB approach extends this interval to lower energies~\cite{Napolitani2013}.
In this energy domain, the BLOB approach can describe the interplay between volume and surface instabilities below nuclear saturation density. As a consequence it is capable of describing extreme situations ranging from multiple breakups in the deep-inelastic regime~\cite{Napolitani2019_quater} to the stream of clusters around Fermi energy~\cite{Napolitani2019_jets}.

At intermediate energies, the inclusion of fluctuations has two opposite effects: on the one hand, it enhances the
fragmentation of the system, and on the other hand, it reduces the directed flow. This effect is seen in Fig.~\ref{fig1} for the $^{197}$Au$+^{197}$Au collision at 100 $A$MeV for an impact parameter of 7 fm.
The simulation is performed with three approaches, Vlasov, SMF and BLOB, using identical parameters for the mean
field and for the two-body collision term, i.e., for a comparable number of attempted and
effective nucleon-nucleon collisions per time interval. The SMF approach describes the outward deflection of the
trajectory imparted by the directed flow, which is absent in the Vlasov description. The BLOB approach exhibits
a reduced directed flow with respect to SMF, because it competes with a more explosive dynamics. This mechanism, due
to the Langevin fluctuations, results in a large variety of very different fragment configurations; two of those
are shown with one where the fragmentation of the quasi-target and the quasi-projectile is observed (bottom, left),
and the other where the emitting source is situated at midrapidity (bottom, right). This example illustrates the main
difference between the SMF approach and the BLOB approach when applied to intermediate energies.

Applications of the BLOB approach to relativistic energies are so far restricted to spallation reactions (heavy nuclei bombarded with protons and deuterons) to describe the prompt emission of intermediate-mass fragments~\cite{Napolitani2015}.

It may be observed that, quite independently of the shapes of the phase-space metric, the fluctuation
amplitude is no longer correct if the occupancy variance in phase space cells is not consistent with the Pauli exclusion principle, with the consequence that the dispersion relation and some of the fragmentation patterns are not described. On the other hand, the phase-space metric (and therefore the shape of the scattering wave packets in configuration space and in momentum space)
affects the collective effects,
and disregarding it would eventually lead to a more transparent dynamics with reduced flow.


\subsection{The BUU@VECC-McGill (BUU-VM) code}
\vskip 0.1in
S. Mallik
\vskip 0.1in
\subsubsection{Code history}
BUU@VECC-McGill is a transport model based on the Boltzmann-Uehling-Uhlenbeck (BUU) equation that
was developed at Variable Energy Cyclotron Centre (VECC), Kolkata, India, in collaboration with
McGill University, Montr{\'e}al, Canada. This model was motivated by the BUU transport model
of Bertsch and Das Gupta \cite{Bertsch:1988ik, DasGupta_book:2019}. The one component version of
the model (i.e. only nucleons are considered) was successfully applied to calculate
the initial conditions for fragmentation reactions \cite{Mallik9:2014, Mallik11:2015}, signatures of nuclear
liquid-gas phase transition \cite{Mallik10:2015,Mallik14:2016,Mallik18:2018}, etc.  Motivated by this, the isospin
degrees of freedom and pion channels were  introduced to treat realistic heavy-ion reactions at intermediate energies \cite{Mallik22:2019,Mallik23:2020}.
Comparisons of model calculations with experimental data are given in Refs.~\cite{Mallik11:2015,Mallik23:2020}.

\subsubsection{Initialization}
The simulation of heavy ion collisions using the BUU@VECC-McGill model is started with two nuclei in their
respective ground states approaching each other with specified velocities. An isospin dependent
Thomas-Fermi solution for ground states is first constructed, and the resulting phase-space distribution is then
modeled by choosing $N_{test}$ test particles for each nucleon with appropriate positions and momenta using the Monte-Carlo method.
For the collision of projectile nucleus
of mass number $A_p$ (=$Z_p+N_p$) with a target nucleus of mass number $A_t$ (=$Z_t+N_t$) with $Z_p$ ($Z_t$) and $N_p$ ($N_t$) being the proton and neutron numbers of projectile (target) nuclei, respectively, the time evolution
of $(A_p+A_t)N_{test}$ test particles is studied. The test particles move in a mean
field $U(\rho(\vec{r}),t)$, generated by the potential energy density and occasionally suffer
two-body collisions when two of them pass close to each other, and the collisions are not blocked by
the Pauli principle. Details on the mean-field propagation and collision are described in next two sections.

\subsubsection{Mean-field propagation}
The propagation of test particles can be described by the Hamilton's canonical equations of motion,
\begin{eqnarray}
\frac{d\vec{p}_i}{dt}&=&-\nabla_r U(\rho(\vec{r}_i),t),\nonumber\\
\frac{d\vec{r}_i}{dt}&=&\vec{v}_i, \qquad i=1,2,.....,(A_p+A_t)N_{test},
\label{Hamilton_equation}
\end{eqnarray}
where, depending on the beam velocity, $\vec{v}_i$ can be calculated relativistically or
non-relativistically. The mean-field potential for a neutron ($n$) or a proton ($p$) in Eq. (\ref{Hamilton_equation}) is given by
\begin{equation}
U_{n,p}(\vec r,t)=A\bigg{\{}\frac{\rho(\vec r,t)}{\rho_0}\bigg{\}}+B\bigg{\{}\frac{\rho(\vec r,t)}{\rho_0}\bigg{\}}^{\sigma}+\frac{C}{\rho_0^{2/3}}\nabla_r^2\bigg{\{}\frac{\rho(\vec r,t)}{\rho_0}\bigg{\}}+\tau_zS_{sym}\bigg{\{}\frac{\rho_n(\vec r,t)-\rho_p(\vec r,t)}{\rho_0}\bigg{\}}+\frac{1}{2}(1-\tau_z)U_c,
\label{Lenk_potential}
\end{equation}
where the first two terms represent the zero-range Skyrme interactions, and the derivative term, which does not affect
nuclear matter properties, is introduced for nuclei. A realistic description of the diffuse surfaces and binding energies of finite nuclei can be achieved with $A=$-356.8 MeV, $B=$303.9 MeV,  $\sigma=$7/6, $\rho_0$=0.16 $\textrm{fm}^{-3}$
and $C$=-6.5 MeV$\textrm{fm}^{5/2}$. $\rho(\vec r,t)$, $\rho_p(\vec r,t)$ and $\rho_n(\vec r,t)$ are, respectively,  the total, proton and neutron densities at position $\vec r$ and time $t$, $\tau_z$ is
the z-component of isospin, which is +1 and -1 for neutrons and protons, respectively, and $U_c$ is the Coulomb
potential. The coefficient of the symmetry potential ($S_{sym}$) is 32 MeV \cite{Mallik23:2020}.\\
\indent
Recently, the bulk part of the isospin-dependent mean-field potential in the BUU@VECC-McGill model has been updated \cite{Mallik22:2019} by a recently proposed meta-functional \cite{Margueron:2018} based on a polynomial expansion in density around saturation and including deviations from the parabolic isospin dependence through the effective mass splitting in the kinetic term. The resulting mean-field potential for neutrons ($n$) or protons ($p$) in Eq. (\ref{Hamilton_equation}) can be expressed as
\begin{eqnarray}
U_{n,p}(\vec r,t)&=&(v^{is}_{0}+v^{iv}_{0}\delta^2)+\sum_{k=1}^{4}\frac{k+1}{k!}(v^{is}_{k}+v^{iv}_{k}\delta^2)x^{k}+\frac{1}{3}\sum_{k=1}^{4}\frac{1}{(k-1)!}(v^{is}_{k}+v^{iv}_{k}\delta^2)x^{k-1}\nonumber\\
&+&2\delta\tau_z(1-\delta\tau_z)\sum_{k=1}^{4}\frac{1}{k!}v^{iv}_{k}x^{k}+\exp\{-b(1+3x)\}\nonumber\bigg{[}(a^{is}+a^{iv}\delta^2)\bigg{\{}\frac{5}{3}x^4+(6-b)x^{5}-3bx^{6}\bigg{\}}\nonumber\\
&+&2\delta\tau_z(1-\delta\tau_z)a^{iv}x^{5}\bigg{]}+\frac{3C}{\rho_0^{2/3}}\nabla^2x+\frac{1}{2}(1-\tau_z)U_c,
\label{Meta_model_potential}
\end{eqnarray}
where $x=(\rho(\vec r,t)-\rho_0)/{3\rho_0}$ and $\delta=(\rho_n(\vec r,t)-\rho_p(\vec r,t))/\rho(\vec r,t)$. The parameters $v_k^{is}$ with $k$= 1 to 4 can be linked with the usual isoscalar empirical parameters of the saturation energy ($E_{sat}$), incompressibility modulus ($K_{sat}$),  isospin symmetric skewness ($Q_{sat}$) and  kurtosis ($Z_{sat}$), respectively, and $v_k^{iv}$ with $k$= 1 to 4 can be linked with  the usual isovector empirical parameters of the symmetry energy ($E_{sym}$), slope  ($L_{sym}$), and associated incompressibility ($K_{sym}$),  skewness ($Q_{sym}$) and  kurtosis ($Z_{sym}$), respectively. These empirical parameters are taken from the SLY5 EoS. The details can be found in Ref. \cite{Mallik22:2019}.\\
\indent
The density dependence of the effective masses induces the following extra term to Eq. (\ref{Meta_model_potential}) for the mean field:
\begin{equation}
U_q^{eff}=\sum_{q=n,p} \tau_q \frac{\partial}{\partial \rho_q}\left ( \frac{m_q}{m^*_q}\right )=\tau_q \frac{\kappa_0+\kappa_{sym}}{\rho_0} + \tau_{q'} \frac{\kappa_0-\kappa_{sym}}{\rho_0}.
\end{equation}
\indent
The numerical method employed to compute trajectories of test particles from Eq. (\ref{Hamilton_equation}) need a high accuracy in energy and momentum conservation. This is achieved by using the Lattice Hamiltonian method, which was proposed in Ref. \cite{Lenk89} and has proven to be
phenomenally accurate. According to this method, the configuration space is divided into cubic lattices with the lattice points $l$ fm apart. Thus, the configuration space is discretized into boxes of
size $l^3$~fm$^3$. The density at lattice point $r_{\alpha}$ is defined by
\begin{equation}
\rho_L(\vec{r}_\alpha)=\sum_{i=1}^{A_tN_{test}}S(\vec{r}_{\alpha}-\vec{r}_i),
\label{LH_lattice_density}
\end{equation}
where $\alpha$ stands for values of the three co-ordinates of the lattice point $\alpha=(x_l,y_m,z_n)$ , $\vec{r}_{\alpha}$ is the position of site $\alpha$ and $S(\vec{r})$ is the
form factor given by
\begin{equation}
S(\vec{r})=\frac{1}{N_{test}(nl)^6}g(x)g(y)g(z),
\label{LH_form_factor}
\end{equation}
\begin{equation}
g(q)=(nl-|q|)\Theta (nl-|q|),
\label{LH_g}
\end{equation}
where $\Theta$ is the Heaviside function and $n$ is an integer which determines the range of $S$. A test particle contributes to the average density $\rho_L$ at exactly $(2n)^3$ lattice
sites, and the movement of a particle results in a continuous change in $\rho_L$ at nearby lattice sites.
In our calculation, we always use $l$=1 fm and $n$=1. Therefore, by knowing test particle positions, the density
at a lattice point can be calculated from Eq. (\ref{LH_lattice_density}), and, if required, the potential at a lattice point can be calculated. Then, the positions and momenta of test particles are modified
using Eq. (\ref{Hamilton_equation}).

\subsubsection{Collisions term}
The effect of collisions in the BUU@VECC-McGill model is included by using the Monte-Carlo methods, which was
formerly used in the intra-nuclear cascade model \cite{Cugnon:1980rb}. In the cascade calculation, the frequency of
collisions is governed by the scattering cross-section only, whereas in the BUU@VECC-McGill model, the Pauli blocking effect
is also included as is described in the next section. The main aim of a cascade program is to decide where and when particles collide. This is achieved by dividing the collision time into small time intervals $\delta t$ and examine all pairs of particles in each time interval to check whether they scatter. The collision between two nucleons is performed in their local center of mass frame. There are two conditions that need to be fulfilled, i.e., the two nucleons must pass the point of closest approach in the time interval and
the distance of closest approach must be less than $b_{max}=\sqrt{\sigma_{NN}^t(\sqrt{s})/\pi}$.  Here $\sigma_{NN}^t(\sqrt{s})$ is the total cross section for the nucleon-nucleon scattering with center of mass energy $\sqrt{s}$, and can be different for proton-proton, neutron-neutron and neutron-proton scattering, i.e., $\sigma_{pp}$, $\sigma_{nn}$ and $\sigma_{np}$. The time interval should be small enough that the probability for the same nucleon being scattered more than once in one time interval is very small. After a collision occurs, the two particles can scatter elastically or inelastically. If the beam energy is 200 MeV/nucleon or
less, the inelastic channel is suppressed and non-relativistic kinematics can be used with considerable simplification. In general, the only inelastic channel in the BUU@VECC-McGill model is the pion channel, and this is included via the $\Delta$ state of the nucleon. If two particles collide and if both elastic and inelastic channels are allowed, a Monte-Carlo decision is
made for the channel. The magnitude of the momentum of the final-state particles is determined according to the energy-momentum conservation, and another Monte-Carlo decision is taken to fix the angle of scattering. The elastic scattering
differential scattering cross-section it is taken from experiments. Isotropic scattering is assumed in the inelastic  channel. Details of the cross-section parametrization in the BUU@VECC-McGill model can be found in Appendix C of Ref.~\cite{DasGupta_book:2019}. After determining the final momenta, one can go back to the global frame of reference of the colliding nuclei and check whether the final states are Pauli unblocked or not.\\
\indent
After a sufficient time, when the collisions are almost over, the nucleons and also pions at higher beam energies are freely streaming, and one can consider it as one event. The BUU@VECC-McGill model includes both a mean field and hard collisions. Instead of
deriving it formally, it will be more useful to consider it as an extension of the cascade model. Each
cascade model run will produce a different result as the positions of nucleons are generated by Monte-Carlo
sampling. To obtain an average answer, many runs are needed. It is advantageous to have $N_{test}$ runs simultaneously.
In the cascade model, different runs do not communicate with each other i.e., nucleus $1$ hits nucleus $1'$, nucleus
$2$ hits nucleus $2'$.... etc. Communication between runs is, however, introduced
in the BUU@VECC-McGill code for the mean field calculation described in previous section. In usual applications of the BUU@VECC-McGill code, different test particles 
 scatter with each other with a reduced cross-section $\sigma_{NN}/N_{test}$. 
 In another way of performing the collisions, described in more detail in Ref.~\cite{Mallik10:2015}, $N_{test}$ neighboring test particles scatter together with the cross section $\sigma_{NN}$, which will reduce substantially the computation time.
 
\subsubsection{Pauli Blocking}
Consider two test particles of any isospin, that  come within the distance of closest approach during a given time step, and due to  the above collision criteria they change their momenta
from ($\vec{r_1}$,$\vec{p_1}$), ($\vec{r_2}$,$\vec{p_2}$) to ($\vec{r_1}$,$\vec{p_1}'$), ($\vec{r_2}$,$\vec{p_2}'$).
Since the colliding particles are Fermions, one has to check
whether the final states are allowed or not, i.e., whether the collision will actually take place or whether it will be Pauli blocked.\\
\indent
To obtain the Uehling-Uhlenbeck term for intermediate energy heavy ion reactions, the phase-space densities about
the final states ($\vec{r_1}$,$\vec{p_1}'$) and ($\vec{r_2}$,$\vec{p_2}'$) are required. Therefore, a radius $r_p$
around $\vec{r_1}$ in configuration space and a radius $p_p$ around $\vec{p_1}'$ in momentum space are selected, such
that $N_p$ similar kinds of test particles inside this phase-space volume imply a complete filling~\cite{Bertsch:1988ik,Aichelin:1985}, i.e.,
\begin{equation}
\frac{2}{h^3}\int_0^{r_p}\int_0^{p_p} d^3rd^3p=\frac{N_p}{N_{test}}.
\label{Blocking}
\end{equation}
$N_p$ should be small so that one is examining the phase-space densities near the collision points, but not so small that fluctuations inherent in the Monte-Carlo sampling become severe. For $N_{test}=100$, past
works indicate that $N_p=8$ is a good choice. From Eq. (\ref{Blocking}), it is clear that specifying $N_p$
does not determine both $r_p$ and $p_p$, and one has to add an extra condition. In the present model, $r_p/p_p=R/P_F$ is
used with $R$ being the hard sphere radius of the static nucleus and $P_F$ being the Fermi momentum of proton or neutron (depending upon the nature of the test particle for which Pauli blocking is checked) at normal
nuclear matter density. The blocking factor at ($\vec{r_1}$,$\vec{p_1}'$) is then given by $f_1=N_1/N_p$,
where $N_1$ is the number of test particles of similar kind (excluding the colliding test particle) at ($\vec{r_1}$,$\vec{p_1}'$)
within a radius $r_p$ around $\vec{r_1}$ in configuration space and a radius $p_p$ around $\vec{p_1}'$ in momentum
space. So, the collision probability factor, is then equal to $1-f_1$.  Similarly, for the second particle the blocking factor
is $f_2=N_2/N_p$ and the collision probability factor is equal to $1-f_2$. Therefore, the probability of successful scattering is taken
to be $(1-f_1)(1-f_2)$, and this is calculated by the Monte-Carlo method \cite{Bertsch:1988ik}.\\

\subsection{DaeJeon Boltzmann-Uehling-Uhlenbeck (DJBUU) code}
\vskip 0.1in
Y. Kim, S. Jeon, M. Kim, C.-H. Lee, K. Kim 
\vskip 0.1in

Though DJBUU in many cases adopts a widely-used or standard method in transport theory, it has also  its distinctive features.
Here, we present a brief introduction to DJBUU.
The main references are Refs.~\cite{DJBUU2016, Kim:2020sjy}.

\subsubsection{Code history}
The primary version of DJBUU was written when Sangyong Jeon from McGill University was visiting RISP (Rare Isotope Science Project) of  IBS (Institute for Basic Science) in Korea.
DJ stands for Daejeon, which is a city in south Korea where the rare isotope beam facility, dubbed RAON, is being constructed.
RAON was one of thrusts for us to develop a new transport code.
Since RAON will provide heavy ion beams in the lab frame energy up to a few hundreds AMeV, currently nucleons, and the nucleon resonances $\Delta$(1232), N*(1440) and N*(1520) are included in the code.

\begin{itemize}
\item
October 2015 - January 2016:

The primary version of  DJBUU was completed.
The DJBUU code is written in C/C++ and programmed in OpenMP (Open Multi-Processing).

\item
 2016 - 2017:

 We reviewed the DJBUU code and performed innumerable test-runs to fix typos and errors.
 Meanwhile, we published a short article  about DJBUU in a domestic journal (New Physics: Sae Mulli).

\item 2018 - present:

We have joined the Transport Code Evaluation project with DJBUU.

\item 2019 - 2020:

To see if we can address the issue of the origin of the nucleon mass in heavy ion collisions, we have added a
parity doublet model to DJBUU.

\end{itemize}

\subsubsection{Initialization}
We adopt the widely used method for initialization:  In coordinate space, we sample from parametrized nucleon distribution functions. In momentum space, the (test) particle momenta are randomly distributed in an isotropic Fermi sphere whose radius is the local Fermi-momentum.

For the reference nucleon density distributions in coordinate space, we  initially  used the
Wood-Saxon parametrization
\begin{equation}
F(r) =\frac{1}{1+\exp [(r-R)/a]}\, .
\end{equation}
After developing a nuclear structure code based on the relativistic Thomas-Fermi (RTF) approximation, we now mainly
use the nucleon density profiles from the RTF calculations.

 The profile function of the test particle is given by the following non-Gaussian type function,
\begin{equation}
g({\bf u}) = g(u) = {\cal N}_{m,n} (1 - (u/a)^m)^n
\ \ \ \hbox{for}\  0 < u/a < 1
\label{eq:profile_func}
\end{equation}
where $u = |{\bf u}|$, ${\cal N}_{m,n}$ is the normalization constant, and
$m > 1$ and $n > 1$ are positive integers. For $|u/a| > 1$, $g(u) = 0$.
We use this profile function instead of the usual truncated Gaussian
because it is not only exactly integrable but also smoothly vanishes at $a$.
That is, $g(a) = 0$ and $g'(a) = 0$.
The default values are $m=2$, $n=3$ and $a_x = 4.2\,\hbox{fm}$ for the position profile
and $a_p = s\hbar/a_x$ for the momentum profile with $s \approx 0.6$.

\subsubsection{Potentials}
We adopt the relativistic mean-field model in Ref.~\cite{Liu:2001iz} with scalar ($\sigma$) and vector ($\rho$, $\omega$) mesons  to obtain a mean-field potential. We take the Thomas-Fermi approximation for nucleons in which nucleon bilinear fields are replaced by scalar or vector
densities. Since the relativistic mean-field model~~\cite{Liu:2001iz} has been widely used in nuclear transport models, we will not explicitly write down the corresponding relativistic Lagrangian density.
We note here  that the $\sigma$-field in Ref.~\cite{Liu:2001iz} is not related to the chiral symmetry in QCD and its vacuum expectation value in free space is zero.

To address the issue about the origin of the nucleon mass and (partial) chiral symmetry restoration in heavy ion collisions, we incorporate a parity doublet model in DJBUU.
In the parity doublet model~\cite{Detar:1988kn}, the nucleon mass comes from chiral symmetry breaking but the chiral-invariant mass is unrelated to the QCD chiral symmetry.

We use an extended parity doublet model in Refs.~\cite{Motohiro:2015taa, Shin:2018axs}. The Lagrangian density is given by
\begin{eqnarray}
{\cal L} =&&~\bar{\psi}_1 i \not\!{\partial} \psi_1 + \bar{\psi}_2 i \not\!{\partial}\psi_2 + m_0 (\bar{\psi}_2 \gamma_5 \psi_1 - \bar{\psi}_1 \gamma_5 \psi_2)
 + g_1\bar{\psi}_1 (\sigma + i \gamma_5 \vec{\tau} \cdot \vec{\pi})\psi_1 + g_2\bar{\psi}_2(\sigma - i \gamma_5 \vec{\tau} \cdot \vec{\pi})\psi_2 \nonumber\\
&& -g_{\omega N\!N} \bar{\psi}_1 \gamma_\mu \omega^\mu \psi_1 -g_{\omega N\!N} \bar{\psi}_2 \gamma_\mu \omega^\mu \psi_2
 -g_{\rho N\!N} \bar{\psi}_1 \gamma_\mu \vec{\rho}^{\,\mu} \cdot \vec{\tau} \psi_1 -g_{\rho N\!N} \bar{\psi}_2 \gamma_\mu \vec{\rho}^{\,\mu} \cdot \vec{\tau} \psi_2 \nonumber\\
&& -e \bar{\psi}_1 \gamma^\mu A_\mu \frac{1-\tau_3}{2} \psi_1 -e \bar{\psi}_2 \gamma^\mu A_\mu \frac{1-\tau_3}{2} \psi_2 + {\cal L}_M\, ,~~
\end{eqnarray}
where the nucleon fields $\psi_1$ and $\psi_2$ transform  as
\begin{eqnarray}
\psi_{1R} \rightarrow R\psi_{1R}\, , \quad \psi_{1L} \rightarrow L\psi_{1L},
\quad \psi_{2R} \rightarrow L\psi_{2R}\, , \quad \psi_{2L} \rightarrow R\psi_{2L}\, .
\end{eqnarray}
Here, $L$ and $R$ denote the elements of  $SU(2)_L$ and  $SU(2)_R$ chiral symmetry group, respectively.
The mesonic part of the Lagrangian reads
\begin{eqnarray}
{\cal L}_M =&&\frac{1}{2} \partial_\mu \sigma \partial^\mu \sigma + \frac{1}{2} \partial_\mu \vec{\pi} \cdot \partial^\mu \vec{\pi}
 - \frac{1}{4} \Omega_{\mu\nu} \Omega^{\mu\nu} -\frac{1}{4} \vec{R}_{\mu\nu} \cdot \vec{R}^{\mu\nu} -\frac{1}{4} F_{\mu\nu} F^{\mu\nu} \nonumber \\
&& + \frac{\bar{\mu}^2}{2} (\sigma^2 + \vec{\pi}^2) -\!\frac{\lambda}{4}(\sigma^2 + \vec{\pi}^2)^2 +\!\frac{\lambda_6}{6}(\sigma^2 + \vec{\pi}^2)^3 + \epsilon\sigma
+ \frac{1}{2}m_\omega^2 \omega_\mu \omega^\mu  + \frac{1}{2}m_\rho^2 \vec{\rho}_\mu \cdot \vec{\rho}^{\,\mu}\, .
\end{eqnarray}
We then take the mean-field approximation: $\sigma\rightarrow \langle \sigma \rangle$,
$\omega_\mu\rightarrow \delta_{\mu 0}\langle \omega_\mu \rangle$, and $\rho_{i \mu}\rightarrow \delta_{i3}\delta_{\mu 0}\langle \rho_\mu^i \rangle$.
We remark here that the $\sigma$-field in this model is the chiral partner of the pion field, which is a Goldstone boson from spontaneous chiral symmetry breaking. The vacuum expectation value of this $\sigma$-field in free space is nonzero, and its value is equal to the pion decay constant.

\subsubsection{Collisions}
For the nucleon-nucleon collision, we adopt the following criterion
\begin{equation}
d \leq \sqrt{\hat\sigma\over \pi}\, ,
\end{equation}
where $d$ is the transverse distance of two particles in their center-of-mass (CM) frame. It is given by
\begin{equation}
d^2 = |\Delta{\bf x}_{\rm CM}\times {\bf n}_{\rm CM}|^2\, ,
\end{equation}
where $\Delta {\bf x}_{\rm CM}$ is the distance between the two particles and
${\bf n}_{\rm CM}$ is the unit vector along the momentum direction in the CM frame.
The time of the collision in the common-time frame is set to
\begin{equation}
t_{\rm coll} = t_{\rm prev} + \Delta t_{\rm coll}\, ,
\end{equation}
where
$t_{\rm prev}$ is time at the end of the previous time step
and
\begin{equation}
\Delta t_{\rm coll} = - {\Delta{\bf x}\cdot\Delta{\bf v} \over |\Delta{\bf v}|^2}.
\end{equation}
For a collision to occur within this time interval, we must have
$0 < \Delta t_{\rm coll} < \Delta t~$\cite{Kodama:1983yk}. We usually take $\Delta t=  0.1,\, 0.2$ fm/c.
If resonances do not undergo a scattering within the time interval, then they could
decay with the total decay rate taken as the Breit-Wigner width. Even though the pairs meet the criterion for collision,
decay could happen when the decay probability is high enough within $\Delta t_{\rm coll}$.

In DJBUU, a collision proceeds as follows.
We first decide which pairs collide or which resonances decay within $\Delta t$.
To identify the pairs, we use the nearest neighbor pair searching method. We remark here that in DJBUU, space is divided into $N_x\times N_y\times N_z$ cells and each cell has about $1$ fm$^3$ in volume.
If the pairs do not scatter or decay, we update the position and momentum of the pairs.
If the pairs satisfy the criterion for collision, they collide with each other.
In case decay happens,  we update  the position and momentum of the daughters after decay.
We then check if the final state is allowed by drawing a random number and comparing it
with the Pauli-Blocking factor $1 - f_a(x,{\bf p})$.
\subsection{The Giessen Boltzmann-Uehling-Uhlenbeck (GiBUU) code}\label{sec:GiBUU}
\vskip 0.1in
J.~Weil, T.~Gaitanos, K.~Gallmeister, U.~Mosel
\vskip 0.1in

\subsubsection{Introduction}

The Giessen-BUU-model (GiBUU) is a hadronic transport framework to handle a
variety of reaction types in a wide energy range. One of its specialties is the ability to describe heavy-ion collisions (AA) and other
hadronic collisions ($pA$, $\pi A$, $\bar{p}A$) as also elementary
reactions on nuclei ($\gamma A$, $eA$, $\nu A$) on the same footing. Other noteworthy features
include the availability of two different mean-field models (non-relativistic
Skyrme potentials as well as relativistic mean fields, RMF) and a treatment of
high-energy collisions via the Pythia string-fragmentation model. The model
is based on hadronic degrees of freedom and currently contains 61 baryon and 22
meson states.
A comprehensive overview of the model can be found in \refcite{Buss:2011mx}. 
In the following we restrict ourselves to the discussion of the features important
for low-energy heavy-ion collisions, as performed for the code comparison.

The origins of the model reach back into the year 1986, initiated by W.~Bauer and U.~Mosel. Many notable scientists contributed during the following decades. It is in order to mention particularly the code versions of S.~Teis and M.~Effenberger (1997) and the new rewrite of O.~Buss (2008). 
For more details of the historical development of the code we refer to \refcite{gibuu_history}.

GiBUU is implemented in a rather modern Fortran dialect (Fortran 95/2003) using modularization and some simple object-oriented techniques. 
The code base is managed through a \texttt{subversion} version control since the year 2008, which is essential for reproducibility of results and the collaboration of multiple authors on the same code base.
The code and its usage is well-documented by means of a wiki \cite{gibuu_wiki} and semi-automatic code documentation \cite{gibuu_docu}.
The code is available on the base of public releases approx.~every second year, superimposed with actual code patches, hosted at and downloadable from the \texttt{hepforge} server \cite{gibuu_url}.

\subsubsection{Initialization}

In order to prepare the phase-space density of the nuclear ground
state, the coordinates of neutrons and protons are sampled according
to density profiles, either (i) taken from
empirical systematics (Woods-Saxon or harmonic oscillator type) or (ii) determined from relativistic
Thomas-Fermi (RTF) calculations with the same mean-fields as those used in
the dynamical propagation afterwards. Method (i) is used in the transport
calculations with non-relativistic Skyrme-type potentials, whereas method (ii)
is used in transport simulations with relativistic mean-fields.

In the default prescription, the particle momenta are distributed according to a local
Thomas-Fermi (LTF) approximation,
\begin{equation}
  f_{n,p}(\bvec{r},\bvec{p})=
  \Theta\left[p_{F,n,p}(\bvec{r})-\left|\bvec{p}\right| \right ],
  \label{eq:oneParticleFSD}
\end{equation}
where the momentum distribution is given by an isotropic Fermi sphere
at each point in space with the radius in momentum space determined by
the local Fermi momentum,
\begin{equation}
  p_{F,n,p}(\bvec{r})=[3\pi^2\rho_{n,p}(\bvec{r})]^{1/3}.  \label{eq:fermimom}
\end{equation}
The normalization is chosen such that the proton and neutron densities (which serve as an input) are retrieved by
\begin{equation}
  \rho_{n,p}(\bvec{r})=g~  \int f_{n,p}(\bvec{r},\bvec{p})~ \frac{\dd^3 p}{(2\pi)^3}~.
\end{equation}

In order to overcome the usual pitfalls of this LTF approach, an additional mode has been implemented, where the density profile of particles is adjusted such, that the binding energy of the particles is a constant for all radii. 

Both methods provide us with the full phase-space information
at the initial time, before starting the propagation according to the
Hamilton equations of motion.  Smooth distributions in coordinate and
momentum space are achieved by using $\sim10^3$ test particles per
nucleon, which is an important issue for the numerical treatment of
mean-field gradients in the Hamilton equations of motion. The smoothness of the test particle distribution
in phase space is also important for numerical evaluation of the
Pauli-blocking factors, $(1-f_{n,p}(\bvec{r},\bvec{p}))$, which enter
the collision term of the BUU equation.

\subsubsection{Potentials: Skyrme potentials}

The first kind of mean-field potential which is available in GiBUU is a
non-relativistic Skyrme-type nucleon potential with momentum dependence of the form

\begin{equation}
  \begin{split}
    \label{U_N}
    U_i(x,\bvec{p})
    =& \mbox{ } A \frac{\rho(x)}{\rho_0} +
    B\left(\frac{\rho(x)}{\rho_0}\right)^\gamma
    + \frac{2C}{\rho_0}
    \sum_{i=n,p} \int \frac{g\, \dd^3 p^\prime}{(2\pi)^3}
    \frac{f_i(x,\bvec{p}^\prime)}{1 +
      (\bvec{p}-\bvec{p}^\prime)^2/\Lambda^2} + d_{\rm symm}\frac{\rho_p(x)-\rho_n(x)}{\rho_0}\tau^3_i~,
 \end{split}
\end{equation}
where $i=p,n$ with $\tau^3_p=1$ and $\tau^3_n=-1$. The first two terms are the
common Skyrme form, while the third one implements an (optional) momentum dependence
according to \refcite{welke} and the fourth one is the asymmetry term, which carries the isospin dependence.

To reduce the computation
time for calculating the momentum-dependent part of the potential,
we approximate in the integral the (dashed) nucleon phase-space distribution by
a Fermi distribution. This allows to evaluate the momentum
integral as an analytic function of
$|\bvec{p}|$ and of the local baryon density $\rho(x)$
(see \refcite{welke} for details).
The six free parameters, $A$, $B$, $\gamma$, $C$, $\Lambda$, and
$d_{\rm symm}$, of the nucleon potential are determined
from various properties of nuclear matter (like the nuclear binding energy and nuclear-matter incompressibility $K$).
For the code comparison, we used the values $A=-209.2\MeV$, $B=156.4\MeV$, $\gamma=1.35$, $C=0$ (i.e.~no momentum dependence) and $d_{\rm symm}=30\MeV$,
corresponding to a rather soft equation of state with $K=240\MeV$.

In order to avoid problems with Lorentz invariance, it is
important to choose a proper frame, where the mean-field potentials should
be calculated. For this purpose, we have chosen the local rest frame
(LRF) of the nuclear medium, where the spatial components of the
baryon four-current, $j^\mu=(j^0,\bvec{j})$, vanish, i.e., $\bvec{j}
=0$ at the space position of the particle under consideration.
This choice is only possible if the antibaryons (contributing with negative
sign to the baryon four-current) are not abundant.

\subsubsection{Potentials: Relativistic mean fields}

The second type of interaction that is available in the GiBUU code is the
Walecka model in the well-known relativistic mean-field (RMF) approximation. Here
the mean-field potential consists of two contributions, a Lorentz-scalar
attractive $S\equiv\Sigma_{s}$ and a Lorentz-vector
repulsive $\Sigma^{\mu}$ ($V\equiv\Sigma^{0}$) selfenergies. Both fields are rather strong
and in the order of the nucleon mass, however, their particular cancellation leads to the
small binding of a nuclear system in its ground state.

The corresponding potential in the relativistic case can be expressed in terms
of the in-medium Schr\"{o}dinger-equivalent potential, which reads as

\begin{align}
U_{opt}(E) = -S + \frac{E}{m}V + \frac{1}{2m}(S^{2}-V^{2})
\label{Uopt}
\end{align}
with the energy $E$ extracted from in-medium dispersion relations of protons and neutrons.

The parameters of the RMF-model are the meson-nucleon coupling constants and the additional
constants of the non-linear self-interactions between the $\sigma$-meson. These parameters
are usually adjusted either to the bulk nuclear matter properties or to properties of
finite nuclei.
Several parameter sets are available in the GiBUU code, which differ from each
other in the stiffness of the nuclear equation of state at densities just above saturation.
For the code comparison we have used parameter set I from \cite{Liu:2001iz}, whose
properties roughly match the Skyrme parametrization described above.

\subsubsection{Collision term}

The GiBUU collision term is divided into two energy domains: A low-energy regime
that is dominated by resonance excitation, and a high-energy regime that is
treated via a string-fragmentation model (PYTHIA 6.4) \cite{Sjostrand:2006za}.

The transition is currently performed at energies of $\sqrt{s}=2.2\pm0.2\GeV$ for
meson-baryon collisions (which corresponds to the mass of the heaviest included
nucleon resonances) and $\sqrt{s}=3.4\pm0.1\GeV$ for baryon-baryon collisions
(because resonance production channels like $NN\rightarrow NR,\Delta R$ start to
fail saturating the total NN cross sections at this point, cf.~\refcite{Weil:2012ji}).

For the purpose of the code comparison, solely a constant isotropic elastic
cross section of $\sigma_{\rm el}=40\mb$ was used. Therefore we refrain here from
discussing the details of all the reactions implemented in the GiBUU collision
term and instead refer to \refcite{Buss:2011mx}.
The calculation of the collision term is speeded up by using the local-ensemble method of Ref.~\cite{lang93}.

\subsubsection{Pauli blocking}

The phase-space distribution, $f_i(\bvec{r},\bvec{p})$, which enters in the
Pauli blocking factor, $1-f_i(\bvec{r},\bvec{p})$, is calculated by counting the
number of test particles in the phase-space volume element composed of small
spherical volumes $\Delta V_r$ with radius $r_r$ centered at $\bvec{r}$ in
coordinate space and $\Delta V_p$ with radius $r_p$ centered at $\bvec{p}$ in
momentum space,
\begin{equation}
\begin{split}
  f_i(\bvec{r},\bvec{p}) &= \sum_{j:~\bvec{p}_j \in \Delta V_p}
  \frac{1}{\kappa(2\pi\sigma^2)^{3/2}} 
  \int\limits_{\Delta V_r, |\bvec{r}-\bvec{r}_j| < r_c} \dd^3 \bvec{r}
  \;\exp\left\{ -\frac{(\bvec{r}-\bvec{r}_j)^2}{2\sigma^2}\right\}, \label{f_i}
  \end{split}
\end{equation}
where
\begin{equation}
  \kappa = \frac{2\, \Delta V_r\, \Delta V_p\, N}{(2\pi)^3}
  \frac{4\pi}{(2\pi\sigma^2)^{3/2}}
  \int\limits_0^{r_c} \dd r \; r^2 \exp\left\{-\frac{r^2}{2\sigma^2}\right\}
  \label{kappa}
\end{equation}
is a normalization factor. In \cref{f_i}, the sum is taken over all test
particles, $j$, of the type $i=p,n$ whose momenta belong to the volume
$\Delta V_p$. In coordinate space, the test particles are represented by
Gaussians of the width, $\sigma$, cut off at the radial distance $r_c$, in a
similar way as done for the folding of the density fields with Gaussians. The
default values of parameters are $r_p=80\MeVc$, $r_r=1.86\fm$, $\sigma=1\fm$,
$r_c=2.2\fm$. This set of parameters is a compromise between the quality of the
Pauli blocking in the ground state and the smallness of statistical fluctuations
in the case of simulations with $N\sim200$ test particles per nucleon.
Typically, this is sufficient in accuracy for modeling heavy-ion collisions at
beam energies above $\sim100\AMeV$.

However, for small-amplitude dynamics near the nuclear ground state, like the
giant monopole resonance vibrations studied in \refcite{Gaitanos:2010fd}, the
accuracy provided by \cref{f_i,kappa} is not sufficient, when the default
parameters are used. The main reason is the constant, i.e., momentum-independent
radius, $r_p$, which introduces a spurious temperature of the order of several
MeV. To reduce this effect, we have introduced a position- and
momentum-dependent radius of the momentum-space volume $\Delta V_p$ by
$r_p(\bvec{r},|\bvec{p}|)=\mbox{max}[20\MeVc, p_{F,i}(\bvec{r}) - |\bvec{p}|]$,
which provides a sharper Fermi surface. This allows us to use the reduced
parameters also in coordinate space ($r_r=0.9-1.86\fm$, $\sigma=0.5\fm$,
$r_c=1.1\fm$).

\subsubsection{Off-shell transport}
Inside the nuclear medium hadronic properties, and in particular their spectral functions, may change. When hadrons are getting knocked out of the nucleus their in-medium spectral functions have to change into the free ones. In GiBUU this change can be achieved by turning on the off-shell transport~\cite{Buss:2011mx}, which is essential for any investigation of observable effects of in-medium changes.

\subsection{The isospin-dependent Boltzmann-Langevin (IBL) code}
\vskip 0.1in
W.J. Xie, F.S. Zhang
\vskip 0.1in

The most important features of the IBL model are presented in the following. The main references
are~\cite{AbeAyik96,Zhang95,Xie13}.

\subsubsection{Code history}

\begin{itemize}

\item 1988-1990:\\
The physics idea of the Boltzmann-Langevin (BL) code was proposed by S. Ayik and C. Gregoire in 1988 ~\cite{Ayik88}. 
That is, the correlated part of the two-body collisions acts as a random force and the correlation function of the 
random force is calculated within the semi-classical approximation.
\item 1990-1993: \\
The BL method was applied to nuclear reactions by E. Suraud and M. Belkacem~\cite{Suraud92,Belkacem93}. An approximate
method to obtain numerical solutions of the Boltzmann-Langevin equation was proposed, in which fluctuations were 
evaluated based on the first two non-vanishing terms of the multipole expansion of the local momentum distribution, namely
the quadrupole and octupole ones. The method was used to describe sub-threshold kaon production.
\item 1993-1995:\\
The BL method was used to describe the multifragmentation phenomenon in heavy-ion collisions by F.S. Zhang and E. Suraud~\cite{Zhang93,Zhang95}.
\item 2000-2001:\\
Isospin effects were incorporated into the BL method (called IBL thereafter) by Z.Y. Ming and L.W. Chen~\cite{Ming00,Ming01}.
\item 2008:\\
The symmetry potential was introduced into the IBL method by B.A. Bian~\cite{Bian08}.
\item 2011-2013:\\
Further improvements of the IBL method through incorporating the various density-dependent symmetry potentials, 
momentum-dependent mean-field potential and pion production inelastic reaction channels were implemented by W.J. Xie and J. Su~\cite{Xie13}.
\end{itemize}

\subsubsection{Initialization}

There are three steps for the initialization of nuclei in the IBL code. The first step is the initialization of the space coordinates
and momenta of nucleons. We use the Skyrme-Hartree-Fock approach to obtain the neutron and proton density distributions and the 
radial coordinates of the nucleons are obtained by using Monte Carlo sampling approach. The local Fermi momenta of nucleons is obtained 
by making use of the following formula:
\begin{equation}
p_{F}^{i}(\mathbf{r}) = \hbar [3\pi^{2}\rho_{i}(\mathbf{r})]^{1/3}, i=n,p.
\end{equation}
Momenta of nucleons are then obtained by using Monte Carlo sampling in the interval $[0,p_{F}^{i}]$.

The second step consists of checking the initialization results. Two requirements need to be simultaneously fulfilled: 
(1) The binding energy of the initialized nucleus is consistent with its experimental value; (2) The root mean square radius
is within certain specified limits. If one of the above two conditions is not satisfied, the initialization is redone.

The third step consists of initializing the system. Firstly, the nuclei are rotated around the x,y,z axes by random Euler angles. 
The projectile and target nuclei are boosted to the desired kinetic energy values. Finally, the initial distance between the nuclei 
and the impact parameter are set.

\subsubsection{Forces}
The test particle method is used to solve the BL equation. For the numerical implementation, the propagation of the particles 
is realized through the simple relation
\begin{equation}
\begin{split}
\mathbf{p}(t+\delta t)&=\mathbf{p}(t)-\delta t \nabla U(\mathbf{r},t+0.5\delta t),\\
\mathbf{r}(t+0.5\delta t)&=\mathbf{r}(t-0.5\delta t)+\delta t\mathbf{p}(t)/m.
\end{split}
\end{equation}
The force field $\nabla U$ is calculated as in the case of the usual BUU method. The single-particle potential in the IBL code is as follows,
\begin{equation}\label{U_tau}
U_{\tau}(\rho,\delta,\mathbf{p})= \alpha\frac{\rho}{\rho_{0}} + \beta(\frac{\rho}{\rho_{0}})^{\gamma} + E_{sym}^{loc}(\rho) \delta^{2} + \frac{\partial E_{sym}^{loc}(\rho)}{\partial\rho}\rho\delta^{2}+E_{sym}^{loc}(\rho)\rho\frac{\partial\delta^{2}}{\partial\rho_{\tau}} + U^{iso}_{MDI},
\end{equation}
where the $E_{sym}^{loc}(\rho)$ is the local part of the symmetry energy. The $U^{iso}_{MDI}$ is the momentum-dependent mean-field potential, which can be written as
\begin{equation}\label{U_tau_2}
U^{iso}_{MDI}=\frac{C_{\tau,\tau}}{\rho_{0}}\int d \textbf{p}^{'} f_{\tau}(\textbf{r},\textbf{p}) \ln^{2}[0.0005(\textbf{p}-\textbf{p}^{'})^{2} + 1]
+\frac{C_{\tau,\tau^{'}}}{\rho_{0}}\int d \textbf{p}^{'} f_{\tau^{'}}(\textbf{r},\textbf{p}) \ln^{2}[0.0005(\textbf{p}-\textbf{p}^{'})^{2} + 1].
\end{equation}
The parameter values of Eqs. \eqref{U_tau} and \eqref{U_tau_2} can be found in Ref.~\cite{Xie13prc}.

\subsubsection{Collision}
The collision term in the IBL code originates from the cascade model~\cite{Bertsch:1988ik}. There are two conditions that are required to be fulfilled
for a two-nucleon collision to take place. Firstly, the two nucleons must pass the point of closest approach within the specified time interval
and secondly, the distance of closest approach must be less than $\sqrt{\sigma_{\rm tot}/\pi}$. The elastic nucleon-nucleon cross section
parametrization of Cugnon {\it et al.} \cite{Cugnon:1996kh} is used. For the inelastic channels involving resonance
($\Delta(1232),~N^{*}(1440)$) production, the parametrization by Huber and Aichelin is used~\cite{Huber:1994ee}. Free space
and in-medium cross sections are considered for the elastic and inelastic channels, respectively. The in-medium effects 
are introduced by a medium correction of $\rho$-meson mass~\cite{Li00}.

\subsubsection{Pauli-blocking}
The Pauli blocking algorithm implemented in the IBL code resembles closely the one of the CoMD model by M. Papa et al.~\cite{Papa01}.
Firstly, we calculate the phase space occupancy of the two scattered nucleons. Secondly, if one of the phase space occupancy of the
two nucleons has the value greater than 1, the collision is blocked.

\subsubsection{Fluctuation}
The fluctuating collision term in the IBL code can be interpreted  as a stochastic force acting on density and is characterized
by a correlation function
\begin{eqnarray}
\langle \delta \textit{K} (\mathbf{r_{1}}, \mathbf{p_{1}}, t_{1}) \delta \textit{K} (\textbf{r$_{2}$}, \textbf{p$_{2}$}, t_{2})\rangle =
 C(\textbf{p$_{1}$}, \textbf{p$_{2}$}) \delta (\textbf{r$_{1}$} - \textbf{r$_{2}$}) \delta (t_{1} - t_{2}).
\end{eqnarray}
A projection method is used in the IBL model. The fluctuations are projected on a set of low-order multipole moments of the momentum distribution. The fluctuations of these multipole moments are characterized by a diffusion matrix
\begin{eqnarray}
 C_{LML'M'} &=& \int d\mathbf{p}d\mathbf{p}' Q_{LM}(\mathbf{p})Q_{L'M'}(\mathbf{p}')C(\mathbf{p},\mathbf{p}') \nonumber\\
&=&\int d\mathbf{p}_{1}d\mathbf{p}_{2}d\mathbf{p}_{3}d\mathbf{p}_{4}\Delta Q_{LM} \Delta Q_{L'M'} W(12,34)f_{1}f_{2}(1-f_{3})(1-f_{4}).
\end{eqnarray}
It is found that only the $Q_{20}$ and $Q_{30}$ are not negligible, that is
\begin{equation}
\begin{split}
\hat{Q}_{20}(\mathbf{r},t+\Delta t)&= {Q}_{20}(\mathbf{r},t+\Delta t)+\sqrt{\Delta t C_{20}(\mathbf{r},t)}W_{2},\\
\hat{Q}_{30}(\mathbf{r},t+\Delta t)&= {Q}_{30}(\mathbf{r},t+\Delta t)+\sqrt{\Delta t C_{30}(\mathbf{r},t)}W_{3}.
\end{split}
\end{equation}
where the $W_{2}$ and $W_{3}$ are Gaussian random numbers with mean 0 and variance 1. $\hat{Q}_{20}$ and $\hat{Q}_{30}$ are the 
fluctuating values, while ${Q}_{20}$ and ${Q}_{30}$ are their averaged values. The last terms on the right side are the fluctuation terms. 
The fluctuations are inserted back into the phase-space through scaling the local momentum distribution to the new values of $Q_{LM}$.
\newpage
\subsection{The isospin-dependent Boltzmann-Uehling-Uhlenbeck (IBUU) code}
\vskip 0.1in
J. Xu, L. W. Chen, B. A. Li
\vskip 0.1in

\subsubsection{About the IBUU model}

The isospin-dependent Boltzmann-Uehling-Uhlenbeck (IBUU) transport model was developed by including the isospin degree of freedom~\cite{Li:1991pq,Li98,Li:1991pq} into the original BUU model by Bertsch {\it et al.}~\cite{Bertsch:1988ik}. Later, an isospin- and momentum-dependent nuclear interaction (MDI) was incorporated into the IBUU transport model~\cite{Li:2003ts,Das:2002fr}. The IBUU transport model together with the MDI interaction has been used to study extensively the isospin dynamics in intermediate-energy heavy-ion collisions, e.g., constrain the symmetry energy at subsaturation densities with isospin diffusion~\cite{Che05,Li05} and explore the symmetry energy at suprasaturation densities with $\pi^-/\pi^+$ yield ratios~\cite{Li:2004cq,Xia09}. Most studies using the IBUU transport model before 2008 can be found in Ref.~\cite{Li08}.

\subsubsection{Initialization}

The initial coordinates of protons and neutrons in the projectile and target nuclei are sampled according to those obtained by a Skyrme-Hartree-Fock (SHF) calculation. With the standard SHF functional form, we can get the desired values of Skyrme parameters from empirical macroscopic quantities~\cite{Che10}. If we want to investigate, for instance, the effects due to the uncertainty of neutron skin thickness in heavy-ion collisions~\cite{Wei:2013sfa}, we use Skyrme forces with different slope parameters of the symmetry energy, while keeping all other macroscopic quantities the same.

Knowing the spatial distribution of nucleons, one can get the initial momenta of nucleons according to the local density approximation, i.e., sampling the nucleon momentum within $[0, p_F^{}]$, where $p_F^{}$ is the Fermi momentum from the local density and is isospin dependent. The local momentum distribution is sampled isotropically. Besides the above standard treatments, the high-momentum tail due to the nucleon-nucleon short-range correlations in nuclei has recently been considered \cite{Li:2018lpy}, and deformed Fermi distributions are also used~\cite{Yong:2017zgg}. 

Finally, the spatial coordinates are shifted according to the collision impact parameter, and modified by the Lorentz contraction according to the given beam energy. The momenta of all nucleons are also boosted according to the collision energy. This finishes the initialization of the IBUU transport model.

\subsubsection{The isospin- and momentum-dependent interaction}

The isospin- and momentum-dependent mean-field potential generally used in the IBUU transport model can be written as~\cite{Das:2002fr}
\begin{eqnarray}
U_\tau(\rho ,\delta ,\vec{p}) &=&A_{u}\frac{\rho _{-\tau }}{\rho _{0}}%
+A_{l}\frac{\rho _{\tau }}{\rho _{0}} +B\left(\frac{\rho }{\rho _{0}}\right)^{\sigma }(1-x\delta ^{2})-4\tau x\frac{B}{%
\sigma +1}\frac{\rho ^{\sigma -1}}{\rho _{0}^{\sigma }}\delta \rho
_{-\tau }
\notag \\
&+&\frac{2C_{\tau,\tau}}{\rho _{0}}\int d^{3}p^{\prime }\frac{f_{\tau }(%
\vec{r}, \vec{p}^{\prime })}{1+(\vec{p}-\vec{p}^{\prime })^{2}/\Lambda ^{2}}
+\frac{2C_{\tau,-\tau}}{\rho _{0}}\int d^{3}p^{\prime }\frac{f_{-\tau }(%
\vec{r}, \vec{p}^{\prime })}{1+(\vec{p}-\vec{p}^{\prime })^{2}/\Lambda ^{2}}.
\label{MDIU}
\end{eqnarray}%
In the above, $\tau=1(-1)$ for neutrons (protons) is the isospin index, $\rho=\rho_n+\rho_p$ is the total number density with $\rho_n$ and $\rho_p$ being the neutron and proton number densities, respectively, $\delta=(\rho_n-\rho_p)/(\rho_n+\rho_p)$ is the isospin asymmetry, and $f_\tau(\vec{r},\vec{p})$ is the phase-space distribution function. $\rho_0$ is the saturation density, and $A_u$, $A_l$, $B$, $\sigma$, $C_{\tau,\tau}$, $C_{\tau, -\tau}$, and $\Lambda$ are parameters to fit empirical nuclear matter properties. This mean-field potential can be obtained from an effective two-body interaction with a density-dependent zero-range term and a density-independent finite-range Yukawa term~\cite{Xu10}, similar to the Gogny interaction. The $x$ parameter is used to vary the density dependence of the symmetry energy by varying the relative contribution of different spin-isospin channels in the zero-range density-dependent interaction. The values of the parameters can be found in Ref.~\cite{Che05}, and Ref.~\cite{Che15} gives an extensive discussion on the studies based the MDI interaction. This interaction was later improved by refitting the mean-field potential at high nucleon momenta according to the nucleon-nucleus scattering data and introducing two additional parameters to vary the momentum dependence of the symmetry potential as well as the symmetry energy at saturation density. The details of the improved MDI interaction can be found in Ref.~\cite{ImMDI1}.

In the IBUU model, the phase-space distribution $f_\tau(\vec{r}, \vec{p})$ as well as the local density $\rho_\tau (\vec{r})$ for nucleons with isospin $\tau$ can be obtained by averaging $N_{TP}$ parallel events, i.e.,
\begin{eqnarray}
f_\tau(\vec{r}, \vec{p})&=&\frac{1}{N_{TP}}\sum_{i \in \tau}^{AN_{TP}} g(\vec{r}-\vec{r}_{i})\delta(\vec{p}-\vec{p}_{i}), \\
\rho_\tau (\vec{r})&=&\frac{1}{N_{TP}}\sum_{i \in \tau}^{AN_{TP}}g(\vec{r}-\vec{r}_{i}),
\end{eqnarray}
where $g$ is a smooth function in coordinate space, and $A$ is the number of real particles, with each represented by $N_{TP}$ test particles. The coordinate space is divided into cells with each cell of volume 1 fm$^3$. In the original IBUU (IBUU\_O), each nucleon contributes a factor of $1/3$ to the density of the local cell and $1/9$ to that of each nearest neighboring cell. In the improved IBUU (IBUU\_L), the form of the smooth function $g$ is taken from that in the lattice Hamiltonian framework~\cite{Lenk89}, i.e., the phase-space distribution function $f_L$ and the density $\rho_L$ at the sites of a three-dimensional cubic lattice are expressed as
\begin{eqnarray}
f_{L,\tau}(\vec{r}_{\alpha},\vec{p}) &=& \sum_{i \in \tau}^{AN_{TP}}S(\vec{r}_{\alpha}-\vec{r}_i)\delta(\vec{p}-\vec{p}_{i}),\\
\rho_{L,\tau}(\vec{r}_{\alpha}) &=& \sum_{i \in \tau}^{AN_{TP}}S(\vec{r}_{\alpha}-\vec{r}_i).
\end{eqnarray}
In the above, $\alpha$ is the site index, $\vec{r}_{\alpha}$ is the position of the site $\alpha$, and $S$ is the shape function describing the contribution of a test particle at $\vec{r}_i$ to the value of the quantity at $\vec{r}_{\alpha}$, i.e.,
\begin{eqnarray}
S(\vec{r})=\frac{1}{N_{TP}(nl)^6}g(x)g(y)g(z)
\end{eqnarray}
with
\begin{eqnarray}
g(q)=(nl-|q|)\Theta(nl-|q|).
\end{eqnarray}
$l$ is the lattice spacing, $n$ determines the range of $S$, and $\Theta$ is the Heaviside function. We generally adopt the values of $l=1$ fm and $n=2$.

In IBUU\_O, the equations of motion for the $i$th test particle are expressed as
\begin{eqnarray}
\frac{d\vec{r}_{i}}{dt}&=& \frac{\vec{p}_i}{\sqrt{\vec{p}_{i}^{2}+m^2}} + \frac{\partial U[\rho(\vec{r}_c),\delta(\vec{r}_c),\vec{p}_i]}{\partial\vec{p}_i}, \\
\frac{d\vec{p}_{i}}{dt} &=& - \frac{\partial U[\rho(\vec{r}_c),\delta(\vec{r}_c),\vec{p}_i]}{\partial\vec{r}_c},
\end{eqnarray}
where $m$ is the bare nucleon mass, $\vec{r}_c$ represents the coordinate of the cell containing the $i$th test particle, and ${\partial U}/{\partial\vec{r}_c}$ is calculated by taking numerical derivatives based on the single-particle potentials $U$ in neighboring cells. In IBUU\_L, the Hamiltonian of the system can be expressed as
\begin{equation}\label{htotal}
H=\sum_{i}^{AN_{TP}}\sqrt{\vec{p}_{i}^{2}+m^2}+N_{TP}\widetilde{V},
\end{equation}
with the total potential energy expressed as
\begin{equation}
\widetilde{V}=l^3\sum_{\alpha}V_{\alpha},
\end{equation}
where $V_{\alpha}$ is the potential energy at the site $\alpha$, with the energy-density functional form consistent with the mean-field potential, e.g., Eq.~(\ref{MDIU}). The canonical equations of motion for the $i$th test particle from the above Hamiltonian can thus be written as
\begin{eqnarray}
 \frac{d\vec{r}_{i}}{dt}&=&\frac{\partial H}{\partial\vec{p}_{i}}
 = \frac{\vec{p}_i}{\sqrt{\vec{p}_{i}^{2}+m^2}}+N_{TP}\frac{\partial\widetilde{V}}{\partial\vec{p}_{i}},  \label{rt}\\
 \frac{d\vec{p}_{i}}{dt} &=&-\frac{\partial H}{\partial\vec{r}_{i}}
 = -N_{TP}\frac{\partial\widetilde{V}}{\partial\vec{r}_{i}}.  \label{pt}
\end{eqnarray}
A time step $\Delta t =0.5$ fm/c is generally used to solve numerically the above differential equations. Details of IBUU\_L can be found in Refs.~\cite{Wan18,Xu20}.

\subsubsection{Nucleon-nucleon scattering and Pauli blocking}

The nucleon-nucleon scattering cross sections in free space are taken from the parametrized form as~\cite{Cha90}
\begin{eqnarray}
&&\sigma_{pp(nn)} = 13.73 - 15.04/v + 8.76/v^2 +
68.67v^4,\label{sigma1}\\
&&\sigma_{np} = -70.67 - 18.18/v + 25.26/v^2 +
113.85v,\label{sigma2}
\end{eqnarray}
where the cross sections are in mb and $v$ is the velocity of the
projectile nucleon with respect to the fixed target nucleon. The parametrization can be further improved by considering the different angular distributions for $pp(nn)$ and $np$ collisions. The in-medium scattering cross sections are calculated from~\cite{Li05}
\begin{equation}
\sigma^{medium}_{NN} =
\sigma_{NN}\left(\frac{\mu_{NN}^\star}{\mu_{NN}}\right)^2,
\end{equation}
where $\mu_{NN}$ ($\mu_{NN}^\star$) is the free-space (in-medium)
reduced mass of colliding nucleons, with the nucleon effective mass defined from the momentum dependence of the mean field potential  as
\begin{equation}
\frac{m_{\tau }^{\ast }}{m}=\left( 1+\frac{m}{p}\frac{dU_{\tau
}}{dp}\right) ^{-1}. \label{Meff}
\end{equation}%

In the original IBUU, a possible attempted collision between two nucleons is determined by Bertsch's prescription~\cite{Bertsch:1988ik}, i.e.,
\begin{equation}
\sqrt{(\Delta \vec{r})^2 + (\Delta \vec{r} \cdot \vec{p}/p)^2 }<\sqrt{\sigma/\pi}\\
\end{equation}
and
\begin{equation}\label{b2}
\left|\frac{\Delta{\vec{r}} \cdot \vec{p}}{p}\right| < \left(\frac{p}{\sqrt{p^2+m_1^2}}+\frac{p}{\sqrt{p^2+m_2^2}}\right)\Delta t/2,
\end{equation}
where $\vec{p}$ and $\Delta \vec{r}$ are the momentum of one of the nucleons and the relative coordinate between two nucleons in their center-of-mass (C.M.) frame, respectively. The first condition indicates that the distance between the two nucleons perpendicular to their C.M. velocity is within the range of the scattering cross section, while the second condition means that the collision can happen right in this time step. In the improved IBUU, the right-hand side of Eq.~(\ref{b2}) is divided by a Lorentz factor from the calculational frame to the C.M. frame of the two colliding nucleons. For elastic scatterings, the nucleon momentum in the C.M. frame is then sampled according to the angular distribution of the cross section $\sigma$ in this trial collision. If the trial collision is not Pauli-blocked, the final momenta of nucleons can then be obtained from the inverse Lorentz transformation. For the more complicated inelastic scatterings in IBUU, we refer the reader to Appendix B of Ref.~\cite{Bertsch:1988ik} and Ref.~\cite{Li1995aa}.

The trial collision can be Pauli-blocked if the local phase-space occupation is too `crowded', and this will be checked after the attempted collision is assumed to happen. The isospin-dependent phase-space density $f_\tau(x,y,z,p_x,p_y,p_z)$ for Pauli-blocking is updated at each time step and after each successful collision. When a collision is attempted, the occupation number of the phase space
\begin{equation}
n_{occup} = \frac{h^3}{d\Delta x \Delta y \Delta z \Delta p_x \Delta p_y \Delta p_z} f_\tau(x,y,z,p_x,p_y,p_z),
\end{equation}
calculated by interpolation, determines whether the collision is Pauli-blocked or not, with $h$ being the Planck constant and $d=2$ being the spin degeneracy. At the current stage, the cell size of the Pauli lattice is set to be $\Delta x=\Delta y=\Delta z = 2$ fm and $\Delta p_x =\Delta p_y= \Delta p_z=100$ MeV/c.

\subsubsection{Current status of the IBUU model}

In the past, the IBUU transport model did a good job in studying isospin dynamics in heavy-ion collisions at intermediate energies~\cite{Li08}. The electromagnetic field was later incorporated into IBUU~\cite{Wei:2017dib,Ou11}. The applications of this IBUU model now follow three lines. In one line, the spin degree of freedom for nucleons, the nucleon spin-orbit coupling~\cite{Xu13}, and the spin-dependent nucleon-nucleon scattering cross sections~\cite{Xia17} are incorporated to study the spin-dependent collective flows~\cite{Xia14} as well as the spin polarization~\cite{Xia20} in intermediate-energy heavy-ion collisions. In a second line, the stability of the initialization is further improved to construct the ground state of the nucleus~\cite{Xu20}, in order to study nuclear giant resonances~\cite{Kon17}. The third line is trying to incorporate approximately the main physics of short-range correlations into the IBUU transport model, see, e.g., Refs. \cite{Guo:2021zcs,PLi:2014vua,Hen:2014yfa}.

\newpage 
\subsection{Solving Boltzmann-Uehling-Uhlenbeck equation with lattice-Hamiltonian method~(LBUU)}
\vskip 0.1in
R. Wang, Z. Zhang, L.-W. Chen
\vskip 0.1in
\subsubsection{Introduction}

The lattice Boltzmann-Uehling-Uhlenbeck (LBUU) transport model was first developed in Ref.~\cite{WRPRC99} by Wang, Chen and Zhang, where only the Vlasov equation is solved with the lattice-Hamiltonian method~(LHV).
The collision integral was then included in Ref.~\cite{WRPLB807} following a full-ensemble stochastic method.
As a newly developed framework, one feature of the LBUU model is that it is implemented with GPU parallel computing, which significantly increases the numerical efficiency and thus allows the LBUU to reach a high accuracy by employing a huge number of test particles (e.g., up to $10^5$ per nucleon for Sn+Sn collisions).
The latter is crucial to ensure the convergence of the results for certain observables, e.g., the width of nuclear giant resonances.
The LBUU model has well reproduced the strength function of the nuclear iso-vector dipole resonances~\cite{WRPLB807}.
Further applications on heavy-ion collisions are in progress.
More details on the LBUU model can be found in Ref.~\cite{WRFiP8}.

The lattice-Hamiltonian~(LH) method of solving the BUU equation, originally proposed in Ref.~\cite{Lenk89}, improves the sample smoothing technique of the usual test particle approach~\cite{Won82}.
In the LH method, the equations of motion of test nucleons are governed by the total Hamiltonian of the system, which is approximated by the lattice-Hamiltonian, i.e.,
\begin{equation}\label{E:HL}
    H = \int {\mathcal{H}}(\vec{r})d\vec{r} \approx l_xl_yl_z\sum_{\alpha}{\mathcal{H}}(\vec{r}_{\alpha})\equiv H_L,
\end{equation}
where $\vec{r}_{\alpha}$ represents the coordinate of a certain lattice site $\alpha$ and $l_x$, $l_y$, and $l_z$ are lattice spacing.
The Hamiltonian density at a given lattice site ${\mathcal{H}}(\vec{r}_{\alpha})$ can be expressed in terms of the Wigner function $f_{\tau}(\vec{r}_{\alpha},\vec{p},t)$.
In the LBUU model, we represent $f_{\tau}(\vec{r}_{\alpha},\vec{p},t)$ by a large number of test nucleons with a form factor $S$, i.e.,
\begin{equation}\label{E:f}
    f_{\tau}(\vec{r}_{\alpha},\vec{p},t) = \frac{1}{g}\frac{(2\pi\hbar)^3}{N_{\rm E}}\sum_i^{\alpha,\tau}S\big[\vec{r}_i(t) - \vec{r}_{\alpha}\big]\delta\big[\vec{p}_i(t) - \vec{p}\big].
\end{equation}
where $g$ is the spin degeneracy factor, and $N_{\rm E}$ is the number of test particles per nucleon introduced in the calculation.
The form factor $S$ is taken to be of a triangular form.
The sum in the above expression runs over all test nucleons with isospin $\tau$ that contribute to the lattice site $\alpha$.

\subsubsection{Mean-field evolution}

In the LBUU model, we employ the Skyrme pseudopotential to calculate the Hamiltonian density ${\mathcal{H}}(\vec{r})$ in Eq.~(\ref{E:HL}).
The Skyrme pseudopotential~\cite{RaiPRC83} generalizes the standard Skyrme interaction~\cite{VauPRC5,ChaNPA627} by including in the latter additional derivative terms up to the next-to-next-to-next-to leading order.
It enables us to reproduce the empirical nuclear optical potential up to about $1~\rm GeV$ in kinetic energy~\cite{WRPRC98}, which standard Skyrme interactions fail to achieve.
The equations of motion for the $i$-th test nucleon in the LBUU model are expressed as
\begin{align}
\frac{d\vec{r}_i}{dt} = &~\frac{\vec{p}_i(t)}{m} + N_{\rm E}l_xl_yl_z\sum_{\alpha\in V_i}\frac{\partial{\mathcal{H}}^{\rm MD}_{\alpha}}{\partial\vec{p}_i}\label{E:ri},\\
\frac{d\vec{p}_i}{dt} = &~- N_{\rm E}l_xl_yl_z\sum_{\alpha\in V_i}\bigg\{\sum_{\tau}^{n,p}\bigg[\frac{\partial(\mathcal{H}^{\rm loc}_{\alpha} + \mathcal{H}^{\rm Cou}_{\alpha} + \mathcal{H}^{\rm DD}_{\alpha})}{\partial\rho_{\tau,\alpha}} + \sum_{n = 0}(-1)^n\nabla^n\frac{\partial{\mathcal{H}}^{\rm grad}_{\alpha}}{\partial\nabla^n\rho_{\tau,\alpha}} \bigg]\frac{\partial\rho_{\tau,\alpha}}{\partial\vec{r}_i} + \frac{\partial{\mathcal{H}}^{\rm MD}_{\alpha}}{\partial\vec{r}_i}\bigg\}\label{E:pi}.
\end{align}
In the above two equations, the subscripts $\alpha$ for various quantities denote their values at lattice site $\alpha$.
The sums run over all lattice sites inside a spatial volume $V_i$, which the form factor of the $i$-th test nucleon covers.
The ${\cal H}_{\alpha}^{\rm loc}$, ${\cal H}_{\alpha}^{\rm Cou}(\vec{r})$, ${\cal H}_{\alpha}^{\rm DD}(\vec{r})$, ${\cal H}_{\alpha}^{\rm grad}(\vec{r})$ and ${\cal H}_{\alpha}^{\rm MD}(\vec{r})$ represent the local, Coulomb, density-dependent, gradient and momentum-dependent terms within the Skyrme pseudopotential, respectively. Detailed expressions are given in Ref.~\cite{WRPRC99}.
The partial derivative of local density $\rho_{\tau,\alpha}$ in Eq.~(\ref{E:pi}) can be calculated in terms of the spatial derivative of $S$, i.e.,
\begin{equation}
    \frac{\partial\rho_{\tau,\alpha}}{\partial\vec{r}_i} = \frac{\partial}{\partial\vec{r}_i}\sum_{\vec{r}_j\in V_{\alpha}}^{\tau_j=\tau}S(\vec{r}_j-\vec{r}_{\alpha})
     = \begin{cases}
         & \frac{\partial S(\vec{r}_i-\vec{r}_{\alpha})}{\partial\vec{r}_i},\quad \tau_i = \tau,\\
         & 0,\quad \tau_i \ne \tau.
        \end{cases}
\end{equation}
The contribution of the momentum-dependent parts of the lattice Hamiltonian to the equations of motion of test nucleons are expressed as the summations of test nucleons, i.e.,
\begin{align}
    \frac{\partial\mathcal{H}^{\rm MD}(\vec{r}_{\alpha})}{\partial\vec{r}_i} = &~2\frac{\partial S\big[\vec{r}_i(t) - \vec{r}_{\alpha}\big]}{\partial\vec{r}_i}\bigg\{\sum_{j\in V_{\alpha}}S\big[\vec{r}_j(t) - \vec{r}_{\alpha}\big]{\mathcal{K}}_{\rm s}\big[\vec{p}_i(t),\vec{p}_j(t)\big] + \sum_{j\in V_{\alpha}}^{\tau_j = \tau_i}S\big[\vec{r}_j(t) - \vec{r}_{\alpha}\big]{\mathcal{K}}_{\rm v}\big[\vec{p}_i(t),\vec{p}_j(t)\big]\bigg\}\label{E:mdr},\\
    \frac{\partial\mathcal{H}^{\rm MD}(\vec{r}_{\alpha})}{\partial\vec{p}_i} = &~2S\big[\vec{r}_i(t) - \vec{r}_{\alpha}\big]\bigg\{\sum_{j\in V_{\alpha}}S\big[\vec{r}_j(t) - \vec{r}_{\alpha}\big]\frac{\partial{\mathcal{K}}_{\rm s}\big[\vec{p}_i(t),\vec{p}_j(t)\big]}{\partial\vec{p}_i} + \sum_{j\in V_{\alpha}}^{\tau_j = \tau_i}S\big[\vec{r}_j(t) - \vec{r}_{\alpha}\big]\frac{\partial{\mathcal{K}}_{\rm v}\big[\vec{p}_i(t),\vec{p}_j(t)\big]}{\partial\vec{p}_i}\bigg\}\label{E:mdp}.
\end{align}
The expression of the iso-scalar(vector) momentum-dependent kernel $\mathcal{K}_{\rm s}$($\mathcal{K}_{\rm v}$) can be found in Ref.~\cite{WRPRC99}.

\subsubsection{Collision integral}

In the LBUU model, we treat the collision integral through a full-ensemble stochastic method~\cite{Danielewicz:1991dh}.
The collision probability of two test nucleons is derived as follows.
Considering test nucleons around lattice site $\vec{r}_{\alpha}$ from two momentum space volumes $V_{\vec{p}_1}$ $=$ $\vec{p}_1\pm\frac{1}{2}\Delta^3p_1$ and $V_{\vec{p}_2}$ $=$ $\vec{p}_2\pm\frac{1}{2}\Delta^3p_2$ we average over the momentum-space volume $V_{\vec{p}_i}$ to obtain their $f(\vec{r}_{\alpha},\vec{p}_i)$ according to Eq.~(\ref{E:f}), i.e.,
\begin{equation}\label{E:fi}
    f(\vec{r}_{\alpha},\vec{p}_i) \approx \frac{1}{\Delta^3p_i}\frac{(2\pi\hbar)^3}{gN_{\rm E}}\sum_j^{\vec{p}_j\in V_{\vec{p}_i}}S(\vec{r}_j - \vec{r}_{\alpha}).
\end{equation}
The number of collisions between nucleons from those two momentum space volumes that happen in time interval $\Delta t$ is
\begin{equation}\label{E:DN}
        \Delta N^{\rm coll}(\vec{r}_{\alpha}, \vec{p}_1,\vec{p}_2) = g\frac{\Delta^3p_1}{(2\pi\hbar)^3}\Big|\frac{df(\vec{r}_{\alpha},\vec{p}_1)}{dt}\Big|_{\vec{p}_2}^{\rm coll}l_xl_yl_z\Delta t = g\frac{\Delta^3p_2}{(2\pi\hbar)^3}\Big|\frac{df(\vec{r}_{\alpha},\vec{p}_2)}{dt}\Big|_{\vec{p}_1}^{\rm coll}l_xl_yl_z\Delta t.
\end{equation}
The $\big|\frac{df(\vec{r}_{\alpha},\vec{p}_1)}{dt}\big|_{\vec{p}_2}^{\rm coll}$ and $\big|\frac{df(\vec{r}_{\alpha},\vec{p}_2)}{dt}\big|_{\vec{p}_1}^{\rm coll}$ are the changing rate of $f(\vec{r}_{\alpha},\vec{p}_1)$ and $f(\vec{r}_{\alpha},\vec{p}_2)$, respectively, caused by the two-body collision between $V_{\vec{p}_1}$ and $V_{\vec{p}_2}$.
These terms can be obtained directly from the collision term of the BUU equation,
\begin{equation}\label{E:df}
    \Big|\frac{df(\vec{r}_{\alpha},\vec{p}_1)}{dt}\Big|_{\vec{p}_2}^{\rm coll} = g\frac{\Delta^3p_2}{(2\pi\hbar)^3}f(\vec{r}_{\alpha},\vec{p}_1)f(\vec{r}_{\alpha},\vec{p}_2)\int\frac{{\rm d}\vec{p}_3}{(2\pi\hbar)^3}\frac{{\rm d}\vec{p}_4}{(2\pi\hbar)^3}|{\mathcal{M}}_{12\rightarrow34}|^2(2\pi)^4\delta^4(p_1 + p_2 - p_3 - p_4).
\end{equation}
By definition, we can replace the integral in Eq.~(\ref{E:df}) with $v_{\rm rel}\sigma_{\rm NN}^*$, where $v_{\rm rel}$ and $\sigma_{\rm NN}^*$ are the relative velocity and in-medium cross section in the \emph{two-nuclei} center-of-mass frame, respectively.
Substituting Eq.~(\ref{E:fi}) and Eq.~(\ref{E:df}) into Eq.~(\ref{E:DN}), we obtain
\begin{equation}
 \Delta N^{\rm coll}(\vec{r}_{\alpha}, \vec{p}_1,\vec{p}_2) = \sum_{i,j}^{\substack{\vec{p}_i\in V_{\vec{p}_1}\\{\vec{p}_j\in V_{\vec{p}_2}}}}\frac{1}{N_{\rm E}^2}v_{\rm rel}\sigma_{\rm NN}^*S(\vec{r}_i-\vec{r}_{\alpha})S(\vec{r}_j-\vec{r}_{\alpha})l_xl_yl_Z\Delta t \equiv \sum_{i,j}^{\substack{\vec{p}_i\in V_{\vec{p}_1}\\{\vec{p}_j\in V_{\vec{p}_2}}}}\Delta N_{ij}^{\rm coll}.
\end{equation}
The  summed over quantity, denoted by $\Delta N_{ij}^{\rm coll}$, is the number of physical collisions from the scattering of the $i$-th and $j$-th test nucleons.
Given that every test nucleon represents $1/N_{\rm E}$ of a physical nucleon, one obtains the scattering probability of the $i$-th and $j$-th test nucleons as
\begin{equation}
    P_{ij} = \frac{\Delta N_{ij}^{\rm coll}}{(1/N_{\rm E})^2} = v_{\rm rel}\sigma_{\rm NN}^*S(\vec{r}_i - \vec{r}_{\alpha})S(\vec{r}_j - \vec{r}_{\alpha})l_xl_yl_z\Delta t.
\end{equation}
One should note the difference from the scattering probability in a typical stochastic approach~\cite{Xu:2004mz}, due to the presence of the form factor $S$ in the LBUU model.
In the full-ensemble scenario, the collision probability is reduced, i.e., $P_{ij}$ $\rightarrow$ $P_{ij}/N_{\rm E}$, by the scaling $\sigma_{\rm NN}^*$ $\rightarrow$ $\sigma_{\rm NN}^*/N_{\rm E}$.
The above stochastic method has been generalized to the production of light nuclei through $3\leftrightarrow2$ and $4\leftrightarrow2$ scatterings, where the internal wave function and finite size of light nuclei have been taken into account~\cite{Sun:2021dlz}.

In the LBUU model, we obtain the in-medium nucleon-nucleon scattering cross section $\sigma_{\rm NN}^*$ through multiplying the free nucleon-nucleon scattering cross section $\sigma_{\rm NN}^{\rm free}$ by a medium-correction factor.
The correction factor depends on both the local density and the center-of-mass kinetic energy of the scattering test nucleons.
For the nucleon-nucleon scattering cross section in free space $\sigma_{\rm NN}^{\rm free}$, we use the parametrization provided in Ref.~\cite{Cugnon:1996kh}.

Since the nucleon-nucleon scattering probabilities are very small within one time step, instead of evaluating probabilities of all possible scatterings of test nucleons, we choose $N'$ pairs randomly out of all possible test nucleon pairs $N_{\alpha}^{22}$ around lattice site $\alpha$, and accordingly amplify the scattering probabilities, i.e., $P_{ij}'$ $=$ $N_{\alpha}^{22}/N'P_{ij}$.
Normally we choose $N'$ $=$ $N_{\alpha}^{22}/2$ in the LBUU model so that one test nucleon involves at most in one scattering event.

If the scattering between the $i$-th and $j$-th test nucleons happens in lattice site $\vec{r}_{\alpha}$ according to $P_{ij}$ or $P'_{ij}$, the direction of their final states $\vec{p}_3$ and $\vec{p}_4$ are sampled according to the differential cross-section given in Ref.~\cite{Cugnon:1996kh}, and then the Pauli blocking factor $[1 - f(\vec{r}_{\alpha},\vec{p}_3)]\times[1 - f(\vec{r}_{\alpha},\vec{p}_4)]$ is employed to determine whether the collision is blocked by the Pauli principle.
The distribution function $f_{\tau}(\vec{r}_{\alpha},\vec{p})$ is calculated by averaging its value in Eq.~(\ref{E:f}) over a momentum-space sphere with radius $R_{\tau}^p(\vec{r}_{\alpha},\vec{p})$ centered at $\vec{p}$.
Typically, $R_{\tau}^p(\vec{r}_{\alpha},\vec{p})$ is taken as a constant of about $0.1~{\rm GeV}$.
For the calculation of nuclear giant resonances, we use an improved form of $R_{\tau}^p(\vec{r}_{\alpha},\vec{p})$, which has been specifically proposed in the GiBUU code~\cite{Gaitanos:2010fd} for small-amplitude nuclear collective dynamics near ground state, i.e.,
\begin{equation}
    R_{\tau}^p(\vec{r}_{\alpha},\vec{p}) = {\rm max}[\Delta p, p_{\tau}^F(\vec{r}_{\alpha}) - |\vec{p}|],
\end{equation}
where $p_{\tau}^F(\vec{r}_{\alpha})$ is the nucleon Fermi momentum at lattice site $\vec{r}_{\alpha}$, and $\Delta p$ is a constant that is normally taken to be less than $0.05~{\rm GeV/c}$.

\subsubsection{Initialization}
In the LBUU model, the initial coordinates of test nucleons are generated according to a nucleon radial density $\rho_{\tau}(r)$, while their initial momenta follow zero-temperature Fermi distribution,
\begin{equation}\label{E:f0}
    f_{\tau}(\vec{r},\vec{p}) = \frac{2}{(2\pi\hbar)^3}\theta\big[|\vec{p}| - p^F_{\tau}(r)\big],
\end{equation}
where $p^F_{\tau}(\vec{r})$ is the local Fermi momentum fulfills $p^F_{\tau}(\vec{r})$ $=$ $\hbar\big[3\pi^2\rho_{\tau}(\vec{r})\big]^{1/3}$.

The nucleon radial density $\rho_{\tau}(r)$ of a ground state nucleus at zero temperature is obtained by varying the total energy, i.e., through the Thomas-Fermi initialization.
Within the local density approximation, the total energy of a ground state spherical nucleus at zero temperature can be obtained by integrating the Hamiltonian density, which is a function of the
$\rho_{\tau}(r)$ and its spatial gradients $\nabla^{n}\rho_{\tau}(r)$, i.e.,
\begin{equation}
    E = \int{\mathcal{H}\big[r,\rho_{\tau}(r),\nabla{\rho_{\tau}(r)},\nabla^2{\rho_{\tau}(r)}\cdots\big]dr}.
\end{equation}
Varying the total energy with respect to $\rho_{\tau}(r)$ we obtain (note that the contribution from the Coulomb interaction should also be included in the Hamiltonian density),
\begin{equation}\label{E:GS}
\frac{1}{2m} \big\{p_{\tau}^F\big[\rho_{\tau}(r)\big]\big\}^2 +U_{\tau}\big\{p_{\tau}^F\big[\rho_{\tau}(r)\big],r\big\} = \mu_{\tau},
\end{equation}
where $\mu_{\tau}$ is the chemical potential of proton (neutron) inside the nucleus to 
obtain the given proton number $Z$ (neutron number $N$).
The $U_{\tau}\big\{p_{\tau}^{\rm F}\big[\rho_{\tau}(r)\big],r\big\}$ is the single-particle
potential of the nucleon at the Fermi surface.
The single-particle potential is defined as the variation of the potential density with respect to
the phase-space distribution function and density gradients, 
and therefore Eq.~(\ref{E:GS}) contains density gradients implicitly.
For the N$3$LO Skyrme pseudopotential used in the LBUU model, the detailed 
expression is given in Ref.~\cite{WRPRC98}.
The nucleon radial density $\rho(r)$ for ground state spherical nuclei is obtained by solving Eq.~(\ref{E:GS}) with the boundary condition
\begin{equation} 
    \frac{\partial\rho(r)}{\partial r}\Big|_{r = 0} = \frac{\partial\rho(r)}{\partial r}\Big|_{r = r_{\rm B}} = 0.
\end{equation}
The $r_{\rm B}$ in the above equation is the boundary of the nucleus, which satisfies $\rho(r_{\rm B})$ $=$ $0$, and it needs to be determined self-consistently when solving Eq.~(\ref{E:GS}).

\subsection{The Pawel's Boltzmann-Uehling-Uhlenbeck (pBUU) code}\label{sec:pBUU}
\vskip 0.1in
P. Danielewicz 
\vskip 0.1in

In this short write-up, the relevant details of the pBUU code are presented. The main references are~\cite{Danielewicz:1991dh,dan00}.

\subsubsection{Code history}

Most of that code's development was done by Pawel~Danielewicz at the Michigan State University in East Lansing, with some of the advances made during his visits to GSI Darmstadt and to INT Seattle.  The abbreviation pBUU stands for Pawel's BUU.  Early on, the code was also known under the abbreviation BEM for the Boltzmann-Equation Model.  Some other contributions are mentioned within the chronology of the code's development below.

\begin{itemize}

\item 1990-91:
Codes incorporating self-consistent mean-fields, prior to pBUU, were built around the concept of extending the cascade model. However, pBUU (BEM) was designed around the idea of a Monte-Carlo solution to the Boltzmann equation.
The code included deuterons as a~degree of freedom, with deuteron production described as a~process inverse to deuteron breakup in collisions with nucleons, i.e., the production taking place in 3-nucleon collisions, based on the formal developments in~\cite{dan90}.

From its nascency, the code included an option of enclosing the nuclear system in a spatial box, for the sake of studying equilibration in that box and verifying whether the expected equilibrium limit could be reached.
Cross-sections for $\Delta$ absorption in collisions with nucleons followed detailed balance relations that accounted for the significant $\Delta$ widths and that led to pion yields in equilibrium consistent with the fireball model.  Details of the code first appeared in a paper co-authored with George Bertsch~\cite{Danielewicz:1991dh}.

\item 1992-93:
Production of $A=3$ clusters, tritons and helions, was added~\cite{dan92}.  Momentum dependence was incorporated into the mean fields~\cite{pan93}.  In~connection with a parallel development of the collision treatment in Ref.~\cite{lang93}, the calculation of collision effects was speeded up.

\item 1994-95:
The lattice-Hamiltonian method of Lenk and Pandharipande~\cite{Lenk89} was adopted for integration of the mean-field effects~\cite{dan95}.

\item 1998-2000:
In the context of application of the code to truly relativistic collisions~\cite{dan98}, the code got more firmly based within the realm of the relativistic Landau theory~\cite{bay76,dan00}.  In the context of experimental investigations of stopping, density-dependent elastic in-medium cross-sections were incorporated~\cite{dan02}.  To tame spurious collective oscillations of initial nuclear states, those states began to be sought as solutions to Thomas-Fermi equations consistent with the energy functional employed in the transport equations~\cite{dan00}.

\item 2002-2003:
Lijun Shi~\cite{shi03} introduced different forms of symmetry energy into the code, in the context of the studies of isospin diffusion~\cite{tsang04}.

\item 2012-2016:
Jun Hong~\cite{Hong14,hong_bulk_2016} introduced provisions for increasing computational statistics of pion production within the subthreshold domain. She also expanded options for interplay of the density and momentum dependence in the nucleonic potentials. Christian Simon~\cite{Simon13} and Yuanyuan Zhang~\cite{Zhang14} made theoretical advances for the sake of a greater flexibility in the energy functional, with only modest computational cost.

\item 2012-2017: Brent Barker expanded the selection of in-medium cross-sections~\cite{Barker13,barker_shear_2019}.

\end{itemize}

\subsubsection{Energy Functional}

In the calculational frame, the energy functional is a sum of three terms:
\begin{equation}
\label{eq:E=}
E = \int \text{d} {\pmb r} \, e + E_\nabla + E_\text{coul} \, .
\end{equation}
Here the first r.h.s.\ term is the volume contribution, treated covariantly. The~last two terms
in Eq.~\eqref{eq:E=} are the finite-range correction,
\be
\label{eq:Enabla=}
E_\nabla = \frac{a_1}{2 \rho_0} \int \text{d} {\pmb r} \, \big( \nabla \rho \big)^2 \, ,
\ee
and the Coulomb term, both treated noncovariantly. In the local rest frame, the volume energy density is
\ba
\label{eq:e=}
e & = &\sum_X g_X \int \frac{\text{d} {\pmb p}}{(2 \pi)^3} \, f_X ({\pmb p}) \, \bigg(m_X + \int_0^p \text{d} p' \, v_X' \bigg)
+ \int_0^\rho \text{d}\rho' \, U(\rho') + 4 \, S(\rho)  \, \rho_T^2/\rho \, .
\ea
Here, $X$ pertains to species $X$, $g_X$ is spin degeneracy, $f_X$ is the phase-space occupation of
species~$X$, $\rho_T$ is isospin density (hence the factor of 4 in the term) and $v_X$ is local velocity of~$X$, parametrized with
\be
\label{eq:vX=}
v_X(p,\rho) = \frac{p}{\sqrt{p^2 + m_X^2/\big(1 +  \frac{m_N}{m_X} \, \frac{A_X \, {\mathcal F }(\rho/\rho_0)}{(1+\lambda p^2/m_X^2)^2} \big)^2}} \, ,
\ee
where $m_X$ and $A_X$ are the mass and mass number, respectively.

The single-particle energies follow from the net energy with
\be
\label{eq:epsilon=}
\epsilon_X({\pmb r},{\pmb p},t) = \frac{(2 \pi)^3}{g_X} \, \frac{\delta E}{\delta f({\pmb r},{\pmb p},t)} \, ,
\ee
and the solved set of Boltzmann equations is of the form
\be
\label{eq:Boltzmann}
\frac{\partial f_X}{\partial t} + \frac{ \partial \epsilon_X}{\partial {\pmb p}} \, \frac{ \partial f_X}{\partial {\pmb r}}
- \frac{ \partial \epsilon_X}{\partial {\pmb r}} \, \frac{ \partial f_X}{\partial {\pmb p}} =
{\mathcal K}_X^< \, (1 \mp f_X) - {\mathcal K}_X^> \,  f_X \, .
\ee
The species $X$ in pBUU include nucleons, deuterons, tritons, helions, $\Delta$~and $N^*$ resonances and pions.
The feeding ${\mathcal K}^<$ and absorption ${\mathcal K}^>$ rates include the effects of elastic collisions,
the effects of absorption of pions on nucleons into resonances and of resonance decays and of inelastic $A=1$
baryon collisions where the resonances are produced or absorbed. Moreover, the rates include collisions
of $A=2$ and 3 clusters with nucleons, where the clusters are broken up, and inverse processes, where 3 or 4
nucleons collide and clusters are formed.

In solving the Boltzmann equation set, the phase-space density is represented in terms of a set
of $\delta$-functions, or test particles:
\be
\frac{g}{(2 \pi)^3} \, f({\pmb r},{\pmb p},t) = \frac{1}{\mathcal N} \sum_k \delta({\pmb r} - {\pmb r}_k(t)) \, \delta({\pmb p} - {\pmb p}_k(t)) \, ,
\ee
where ${\mathcal N}$ is the number of test particles per particle.
The~phase-space derivative terms on the l.h.s.\ of \eqref{eq:Boltzmann} are accounted for by making the
test-particles follow equations of the Hamiltonian form:
\be
\frac{\text{d}{\pmb r}_k}{\text{d}t} = \frac{\text{d}\epsilon}{\text{d}{\pmb p}_k} \, , \hspace*{.14\textwidth}
\frac{\text{d}{\pmb p}_k}{\text{d}t} = - \frac{\text{d}\epsilon}{\text{d}{\pmb r}_k} \, .
\ee
Numerical algorithms, for calculating the single-particle energies and their  derivatives in integrating
the Boltzmann equation, are made consistent by tying them to a lattice Hamiltonian for the system~\cite{Lenk89}.
Specifically, the space is divided into cartesian cells of volume $\Delta V$. The numerical system energy
$\overline{E}$ is constructed as a sum of contributions from cell corners, or nodes, following
Eqs.~\eqref{eq:E=}--\eqref{eq:vX=}.  Associated with each node is a piecewise differentiable form factor
$S({\pmb r})$, where ${\pmb r}$ is the distance from the node, normalized according to
$\int \text{d}{\pmb r} \, S({\pmb r}) = \Delta V$. Then, the particle distribution functions for a~node
$\alpha$ at ${\pmb r}_\alpha$, for calculating that node's contribution to $\overline{E}$, are
\be
\overline{f}({\pmb r}_\alpha) = \frac{1}{{\mathcal N} \, \Delta V} \sum_k S({\pmb r}_\alpha - {\pmb r}_k) \, \delta({\pmb p} - {\pmb p}_k) \, .
\ee
For the sake of the numerical calculation, the numerical energy~$\overline{E}$ from $\overline{f}$ replaces~$E$
in Eq.~\eqref{eq:epsilon=} in determining the single-particle energy and its derivatives.

\subsubsection{Initialization}

Ahead of a transport calculation, a set of Thomas-Fermi (TF) equations is solved for the neutron $\rho_n (r)$
and proton $\rho_p (r)$ density profiles, consistent with the energy functional Eqs.~\eqref{eq:E=}-\eqref{eq:vX=}
employed for the transport calculations. For a~spherically symmetric nucleus, that equation set is conveniently
transformed into:
\ba
\frac{1}{r^2} \, \frac{\text{d}}{\text{d}r} \, r^2 \,  \frac{\text{d}}{\text{d}r} \, \rho & = & \frac{\rho_0}{2 a_1} \, \big[ \epsilon_n^F + \epsilon_p^F - \mu_n - \mu_p \big] \, , \\[.5ex]
\mu_n - \mu_p & = & \epsilon_n^F - \epsilon_p^F \, ,
\ea
where $\epsilon^F$ are Fermi energies that include here the Coulomb, symmetry-energy and momentum-dependent
contributions. The density difference $\rho_n - \rho_p$ is implicitly contained in the Fermi energies. The
set is solved by assuming $\mu_n$ and $\mu_p$ and $\rho(r=0)$.  The acceptable $\rho(r=0)$ is one for
which $\rho=0$ is reached together with $\text{d}\rho/\text{d}r=0$, ensuring a parabolic termination for
the density distribution.  The chemical potentials $\mu_n$ and $\mu_p$ are adjusted until the required neutron
and proton numbers are achieved for the nucleus.

At the start of a transport calculation the test-particles are distributed in space according to the
densities $\rho_n(r)$ and $\rho_p(r)$ from the TF calculation, and in momentum according to the local
momentum distribution, consistently with the physical content of the TF equations. The distribution is
then Lorentz boosted to give the nucleus desired velocity in the chosen calculational frame.

\subsubsection{Feeding and Absorption Rates}

For the sake of numerical calculations, the rates on the r.h.s.\ of the Boltzmann equations \eqref{eq:Boltzmann}
are averaged out over the volume elements $\Delta V$. The integrals in the rates are evaluated within a
Monte-Carlo procedure relying on the test particles there. This in practice makes the test-particles interact
across the content of the cell $\Delta V$.

The averaging over the volume elements is particularly beneficial in calculating the effects of 3- and
4-nucleon collisions where clusters are produced. Physically, the production and breakup of clusters should
take away probability flux from elastic scattering, but that is not done in the transport calculations.
Given that clusters are unlikely to survive when they are produced at excessive density and the repeated
processes of cluster breakup and reformation would yield excessively strong dissipation, cluster formation
is suppressed, in a tempered fashion, above a subnormal density cut-off. That cut-off is applied in addition
to the Pauli suppression mentioned below.

\subsubsection{Statistical Factors}

In calculating the final-state statistical factors, an averaging is employed for the distributions of nucleonic species over that cell of volume $\Delta V$, where the process occurs, and over the adjacent cells with a reduced, gradually changing, weight.  The aim of the tempering is a reduction of fluctuations for a given number ${\mathcal N}$ of test-particles per particle.  Beyond that, the value of the distribution for nucleonic species at a phase-space location is calculated with one of the following two methods.  In one method, the local momentum distribution is modeled in terms of a superposition of two deformed Fermi distributions, one for the species originally from the projectile and the other from the target.  Moments of the distributions, following the above averaging, are used to determine the distribution parameters, with the number of parameters trimmed if the number of test particles representing the species is low.  In the other method, the phase-space distribution functions are averaged over a volume in momentum space, in addition to the aforementioned spatial averaging.

Cluster states dissolve when the cluster wavefunction is significantly impacted by antisymmetrization at the nucleonic level with the exterior \cite{Danielewicz:1991dh,Beyer04}.  Whenever phase space occupation averaged over cluster phase-space volume exceeds a critical value, the cluster is prevented from forming.  This eliminates clusters from the phase-space areas where nucleonic occupations are high.  Notably, if this were not employed, the nucleonic Fermi spheres would undergo conversion into clusters.

\subsection{The Parton-Hadron-String-Dynamics (PHSD) approach} \label{phsd}
\vskip 0.1in
E. Bratkovskaya, W. Cassing
\vskip 0.1in

\subsubsection{Introduction}

The Parton-Hadron-String Dynamics (PHSD)
~\cite{Cassing:2009vt,Bratkovskaya:2011wp,Linnyk:2015rco,Moreau:2019vhw}
is a microscopic covariant  transport approach for the dynamical  description of strongly interacting
hadronic and partonic matter created in heavy-ion collisions.
It is based on a solution of the Cassing-Juchem generalized off-shell transport equations for test particles \cite{Cassing:1999wx,Cassing:1999mh}, which are derived from the  Kadanoff-Baym equations \cite{KadanoffBaym} in first-order gradient expansion \cite{Juchem:2004cs,Cassing:2008nn}.
This quantum field theoretical basis
distinguishes the PHSD from the semi-classical BUU based models, since the PHSD  propagates Green's functions
(in phase-space representation) which contain information not only on the occupation probability
(in terms of the phase-space distribution functions), but also on the properties of hadronic and partonic degrees-of-freedom via their spectral functions.
The PHSD approach consistently describes the full evolution of a relativistic heavy-ion collision
from the initial hard scatterings and string formation through the dynamical deconfinement phase transition
to the strongly-interacting quark-gluon plasma (sQGP) as well as hadronization and the subsequent
interactions in the expanding hadronic phase in line with the Hadron-String-Dynamics (HSD) transport approach \cite{Ehehalt:1996uq,Cassing:1999es}. The partonic dynamics is realized within the Dynamical Quasiparticle Model (DQPM) that
is constructed to reproduce lattice QCD (lQCD) results for a non-perturbative quark-gluon plasma in thermodynamic equilibrium at finite temperature $T$ and baryon (or quark) chemical potential $\mu_B$ on the basis of
effective propagators for quarks and gluons. Since the QGP dynamics is
not of relevance for the studies addressed here, we abstain from a further explicit description and refer the
interested reader to the review \cite{Linnyk:2015rco} and Ref. \cite{Moreau:2019vhw}.

The hadronic part is governed by the Hadron-String-Dynamics (HSD) part of the transport approach
\cite{Ehehalt:1996uq,Cassing:1999es} incorporating (optionally) self-energies for hadrons \cite{Cassing:2003vz}. The hadronic degrees-of-freedom include the baryon octet and decouplet, the ${0}^{-}$ and ${1}^{-}$ meson nonets as well as some higher resonances.
The low energy hadronic reactions are calculated based on the corresponding hadron-hadron cross sections -
either parametrized to the available experimental data or based on some microscopic chiral or one-boson exchange models (OBE).
With increasing energy the description of multi-particle production in elementary baryon-baryon ($BB$),
meson-baryon ($mB$) and meson-meson ($mm$) reactions is realized within the Lund model \cite{NilssonAlmqvist:1986rx}, in terms of the event generators FRITIOF~7.02 \cite{NilssonAlmqvist:1986rx,Andersson:1992iq} and PYTHIA~6.4 \cite{Sjostrand:2006za},  which are "tuned", i.e., adjusted, to get a better agreement with experimental data
on elementary $p+p$ collisions, especially at intermediate energies (cf. Ref.~\cite{Kireyeu:2020wou}).  
Furthermore,  the PHSD  incorporates the description of the chiral symmetry restoration via the Schwinger mechanism for the string decay \cite{Cassing:2015owa,Palmese:2016rtq} in a dense medium,
as well as in-medium effects such as a collisional broadening of the vector-meson spectra functions \cite{Bratkovskaya:2007jk} and the modifications of strangeness degrees-of-freedom in line with many-body G-matrix calculations \cite{Cassing:2003vz,Song:2020clw}.
Moreover, the implementation of detailed balance on the level of  $2\leftrightarrow 3$
reactions is realized for the main channels of strangeness production/absorption
by baryons ($B=N, \Delta, Y$) and pions \cite{Song:2020clw}, as well as for the multi-meson fusion reactions $2\leftrightarrow n$ for the formation of $B +\bar B$  pairs  \cite{Cassing:2001ds,Seifert:2017oyb}.
We note that although the excitation of strings and multi-particle production become of relevance above about 1 AGeV, they are not important in the energy range of a few hundred AMeV, where the scattering of protons and neutrons, the excitation of $\Delta$-resonances and their decay as well as the dynamics of pions are of primary importance.


The history of the (P)HSD code, similar to that of the GiBUU code, started in 1986
since both models emerged from the same origin, cf. Ref. \cite{Bauer:1986zz}, however, with a different focus on
reactions and energy range. The implementation of string formation and decay was carried out (in HSD) in 1996 \cite{Ehehalt:1996uq} and the implementation of off-shell propagation for selected hadrons in the year 2000 \cite{Cassing:1999wx}.
In 2009 the HSD approach was extended to the partonic sector (PHSD).
The propagation of hadrons and partons  in self-generated electromagnetic fields was included in 2011 \cite{Voronyuk:2011jd}.
The non-perturbative incorporation of charm and bottom degrees-of-freedom followed in 2016 \cite{Song:2015sfa}, and in 2019 the description of the QGP phase
was upgraded by introducing the explicit $\mu_B$ dependence of partonic complex self-energies
and cross sections \cite{Moreau:2019vhw}.
Moreover, in 2020 the PHSD gave birth to the PHQMD  (Parton-Hadron-Quantum-Molecular-Dynamics) approach \cite{Aichelin:2019tnk}, a novel branch of the PHSD, where the mean-field propagation of baryons is replaced by  Quantum Molecular Dynamics (QMD) to allow to follow the $n$-body dynamical evolution (similar to the IQMD approach), which is of relevance for the study of cluster and hypernuclei formation.

The PHSD approach is applicable to  $p+p$, $p+A$, $A+A$ collisions  as well as to $\pi +p$, $\pi+A$ reactions
from  SIS to LHC energies. It has been extensively used to study the dynamics of hadronic and partonic degrees-of-freedom as well as electromagnetic probes - dileptons and photons \cite{Linnyk:2015rco}.
PHSD also provides the possibility to study the thermal properties of the system in equilibrium by initializing
the system  in a 'box' with periodic boundary conditions or in a static 'brick'.  Furthermore, a `coarse-grained' calculation of the time evolution of the 
energy-momentum tensor $T_{\mu\nu}$ is implemented, which can be used  as 'initial conditions'
for hydrodynamic models or for the time evolution of fireball-like models.
The PHSD source code is publicly available since 2004 via registration from the PHSD web-page \cite{PHSDweb}.

\subsubsection{Initialization}
The phase-space densities of the two nuclei are described in the local Thomas-Fermi (LTF) approximation with density profiles taken from experimental data, i.e. of harmonic oscillator type for masses up to A=16 and Woods-Saxon type for larger masses in case of non-relativistic systems. In case of relativistic systems, the nuclei are initialized in the relativistic Thomas-Fermi approximation employing the same scalar and vector self-energies as used for the propagation of baryons. In the present case, the LTF is used,
\begin{equation}
f_{n,p}({\bf r}, {\bf p}) = \Theta(p_F^{n,p}({\bf r})-|{\bf p}|),
\end{equation}
for protons and neutrons with the local Fermi momentum
\begin{equation}
p_F^{n,p}({\bf r}) = \left( 3 \pi^2 \rho_{n,p}({\bf r}) \right)^{1/3} .
\end{equation}
 The two nuclei are shifted in space according to the actual impact parameter $b$ (in $x$-direction) and boosted towards each other (in $z$-direction) according to the bombarding energy of interest.

The nucleons are then propagated in time according to the Hamilton equations of motion employing the parallel ensemble method, i.e., solving for typically $\sim 10^3$ ensembles simultaneously. Particles are only allowed to scatter within the same ensemble but their mean fields are evaluated by ensemble averaging. This leads to rather smooth density distributions and mean fields with 'soft' gradients in space, which is mandatory for the time integration of the equations of motion.

\subsubsection{Potentials: non-relativistic}
In the case of non-relativistic systems, a Skyrme-type mean field is employed for baryons.
The Coulomb potential for the charged particles is also incorporated. For the study here, the asymmetry energy is included by using the following mean-field potentials for protons and neutrons,
\begin{equation}
\label{meanfieldsym}
U_{n,p}({\bf r}) = A \frac{\rho({\bf r})}{\rho_0} + B \left( \frac{\rho({\bf r})}{\rho_0} \right)^\lambda \pm
\left(\frac{\rho_n({\bf r})-\rho_p({\bf r})}{\rho_n({\bf r})+\rho_p(\bf r)} \right) 
U_{sym}\left(\frac{\rho_n({\bf r})+\rho_p(\bf r)}{\rho_0}\right)
\end{equation}
with
\begin{equation}
U_{sym}(\frac{\rho}{\rho_0}) = 12.3 \left( \frac{\rho}{\rho_0} \right)^{2/3} + 20 \left( \frac{\rho}{\rho_0} \right)^{\gamma}
\hspace{1cm} {\rm MeV}.
\end{equation}
Here the exponent $\gamma$ is taken as $\gamma=0.5$ for the 'isosoft' case and $\gamma =2$ for the 'isohard' version; the default version uses $\gamma=1$. In Eq. (\ref{meanfieldsym}) the (+)-sign holds for neutrons and the (-)-sign for protons.
The $\rho({\bf r})$ is the local baryon density calculated in the corresponding grid cell by averaging over all parallel ensembles.
Resonances are not incorporated in the calculation of $\rho_n$ and $\rho_p$ but are propagated in the isospin symmetric mean field, which is  reduced by a factor 2/3.
The parameters $A,B$ and $\lambda$ are determined by the binding energy per nucleon of -16 MeV at saturation density $\rho_0 = 0.166 fm^{-3}$ and the nuclear incompressibility $K$, which can be chosen in the range 210 MeV $\leq K \leq$ 380 MeV to explore the sensitivity of observables on the nuclear equation of state (EoS). The default value for $K$ is 300 MeV in the HSD code. The kinematics is relativistic throughout.

\subsubsection{Potentials: relativistic}
In the case of relativistic systems, the covariant mean-field (RMF) model is employed with an attractive scalar self energy $\Sigma_s({\bf r};t)$ and a repulsive 4-vector self energy $\Sigma_\mu ({\bf r};t))$ which is taken proportional to the baryon 4-current $\j_\mu$. The scalar self energy incorporates self interactions up to 4th order and is obtained by iteration of a gap-equation in each space-time cell. The parameters of the RMF model are fixed by the binding energy per nucleon at $\rho_0$, the choice of the incompressibility $K$, the effective nucleon mass $m^*$ at density $\rho_0$ and proton-nucleus Schr\"odinger equivalent potential
\begin{equation}
U_{SEP}(E) = -\Sigma_s + \frac{E}{m} \Sigma_0 + \frac{1}{2m}(\Sigma_s^2- \Sigma_0^2),
\end{equation} where $m$ denotes the nucleon mass and the energy $E=\sqrt{{\bf p}^2 +(m-\Sigma_s)^2}$.
Actual parameter-sets are listed in Ref. \cite{Palmese:2016rtq}. Due to the numerical complexity and large scalar and vector self energies of opposite sign, high statistics calculations are mandatory to achieve stable nuclei and robust results in particular for the space-time gradients of the self energies. The actual version of (P)HSD does not incorporate isospin-dependent self energies in the covariant mode and thus is not employed for the studies in this work.

\subsubsection{Collisions}
Baryon-baryon collisions with invariant energies below $\sqrt{s}$ = 2.6 GeV are treated by resonance excitation and decays; the same holds for meson-baryon collisions below $\sqrt{s}$ = 2.35 GeV. Above these thresholds, the string excitation model \cite{Andersson:1992iq} is incorporated producing 'leading' hadrons and 'pre-hadrons'. The 'pre-hadrons' are only allowed to scatter after a formation time $\tau_F$=0.8 fm/c (in their rest frame), while the 'leading' hadrons may scatter instantly with a reduced cross section in line with their fractional number of constituent quarks/antiquarks. The multitude of hadronic cross sections and (optional) mesonic self energies are described in the review \cite{Cassing:1999es} but are not of central interest here.

\subsubsection{Pauli blocking}
The Pauli-blocking factors $(1-f_{n,p}({\bf r}, {\bf p}))$ require a proper determination of the phase-space occupation $f_{n,p}$, which is described here in the non-relativistic version. Furthermore, neutrons and protons are not separated explicitly and treated as nucleons. Since the computational phase-space is a sum over $\delta$-distributions the average occupation factor ${\tilde f}({\bf r},{\bf p})$ has to be calculated in a suitable phase-space volume.  The evaluation proceeds by counting all nucleons in a spherical volume $\Delta V_R({\bf r},r_R)$ around ${\bf r}$ with radius $r_R$ and the spherical momentum-space volume $\Delta V_P({\bf p},r_P)$ around ${\bf p}$ with radius $r_P$,  \begin{equation}
\label{Pauli1}
{\tilde f}({\bf r}, {\bf p}) = \frac{1}{{\cal N}} \sum_{j \epsilon \Delta V_P\cdot \Delta V_R} 1
\end{equation}
with the normalization
\begin{equation}
{\cal N} = \frac{g \Delta V_R \Delta V_P N}{(2 \pi)^3} ,
\end{equation}
where $g=4$ denotes the degeneracy factor for spin and isospin.
Typical parameters are $r_P= $90 MeV/c and $r_R$ = 1.8 fm, which works reasonably for a large number of ensembles $N \geq $ 500 and heavy-ion collisions above about 50 AMeV. However, for lower bombarding energies this recipe does not work well due to the finite size of the phase-space cell in Eq. (\ref{Pauli1}) and the surface effects have to be appropriately accounted for.

\subsection{The relativistic Boltzmann-Uehling-Uhlenbeck (RBUU) code}
\vskip 0.1in
T. Gaitanos 
\vskip 0.1in

In this short write-up, we provide information on the RBUU code. The main references are~\cite{gaitanos01,Ferini1}.

\subsubsection{Code history}

\begin{itemize}

\item Munich, 1995:\\
The RBUU code was developed in Munich by C. Fuchs in the middle nineties~\cite{fuchs95},
where the Landau-Vlasov method was introduced.
\item Munich, 1996-2000: \\
Implementation of Density-Dependent RMF models and of DBHF approaches within the Local
Configuration Approximation (LCA) by Gaitanos and Fuchs~\cite{gaitanos99,gaitanos01}.
\item Munich/Catania, 2002-2005:\\
Implementation of iso-vector mean-fields by Gaitanos~\cite{gaitanos04}. Implementation of
isospin effects in the production thresholds in the collision integral by Ferini
and Gaitanos~\cite{Ferini1,Ferini2}.
\item Munich/Catania/Thessaloniki, 2005-2010:\\
Further improvement of kaon potentials. Implementation of in-medium isospin effects in cross
sections and kaon potentials by Prassa and Gaitanos~\cite{prassa}.
\end{itemize}

\subsubsection{Initialization}

The initial distribution of neutrons and protons of a nucleus of interest is performed
by using the Fermi function in coordinate space and by a theta function in momentum space.

In coordinate space, the nucleus is initialized by  fitting to the Fermi function,
\begin{equation}
F(r) =\frac{1}{1+\exp [(r-R)/a]}\, ,
\end{equation}
where the parameters, $R$ and $a$, are obtained from the relativistic Thomas-Fermi (RTF) calculations.
In the (relativistic) Thomas-Fermi calculations, we first express the total energy of the system in terms of
densities. For a specific nucleus with $Z$ protons and $N$ neutrons, we obtain RTF equations for mean fields by minimizing
the total energy with fixed number of protons and neutrons,
\begin{equation}
\delta \int d^3{\bf r} [\epsilon (\rho_p({\bf r}),\rho_n({\bf r})) -\mu_p\rho_p ({\bf r})-\mu_n\rho_n ({\bf r})]=0\, .
\end{equation}

The (test) particle momenta are randomly distributed in an isotropic Fermi sphere, $f_{n,p} ({\bf r}, {\bf k}) =\Theta \left( k_{{\rm F},n,p}({\bf r}) -|{\bf k}| \right)$, whose radius is determined by the local Thomas-Fermi approximation, $k_{{\rm F},n,p}({\bf r}) = \left(3\pi^2 \rho_{n,p} ({\bf r}) \right)^{1/3}$.
The test particle in the RBUU code has $1.4$ fm gaussian width in coordinate space and $0.346$ fm$^{-1}$ width in momentum space.

To ensure the stability of initialized nuclei, we used the same energy density functional both in the RTF calculations and in the dynamical evolution.

\subsubsection{Forces}
In the RBUU code, a relativistic mean field model is used to obtain a mean-field potential under which
particles propagate as described by the RBUU equation.
The relativistic Lagrangian density adopted in the RBUU code is given by,
\begin{eqnarray}
{\mathcal L}&=&\bar\psi [\gamma_\mu\partial^\mu -(M_N-g_\sigma\sigma-g_\delta\vec\tau\cdot \vec\delta)
-g_\omega\gamma_\mu\omega^\mu-g_\rho\gamma^\mu\vec\tau\cdot\vec\rho_\mu]\psi
+ \frac{1}{2}(\partial_\mu\sigma\partial^\mu\sigma-m_\sigma^2\sigma^2) -U(\sigma) \nonumber\\
&&+\frac{1}{2}m_\omega^2\omega^2  +\frac{1}{2} m_\rho^2\vec\rho_\mu\cdot{\vec\rho}^\mu
+\frac{1}{2}(\partial_\mu\vec\delta\cdot\partial^\mu\vec\delta-m_\delta^2{\vec\delta}^2) -\frac{1}{4}F_{\mu\nu}F^{\mu\nu}-\frac{1}{4}\vec G_{\mu\nu}\cdot \vec G^{\mu\nu}\, .\label{MFT}
\end{eqnarray}
Here $F_{\mu\nu}\equiv \partial_\mu \omega_\nu -\partial_\nu\omega_\mu$,
 $\vec G_{\mu\nu} \equiv \partial_\mu\vec\rho_\nu-\partial_\nu\vec\rho_\mu$,
 and $U(\sigma)=\frac{1}{3}a\sigma^3+\frac{1}{4}b\sigma^4$.
The propagation of the phase-space density $f({\bf x},{\bf k})$
under the influence of the mean-field potential obtained from Eq. (\ref{MFT}) and two-body collisions are described by
the relativistic Boltzmann-Uehling-Uhlenbeck (RBUU) equation,
\begin{eqnarray}
&&\left[ p_\mu^\star \partial_x^\mu +\left(p_\nu^\star F^{\mu\nu} +m^\star (\partial_x^\mu m^\star) \right)\partial_\mu^{p^\star} \right]
f({\bf x},p^\star) \\ \nonumber
&& = \frac{1}{2} \int \frac{d^3p_2}{E^\star_{p_2} (2\pi)^3}  \frac{d^3p_3}{E_{p_3}^\star (2\pi)^3}  \frac{d^3p_4}{E_{p_4}^\star (2\pi)^3}
W(pp_2|p_3p_4)\\ \nonumber
&& ~~~\times [f({\bf x},{\bf p}_3)f({\bf x},{\bf p}_4) (1-f({\bf x},{\bf p})) (1-f({\bf x},{\bf p}_2)) - f({\bf x},{\bf p})f({\bf x},{\bf p}_2) (1-f({\bf x},{\bf p}_3)) (1-f(x,{\bf p}_4))]\, .
\end{eqnarray}
$W$ is the transition probability for a scattering process $p+p_{2}\to p_{3}+p_{4}$, and it is
defined by
$$
W(pp_{2}|p_{3}p_{4})=(p+p_{2})^{2}\frac{d\sigma}{d\Theta}\,
\delta^{(4)}(p+p_{2}-p_{3}-p_{4})
\,.
$$
Furthermore, $m^\star=M_N-g_\sigma\sigma$ and $p_\mu^\star=p_\mu-g_\omega\omega^\mu$, where
$\sigma$ and $\omega^\mu$ denote the mean fields.

\subsubsection{Collision}
The collision term in the RBUU code originates from T{\"u}bingen (R)QMD, however, it has been
further developed by including all possible isospin channels of strangeness production and
isospin effects in the particle thresholds~\cite{Ferini1,Ferini2}.
Two particles collide if the  minimum relative distance is equal to or less than
$\sqrt{\sigma_{\rm tot}/\pi}$  within the time interval used in the propagation. $\sigma_{\rm tot}$,
which depends on the center of mass energy $\sqrt{s}$, is the total cross section for  a two-body process.
In the RBUU approach, the Lorentz invariant distance is used to compare with $\sqrt{\sigma_{\rm tot}/\pi}$.
The elastic nucleon-nucleon cross section is taken from a
parametrization by Cugnon et al. \cite{Cugnon:1996kh}. For the inelastic channels involving
resonance ($\Delta(1.232),~N^{*}(1.445)$) production, the analysis of Huber and Aichelin is
used~\cite{Huber:1994ee}. There is also an option to use in-medium reduced elastic and inelastic cross sections
by using the parametrizations taken from Dirac-Brueckner-Hartree-Fock calculations~\cite{Fuchs:2001fp,TerHaar:1987ce}.

The collision processes take place in the local c.m. system of the two colliding particles.
Once the elastic or inelastic scattering occurs, the modulus of the momentum of the final state
is calculated  according to energy-momentum conservation. The direction of the final state momentum
is distributed according to experimental differential cross sections, when available. For those
processes without any empirical information on differential cross sections, the momenta of
the final particles are distributed isotropically. After the calculation of the final state one
returns to the calculational frame, i.e., to the global c.m. system of the two colliding
nuclei, where the Pauli-blocking is checked.

\subsubsection{Pauli-blocking}
The Pauli-blocking factor is defined by $1-f({\bf x},{\bf p})$, where $f({\bf x},{\bf p})$
is calculated by counting the number of test particles within a sphere centered at
(${\bf x},{\bf p}$) in the phase space, which is fixed by the width of the Gaussian functions
in coordinate and momentum space.
If the blocking factor is larger than a generated random number, the collision is allowed.
If the collision is blocked, the associated particles take back their original momenta.

\subsection{The relativistic Vlasov-Uehling-Uhlenbeck (RVUU) code\label{Sec:RVUU}}
\vskip 0.1in
Z. Zhang, C. M. Ko, T. Song
\vskip 0.1in

\subsubsection{{Code history}}

\begin{itemize}
    \item 1987-1989{:} Ko, Li and Wang developed the RVUU model based on the Walecka model with scalar-isoscalar $\sigma$ meson and vector-isoscalar $\omega$ meson~\cite{Ko:1987gp,Li:1989zza}. 
    
    \item 1993-1998{:} Ko and Li extended the RVUU model to study particle production in high-energy heavy ion collisions~\cite{Ko:1996yy}.
    
    \item 2012{:} Li, Chen, Ko, and Lee included hyperon-hyperon scattering in RVUU to study the threshold cascade production in heavy-ion collisions~\cite{LiChenKo2012}.  

    \item 2015: Song and Ko extended the RVUU model to include the isospin degrees of freedom and the threshold effect on pion production, which was then used to study the symmetry energy effect on the charged pion ratio in medium-energy heavy ion collisions~\cite{Song:2015hua}. 
   
    \item 2016-2018: Zhang and Ko introduced the pion s-wave and p-wave potentials in the RVUU model and studied their effects on pion production in medium-energy heavy ion collisions~\cite{Zhang:2017mps}. Effects of mean-field potentials on the detailed balance relations in $N+N\leftrightarrow N+\Delta$ and $\Delta \leftrightarrow N+\pi$ reactions were validated in a box calculation~\cite{Zhang:2017nck}.

\end{itemize}
\subsubsection{Initialization}

In the RVUU model, the initial nucleon positions in a nucleus are distributed according to the density profile of the Wood-Saxon form,
\be
\rho(\bm{r}) \sim \frac{1}{1+\exp [(r-c) / a]}
\ee
with the parameters $c=1.1 A^{1 / 3}~ \mathrm{fm}$ and $a=0.535 ~\mathrm{fm}$. Whether a nucleon is a proton or a neutron is determined from the probability given by the ratio of the atomic to the mass number of a nucleus. Ordering protons and neutrons randomly helps to avoid possible artifacts due to labeling protons before neutrons or vice versa when checking their scatterings. The nucleon momenta in a nucleus are determined by assuming that nucleons in a local cell forms a Fermi gas with the Fermi momentum determined by the local density. Nucleons in the two colliding nuclei are then boosted to their center-of-mass frame by using the velocity $p / E$ for projectile nucleons and $-p / E$ for target nucleons, where $p$ and $E$ are, respectively, the total momentum and energy of the colliding system in the laboratory frame.

\subsubsection{Mean-field potentials}

The nucleon mean-field potentials in the RVUU model are taken from the non-linear relativistic mean-field (RMF) model with isoscalar scalar $\sigma$ and vector $\omega$ mesons as well as isovector scalar $\delta$ and vector $\rho$ mesons.  With the plus and minus signs for the proton and neutron, respectively, the single-nucleon energy or potential is given by
\be
U\equiv p_i^0=\sqrt{m_N^{*2} + {\bf p}^{*2}} + g_\omega \omega^0 \pm g_\rho\rho_3^0.
\ee
where $m_N^*=m_N-(g_\sigma\sigma\pm g_\delta\delta_3)$ and ${\bf p}^*={\bf p}+g_\omega{\boldsymbol\omega}\pm g_\rho{\boldsymbol\rho_3}$ are the nucleon in-medium mass and kinetic momentum, respectively, and the meson fields $\sigma$, $\delta_3$, $\omega^\mu$ and $\rho_3^\mu$ are related to the proton and neutron scalar densities and currents. 
The default parameters of the non-linear RMF model in the RVUU is the parameter set I 
in Ref.~\cite{Liu:2001iz}, which gives the nuclear matter saturation density $\rho_0 = 0.16~\mathrm{fm}^{-3}$, binding energy per nucleon $E_0(\rho_0)=-16$ MeV, incompressibility $K_0=240$ MeV, and the symmetry energy $E_{\mathrm{sym}}(\rho_0)=30.5$ MeV and its slope parameter $L=84$ MeV at the saturation density. The potential of a $\Delta$ resonances has a similar form in terms of its in-medium mass $m_i^* =~m_i-g_{\sigma}\sigma - x_i g_{\delta}\delta$ and kinetic momentum $p_i^{\mu*} =~p^{\mu}-g_{\omega}\omega^{\mu}-x_i g_{\rho}\rho^{\mu}$, where $x_i =1$, $1/3$, $-1/3$ and $-1$ for $\Delta^{++}$, $\Delta^+$, $\Delta^0$ and $\Delta^-$, respectively, from their isospin structures in terms of nucleon and pion. 

By default, pions are treated as free particles. The option of including the pion s-wave and p-wave potentials from calculations based on the chiral perturbation theory and the $\Delta$-hole model, respectively, is also available in the RVUU code~\cite{Zhang:2017mps}. 

Solving the RVUU equation for the baryon distribution function $f(\bm{x},\bm{p})$, i.e.,
\be
\frac{\partial}{\partial t}f+\bm{v}\cdot\bm{\nabla}_rf-\bm{\nabla}_r U\cdot\bm{\nabla}_pf=\mathcal{C},
\ee
with $\mathcal{C}$ denoting the collision integral, is carried out by the test particle method~\cite{Bertsch:1988ik} with $N_{\rm TP}$ test particles and including also the effect of the Coulomb potential on protons via the electric and magnetic fields. The test particle $i$, with electric charge $q_i$, then obeys the classical equations of motion, 
\ba
\dot{\bm{r}}_i &=&\frac{\bm{p}^{\ast}_i}{p_i^{0\ast}}, \notag\\
\dot{\bm{p}}_i &=& - \bm{\nabla} p_i^0+q_i(\bm{E}_i+\dot{\bm{r}}_i\times \bm{B}_i).
\ea
In the above, the electric and magnetic fields acting on particle $i$ are evaluated according to
\ba
\bm{E}_i &=&\frac{1}{4\pi N_{\rm TP}}\sum_{i\neq j}q_j\frac{\bm{r}_{ij}}{r_{jk}^3}, \notag\\
\bm{B}_i & = & \frac{1}{4\pi N_{\rm TP}}\sum_{i\neq j } q_j
\frac{\dot{\bm{r}}_{j} \times \bm{r}_{ij} }{r_{ij}^3},
\ea
with  $\bm{r}_{ij}=\bm{r}_i-\bm{r}_j$, and the sum running over all test particles.

\subsubsection{Two-body scatterings}

The RVUU code implements the geometrical minimum distance criterion for a two-body scattering. The two particles are first transformed to their center-of-mass frame where the relative distance is separated into components $r_{\parallel}$ parallel and $r_{\bot}$ perpendicular to their relative velocity $\bm{v}_{\mathrm{rel}}$. A trial collision occurs  if 
\be
    r_{\bot}  < \sqrt{\frac{\sigma_{\mathrm{tot}}}{\pi}},\quad
    r_{\parallel} < v_{\text {rel}} \frac{\Delta t }{ 2},
\ee
where $\sigma_{\mathrm{tot}}$ is the total cross section of the two scattering particles, and $\Delta t$ is the time step. A random number is then generated to determine the reaction channel based on the ratio $\sigma_i/\sigma_{\mathrm{tot}}$ with $\sigma_i$ being the cross section for channel $i$. Finally, after determining the momenta of the outgoing two particles according to the total energy and momentum conservation, the Pauli blocking is checked in the computational frame.

Unlike normal transport models that use vacuum kinematics in collisions (and decays), i.e., 
only the kinetic energy is conserved, the RVUU takes into account the conservation 
of total energy by including also the nucleon and $\Delta$ potentials, and the effect of modified threshold energy due  to the potentials~\cite{Song:2015hua}. 
Taking a reaction $1+2\rightarrow 3+4$ for example, the RVUU guarantees 4-momentum conservation, $p_1+p_2 = p_3+p_4$, and the possible minimum invariant mass $\sqrt{s}_{\mathrm{th}}$ of the final two particles (i.e., the threshold energy in inelastic scatterings) is given by ~\cite{Song:2015hua} 
$$
\sqrt{s}_{\mathrm{th}}=\sqrt{(m_3^*+m_4^*+\Sigma_3^0+\Sigma_4^0)^2-
\vert \bm{\Sigma}_3+\bm{\Sigma}_4 \vert^2},
$$
where $m_3^*$ and $m_4^*$ are the effective masses of the two final particles, and $\Sigma_3^{\mu}$ and $\Sigma_4^{\mu}$ are their mean-field potentials in the frame where the total kinetic momentum vanishes, i.e., $\bm{p}_3^{\ast}+\bm{p}_4^{\ast}=0$.

The inclusion of mean-field potentials in collision kinematics also affects 
the detailed balance relations in $N+N\leftrightarrow  N+\Delta$ and $\Delta \leftrightarrow  N+\pi$ reactions. 
Specifically, the mass of the produced $\Delta$ resonance is determined according to 
 \be
 P\left(m^{*}\right)=\frac{\mathcal{A}(m^*) p^{*}}{\int_{m^*_{\min }}^{m^*_{\max }} d m^{*\prime} \mathcal{A}\left(m^{*\prime}\right) p^{*}\left(m^{*\prime}\right)},
 \ee
where $m^{*}$ is the effective mass of $\Delta$, and $m^*_{\min }$ and $m^*_{\text {max }}$ are, respectively,  the minimum and maximum allowed effective masses of $\Delta$ in the $NN\to N\Delta$ reaction. The  $\mathcal{A}(m^*)$ is the in-medium $\Delta$ spectral function, given by
 \begin{eqnarray}
 \mathcal{A}(m^*)=\frac{1}{\mathcal{N}} \frac{4 m_{0}^{*2} \Gamma}{\left(m^{*2}-m_{0}^{*2}\right)^{2}+m_{0}^{*2} \Gamma^{2}},
 \end{eqnarray}
 where  $\mathcal{N}$ is the normalization factor, $m^*_0$ is the $\Delta$ pole mass of $1.232$ GeV shifted by the scalar potential, and $\Gamma$ is the $\Delta$ total decay width. The $\Delta$ absorption cross section $\sigma_{N\Delta\rightarrow NN}$ is related to the $\Delta$ production cross section $\sigma_{NN\rightarrow N\Delta}$ by
 $$
 \begin{aligned}
    \sigma_{N\Delta\rightarrow NN}= \frac{\sigma_{NN\rightarrow N\Delta}}{2\left(1+\delta_{12}\right)} \frac{2 \pi p_{1}^{* \prime \prime}}{\int d m \mathcal{A}(m) p_{4}^{* \prime}(m)} \frac{E_{3}^{* \prime}+E_{4}^{* \prime}}{E_{1}^{* \prime \prime}+E_{2}^{* \prime \prime}} \frac{\left\vert E_{1}^{*^{\prime}} p_{2}^{*^{\prime}}-E_{2}^{*^{\prime}} p_{1}^{*^{\prime}}\right\vert}{\left\vert E_{3}^{* \prime \prime} p_{4}^{* \prime \prime}-E_{4}^{* \prime \prime} p_{3}^{*^{\prime \prime}} \right\vert}
    \end{aligned},
 $$
where $1$, $2$, $3$ and $4$ indicate the final two nucleons and the initial nucleon and $\Delta$, respectively.  The factor $1/(1+\delta_{12})$ takes into account the case that $1$ and $2$ are identical particles, and the single  and  double primes indicate quantities in the frames of $\bm{p}_3^*+\bm{p}_4^*=0$ and $\bm{p}_1^*+\bm{p}_2^*=0$, respectively.

The pion absorption cross section $\sigma_{N\pi\rightarrow \Delta}$ is related to the partial decay width of $\Delta$ by 
$$
\sigma_{N \pi \rightarrow \Delta}=\frac{2 \pi}{p_{N}^{* 2}} \mathcal{A}\left(m_{\Delta}\right) \Gamma\left(m_{\Delta}\right),
$$
with $\bm{p}_N^{\ast}$ being the kinetic momentum of the scattering nucleon in the frame of  $\bm{p}_N^{\ast}+\bm{p}_{\pi}=0$.  Note that both $\sigma_{N\pi\rightarrow\Delta}$ and $\Gamma\left(m_{\Delta}\right)$ refer to the same isospin channels of the process $\Delta \leftrightarrow N\pi$, while the $\Delta$ width in $\mathcal{A}(m_{\Delta})$ refers to the total width of the $\Delta$ resonance. The detailed balance conditions introduced above have been validated in a box calculation in Ref.~\cite{Zhang:2017nck}.

\subsubsection{Pauli blocking}

Considering a trial collision of two particles with their final positions and momenta given by $(\bm{r}_1,\bm{p}_1)$ and $(\bm{r}_2,\bm{p}_2)$, the blocking probability due to the Pauli principle is $1-[1-f_1(\bm{r}_1,\bm{p}_1)][1-f_2(\bm{r}_2,\bm{p}_2)]$. To determine the phase-space density $f_i(\bm{r}_i,\bm{p}_i)$, the RVUU code counts the number $N_i$ of test particles of the same particle species $i$ inside a spherical cell of the phase-space centering at $(\bm{r}_i,\bm{p}_i)$.  The $f_i(\bm{r}_i,\bm{r}_p)$  is then calculated according to 
\be
f_i = \frac{N_i}{gN_{\mathrm{TP}}}\frac{h^3}{(4\pi R_r^3/3)(4\pi R_p^3/3)},
\ee
where $g$ is the spin degeneracy and $R_r$ ($R_p$) is the radius of the sphere in coordinate (momentum) space. For $N_{\mathrm{TP}}=100$, $R_r$ and $R_p$ are taken to be $2$ fm and $100~\mathrm{MeV/c}$, respectively. The accuracy of  the Pauli blocking calculation can be improved with a larger $N_{\mathrm{TP}}$ and smaller $R_r$ and $R_p$ in the treatment of the collision integral in the RVUU model.



\subsection{SMASH (Simulating Many Accelerated Strongly-interacting Hadrons)} \label{sec:SMASH}
\vskip 0.1in
D. Oliinychenko, H. Elfner, A. Sorensen
\vskip 0.1in
\subsubsection{Code history}

SMASH \cite{Weil:2016zrk} is a new C++ relativistic transport code currently developed by the
group of Prof.\ Elfner (n\'{e}e Petersen) at the Frankfurt Institute for Advanced Studies, and it is partly inspired by features and implementation ideas from UrQMD \cite{Bass:1998ca,Bleicher:1999xi} and GiBUU (see Section \ref{sec:GiBUU}). The degrees of freedom evolved in SMASH are hadrons and strings. The list of hadrons includes all hadrons with *** and **** status from the Particle Data Group \cite{Olive:2016xmw}, while string fragmentation is performed within Pythia 8 \cite{Sjostrand:2014zea}. Hadronic reactions include $2 \to 1$ resonance formations, $1\to 2$ decays, as well as $2\to 2$ reactions: $NN \leftrightarrow NN^*$,
$NN\leftrightarrow N \Delta^*$, and strangeness exchange reactions. All hadronic reactions fulfill the principle of detailed balance, which in practice means that $1 \to 3$ decays are replaced by a chain of two $1 \to 2$ decays. Recently, $2 \leftrightarrow 3$ particle scatterings \cite{Staudenmaier:2021lrg} and $p\bar{p} \leftrightarrow 5\pi$ annihilations \cite{Garcia-Montero:2021haa} were also introduced. In addition to hadrons, SMASH can also simulate photon and dilepton emission \cite{Schafer:2019edr,Staudenmaier:2017vtq} and deuteron production \textit{via} explicit reactions \cite{Oliinychenko:2018ugs}.

SMASH has four simulation modi: ``collider'' for collisions of hadrons and nuclei, ``box'' for infinite matter simulations, ``sphere'' for simulations of spherically initialized expanding matter as well as comparisons with an analytical solution of the Boltzmann equation \cite{Tindall:2016try}, and ``list'' for operating as an afterburner for hydrodynamic simulations. Mean-field potentials can be optionally included in all modi. A treatment of the test-particle ansatz known as the full ensemble is adopted: the number of test particles sampled to describe a system of $A$ particles is increased by a factor of $N_{\rm test}$, resulting in $N = AN_{\rm test}$ test particles evolved in the simulation, while all cross sections are scaled by the factor $N_{\rm test}^{-1}$, so that the average number of scattering events per test particle is the same as in a system of $A$ particles with unscaled cross sections. Each of the $N$ test particles contributes to the calculation of mean fields with a scaling factor of $N_{\rm test}^{-1}$, which in particular ensures that the total charge evolved in the simulation corresponds to the charge of $A$ particles. Recently, the development version of SMASH has been extended by the possibility to use the parallel-ensembles technique, within which $N_{\rm test}$ separate systems of $A$ particles are evolved simultaneously. In this method, scatterings are performed only between test particles that belong to the same system, so that cross sections need not be scaled, while at the same time, the mean field is calculated based on all $N_{\rm test}$ instances of the system, where the contribution of each test particle is again scaled by a factor of $N_{\rm test}^{-1}$. Using a large number of test particles per particle, $N_{\rm test}$, is necessary in the mean-field calculation, both in the full ensemble and parallel ensemble approach, to suppress fluctuations due to the finite number of sampled particles.

The current applicability range of the code is for collisions satisfying $E_{\rm lab} \gtrsim  0.5 \AGeV$. The improvements to the description of mean fields necessary to operate at lower energies are currently in development.

\subsubsection{Initialization}

The initialization of the nucleus is achieved as follows. First, the coordinates of a given nucleon $\bm{r}$ are
sampled from the Woods-Saxon distribution, $\rho(r') = \left( 1 + e^{\frac{r' - R}{d}}\right)^{-1}$. Then, the local Fermi momentum at a point $\bm{r}$ is computed based on the value of the local density given by $\rho(|\bm{r}|)=\rho(r)$, $p_F(r) = \hbar c \big(3 \pi^2 \rho(r) \big)^{1/3}$, and the nucleon momentum in the rest frame of the nucleus $p^i_{f}$ is sampled from the Fermi sphere of radius $p_F(r)$. This process is repeated for all initialized nucleons. Finally, the boost of momenta $p_{f}^i$ to the computational frame is performed using $p^i_{\rm comp} = p_{\rm beam} + \gamma p^i_{f}$, where $p_{\rm beam}$ is the beam momentum per nucleon and $\gamma$ is the corresponding relativistic gamma factor. No additional effort is taken to obtain the nucleus in the ground state. The effects of nucleon-nucleon correlations and neutron skin are also neglected. Recently, a nucleus initialization from an external list of particles and their properties was implemented; in particular, such a list can be generated by an approach that takes into the account the physical effects neglected in the SMASH nucleus initialization.

The box can be initialized in the canonical or the grand-canonical ensemble. The distribution of coordinates in the box is uniform and the momenta are sampled from the Boltzmann, Fermi, or Bose distributions. Multiplicities in the grand-canonical ensemble are always sampled from the Poisson distribution, regardless of the chosen Boltzmann, Fermi, or Bose statistics.

\subsubsection{Mean-field potentials}

When mean-field potentials are used, SMASH samples and evolves the kinetic momenta of the test particles \cite{Blaettel:1993uz} (an alternative approach, in which the canonical momenta are propagated, is possible \cite{Ko:1988zz}). By default, SMASH is using the same nuclear Skyrme potential that was suggested in comparison 2 (Sec.~\ref{HIC_100AMEV}). The potential is calculated as a function of the local density,
\begin{equation}
U = a \left(\frac{\rho_B}{\rho_0}\right) + b \left(\frac{\rho_B}{\rho_0}\right)^{\tau} \pm 2 S_\text{pot} \left(\frac{\rho_{I3}}{\rho_0}\right) \,.
\label{SMASH_Skyrme}
\end{equation}
Here, $\rho_B$ is the Eckart rest frame baryon density, $\rho_0$ is the saturation density of nuclear matter, and $\rho_{I3}$ is the Eckart rest frame baryon
isospin density of the relative isospin projection $I_3/I$. The default values of the parameters are
$\rho_0=0.168\ \text{fm}^{-3}$, $a=-209.2$ MeV, $b=156.4$ MeV, $\tau = 1.35$, and $S_\text{pot} = 18$ MeV.
The potential is only exerted on baryons. No Coulomb or momentum-dependent potentials are implemented.

The Eckart rest frame density is obtained from the four-current
$j^{\mu}$ as $\rho = \rho^+ - \rho^- =  \sqrt{j^{+ \mu}j_{\mu}^+} - \sqrt{j^{- \mu}j_{\mu}^-}$, where~$+$
corresponds to a positive baryon number or isospin projection and $-$ corresponds to negative ones; a naive calculation from $\rho = \sqrt{j^{\mu}j_{\mu}}$ would fail for $j^{\mu}j_{\mu} < 0$.
The four-current is calculated as follows:
\begin{align}
j_B^{\mu}(\bm{r})    &= \frac{1}{N_\text{test}} \sum_i B_i  \frac{\Pi^{\mu}_i}{\Pi^0_i} K(\bm{r} - \bm{r_i}, \Pi_i) \,,
\end{align}
where $B_i$ is the baryon number of the $i$-th particle, $\Pi^{\mu}_i$ is its kinetic four-momentum, and $K$ is a relativistic
smearing kernel, shown in \cite{Oliinychenko:2015lva} to be a kernel with correct properties under the Lorentz-transformation,
\begin{eqnarray}
K(\bm{r} - \bm{r_i}, u, \sigma) = \frac{u_0}{(2\pi \sigma^2)^{3/2}}
\exp\bigg[-\frac{(\bm{r} - \bm{r_i})^2 +
	\big(\bm{u} \cdot (\bm{r} - \bm{r_i})\big)^2}{2\sigma^2}\bigg] \,, 
\end{eqnarray}
where $u^{\mu} = (u^0, \bm{u}) = \Pi^{\mu}/m$ should not be confused with a local collective velocity of a fluid.

In the development version of the code, different types of smearing kernels as well as the Coulomb potential were added. The equations of motion with relativistic particles but a non-relativistic potential as introduced in SMASH 1.0,
\begin{eqnarray}
&&\frac{d\bm{r}}{dt} = \frac{\bm{\Pi}~}{\Pi_0} \,, \\
&&\frac{d\bm{\Pi}}{dt} = - \bm{\nabla} U \,,
\end{eqnarray}
were updated to relativistic ones \cite{Mohs:2020awg} including magnetic-type forces,
\begin{align}
\frac{d\bm{\Pi}}{dt} = \frac{\partial U}{\partial \rho_B} \bigg[-\bigg(\bm{\nabla} j^0 + \frac{ \partial \bm{j}}{\partial t} \bigg)+  \frac{d\bm{x}}{dt} \times \big(\bm{\nabla} \times \bm{j}\big)\bigg] \,,
\label{Feng_Li's_EOMs_p}
\end{align}
and recently corrected to be fully Lorentz-covariant \cite{Sorensen:2020ygf},
\begin{align}
\frac{d \bm{\Pi}}{dt} =  \sum_{k=1}^K \left\{ - \Big(  \bm{\nabla} A_k^{0}  + \partial^{0} \bm{A}_k   \Big) + \frac{\bm{\Pi}~}{\Pi_0} \times \big( \bm{\nabla} \times \bm{A}_k  \big)  \right\}   ~,
\label{VDF_EOM}
\end{align}
where the sum is performed over a chosen number $K$ of vector fields $A_k^{\mu}$, 
given by
\begin{align}
A_k^{\mu}(x; \tilde{C}_k, b_k) = \tilde{C}_k \frac{\rho_B^{b_k - 2}}{\rho_0^{b_k -1}} j^{\mu}(x) ~,
\end{align}
and the parameters of the interaction, $\tilde{C}_k$ and $b_k$, can be fitted to reproduce the chosen properties of dense nuclear matter (e.g., the saturation properties and the critical point of nuclear matter for $K=2$, or these properties and in addition the characteristics of a postulated QCD critical point at high baryon number density for $K=4$, etc.). In particular, taking
\begin{align}
K = 2~, \hspace{10mm}   \tilde{C}_1 = a ~, \hspace{10mm} b_1 = 2 ~,  \hspace{10mm} \tilde{C}_2 = b ~, \hspace{10mm} b_2 = \tau + 1 ~, \hspace{10mm} \rho_0 = 0.168\ \textrm{fm}^{-3}
\end{align}
reproduces the isospin symmetric part of the original SMASH potential, Eq.\ \eqref{SMASH_Skyrme}.

\subsubsection{Collision term}

The same geometrical collision criterion as in  UrQMD \cite{Bass:1998ca,Bleicher:1999xi} is employed,
\begin{eqnarray}
&& d_{\rm trans} < d_{\rm int} = \sqrt{\frac{\sigma_{\rm tot}}{\pi}} \,, \\
&& d_{\rm trans}^2 = (\bm{r}_a-\bm{r}_b)^2-\frac{\big((\bm{r}_a-\bm{r}_b)\cdot(\bm{p}_a-\bm{p}_b)\big)^2}{(\bm{p}_a-\bm{p}_b)^2} \,,
\end{eqnarray}
where $\bm{r}_{a,b}$ and $\bm{p}_{a,b}$ are the coordinates and momenta of the colliding particles $a$ and $b$ in
their center-of-mass frame. The time of the collision is set to the time of the closest approach in
the computational frame,
\begin{equation}
t_{\rm coll}=-\frac{(\bm{r}_a-\bm{r}_b)\cdot (\bm{p}_a/E_a-\bm{p}_b/E_b)}{(\bm{p}_a/E_a-\bm{p}_b/E_b)^2} \,,
\end{equation}
where $E_{a,b}$ are energies of particles $a$ and $b$, and all coordinate and momentum vectors are taken in the computational frame. Within a single time step $\Delta t$, collisions and decays (``actions'') are ordered according to the time at which they occur in the computational frame, and particles are propagated along straight lines from action to action. After an action is performed, some new collisions or decays might become possible while others might become obsolete, so that the list of the time-ordered actions is updated at each action. If potentials are not employed, then the time step $\Delta t$ can be arbitrarily large and it was proven that the collision rate does not depend on it. On the other hand, if potentials are used, then the time step must be small, as changes in momentum due to potentials are performed after each $\Delta t$; unless it is specified by the user, $\Delta t = 0.1\ \text{fm}$ by default. 

There are two recent developments concerning the collision criterion: (i) the UrQMD collision criterion was replaced by its covariant generalization \cite{Hirano:2012yy}, and (ii) a stochastic collision criterion similar to pBUU (see Section \ref{sec:pBUU}) was implemented \cite{Staudenmaier:2021lrg,Garcia-Montero:2021haa}, which can optionally replace the geometric criterion.

The cross sections and resonance treatment in SMASH are described in detail in \cite{Weil:2016zrk}, while \cite{Steinberg:2018jvv} focuses on reactions related to strange hadrons. For the Transport Model Evaluation Project, the default SMASH reaction treatment is modified, as some cross sections are set to particular values or switched off entirely according to the comparison assignment.

\subsubsection{Pauli blocking}

To account for effects due to Pauli blocking, reactions with baryons in the final state are rejected with probability $1 - \prod_i (1-f_i)$, where the product is running over the outgoing baryons and $f_i$ is the phase-space density at their positions. The calculation of the phase-space density at a point $(\bm{r}_i,\bm{p}_i)$ follows the method used in the GiBUU model (see Section D.4.3 in \cite{Buss:2011mx}), according to which $f_i$ is given by
\begin{align}
f_i(\bm{r}_i, \bm{p}_i) = \sum_{j} w_r(\bm{r}_i
- \bm{r}_j) w_p (\bm{p}_i - \bm{p}_j)  \,,
\end{align}
where the sum formally goes over all particles and $w_r$ and $w_p$ are weights in coordinate and momentum space defined as
\begin{eqnarray}
&& w_r(\bm{r}) = \mathcal{N}~ \Theta(r_r - |\bm{r}\hspace{1pt}|) \int d^3r'~ \Theta\big(r_c - |\bm{r} - \bm{r}'|\big) \exp \left( - \frac{(\bm{r} - \bm{r}')^2}{2 \sigma^2} \right) \,, \\
&& w_p(\bm{p}) = \Theta(p_0 - |\bm{p}\hspace{1pt}|)  / \left(\frac{4}{3} \pi p_0^3 \right)  \,.
\end{eqnarray}
Here, the parameters are chosen to be $r_r = 1.86$ fm, $r_c = 2.2$ fm, $p_0 = 80$ MeV/c, and the normalization $\mathcal{N}$ is chosen to satisfy the condition $N_{\rm test} \int d^3r \, w_r(r) = 1$ (see Appendix A of \cite{Weil:2016zrk} for an analytical expression for $\mathcal{N}$).

\subsection{The Stochastic Mean Field (SMF) code} 
\label{sec:SMF}
\vskip 0.1in
M. Colonna, P. Napolitani
\vskip 0.1in

\subsubsection{Introduction}



The SMF model \cite{Colonna1998} can be considered as an approximate tool to solve the
so-called Boltzmann-Langevin (BL) equation \cite{Ayik88}:
\begin{equation}
{\frac{df}{dt}} = {\frac{\partial f}{\partial t}} + \{f,H\} = I_{coll}[f]
+ \delta I[f],
\label{eq:BL}
\end{equation}
 where 
$\delta I[f]$ is the fluctuating 
part of the collision integral $I_{coll}$. 
The coordinates of isospin are not shown for brevity.

It should be noticed that in the BUU or Boltzmann-Nordheim-Vlasov (BNV) models, the fluctuating term $\delta I[f]$
is neglected \cite{buu,bonasera1994}.  
In the present SMF treatment we project the fluctuations of the distribution
function, generated by the stochastic collision integral in Eq.~(\ref{eq:BL}),
on the coordinate space and consider local density fluctuations, which can be implemented
as such in a numerical calculation. We make the further assumption of local thermal equilibrium,
thus being able to derive  analytical expressions for the density fluctuations.

Within our framework, the system is described in terms of the one-body distribution function $f$, but this function
may experience a stochastic evolution in response to the action of the fluctuating term. 
Then, the model is suitable to treat the occurrence of instabilities and bifurcations of trajectories in
nuclear dynamics \cite{rep}. When instabilities are encountered along the reaction path,
the evolution of the fluctuation ``seeds" introduced by the SMF method is then
determined by the dissipative dynamics of the BNV evolution,
allowing the system to choose its trajectory through the fragmentation configuration. In this way, we create a series of
``events" in a heavy-ion collision, which can then be analyzed and sampled in various ways. In the following,
we will give the details of the implementation of the different terms of 
Eq.(\ref{eq:BL}).

\subsubsection{Mean-field propagation}

Eq.(\ref{eq:BL}) is solved by adopting the test-particle method.  Then, the
one-body distribution function is parametrized as follows:
\begin{equation}
f(\vecr,\vecp,t) = {\frac{Ch^3}{4}} \sum_i g_r(\vecr - \vecr_i)g_p(\vecp - \vecp_i),
\label{eq:f}
\end{equation}
where the sum runs over $N_{tot} = N_{test} \cdot A$ test particles, with
$N_{test}$ being the number of test particles per nucleon and $A$ the nucleon number
of the system considered. $C$ is a normalization factor.
In the SMF model, we adopt triangular functions
in $\vecr$ space (for $g_r$) and $\delta$ functions in
momentum space (for $g_p$) \cite{TWINGO}.

The coordinate space is discretized by introducing a lattice of mesh size $l$.
Then, the function $g_r(\vecr - \vecr_i)$ is defined, at each lattice site, as
the product of three triangular functions (to account for the three spatial dimensions)
of the type
$g(x^j-x_i^j) = 2l - |x^j-x_i^j|$, where $x^j$ ($j$ = 1,2,3) denotes the
spatial coordinate \cite{Lenk89}. We note that the nucleon density simply reads:
\begin{equation}
\rho(\vecr,t) = C \sum_i g_r(\vecr - \vecr_i).
\label{eq:ro}
\end{equation}
Within this framework, the total energy of the system, for the
$N_{tot}$ test particles, can be written as:
\begin{equation}
E_{tot} = \sum_i p_i^2/(2m) + N_{test}\big[\int d\vecr~\rho(\vecr) E_{pot}(\rho_n,\rho_p)
+\int d\vecr~\rho_p(\vecr) E_{pot}^{Coul}(\rho_p)/2\big] ,
\label{eq:etot}
\end{equation}
where $\rho_n$ and $\rho_p$ denote neutron and proton densities, respectively, $E_{pot}$ is the potential
energy per nucleon, connected to the (momentum independent) mean-field interaction, and  $E_{pot}^{Coul}$
denotes the Coulomb potential.

Effective interactions, associated with a given Equation of State (EOS),
can be considered as an input of all transport codes.
We adopt a soft isoscalar EOS (compressibility $K = 200$ MeV).
We notice that the considered compressibility value is favored, e.g.,
from flow, monopole oscillation and multifragmentation studies \cite{rep,Borderie}.
The choice considered corresponds
to a Skyrme-like effective interaction, namely $SKM^*$, for which we take the effective
mass as being equal to the nucleon bare mass. 
Then $E_{pot}$ can be written as:
\begin{equation}
E_{pot}(\rho)  = {\frac{A}{2}} {\tilde \rho} +  {\frac{B}{\sigma+1}}
{\tilde \rho}^\sigma + \frac{C_{surf}}{2\rho} (\nabla\rho)^2
+ \frac{1}{2} C_{sym}(\rho){\tilde \rho} \beta^2,
\label{eq:epot}
\end{equation}
where ${\tilde \rho} = \rho / \rho_0$ ($\rho_0$ denotes the saturation density), $A = -356~MeV$,
$B = 303~ MeV$, $\sigma = 7/6$.
We notice that surface effects are automatically introduced
in the dynamics when considering finite width wave packets
for the test particles employed in the numerical resolution.
An explicit surface term is also added (third term of Eq.(\ref{eq:epot})) and
tuned in such a way that the total surface energy reproduces
the surface energy of nuclei in the ground state
\cite{TWINGO}.  This procedure yields $C_{surf} = -6/\rho_0^{5/3}$~MeV$\,$fm$^5$.
The fourth term of Eq.(\ref{eq:epot}) represents the potential part of the symmetry energy
per nucleon, with $\beta = (\rho_n - \rho_p) / \rho$.

For some of the reaction mechanisms analyzed with SMF,
the sensitivity of the simulation results is tested against
different choices of the density dependence of $C_{sym}$:
the asysoft,  $\displaystyle C_{sym}(\rho) = 
(77.1-41.9 \, \tilde{\rho})$ MeV,
the asystiff, $\displaystyle C_{sym}(\rho) = 36$ MeV,
and the asysuperstiff, $\displaystyle C_{sym}(\rho) =72 \,\tilde{\rho}/(\tilde{\rho}+1)$ MeV 
\cite{epja}.
The value of the symmetry energy,
$\displaystyle E_{sym}/A = \epsilon_F/3+(1/2) C_{sym}(\rho){\tilde \rho}$,
at saturation, as well as
the slope parameter, $\displaystyle L = 3 \rho_0 \, (dE_{sym}/A)/d \rho |_{\rho=\rho_0}$,
are given in Table~\ref{table2} for each of these asy-EOSs
($\epsilon_F$ denotes the Fermi energy).
Just below the saturation density,
the asysoft parametrization exhibits a weak variation with density, while the asysuperstiff shows
a rapid decrease.
\begin{table}
\begin{center}
\begin{tabular}{|l|r|r|r|r|r|} \hline
asy-EoS       &E$_{sym}$/A (MeV)    &~~~L(MeV)  \\ \hline
asysoft       & 30                  & 14   \\ \hline
asystiff      & 28                  & 73   \\ \hline
asysuperstiff   & 28                 & 97   \\ \hline
\end{tabular}
\caption{The symmetry energy at saturation (MeV) and the slope parameter (MeV)
for the three asy-EOS considered (see text).}
\label{table2}
\end{center}
\end{table}
Finally, the Coulomb potential is determined from solving the Laplace equation:
\begin{equation}
\nabla^2 E_{pot}^{Coul} = -4\pi e^2 \rho_p = -18.1 \rho_p,
\label{eq:surf}
\end{equation}
on the lattice sites.
Momentum-dependent effective interactions may also be implemented into 
Eq.(\ref{eq:BL}) \cite{baranPR,Joseph,Zheng_2016}.

The ground state configuration of nuclei is obtained by distributing the test
particle positions inside a sphere of radius $R_{gs}$ and their momenta
inside the corresponding local Fermi sphere, with a density-dependent Fermi
momentum. $R_{gs}$ is determined by searching for the minimum
 of the total energy i.e., Eq. (\ref{eq:etot}).

The dynamics is followed by solving the Hamilton equations for the test
particle positions and momenta, which are derived from the expression
of the total energy, Eq.(\ref{eq:etot}).

\subsubsection{Collision integral}

Two-body correlations are taken into account through the collision integral in Eq. (\ref{eq:BL}), which is
evaluated from
considering collisions between pairs of test particles.
We adopt the mean free path method to determine the collision probability, as in the standard BNV approach \cite{bonasera1994}.
Each test particle $k$ looks for the closest test particle $l$ and the mean free path
$\lambda = 1 / (\rho(\vecr) \sigma_{NN})$ is evaluated.
The  associated collision time for the two particles considered is:
\begin{equation}
\tau_{col} = \frac{\lambda}{v_{kl}}=\frac{1}{\rho\sigma_{NN} v_{kl}},
\label{eq:tau}
\end{equation}
where $v_{kl}$ is the relative velocity between particles $k$ and $l$.
The main ingredient entering this process is the nucleon-nucleon cross section $\sigma_{NN}$, for which
we use the free isospin, angle and energy dependent values.

The probability for the collision to happen in the time step $\Delta t$ can then be expressed as:
$P_{col}(\Delta t)  \approx \Delta t / \tau_{col} =  \rho\sigma_{NN} v_{kl} \Delta t$.
It should be noticed that, since the collision process involves pairs of particles, in order to
avoid double counting, the probability $P_{col}$ has to be divided by 2.
If $P_{col}$ is larger than a random number (selected in the interval between 0 and 1)
new momenta are chosen for the two test particles, with the requirement of energy
and momentum conservation.

In order to finally accept the collision between the two considered particles, the Pauli blocking has to be checked
at the final positions in phase space.  To do so, the occupation number $f(\vecr,\vecp)$ has to be evaluated at the
two considered phase-space points, associated with the new momenta of particles $k$ and $l$.
To optimize the calculation of the Pauli blocking factor, instead of considering
the $g_r$ and $g_p$ functions introduced above (see Eq. (\ref{eq:f})), we now take a $\Theta$ function in $\vecr$ space and
a gaussian function, with $\sigma = 29~ MeV/c$, in momentum space. The $\Theta$ function is defined as:
$\Theta(\vecr-\vecr_i) = \Theta(R - |\vecr-\vecr_i|)$, with $R = 2.53 fm$.
The new definition makes the occupation number smoother (though less local), reducing fluctuations which may induce
spurious collisions.
The Pauli blocking factors, defined as $P_{Pauli} = (1-f_l)(1-f_k)$, is confronted to a random number and, if
it is larger than it, the collision is finally accepted.

It should be noticed that the procedure described above, which employs random
numbers, is stochastic.  However, owing to the fact that collisions are treated
for pairs of test particles, fluctuations are reduced by $1 / N_{test}$.
Thus an explicit fluctuation term is needed, as indicated in Eq.(\ref{eq:BL}), to account for the stochastic nature of the nucleon-nucleon collision process.

\subsubsection{Fluctuations}
It would seem to be attractive to introduce the fluctuations directly in
the phase space, i.e., to use $\sigma^2_f = {f}(1 - {f}) $
locally in the phase space. 
However, this is difficult numerically because of the high dimension of the phase space. In the
present application of the method, we therefore project on density fluctuations, in a volume $V$, by
\begin{equation}
\sigma^2_\rho(\vecr, t) = \frac{1}{V} \int \frac{d\vecp}{h^3/4} \sigma^2_f
(\vecr,\vecp,t).
\label{eq:sigma}
\end{equation}

This variance can be directly calculated from the BNV simulation, and fluctuations
be introduced accordingly. However, it is more practical to have explicit analytical expressions
for the density fluctuations. Within our assumption of local thermal equilibrium,
the mean distribution function can be parametrized by the expression
${f} ( \vecr , \vecp , t ) = 1/(1 + exp(\epsilon - \mu( \vecr , t ) )
/ T ( \vecr , t ) ) $
with a local chemical potential and temperature $\mu(\vecr, t)$ and $T(\vecr, t)$,
respectively, and with $\epsilon = p^2/2m$. The determination of the temperature will be discussed
below. Introducing the expression for the fluctuation variance into 
Eq.(\ref{eq:sigma}) and converting the
$\vecp$-integration into an $\epsilon$-integration,
we obtain, after integrating by parts:
\begin{equation}
\sigma^2_\rho =  \frac{1}{V} {\frac{2\pi m \sqrt{(2m)}~T}{h^3/4}}
\int \frac{1}{\sqrt \epsilon} \frac{1}{{1+\exp(\epsilon-\mu)/T}} d\epsilon.
\label{eq:sigma2}
\end{equation}
We note that Eq.(\ref{eq:sigma2}) is consistent with the thermodynamical relation
for the variance of the particle number in a given volume. To obtain
a more explicit expression and to eliminate the chemical potential, we can
use the Sommerfeld expansion for the function ${f}$ around $\epsilon =\mu$
for small $T/\epsilon_F$.  We then obtain
\begin{equation}
\sigma^2_\rho =\frac{{16\pi m \sqrt{(2m)}}}{h^3 V} \sqrt{\epsilon_F} T
\left [1 - \frac{\pi^2}{12} \left (\frac{T}{\epsilon_F} \right )^2 + . . . \right ].
\label{eq:sigma21}
\end{equation}
The procedure can be considered and implemented separately for neutrons and protons.

As already mentioned above, in order to use the explicit analytical expression for the
density fluctuations (Eq. (\ref{eq:sigma21})), we make the assumption of local thermal equilibrium.
As a consequence, the implementation of fluctuations can only be considered starting
from the moment when the nuclear system
is locally equilibrated in the dynamical evolution of the collision. This corresponds roughly to the time when the maximum of the
entropy is reached and, in central collisions for instance, a composite nuclear source is
formed. In more peripheral collisions, this is associated with the formation of a locally
equilibrated di-nuclear system.
To introduce the local fluctuations (Eq. (\ref{eq:sigma21})),
local density and temperature are evaluated at each site of the lattice introduced in coordinate space.
The collective momentum ${\bar \vecp}$ of the cell is calculated by
averaging over the momenta of the test particles that belong to the considered cell. 
In the same manner, the excitation energy per nucleon $E^*$ is obtained by averaging over
the kinetic energies (calculated in the frame of the cell) of the test particles and
subtracting the mean energy per nucleon associated with a Fermi gas at zero temperature
and at the density considered. It is then possible to extract the temperature.
Once the temperature and the density are calculated, we derive the value of the
density fluctuation correlation using Eq.(\ref{eq:sigma21}). In the cell being considered,
the density fluctuation $\delta\rho$ is selected randomly according to the gaussian
distribution. This determines the variation of the number of particles
contained in the cell. A few left-over particles are finally randomly distributed
again in order to ensure the conservation of mass.
Once the new density value in the cell has been defined, the excitation energy and
temperature (and chemical potential) are redefined to enforce energy conservation. Then, the
momenta of the test particles are redistributed according to the Fermi-Dirac function
associated with the new values of chemical potential and temperature.

Finally, we note that, as shown in Ref.\cite{Malgor},
for reactions at Fermi energies, the SMF method essentially gives the same results
as a simpler and
computationally much easier approach, based on the introduction
of density fluctuations by a random sampling of the phase space \cite{colonna_new,standard,rep},
where the amplitude of the noise is gauged to reproduce the dynamics of the most unstable modes.
The equivalence of the two methods in the description of the
collision dynamics was checked along the complete evolution, from fast particle emissions to the fragment
production \cite{Malgor}.

\subsection{ Boltzmann-Uehling-Uhlenbeck transport code based
on the energy density functional from chiral effective theory ($\chi$BUU)}
\vskip 0.1in
Z. Zhang, C. M. Ko
\vskip 0.1in
\subsubsection{Introduction}

The $\chi$BUU transport model is developed based on a Skyrme energy density functional,  Sk$\chi$m$^*$, that is constructed by fitting the equation of state of asymmetric nuclear matter and nucleon effective mass from many-body perturbation theory with chiral two- and three-body forces as well as the binding energy of finite nuclei~\cite{Zhang:2017hvh}.  It builds a bridge between the energy density functional from \textit{ab initio} calculations based on chiral effective theory to the observables measured in heavy-ion collisions~\cite{Zhang:2018ool}.   Given the complexity of the transport model description of heavy-ion collisions, the philosophy of the $\chi$BUU model is to fix its ingredients as many as possible by ab-initio calculations and then use it to study various phenomena in heavy-ion collisions and their physics implications. 

The $\chi$BUU was developed by modifying the RVUU code. Its treatment of the collision criteria, Pauli blockings and Coulomb interactions on charged particles is thus similar to those in the RVUU code. It is, however, a non-relativistic code because of the non-relativistic mean-field potentials of  nucleons and $\Delta$ resonances.  The initialization, mean-field potentials, particle propagations, and also detailed balance relations therefore differ from those in the RVUU.

\subsubsection{Initialization}

In the $\chi$BUU code, the initial positions of nucleons in a nucleus are distributed according to the density distribution obtained from Hartree-Fock calculation with the Sk$\chi$m* interaction. As in the RVUU code, their momenta are obtained by assuming that nucleons in a local cell forms a Fermi gas with the Fermi momentum determined by the local density. Nucleons in the two colliding nuclei are then boosted to the two-nuclei center-of-mass frame using the velocity $p / E$ for projectile nucleons and $-p / E$ for target nucleons, where $p$ and $E$ are, respectively, the total momentum and energy of the two colliding nuclei in the laboratory frame.

\subsubsection{Mean-field Potentials}

The potential of a nucleon is obtained from the Sk$\chi$m* energy density functional, 
which predicts the nuclear matter saturation density  $\rho_0 = 0.1651~\mathrm{fm}^{-3}$, 
binding energy per nucleon $E_0(\rho_0)=-16.07$ MeV, incompressibility $K_0=230.4$ MeV, and the symmetry energy $E_{\mathrm{sym}}(\rho_0)=30.9$ MeV 
and its slope parameter  $L=45.6$ MeV at the saturation density. Because of its  quadratic momentum dependence, the single-nucleon potential  in the $\chi$BUU model can be expressed as 
\be\label{Eq:snp}
U_q(\bm{r},\bm{p})= a_q\bm{p}^2-\bm{b}_q\cdot\bm{p}+c_q,
\ee
where $q= p,n$, and the coefficients $a_q$, $\bm{p}_q$ and $c_q$ are given by
\be\label{mean}
    \begin{aligned}
        a_{q}=& 2 C \rho+2 D \rho_{q}, \\
        \boldsymbol{b}_{q}=& 4 C \int d^{3} \boldsymbol{p}^{\prime} \boldsymbol{p}^{\prime} f\left(\boldsymbol{r}, \boldsymbol{p}^{\prime}\right)+4 D \int d^{3} \boldsymbol{p}_{q}^{\prime} \boldsymbol{p}_{q}^{\prime} f_{q}\left(\boldsymbol{r}, \boldsymbol{p}_{q}^{\prime}\right), \\
        c_{q}=& 2 A_{0} \rho-2 A_{1} \rho_{q}+B_{0}(\alpha+2) \rho^{\alpha+1} \\
        &-B_{1} \alpha \rho^{\alpha-1}\left(\rho_{n}^{2}+\rho_{p}^{2}\right)-2 B_{1} \rho^{\alpha} \rho_{q} \\
        &+2 C \int d^{3} \boldsymbol{p}^{\prime} \boldsymbol{p}^{\prime 2} f\left(\boldsymbol{r}, \boldsymbol{p}^{\prime}\right)+2 D \int d^{3} \boldsymbol{p}_{q}^{\prime} \boldsymbol{p}_{q}^{\prime 2} f_{q}\left(\boldsymbol{r}, \boldsymbol{p}_{q}^{\prime}\right).
        \end{aligned}
\ee
The above coefficients  $A_0$, $A_1$, $B_0$, $B_1$, $C$ and $D$ are defined as 
\be
    \begin{aligned}
        &A_{0}=\frac{1}{4} t_{0}\left(2+x_{0}\right), \quad A_{1}=\frac{1}{4} t_{0}\left(1+2 x_{0}\right), \\
        &B_{0}=\frac{1}{24} t_{3}\left(2+x_{3}\right), \quad B_{1}=\frac{1}{24} t_{3}\left(1+2 x_{3}\right), \\
        &C=\frac{1}{16}\left[t_{1}\left(2+x_{1}\right)+t_{2}\left(2+x_{2}\right)\right], \\
        &D=\frac{1}{16}\left[t_{2}\left(2 x_{2}+1\right)-t_{1}\left(2 x_{1}+1\right)\right],
        \end{aligned}
    \ee
with $t_i, x_i, (i=0,1,2,3)$ and $\alpha$ being conventional Skyrme parameters. Values of the corresponding Skyrme parameters in the Sk$\chi$m* interaction can be found in Ref.~\cite{Zhang:2017hvh}.

For the potentials of $\Delta$ resonances, they are determined according to the extensively
used relations in literature~\cite{Li:2008gp},
\be
U_{\Delta^{++}} = U_p,~~U_{\Delta^{+}} = \frac{2}{3}U_p+\frac{1}{3}U_n,~~
U_{\Delta^0} = \frac{1}{3}U_p+\frac{2}{3}U_n,~~U_{\Delta^-} = U_n.
\ee

Based on Eq.~(\ref{Eq:snp}), the effective mass $m_q^*$, kinetic energy $E_q^*$  and kinetic momentum $\bm{p}_q^*$ of baryons can be defined as
\begin{eqnarray}
    \frac{1}{m_q^*} &=& \frac{1}{m_q}+a_q, \notag\\
    E_q^* &=& E_q-\Sigma_q^0, \notag\\
    \bm{p}_q^* &=&\bm{p}_q-\bm{\Sigma}_q,
\end{eqnarray}
with $\Sigma_q^0 = c_q-\frac{m_q^*b_q^2}{2}+m_q-m_q^*$,  $\bm{\Sigma}_q = m_q^*\bm{b}_q$ 
and $E = p_q^2/(2m)+U_q$.

The equations of motion of nucleons and $\Delta$ resonances are then given by 
\begin{eqnarray}
    \dot{\bm{r}} = \frac{\bm{p}^*}{m^*}, \quad \dot{\bm{p}} &=& -\nabla_r E_q.
\end{eqnarray}
Pions are treated as if they are in free space. Because  of their small mass ($138$ MeV), they are treated as relativistic particles moving with a constant velocity of $\bm{p}/\sqrt{m_{\pi}^2+p^2}$.

\subsubsection{Two-body scatterings}

Although $\chi$BUU uses the same geometrical minimum  distance criterion for binary collisions as in the RVUU code, it implements the criterion in the computational frame, i.e., the center of mass frame of the two colliding nuclei, instead of the center of mass frame of two scattering particles. The cross sections for baryon-baryon elastic scattering and $\Delta$ production, the  $\Delta$ decay width  and the $\Delta$ spectral function in $\chi$BUU are the same as those in the RVUU.  Because of the non-relativistic baryon potentials, the detailed balance relations in $N+N \leftrightarrow N+\Delta$ and $\Delta \leftrightarrow N+\pi$ reactions in $\chi$BUU differ from those in RVUU. Specifically, in a $N+N\rightarrow N+\Delta$ reaction with the initial two nucleons, the final nucleon and the final $\Delta$ labeled by 1, 2, 3 and 4, respectively, the mass of produced $\Delta$ is sampled according to 
\be
P(m_{\Delta})= \frac{\mathcal{A}(m_{\Delta})k_f^*\mu_f}{\int dm_{\Delta}\mathcal{A}(m_{\Delta})k_f^*\mu_f},
\ee
where $\mathcal{A}(m_{\Delta})$ is the $\Delta$ spectral function normalized by 
$\int \frac{dm_{\Delta}}{2\pi}\mathcal{A}(m_{\Delta})=1$, $\mu_f= m_3^*m_{4}^*/(m_3^*+m_{4}^*)$ is the reduced effective mass of the final nucleon and $\Delta$, and $\bm{k}_f^*$ is the relative kinetic momentum in the final state defined as
\be
\boldsymbol{k}_{f}^{*}=\frac{\boldsymbol{p}_{3}^{*}}{m_{3}^{*}}
-m_{3}^{*} \frac{\boldsymbol{p}_{3}^{*}+\boldsymbol{p}_{4}^{*}}{m_{3}^{*}+m_{4}^{*}}
=-\left(\frac{\boldsymbol{p}_{4}^{*}}{m_{4}^{*}}
-m_{4}^{*} \frac{\boldsymbol{p}_{3}^{*}+\boldsymbol{p}_4^{*}}{m_{3}^{*}+m_{4}^{*}}\right).
\ee
The  cross section of its inverse reaction $N+\Delta \rightarrow N+N$  is then determined 
by the detailed balance relation as 
\begin{eqnarray}
\sigma_{N \Delta \rightarrow N N}=\frac{\sigma_{NN\rightarrow N\Delta}}{2\left(1+\delta_{12}\right)} 
\frac{k_{i}^{* 2}}{k_{f}^{*}} \frac{\mu_{f}}{\int \frac{d m^{\prime}}{2 \pi} k_{f}^{* \prime} \mu_{f}^{\prime} \mathcal{A}\left(m^{\prime}\right)},  
\end{eqnarray}
where $k_i$ is the relative kinetic momentum of the scattering two nucleons and the 
factor $1/(1+\delta_{12})$ takes into account the case that nucleons 1 and 2 are identical.

For the pion absorption reaction $N+\pi\rightarrow \Delta$, the cross section $\sigma_{N\pi}$ is related to the $\Delta$ decay width by 
\begin{eqnarray}
\sigma_{N \pi \rightarrow \Delta}=\frac{4 \pi P^{*}}{m_{N}^{*}\left|\boldsymbol{v}_{N}-\boldsymbol{v}_{\pi}\right|\left|1-\frac{p_{\Delta}^2}{2m^2_{\Delta}}\right|} \frac{\mathcal{A}\left(m_{\Delta}\right) \Gamma\left(m_{\Delta}\right)}{\omega\left(p_{\pi}\right)\left[\omega\left(p_{\max }\right)-\omega\left(p_{\min }\right)\right]}.\notag\\
\end{eqnarray}
In the above, $\bm{P}^{\ast} =\bm{p}_N^{\ast}+\bm{p}_{\pi}$, $\omega(p_{\pi}) = \sqrt{m_{\pi}^2+p_{\pi}^2}$ is the pion energy, $\bm{v}_N=\bm{p}_N^{\ast}/m_N^{\ast}$ and $\bm{v}_{\pi} =\bm{p}_{\pi}/\omega(p_{\pi})$ are, respectively, the velocities of the colliding nucleon and pion, $\Gamma(m_{\Delta})$ is the $\Delta$ partial decay width, $\bm{p}_{\Delta} = \bm{p}_N+\bm{p}_{\pi}$ is the momentum of the final $\Delta$, and $p_{\mathrm{max}}$ and $p_{\mathrm{min}}$ are, respectively, the allowed maximum and minimum pion momenta from the decay of the $\Delta$ in the computational frame. The $\Delta$ mass is determined by 
\be
m_{\Delta} = \frac{p_{\Delta}^2}{2(E-a_{\Delta}p_{\Delta}^2+\bm{b}_{\Delta}\cdot \bm{p}_{\Delta}-c_{\Delta})},
\ee
with $E$ being the total energy of the colliding 
nucleon and pion. 



\newpage
\section{Quantum molecular dynamics (QMD)-like codes} \label{sec:qmd}

As in the previous section, we collect the descriptions of the codes of QMD type here. The format of the code description is the same as descriptions of the BUU-type of codes (see the paragraph at the beginning of Sec.~\ref{sec:buu}).

\subsection{Antisymmetrized molecular dynamics (AMD) code\label{sec:AMD}}
\vskip 0.1in
A. Ono
\vskip 0.2in
\subsubsection{About the code}

The comparison calculations with the antisymmetrized molecular dynamics
approach (AMD) have been performed by employing a recently developed AMD
code written in Fortran90 \cite{Ono:NN12,Ikeno}.  This code includes the
feature to efficiently calculate the force with the Skyrme
parametrization and an option to take into account cluster correlations
in the final states of two-nucleon collisions.  However, in the comparison calculations of Ref.~\cite{Xu2016}, the option of
cluster correlations was turned off, so
that the code works equivalently to the old version of the AMD code
\cite{ONOb}, except for the detailed points described in the following
sections. In the comparison of Ref.~\cite{SpRIT:2020blg} for the prediction of pion production, the AMD with cluster correlations was used in combination with the JAM code.

\subsubsection{AMD wave function}

AMD employs a Slater determinant of Gaussian wave packets
$\varphi_i=\exp[-\nu(\mathbf{r}-\frac{1}{\sqrt{\nu}}\mathbf{Z}_i)^2]\chi_i$
($i=1,2,\ldots,A$), where the centroid variable $\mathbf{Z}_i$
contains the information of the position and the momentum in its real
and imaginary parts, respectively.  The width parameter $\nu$ is
chosen so that the position and momentum uncertainties are $\Delta
x=\frac{1}{2\sqrt{\nu}}=1.25$~fm and $\Delta
p=\hbar\sqrt{\nu}=78.9$~MeV/$c$.  The spin-isospin states $\chi_i$ are
fixed to be p$\uparrow$, p$\downarrow$, n$\uparrow$ or n$\downarrow$.

Due to the antisymmetrization, the variables $\{\mathbf{Z}_i\}$ do not
have a simple meaning.  In fact, the equations of motion derived from
the time-dependent variational principle show that these are not
canonical variables.  The Wigner transform of the one-body density
matrix is written in a complicated way as
\begin{equation}
f(\mathbf{r},\mathbf{p})=
8 \sum_{i=1}^A\sum_{j=1}^A 
\exp\bigr[-(\mathbf{r}-\mathbf{R}_{ij})^2/2\Delta x^2\big]\,
\exp\bigr[-(\mathbf{p}-\mathbf{P}_{ij})^2/2\Delta p^2\big] \, B_{ij}B^{-1}_{ji},
\label{eq:AMDwigner}
\end{equation}
where
$\mathbf{R}_{ij}=\frac{1}{2\sqrt{\nu}}(\mathbf{Z}_i^*+\mathbf{Z}_j)$,
$\mathbf{P}_{ij}=i\hbar\sqrt{\nu}(\mathbf{Z}_i^*-\mathbf{Z}_j)$, and
$B_{ij}=\langle\varphi_i|\varphi_j\rangle$.  Any one-body quantity can
be calculated precisely from this Wigner function.  However, it is possible to introduce an approximated distribution function
\begin{equation}
f(\mathbf{r},\mathbf{p})\approx
8 \sum_{k=1}^A 
\exp\bigr[-(\mathbf{r}-\mathbf{R}_{k})^2/2\Delta x^2\big] \,
\exp\bigr[-(\mathbf{p}-\mathbf{P}_{k})^2/2\Delta p^2\big]
\label{eq:physcoord}
\end{equation}
by using the so-called physical coordinates \cite{ONOb}, which is often employed in the two-nucleon collision process (see below). More recently, a numerical method has been developed to randomly generate test particles according to the precise Wigner distribution function of Eq.~(\ref{eq:AMDwigner}), as described in Appendix C of Ref.~\cite{Ikeno}. This method is used for many purposes, such as to generate output data in a format similar to other transport codes, to calculate the two-nucleon collision probability, and to combine AMD with the JAM code to predict pion production \cite{Ikeno,SpRIT:2020blg}.

The momentum width of each wave packet has a large contribution
$3\Delta p^2/2M=10.0$ MeV (per nucleon) to the kinetic energy.  This
zero-point energy is regarded as a part of the physical energy, when a many-body wave function of a ground state nucleus is prepared for the initial condition of heavy-ion collisions \cite{ONOb} as well as for any other nuclear structure studies \cite{kanada2012}. However, the zero-point energy for the center of mass of a nucleus or each of isolated fragments is regarded as spurious and is subtracted from the Hamiltonian by using the method of Ref.~\cite{ONOb}.

\subsubsection{Effective interaction}
This AMD code uses the Skyrme force parametrized as
\begin{eqnarray}
v_{ij}&=&t_0(1+x_0P_\sigma)\delta(\mathbf{r})
+\frac{1}{2}t_1(1+x_1P_\sigma)
[\delta(\mathbf{r})\mathbf{k}^2 \nonumber\\
&&+\mathbf{k}^2\delta(\mathbf{r})]+t_2(1+x_2P_\sigma)
\mathbf{k}\cdot\delta(\mathbf{r})\mathbf{k}
+t_3(1+x_3P_\sigma)[\rho(\mathbf{r}_i)]^{\alpha}\delta(\mathbf{r}),
\end{eqnarray}
where $\mathbf{r}=\mathbf{r}_i-\mathbf{r}_j$ and
$\mathbf{k}=\frac{1}{2\hbar}(\mathbf{p}_i-\mathbf{p}_j)$.  In the calculations of Ref.~\cite{Xu2016}, the terms
of $t_1$ and $t_2$ were ignored in order to have no surface and
momentum-dependent terms as specified by the set-up of the comparison.  The other
parameters were uniquely fixed by the set-up condition: $t_0=-1743.33\ \text{MeV}\cdot\text{fm}^3$, $x_0=-0.2419$, $t_3=12639.4\ \text{MeV}\cdot\text{fm}^{3(1+\alpha)}$, $x_3=-0.5$ and $\alpha=0.35$.

More realistic parametrizations were used in the calculations of Ref.~\cite{SpRIT:2020blg} for the prediction of pion production. We use the Skyrme SLy4 effective interaction \cite{chabanat1998} but the spin--orbit interaction is ignored.  The corresponding nuclear-matter incompressibility is $K=230$ MeV at the saturation density $\rho_0=0.160\ \text{fm}^{-3}$.  The nuclear-matter symmetry energy at $\rho_0$ is $S_0=32.0$ MeV with the slope parameter $L=46$ MeV.  In order to study the effect of the density dependence of the symmetry energy, we perform calculations with interactions obtained by changing the density-dependent term in the SLy4 interaction
\begin{equation}
  v_\rho^{\text{SLy4}}=\tfrac16 t_3(1+x_3P_\sigma)\rho(\bm{r}_1)^\alpha\delta(\bm{r}_1-\bm{r}_2)
\end{equation}
to
\begin{equation}
v_\rho' = \tfrac16 t_3(1+x_3'P_\sigma)
\rho(\bm{r}_1)^\alpha\delta(\bm{r}_1-\bm{r}_2)
+ \tfrac16 t_3(x_3-x_3')\rho_0^\alpha P_\sigma\delta(\bm{r}_1-\bm{r}_2).
\end{equation}
We choose $x_3'=1.1$, $0.3$ and $-1.8$, which correspond to $L=55$, $82$ and $152$ MeV, respectively.

In collision calculations at 270 MeV/nucleon, the momentum dependence is modified by the method described in Appendix B of Ref.~\cite{Ikeno}.
The Coulomb interaction between protons is included.

\subsubsection{Preparation of initial nuclei}

To prepare an initial state of a collision, AMD uses ground state
nuclei which are usually obtained by the frictional cooling
method to search a wave packet configuration that minimizes the
energy.  However, the Skyrme force specified for the calculations of Ref.~\cite{Xu2016} is not a good interaction to describe nuclear
structure, and the configuration that minimizes the energy does not
reproduce the properties of the ground state nucleus.  Therefore, the
state is chosen when the experimental binding energy is reached in the
course of the frictional cooling calculation.  Fortunately, this state
of the $^{197}\mathrm{Au}$ nucleus happened to have a density
distribution that is very close to the specification of the comparison, so it
was adopted as the projectile and target nuclei in the initial states
of collisions.

\subsubsection{Two-nucleon collisions and Pauli blocking}

First, the two-nucleon collision process used in the calculations of Ref.~\cite{Xu2016} is described here. This method has not been changed much since Refs.~\cite{ONOb,ONOc}. It uses
physical coordinates in Eq.\ (\ref{eq:physcoord}).  The possibility of collisions is tested for all pairs
of wave packets at every time step.  Let us consider the collision of
the wave packets 1 and 2, without losing generality, in the time step
between $t-\Delta t$ and $t$, where $\Delta t$ is the time step to
solve the equation of motion, which is usually chosen to be $\Delta
t=1.67$ fm/$c$, in heavy-ion collisions below 100 MeV/nucleon.

The probability of the collision (without considering the Pauli
blocking) is assumed to be proportional to the density overlap of the
two wave packets and to the relative velocity, so that
$P(\mathbf{r})|\Delta\mathbf{r}|=\alpha \,
\exp(-\nu\mathrm{r}^2)|\,\Delta\mathbf{r}|$, where
$\mathbf{r}=\mathbf{R}_1-\mathbf{R}_2$ and
$\Delta\mathbf{r}=\mathbf{r}(t)-\mathbf{r}(t-\Delta t)$.  The
proportionality constant $\alpha$ is determined by the condition that
the collision cross section should be the given value of
$\sigma_{\text{NN}}$ when the two nucleons are assumed to move on
straight lines until they collide
\begin{equation}
\int_0^\infty 2\pi bdb
\biggl[1-\exp\big(-\int_{-\infty}^{\infty}P(\mathbf{b}+\mathbf{z})dz\big)\biggr]
=\sigma_{\text{NN}}.
\end{equation}
This allows us to express $\alpha=f(\nu\sigma_{\text{NN}})$ with a
universal function $f(x)$ \cite{ONOc}.

Based on the calculated probability $P(\mathbf{r})|\Delta\mathbf{r}|$, whether
the wave packets 1 and 2 collide or not in this time
step, can then be decided.  If they are to collide, their relative momenta are changed
according to the angular distribution of the two nucleon collision
(which is chosen to be isotropic in this comparison).  This means that
only the two momentum coordinates are changed
$\mathbf{P}_1\rightarrow\mathbf{P}_1'$ and
$\mathbf{P}_2\rightarrow\mathbf{P}_2'$ while the other position and
momentum coordinates do not change.  The magnitude of the relative
momentum should be adjusted for the energy conservation.

The collision is assumed to be Pauli blocked if there is another wave
packet (with the same spin and isospin as the colliding wave packets)
near the phase-space point $(\mathbf{R}_1,\mathbf{P}_1')$ or
$(\mathbf{R}_2,\mathbf{P}_2')$.  The distance condition for blocking
is determined by requiring the blocking region around each point to have a
phase-space volume $(2\pi\hbar)^3$, i.e.,
\begin{equation}
\sqrt{\frac{(\mathbf{R}_k-\mathbf{R}_1)^2}{4\Delta x^2}
+\frac{(\mathbf{P}_k-\mathbf{P}_1')^2}{4\Delta p^2}} < 6^{1/6}
\quad\mbox{or}\quad
\sqrt{\frac{(\mathbf{R}_k-\mathbf{R}_2)^2}{4\Delta x^2}
+\frac{(\mathbf{P}_k-\mathbf{P}_2')^2}{4\Delta p^2}} < 6^{1/6},
\end{equation}
for any $k\ \ne 1,2$.
Furthermore, the physical coordinates of the final state
$\{\mathbf{R}_1,\mathbf{P}_1',\mathbf{R}_2,\mathbf{P}_2',\mathbf{R}_3,\mathbf{P}_3,\dots\}$
are transformed back to the original coordinates
$\{\mathbf{Z}_1',\mathbf{Z}_2',\mathbf{Z}_3',\ldots\}$.  Such a back
transformation does not exist for some region of physical coordinates, and then
the two-nucleon collision is regarded as Pauli blocked.

The two-nucleon collision process in AMD is a collision of two wave
packets rather than of two test particles with definite momenta.  In
the output data for the comparison, it is reported as if two nucleons
with momenta $\mathbf{p}_1$ and $\mathbf{p}_2$ have collided, where
$\mathbf{p}_1$ and $\mathbf{p}_2$ are the sample values taken from the
Gaussian distributions $e^{-(\mathbf{p}_1-\mathbf{P}_1)^2/2\Delta
  p^2}$ and $e^{-(\mathbf{p}_2-\mathbf{P}_2)^2/2\Delta p^2}$,
respectively.

\subsubsection{Two-nucleon collisions with cluster correlations}

Next, the collision process including cluster correlation is described. This method is used in the calculations of Ref.~\cite{SpRIT:2020blg}. Cluster correlations are taken into account in the AMD code  as in Refs.~\cite{Ikeno,ono2020nn18}.  When two nucleons $N_1$ and $N_2$ collide, each of them may form a cluster with other particles around it, so that a general process can be expressed as $N_1+N_2+B_1+B_2\to C_1+C_2$. 
For $B_1$ or $B_2$ being empty this corresponds to a collision where only one cluster is produced, and for both being empty to a usual two-nucleon collision.  The collision probability to a specific cluster configuration $(C_1,C_2)$ and a scattering angle $\Omega$ is
 \begin{equation}
vd\sigma=\frac{2\pi}{\hbar}
P(C_1,C_2,p_{\text{rel}},\Omega)
|M(p_{\text{rel}}^{(0)},p_{\text{rel}},\Omega)|^2
\frac{p_{\text{rel}}^2d\Omega}{\partial E/\partial p_{\text{rel}}},
\end{equation}
where the relative momentum $p_{\text{rel}}$ between the scattered $N_1$ and $N_2$ is determined by the energy conservation.  The total energy $E$ is calculated for the antisymmetrized wave function with the effective interaction.  The probability factor $P(C_1,C_2,p_{\text{rel}},\Omega)$ is essentially the overlap probability between the initial and final internal states of the cluster, but the non-orthogonality of the final states is taken into account to ensure that the sum of $P(C_1,C_2,p_{\text{rel}},\Omega)$ over the cluster configurations $(C_1,C_2)$ adds up to 1 for given $p_{\text{rel}}$ and $\Omega$.

The matrix element $|M(p_{\text{rel}}^{(0)},p_{\text{rel}},\Omega)|^2$ for the two-nucleon scattering is directly related to the assumed in-medium two-nucleon cross sections $\sigma_{NN}$, averaged in some way for the initial and final relative momenta. In the calculation of Ref.~\cite{SpRIT:2020blg} we use the parametrization
\begin{equation}
\sigma_{NN}=\sigma_0\tanh(\sigma_{NN}^{\text{(free)}}/\sigma_0),\quad
\sigma_0=0.5\times(\rho')^{-2/3}.
\end{equation}
Here we assume that it depends on a kind of phase-space density $\rho'=(\rho_1^{\prime\text{(ini)}}\rho_2^{\prime\text{(ini)}}\rho_1^{\prime\text{(fin)}}\rho_2^{\prime\text{(fin)}})^{1/4}$, where
\begin{equation}
\rho_i^{\prime(\text{ini/fin})}=\Bigl(\frac{2\nu}{\pi}\Bigr)^{3/2}
\sum_{k(\ne i)}
\theta\bigl(p_{\text{cut}}>|\bm{P}_i^{\text{(ini/fin)}}-\bm{P}_k|\bigr)
\times \exp\big[-2\nu(\bm{R}_i-\bm{R}_k)^2\bigr],
\end{equation}
are introduced for the initial and final momenta, $\bm{P}_i^{\text{(ini)}}$ and $\bm{P}_i^{\text{(fin)}}$, of the scattered two nucleons $i=1$ and $2$, with a momentum cut of $p_{\text{cut}}=(375\ \text{MeV}/c)\times e^{-\epsilon/(225\ \text{MeV})}$, which has a weak dependence on the collision energy $\epsilon$ in the two-nucleon center-of-mass system.

In the calculation of Ref.~\cite{SpRIT:2020blg}, cluster correlations are suppressed in medium, by allowing clusters to form only in low phase-space density regions of $\rho'<0.125\ \text{fm}^{-3}$. A two-nucleon collision is allowed only when the backward transformation from the physical coordinates to the original wave-packet coordinates exists. In addition, a condition $|\bm{R}_i-\bm{R}_k|^2/(4\Delta x^2)+|\bm{P}'_i-\bm{P}_k|^2/(4\Delta p^2)<1.50^2$ is imposed for the physical coordinates of the scattered nucleons $i=1$ and $2$, and for all the other nucleons $k$ with the same spin-isospin state as $i$.

\subsection{AMD+JAM code}
\vskip 0.1in
N. Ikeno, A. Ono
\vskip 0.2in
\subsubsection{Code history}
The AMD+JAM model~\cite{Ikeno} is a transport model which uses the antisymmetrized molecular dynamics (AMD) and a hadronic cascade model (JAM) in combination.
As described in Sec.~\ref{sec:AMD}, the AMD model solves the many-body system of nucleons with effective mean-field interactions and two-nucleon collisions. Antisymmetrization is exactly treated and cluster correlations can also be included. However, $\Delta$ resonances and pions have not been incorporated. On the other hand, as described in Sec.~\ref{sec:JAM}, JAM is a cascade model that treats production of various hadrons, but it does not include the mean-field interaction. To make use of the advantages of both models, we developed a wrapper code, which accepts input from the AMD code and controls the JAM code, to study pion production and the symmetry energy effect in heavy-ion collisions.

The dynamics of neutrons and protons is solved by AMD, and then pions and $\Delta$ resonances are handled by JAM.  In AMD+JAM, a basic assumption is that $\Delta$ and pion production can be treated perturbatively in heavy-ion collisions at around 300~MeV/nucleon, because the number of $\Delta$ resonances and pions existing at any intermediate time is small compared to the total number of nucleons in the system.

In Ref.~\cite{Ikeno:2019mne}, we improved the Pauli blocking treatment in AMD+JAM to use the precisely calculated Wigner function in AMD as the blocking probability.
In our computation, the $NN \leftrightarrow N \Delta$ and $\Delta \to N \pi$ processes take place always in the JAM code. However, it communicates bidirectionally with an AMD code that calculates the value of occupation probability upon every request from the JAM code, using the information on the AMD.

\subsubsection{Sending test particles from AMD to JAM}
Nucleons in the JAM calculation are always replaced by nucleon test particles calculated by AMD. Namely, particle production is calculated by JAM based on the nucleon dynamics calculated by AMD. The information of nucleons is sent from AMD to JAM at every 1 fm/$c$. This treatment violates some conservation laws in the higher orders. Corrections are introduced for the conservation laws of baryon number, charge and energy, by modifying the nucleon information.


\subsubsection{Pauli blocking}
	
Pauli blocking in hadron--hadron reactions in JAM is considered only for the nucleon(s) in the final state.  As we saw in the calculations of Ref.~\cite{Zhang2017}, the standard method for Pauli blocking in the JAM code has the problem of incomplete blocking due to fluctuation and smearing of the blocking factor, which is commonly observed in QMD codes.  In the calculations of Ref.~\cite{SpRIT:2020blg}, we employ a better method by using the Wigner function $f$ calculated in the AMD code from the antisymmetrized wave function~\cite{Ikeno:2019mne}.


\subsection{The quantum molecular dynamics (BQMD) and the isospin-dependent quantum molecular dynamics (IQMD) code}
\label{sec:BQMD}
\vskip 0.1in
A. Le Fèvre, J. Aichelin, C. Hartnack, R. Kumar 
\vskip 0.2in
In this short write-up, we provide information on the BQMD/IQMD codes. The main references are~\cite{Horst,Aic91,firstIQMD,Hartnack:1997ez}.

\subsubsection{Code history}

\begin{itemize}
\item Frankfurt \& Heidelberg, 1987-1990:\\
The original QMD has been further developed by the Frankfurt/Heidelberg collaboration \cite{Peilert:1989kr,Aichelin:1988me}. This version has later been dubbed BQMD and was used also to include Brueckner G-matrix results \cite{Bohnet:1989dzk}. In parallel,  IQMD has been developed by C. Hartnack as the first Quantum Molecular Dynamics code including isospin. It is based on the collision dynamics of the VUU model \cite{Kruse,Horst}.  Both QMD versions have been used extensively for explaining Plastic Ball data on fragment flow \cite{Peilert:1989kr} and fragment production. It successfully described pion flow data of Diogene \cite{piDiogene} and squeeze out at Plasticball \cite{squeezePB}. One of the major results was that, if the momentum dependence of the optical potential is properly taken into account, soft momentum dependent interactions give similar values for the flow observables as a static hard equation of state \cite{Aichelin:1987ti}. This reconciled the results obtained for giant monopole resonances and for heavy ion collisions. Steffen Bass and Christoph Hartnack refined the IQMD model with
respect to the $N-\Delta-\pi$ cycle \cite{Bass:1995pj,hartnack_pions}. They performed first perturbative kaon calculations \cite{pertKaon} and calculations for equations of state with secondary minima \cite{densityisomer}. \\ 

Zhuxia Li, who joined the Frankfurt HIC group headed by H. St\"ocker worked on Pauli potentials and damping procedures in the initialization of the nucleus with the aim for the application in QMD~\cite{firstIQMD}. She exported the IQMD model to China where it was the seed of several QMD models developed by different Chinese groups.

\item Nantes \& GSI  1991-today: \\
After J. Aichelin and C. Hartnack had been appointed by SUBATECH in  Nantes, the QMD activities moved to there. They extended IQMD by including virtual propagation, which allows to describe the dynamical observables of kaons, antikaons and hyperons \cite{Hartnack:2011cn}. They  proposed,  together with H. Oeschler, to use scaling laws for determining the nuclear equation of state for symmetric matter~\cite{Hartnack:2005tr}. The result of these studies is summarized in  \cite{FOPI:2010xrt,Hartnack:2011cn}.\\
In collaboration with Y. Leifels and A. Le Fèvre from GSI, IQMD was implemented into the FOPI data analysis environment, i.e., IQMD events (or events from any other event generator) undergo completely the same treatment as experimental events. This triggered many comprehensive direct comparisons between results from IQMD and FOPI data with identical conditions on the experimental acceptance (see e.g., \cite{FOPI-Hong,FOPI-pion,LeFevre:2016vpp}). \\
BQMD has been extensively used to describe the fragment data of the INDRA collaboration at GSI and also the ALADIN results at GSI \cite{Zbiri:2006ts,LeFevre:2009er,Fevre:2007pr}.
Later, A. Le Fèvre coupled IQMD to the new fragmentation algorithm FRIGA \cite{LeFevre:2019wuj}, which allows to describe the relation between fragment production and the equation of state \cite{LeFevre:2015paj} as well as the production of hypernuclei \cite{Hartnack:2015vzc}.\\

Rajeev Puri joined the group in late 2000 to develop the SACA algorithm for fragment identification \cite{Puri:1996qv,Puri:1998te,Gossiaux:1997hp}. Returning to the
Panjab University, Chandigarh he continued to work on  the onset of flow \cite{Gautam:2010ak}, on multifragmentation \cite{Vermani_2009,Kumar:2014ewa,sharma2015,rkumar18}, and the influence of isospin asymmetry \cite{preeti2018,sharma2021,sakshi2011,Sharma:2021wtj} using IQMD, together with Sakshi Gautam and Rohit Kumar.
\end{itemize}

\subsubsection{Initialization}

The nucleons are represented by Gaussian type distribution
functions (see eq.~\ref{fdefinition}). The centroids of these Gaussians are initially distributed in a nucleus 
by using a theta function in coordinate space and in momentum space, 
\begin{equation}
r_i < R;~~R= R_0 A^{1/3};~~p_i < P_F, 
\end{equation} 
where $ R_0=1.12~{\rm fm}$ and $ P_F = 268~{\rm MeV/c}$ are standard initialization values. 

\subsubsection{Forces}
In IQMD, a particle is represented by the single-particle Wigner density given by
\begin{equation} \label{fdefinition}
 f_i ({\bf r, p},t) = \frac{1}{\pi^3 \hbar^3 }
 \,\exp\big[-\frac{2}{L} ({\bf r} - {{\bf r}_i} (t) )^2\bigr]
 \,\exp\big[-\frac{L}{2\hbar^2} ({\bf p} - {{\bf{p}}_i} (t) )^2\bigr]
\end{equation}
The total one-body Wigner density is the sum of the Wigner densities of
all nucleons. The time evolution of the wave function is given by the 
Dirac-Frenkel-McLachlan equations \cite{broeck:1988,raab:2000}
which yield, for Gaussian wave functions with a fixed width,
\begin{equation}
\dot{r_i}=\frac{\partial\langle H \rangle}{\partial p_i}; \qquad
\dot{p_i}=-\frac{\partial \langle H \rangle}{\partial r_i},
\end{equation}
where the expectation value of the total Hamiltonian is
\begin{eqnarray}
\langle H \rangle &=& \langle T \rangle + \langle V \rangle= \sum_i \frac{p_i^2}{2m_i} +
\sum_{i} \sum_{j>i} \int f_i({\bf r, p},t) \,{\rm d{\bf p}}\, {\rm d{\bf p\,'}}
\,V({\bf r, r\,', p, p\,'})  f_j({\bf r\,', p\,'},t)\, \rm d{\bf r}\, \rm d{\bf r\,'}.
\end{eqnarray}
For Gaussian wave functions, the QMD equations are thus similar to the classical Hamilton equations, with the classical Hamiltonian replaced by the quantal expectation value.

In the QMD model, particles can interact by two-particle potentials and by collisions.
The two-particle potentials can be described by the convolution of the distribution
functions $f_i$ and $f_j$ with the interactions of Skyrme-, Yukawa- and Coulomb
types, including isospin and  momentum-dependent interactions: 
\begin{eqnarray}
V({{\bf r}_i, {\bf r}_j, {\bf p}_i, {\bf p}_j}) &=& G + V_{\rm Coul} = V_{\rm Sky} + V_{\rm Yuk} + V_{\rm mdi} + V_{\rm sym} + V_{\rm Coul}  \nonumber \\
       &=& t_1 \delta ({{\bf r}_i} - {{\bf r}_j}) +
           t_2 \delta ({{\bf r}_i} - {{\bf r}_j}) \rho^{\gamma-1}({{\bf r}_i}) +
           t_3 \frac{\hbox{exp}\{-|{{\bf r}_i}-{{\bf r}_j}|/\mu\}}
               {|{{\bf r}_i}-{{\bf r}_j}|/\mu} \\
       & & +t_4 {\rm ln}^2 (1+t_5({{\bf{p}}_i}-{{\bf{p}}_j})^2)
               \delta ({{\bf{r}}_i} -{{\bf{r}}_j}) + t_6 \frac{1}{\rho_0}
                T_{3}^i T_{3}^j \delta({{\bf{r}}_i} - {\bf{r}_j}) + 
        \frac{Z_i Z_j e^2}{|{{\bf r}_i}-{{\bf r}_j}|} \nonumber,
\label{Vij}
\end{eqnarray}
where the different contributions can be related to different terms in the
Bethe-Weizsaecker mass formula:
\begin{itemize}
\item Skyrme type forces and momentum-dependent interactions corresponding to
the volume energy. Its density dependence leads directly to the nuclear equation
of state of symmetric matter.
\item Yukawa forces corresponding to the surface energy
\item Coulomb forces corresponding to the Coulomb energy
\item Isospin-dependent forces corresponding to the asymmetry energy and thus leading
to the nuclear equation of state of asymmetric matter. We use for most calculations a linear 
dependence in baryonic density, but other density dependencies can be chosen
optionally.
\end{itemize}

The single-particle potential corresponding to the volume energy and resulting 
from the convolution of the distribution functions $f_i$ and $f_j$ with the interactions
$V_{\rm Skyrme}+ V_{\rm mdi}$ (local interactions including their momentum dependence) is 
for symmetric nuclear matter:
\begin{eqnarray} \label{eosinf}
U_i({{\bf r}_i},t)&=&\alpha
\left(\frac{\rho_{int}}{\rho_0}\right) +
\beta \left(\frac{\rho_{int}}{\rho_0}\right)^{\gamma}+\delta \,\textrm{ln}^2 \left( \varepsilon
\left( \Delta {\bf p} \right)^2 +1 \right)
\left(\frac{\rho_{int}}{\rho_0}\right),
\end{eqnarray}
where $\rho_{int}$ is the interaction density obtained by
convoluting the distribution function of a particle with the
distribution functions of all other particles of the surrounding
medium. Here, $\Delta {\bf{p}}$ is the relative momentum of a particle
with respect to the surrounding medium. The momentum-dependent part of 
the nucleon-nucleon (NN) interaction has been fitted to proton induced
experimental data on the real part of the nucleon optical 
potential \cite{Bertsch:1988ik,aichelin87}. For our approach, we used a parametrization of the optical potential data collected by
Passatore \cite{passatore67,Hartnack:1994zz}. 

Several Skyrme type parametrizations are available in the code with the most frequently used being: $H$ (`stiff'), $S$ (`soft'),
$HM$ (`stiff momentum-dependent'), $SM$ (`soft momentum-dependent'). 
The corresponding coefficients $\alpha$, $\beta$, $\gamma$, $\delta$, and
$\epsilon$ in Eq.~(\ref{eosinf}) are tabulated in Tab.~\ref{eostab}.
\begin{table}[hbt] 
\caption{Parameters of the different parametrizations of the EoS used in the IQMD model}
\label{eostab}
\centering
\begin{tabular}{|l|cccccc|}
\hline
 &$\alpha$ (MeV)  &$\beta$ (MeV) & $\gamma$ & $\delta$ (MeV) &$\varepsilon \,
 \left(\frac{c^2}{\mbox{GeV}^2}\right) $ & $K$ (MeV) \\
\hline
 S  & -356 & 303 & 1.17 & ---  & ---  & 200   \\
 SM & -390 & 320 & 1.14 & 1.57 & 500  & 200 \\
 H  & -124 & 71  & 2.00 & ---  & ---  & 376  \\
 HM & -130 & 59  & 2.09 & 1.57 & 500  & 376 \\
\hline
\end{tabular}
\end{table}
It should be noted that the parameters of $\alpha, \beta, \gamma$ for the
equations of state including momentum-dependent interactions (HM, SM) are adjusted
in such a way that they yield the same volume energy $E/A (\rho)$ in the infinite
nuclear matter as the corresponding equations of state without MDI (H, S).
Furthermore, it should be underlined that the use of momentum-dependent
interactions causes - following the Hamiltonian equations of motion 
$\partial H / \partial p = \dot{q}$ - the particles to propagate with an
effective mass given by
\begin{equation}
\frac{m_{\rm eff}}{m}= \frac{p}{m}\frac{1}{\frac{\partial E}{\partial p}}
\end{equation}

\subsubsection{Collisions}

In addition to the propagation of particles in the simulation,
collisions can take place. We use the common description that two
particles collide if their minimum distance $d$ in their
c.m.-frame, i.e.,\ the distance of the centroids of the
Gaussians, fulfills the requirement: $d\le\sqrt{\frac { \sigma_{\rm tot} } {\pi}}$ with 
 $\sigma_{\rm tot} = \sigma(\sqrt{s},\rm{type})$,
where ``type" denotes the collision type considered (e.g., ${\rm
NN}$, $N\Delta$, \ldots). The total cross section is the sum of
the elastic and all inelastic cross sections
\begin{equation}
 \sigma_{\rm tot} = \sigma_{\rm el}+\sigma_{\rm inel}
 = \sigma_{\rm el}+\sum_{\rm channels} \sigma_i \,.
\label{tot-el-inel}
\end{equation}
For instance, for a pp collision the important contributions are
\begin{equation}
 \sigma_{\rm tot} = \sigma_{\rm el}+ \rm \sigma(pp \to p \Delta^+) +
                  \sigma(pp \to n \Delta^{++}) \, .
\label{pp-example}
\end{equation}
We systematically use free cross sections as given by experiments (however, IQMD also has the option of using density-dependent cross sections), 
with the exception of the NN $\to K^+ \Lambda$N channel, where
experiments have revealed a strong final-state interaction which
is not present in matter. Different isospin channels are weighted
by isospin coefficients, e.g., for a pp collision we have
\begin{equation} \rm
\sigma(pp \to n \Delta^{++}) = 3 \sigma(pp \to p \Delta^+)
   =\frac{3}{4}\sigma_{\rm inelastic} .
\label{pp-delta}
\end{equation}
Experimentally inaccessible cross sections like $\Delta {\rm N}\to {\rm
NN}$ are calculated from their reverse reactions  (here ${\rm NN}
\to\Delta N$) using detailed balance. For reactions involving
unstable particles with a finite width, the form derived in
Ref.~\cite{Danielewicz:1991dh} is used. The probability that a collision leads to a particular 
channel is given by the contribution of this channel to the total cross section:
$P_{\rm channel} = \sigma_{\rm channel}/{\sigma_{\rm tot}}$. 
In the numerical simulation, the channel is chosen randomly
according to the probability of the channel, e.g., in the ${\rm pp \to p \Delta^+}$
case, there will be a 25\% chance to obtain a $\Delta^+$ in an
inelastic pp-collision.

\subsubsection{Pauli-blocking}
The Pauli-blocking is done by checking the phase-space distributions $f_i({\bf{x}}_i,{\bf{p}_i})$, 
where ${\bf x}_i$ is the position of the scattering particles $i$ (typically $i=1, 2$)
and ${\bf p}_i$ is the final state momentum of the corresponding particles.
This can be done in an isospin-explicit or in an 
isospin-averaged mode. For each particle $i$, we choose a random number $r_i$ with $0<r_i<1$ and define the collision to be
allowed if 
\begin{equation}
  r_i < (1-f_i), {\rm for~all~scattering~partners~} i.
\end{equation}
This corresponds to a Monte Carlo integration of the Uehling-Uhlenbeck factor 
$(1-f_1^{\rm final}) (1-f_2^{\rm final})$. For details see \cite{Aic91}.

\subsection{The constrained molecular dynamics (CoMD) code} \label{comd}
\vskip 0.1in
M. Papa
\vskip 0.2in
\subsubsection {Code history}
\begin{itemize}

\item 2000:2004\\
The starting hypothesis of the model \cite{Papa01} was the decomposition of the many-body wave function of nuclei as a direct product of Gaussian wave packets in phase-space \cite{Aic91} with centroids
$\overrightarrow{r}_{i}$,  $\overrightarrow{p}_{i}$ and fixed widths
(see also Eqs.~\ref{eq:QMDwf} and \ref{eq:wignerfxn}). The new idea underlying the model
was the imposition of constraints to the solution of the system of coupled semi-classical single-particle
equations of motion. In this first stage, this constraint was related to the quantal nature of the problem, i.e., to the Pauli principle for Fermionic systems. 
\item 2005:2011\\
An implementation of the code was obtained by introducing 
a further constraint related to the non-conservation of the total angular momentum in  nucleon-nucleon collision processes ~\cite{Papa02}. 
 In these first two stages of development, the model has been used to investigate processes induced by heavy-ion collisions at  Fermi energies with particular reference to the excitation
of the  fluctuating and coherent dipolar modes, cluster production, dynamical fission processes and incomplete fusion
processes. Examples of these studies are \cite{Papag1,Papag2,Papambreakup,Paparap1,Papalim,kohley1,sol}).
\item 2012:2019\\
During this period of time, a particular study was performed
with the aim to highlight typical  many-body correlations characterizing the model relative to a mean-field approach  \cite{Papaskyrme}.
CoMD calculations have been performed  to describe the dynamics of the isospin equilibration phenomenon and the connection to the density dependence of the symmetry energy \cite{Paparap2,Papaj1,Papanc1,Papanc2}.
\item 2020:2021\\
In this last period of time, further modifications have been introduced into the code  to study effects related to the use of finite-range effective interactions. This investigation is still in progress.
\end{itemize}
\subsubsection {Initialization}
"Ground state" (GS) configurations for different nuclei are obtained by coupling a cooling-warming procedure with constraints \cite{Papa01,Papa02}. In particular:

- at the very beginning of the procedure, the spatial coordinates of the A nucleons (centers of the
wave packets) are distributed uniformly inside a sphere of radius $R=r_{0}A^{\frac{1}{3}}+\Delta r$
with $ r_{0}$ =1.12 fm and $\Delta r$  set to obtain a value of $R$  near the experimental value.
The corresponding momenta are distributed inside a  sphere with a radius of about 300 MeV/c.

- in many cases follows a short phase of pre-cooling in which the system  is left to evolve in a spherical box.

- the cooling-warming procedure coupled with the constraints is performed until minimum and stationary energy conditions are reached on average.

- checks on the stability are performed in the final stage for time intervals typically ranging from 200 to several hundreds of fm/c.

According to the particular requirements, the following procedure can be adjusted case-by-case together
with the values of  some parameters such as the strength factor $C_{s}$   of the surface term. In general, the final phase-space distribution of these stable configurations
are different from the initial one. We note also that by using this dynamical method, the possibility
to get "good" GS configurations (charge radius, binding energies, average kinetic contribution, stability time) strongly depends on the used effective microscopic interaction and on the widths of the wave-packets.
\subsubsection{Interaction}
The microscopic nuclear effective interaction of the first versions is  of the Skyrme type. In particular, in the mean-field
limit, it has the following density ($\rho$) dependent  expression:
\begin{eqnarray}\label{Vpap:1}
V(\overrightarrow{r},\overrightarrow{r}')&=&\left[\frac{T_{0}}{\rho_{0}}+
\frac{2T_{3}\rho^{\sigma-1}}{(\sigma+1)\rho_{0}^{\sigma}}+\frac{T_{4}}{\rho_{0}}F'_{\gamma}(\rho)(2\delta_{\tau,\tau'}-1)
+\frac{C_{s}}{2\rho_{0}}\nabla^{2}\right]\delta(\vec{r}-\vec{r\,}')
\end{eqnarray}
with $F'_{\gamma}=(\frac{\rho}{\rho_{0}})^{\gamma-1}$ \cite{Papaskyrme}.
The effective interaction energy $U$ is obtained by convoluting  the above expression with the nucleon wave packets and by
substituting the explicit density dependence with the Gaussian's nucleon-nucleon overlap distribution \cite{Papaskyrme}.
The terms in the above expression represent: the two-body, three-body, isovector, and surface
interaction, respectively. The Coulomb interaction is  added according to Ref. \cite{Aic91}.
Therefore, the total  Hamiltonian is:
\begin{equation}\label{Vpap:2}
H=U_{twb}+U_{trb}+U_{isv}+U_{sur}+U_{Cou}+
\sum_{i=1,N}\frac{\vec{p}_{i}^{2}}{2m}, 
\end{equation}
where the last term is the total kinetic energy. For $\gamma=1$, we get:
\begin{equation}\label{Vpap:3}
\rho_{ij}=\frac{1}{(4\pi\sigma_{r}^{2})^{3/2}}exp\left[ \frac{(\vec{r}_{i}-\vec{r}_{j})^{2}}{4\sigma_{r}^{2}}\right];
\hskip 15pt
U_{twb}=\frac{T_{0}}{2\rho_{0}}\sum_{i,j\neq i}\rho_{ij};
\hskip 15pt
U_{trb}=\frac{T_{3}}{(\sigma+1)\rho_{0}^{\sigma}}\sum_{i}(\sum_{j\neq i}\rho_{ij})^{\sigma}
\end{equation}
\begin{equation}\label{Vpap:4}
U_{sur} =\frac{C_{s}}{2\rho_{0}}\nabla^{2}_{i}\sum_{j\neq i}\rho_{ij}; \hskip 15pt
U_{isv}=\frac{T_{2}}{2\rho_{0}}\sum_{i}\sum_{j\neq i}(2\delta_{\tau_{i},\tau'_{j}}-1)\rho_{ij},
\end{equation}
where $\rho_{ij}$ represents the generic Gaussian overlap between the Wigner distributions of the nucleon  wave-packets.
$\sigma_{r}$ is the related standard deviation. 
We note that in Eq.~(\ref{Vpap:2})  $H$ does not include the constant kinetic contribution $ \frac{3\sigma_{p}^{2}N}{2m}$
 related to the width of wave packets in momentum space.
 Therefore,  the kinetic contribution is associated to the real motion of the wave packets  and then it is available to be transformed and exchanged according to the dynamical evolution.
As an example, in Ref. \cite{Papa01} it was shown that with $T_{0}=-356$ MeV, $T_{3}=303$ MeV, $\sigma=1.166$ and $C_{s} \simeq 1.5$ MeVfm$^{2}$,  it is possible
to reproduce, at an acceptable level, the binding energy and charge radius of "ground state" (GS) configurations for nuclei with mass
number A=30-200. This is obtained by setting the width of the wave packets to 
$\sigma_{r}=1.15 - 1.3$ fm.
The equations of motion, which determine the time evolution of the Gaussian centroids, are obtained according to the  Hamilton equations, which are solved numerically by means of the fourth-order Runge-Kutta method.

In Ref. \cite{Papaskyrme} (CoMD-III), a study has been performed in the
symmetric and asymmetric Nuclear Matter (NM) limit. In this work, the limit has been simulated by large spherical systems whose minimum energy configurations were obtained by means of the cooling-warming procedure.
In particular, after correction for surface effects, we have obtained strength parameter values for the effective interaction that are able to produce the commonly accepted values for the nuclear matter saturation properties. The finite spatial correlation length introduced by the wave-packets and by the constraint, which is associated to the  Pauli principle,  produces  a global repulsive effect that is not balanced by the standard Skyrme effective interactions.
As shown in some detail, the set of the obtained parameter values differ from the usual ones (the ones obtained from the semi-classical mean-field theory).
In particular, in the case of  symmetric NM simulations, the new sets of obtained parameter values for the isoscalar effective interactions  depend on the parameter values describing the isovector interaction (strength and ``stiffness"). This reveals the existence of a  coupling between the two kind of interactions arising from the self-consistent dynamics.
In \cite{Xu2016},  these assigned values of the parameters have been used.


\subsubsection{Collision term}
The hard-core repulsive interaction between nucleons is simulated through nucleon-nucleon elastic scattering
processes with a method similar to the one described in Ref.~\cite{bonasera1994}. In particular:

- at each time step, a check is performed on all the nucleon pairs within a distance of $d$= 2 fm.
These nucleons are candidates for a collision only if they have not attempted a collision process in the
considered time step (no triple collisions),

- the collision is attempted with a probability $P_{c}=1-e^{-\frac{dt}{\tau}}$   where $dt$  is the
integration  time step and where $\tau=\frac{\lambda}{|v|}=\frac{1}{\rho\sigma_{nn}|v|}$, with  $|v|$, $\rho$, and $\sigma_{nn}$
being the relative velocity, average density around the chosen particle, and the assumed nucleon-nucleon
cross-section, respectively. The above relations are inspired from the  classical kinetic theory of gases. Velocity
and isospin dependencies of the cross-section are described according to Ref.~\cite{bonasera1994} with a cut at 50 mb. (In Ref.~\cite{Xu2016} the suggested value has been used).
The described method, based on the evaluated mean-free path $\lambda$, can produce in QMD-like models slight enhancements (10\%-15\%)
of the attempted collision rates compared to the case in which the same method is applied to mean-field BUU-like models in which the density is smooth. This enhancement is generated by the correlation existing  between mean-free path fluctuations (local density fluctuations) and the proximity of the interacting nucleons checked in the first step of the method (see also 
Fig. 20 in~\cite{Zhang2017}).  

- the attempted collision is  accepted if the occupation probabilities $f'_{i}$, $f'_{j}$ after the two-body scattering are both less or at most equal to a reference value $f_{max}$ . $f_{max}$ is set equal to the
average occupation probability $\overline{f}=\sum_{i}f_{i}/N$ as obtained
during the initialization stage (simulating  NM or large systems in GS configurations) through the numerical constraint procedure until minimum stationary values are obtained (see also the following paragraph).
The sharp condition on the occupation probabilities  is consistent with the interpretation of the generic evolving wave-packet as a single coherent-state, and it determines, together with the effectiveness of the constraint
(see Fig. 2 in Ref.~\cite{Papa01}), the behavior of the CoMD blocking probability shown
in Fig. 8  and the lower value of the slope parameter shown in Fig. 4  of  Ref.~\cite{Papa01}). The slow decrease of the CoMD blocking  probability at higher momentum in Fig. 8 reveals the co-volume in momentum space associated to the coherent state.

\subsubsection{Pauli blocking}
\textit{Occupation in phase-space (method-1):}
At each time step, for each particle $i$, we evaluate the occupation probability in phase space as:
\begin{equation}
f_{i}=\sum_{j}\delta( \tau_{i},\tau_{j} )\delta( m_{i},m_{j} )
\int_{V_{i}}F_{j}(\vec{r},\vec{p})d^{3}rd^{3}p,
\end{equation}
where $\tau_{i}$ and $m_{i}$  are the third components of the isospin and spin  quantum numbers of the nucleon, and $F_{j} $ is the Wigner transform of the generic wave-packet with centers at $\vec{r}_{j}$
and $\vec{p}_{j}$  in phase-space. The integral is performed over a hyper-cube
 $V_{i}$
equal to $h^{3}$. The sides are proportional to the wave-packet widths $\sigma_{r}$  and $\sigma_{p}$.
In this case, the generic wave packet is interpreted as a coherent sum of plane-waves. Each
plane wave contributes to the occupation probability with a weight given by the Gaussian.

\textit{Occupation in phase-space (method-2)}:
In this case, in principle more restrictive than the previous, each wave packet is interpreted as a "state" occupying uniformly a hyper-sphere
(or hyper-cube) with volume (1$\div$ 0.75)$h^{3}$ . The radii/or sides are always proportional to the widths
of the wave packets. In this case, the occupation in phase-space around each wave packet is 1 plus
a normalized quantity, which is proportional to the total overlap volume between identical particles.
It better reproduces the change, as a function of the density, of the average kinetic energy related
to the Fermi motion in GS configurations of large slabs of nuclear matter.
In both the cases, the Pauli principle can be fulfilled at the level of the order of 10\% with dedicated
algorithms for treating elastic  multi-scattering processes between identical nucleons that are close
in phase-space \cite{Papa01,Papa02}. In practical cases, the average typical value of the occupation
numbers in GS nuclear matter calculations is approximately in the range $f_{max}$=1.04-1.15, depending  on the effective interaction used and on
details of the numerical procedure related to the constraint.
We observe that in both cases the average stationary value $\overline{f}$  obtained with the numerical procedure related to the Pauli principle constraint is reached after several  time steps (see also the initialization paragraph). The phase-space fluctuations of overlap volumes in CoMD represented in Fig.8 of  Ref.~\cite{Papa01} with a gray line  were instead evaluated in the first step of the calculation.

\subsubsection {Constraint on rotational degrees of freedom}

In calculations with a high collision rate as the one experienced in the dynamics of hot and dense astrophysical objects, non-negligible violation of the total angular momentum
conservation law is seen. In fact, the procedure to simulate the hard core repulsive interaction
at high momenta (nucleon-nucleon residual interaction) including the one related to the constraint on the Pauli
prescription is in general performed with Monte Carlo techniques, which avoid the full
dynamical treatment of hard-core potential interaction. This conservation rule, fundamental also in heavy-ion multi-break-up
processes \cite{Papambreakup}, is restored by imposing further constraints on the dynamical evolution of
the wave-packets. For a given sub-system $C$ of nucleons  in compact configuration (belonging for
example to a cluster) which have undergone a collision in the time interval $dt$, we can
evaluate the dissipated angular momentum
$\Delta \vec{L}$:
\begin{equation}
\Delta \vec{L}=I\, \Delta\vec{\omega}
\end{equation}
where $I$ and $\vec{\omega}$ are the inertia tensor and the collective angular velocity
of the sub-system $C$.
Subsequently, we perform a series of transformations on the momenta
centroids of the $N_{C}$ wave-packets
according to the following relations:
\begin{equation}
\vec{p'}_{k}=\vec{p}_{k}+\vec{r}_{k}\times\vec{\Delta\omega};
\hskip 15pt
\vec{p''}_{k}=\vec{p'}_{k}+\alpha\frac{ (\vec{r}_{k}\vec{p'}_{k})r_{k}}{r_{k}^{2}}
\end{equation}

\begin{equation}
\vec{p'''}_{k}=\vec{p''}_{k}-\sum_{k \subset C}\frac{\vec{p''}_{k}}{N_{C}};
\hskip 15pt
\sum_{k \subset C}\frac{\vec{p'''}_{k}^{2}}{2m}-T_{C}=\epsilon_{min}
\end{equation}
where $T_{C}$ is the initial kinetic energy contents.
The system of the above equations is therefore solved  within the numerical precision $\epsilon_{min}$.
More details are given in Ref.~\cite{Papa02} (CoMD-II).

\subsection{The improved quantum molecular dynamics (ImQMD) code}
\vskip 0.1in
Y. X. Zhang, N. Wang, Z. X. Li 
\vskip 0.2in
The comparison calculations with the ImQMD code have been performed employing a recently-developed code~\cite{zhang01aa,zhang02aa,zha06,zhang04aa,zhang05aa,zhang06aa,Zhang:2014sva}. However, the initialization, the nucleonic mean field, the nucleon-nucleon cross sections, and the width of wave packet are set as in the comparison requirements.

\subsubsection{Parentage of ImQMD}

The ImQMD was developed in the group of Prof. Zhuxia Li at China Institute of Atomic Energy (CIAE) based on the QMD code.
The original version of QMD code was imported from Frankfurt in 1989 by Zhuxia Li, because she was involved in the work
with QMD in the group of Prof. H. St\"ocker.  When she came back, she implemented the symmetry energy term into the mean-field part and made improvements on the sampling algorithm of initial nuclei.

In 1997, Qingfeng Li joined our group in CIAE and started to work using the QMD code. He introduced the 1) Uehling-Uhlenbeck
factor into the Pauli-blocking part of collision term, and 2) isospin-dependent nucleon-nucleon cross sections to replace the original ones.

In 1998, Ning Wang joined our group and made a lot of improvements on the QMD code aiming to its application to reactions
at lower energies such as fusion reactions near the Coulomb barrier. He included the following features: 1) replaced the Yukawa term by density-gradient terms
(surface term and surface asymmetry term) and introduced a $\rho^{5/3}$ term based on the Skyrme energy density functional,
2) readjusted the force parameters by fitting the nuclear ground state properties of $\beta$-stable nuclei and excitation
functions of fusion cross sections, 3) implemented the phase-space constraint, and 4) introduced a phenomenological formula for the width of the wave packet, which depends on the size of the reaction system. This version of QMD was named ``improved QMD'' model, i.e., ImQMD (ImQMD, version: IQ1, IQ2, IQ3~\cite{Wang01,Wang02,Wang03,Wang04,Wang05,Wang06}). It has been mainly applied to heavy-ion reactions at lower energies, i.e., from
energies around the Coulomb barrier to about 50 AMeV.

In 2003, Yingxun Zhang made major improvements on the ImQMD model~\cite{zhang01aa,zhang02aa,zha06,zhang04aa,zhang05aa,zhang06aa,Zhang:2014sva}. These included: 1) the structure of the code was completely upgraded, 2) the parameters of the Skyrme potential energy functional were introduced and the explicit momentum-dependent interaction term as in
original QMD code was adopted, 3) for the study of the density dependence of symmetry energy, different forms of the density dependence
can be chosen in this version of the code, and 4) a new Cugnon parametrization of differential nucleon-nucleon cross sections were
implemented~\cite{Cugnon:1996kh}. This corresponds to version 05 of the model, i.e., ImQMD05. In 2012, the full Skyrme potential energy-functional with
explicit Skyrme-type momentum-dependent interactions (MDI) was introduced in ImQMD. This version is known as ImQMD-Sky~\cite{Zhang:2014sva}. It can be used for the study of the Skyrme
interaction, symmetry energy, nucleon effective mass splitting and other properties in the QMD model. The ImQMD model, versions 05 and Sky,
is appropriate for the study of heavy-ion reactions in the incident energy range of 20 AMeV $\le$ E$_{beam}\le$ 400 AMeV. In 2018, the accurate calculation of the three-body force term in the ImQMD was introduced, and some bugs were corrected. This version is known as ImQMD-L (L means the lattice method)~\cite{yang2021}. 

\subsubsection{Initialization}

In versions 05 and Sky, the width of the nucleon Gaussian wave packet in a reaction system is determined from a phenomenological formula,
$\sigma_r^2=(\sigma_{r,prj}^2+\sigma_{r,tar}^2)/2$, $\sigma_{r,A}^2=(0.16 A^{1/3}+0.49)^2$ $fm^2$.
The nucleon positions are sampled within a hard sphere with radius $R_n$ and $R_p$. A quadrupole deformation of
nuclei is taken into account, for example, $R_{n/p} (\theta)=R_{0n/p} (1+0.631\beta \frac{3 cos^2 \theta-1}{2})$, where  
$R_{0p}=1.18A^{1/3}-0.6$ \text{fm} and $R_{0n}= R_{0p}+\Delta R_{np}$, with $\Delta R_{np}$ being the thickness of neutron skin. The
neutron and proton have different masses of $m_n=0.93957~GeV$ and $m_p=0.93827~GeV$. With the sampled nucleon positions, one
can get the nuclear potential energy by using the interaction that is used in the code.

In version ImQMD-L~\cite{yang2021}, the nucleon positions are sampled within a hard sphere with radius $R_n$ and $R_p$ which are calculated based on the restricted density variation method (RDV). In the calculation of $R_n$ and $R_p$ with RDV, the Hamiltonian is same as that in the mean field propagation in the ImQMD-Sky. 

The momentum of $i^{th}$ nucleon is sampled within the local Fermi sphere with radius $P_f^i-\mathit{pfc}$, where
$P_f^i=(3\pi^2\rho_i)^{1/3}$ and the local density $\rho_i=\sum_j\rho_j(\mathbf{r}_i)$, from which the kinetic energy of system can be calculated, and $\mathit{pfc}$ is related to the width of the wave-packet, and is adjusted to fit the experimental binding energy of the sampled nucleus. If the energy per nucleon of the sampled nucleus falls into the range
of $BE/A\pm0.5MeV$, the sampled nuclei will be used for simulations.

\subsubsection{Force}

The mean field acting on nucleons is derived from the Skyrme potential energy density functional. The equations
of motion of the centroids of the nucleon wave packet are given by:
\begin{equation}
\dot{\vec{r}}_i=\frac{\partial H}{\partial p_i}, \qquad \dot{\vec{p}}_i=-\frac{\partial H}{\partial r_i},
\end{equation}
where the Hamiltonian is written as
\begin{equation}
H=T+U=\sum \frac{p_i^2}{2m}+\int u d^3r +U_{Coul}, \qquad u=u_{\rho}+u_{md}
\end{equation}

1) In version ImQMD-Sky, the real Skyrme potential energy density functional is used:
\begin{eqnarray}
u_{\rho}&=&\frac{\alpha}{2}\frac{\rho^{2}}{\rho_{0}}+\frac{\beta}{\eta+1}\frac{\rho^{\eta+1}}{\rho^{\eta}_{0}}
+\frac{g_{sur}}{2\rho_{0}}(\nabla \rho)^2+\frac{g_{sur,iso}}{\rho_{0}}(\nabla (\rho_{n}-\rho_{p}))^2+A_{sym}\left (\frac{\rho}{\rho_{0}}\right )\delta^{2}\rho+B_{sym}\left (\frac{\rho}{\rho_{0}}\right )^{\gamma_i}\delta^{2}\rho \label{urho3}
\end{eqnarray}

\begin{eqnarray}
u_{md}=C_0 \int d^3p d^3p' f(\vec r,\vec p)f(\vec r,\vec p')(\vec p-\vec p')^2+D_0 \int d^3p d^3p' \sum_{q=n,p} [f_q(\vec r,\vec p)f_q(\vec r,\vec p')(\vec p-\vec p')^2]
\end{eqnarray}
where $f(\vec r,\vec p)$ is the nucleon phase-space density, and is given by $f(\vec r,\vec p)=
\sum_i \frac{1}{(\pi\hbar)^3}\exp[-(\vec r-\vec r_i)^2/2\sigma_r^2-(\vec p-\vec p_i)^2/2\sigma_p^2]$ in the QMD approach.  The coefficients $\alpha$, $\beta$, $\eta$, $g_{sur}$, $g_{sur,iso}$, $A_{sym}$, and $B_{sym}$ are related to the standard Skyrme parameters as in Ref.~\cite{zhang02aa,zha06}. The coefficients $C_0$ and $D_0$ can be determined from the following expressions,
\begin{eqnarray}
C_0 &=&\frac{1}{16\hbar^2}[t_1(2+x_1)+t_2(2+x_2)]\\
D_0 &=&\frac{1}{16\hbar^2}[t_2(2x_2+1)-t_1(2x_1+1)].
\end{eqnarray}
In the ImQMD-Sky, the three-body term is calculated approximately and more details can be found in Ref.~\cite{yang2021}.

2) In version ImQMD-L, the potential energy density functional is the same as in ImQMD-Sky. However, we exactly evaluate the three-body term by a numerical quadrature method. Specifically, the force acting on particle $i$ due to the three-body term is calculated as
\begin{equation}
\label{cal3f}
\dot{\vec{p}}_i=-\frac{\partial U_3}{\partial \vec {r}_i}=-\beta\rho_0 \int \frac{\rho^\eta}{\rho_0^\eta}\frac{\rho_i}{\rho_0} \frac{\vec{r}-\vec{r}_i}{\sigma_r^2}d^3 r.
\end{equation}
The integral in Eq.~(\ref{cal3f}) is solved by using a 11-point Gauss-Legendre quadrature method. This results in a stronger three-body force in ImQMD-L than in ImQMD. To distinguish this from the previous version of ImQMD model, we named it as the ImQMD-L (-L, means the lattice method) in the following discussions.

3) In the versions ImQMD-IQ2 and ImQMD-IQ3, the energy density is taken to be
\begin{eqnarray}
u_{\rho}=&&\frac{\alpha}{2}\frac{\rho^{2}}{\rho_{0}}
+\frac{\beta}{\eta+1}\frac{\rho^{\eta+1}}{\rho^{\eta}_{0}}+\frac{g_{sur}}{2\rho_{0}}(\nabla \rho)^2
+\frac{C_s}{2\rho_0}[\rho^2-\kappa_s (\nabla\rho)^2]\delta^2.
\end{eqnarray}


The model parameters together with the width of the wave packet in coordinate space are determined by the properties
(including the stability) of ground state nuclei, the fusion excitation functions of a number of heavy-ion fusion reactions
at energies around the Coulomb barrier, and the charge distributions in multi-fragmentation process at Fermi energies. They can be found in Ref.~\cite{Wang01,Wang02,Wang03,Wang04,Wang05,Wang06}

\subsubsection{Collision term}

In the ImQMD code, nucleon-nucleon collisions are determined as follows:
firstly, only nucleon pairs with relative distance $r_{ij}<3.5$ fm and energy
$s=(p_i+p_j )^2> $3.556 GeV$^2$, where $p_i=(E_i,\vec{p}_i)$, are considered in order to speed up simulations; then,
the attempted collisions are determined by using the transverse and longitudinal distances of the colliding pairs.
In the center of mass of the colliding pair, if their transverse distance $b_{ij}$ is less than 
$\sqrt{\sigma_{total}^*/\pi}$, where $\sigma_{total}^*$ is the total nucleon-nucleon cross section with medium correction,
$\sigma_{total}^*=(1-\eta(E_{beam}))\sigma_{total}^{free}$, and their longitudinal distance 
$\vec{r}_{ij}^{\prime}\cdot \vec{p}_{ij}^{\prime}/|\vec{p}_{ij}^{\prime}|$ is less than
$v_{ij}^{\prime} \gamma \delta t/2$
with $\delta t$ being the time step, they undergo attempted collisions. The quantities used above are given by the following 
expressions
\begin{equation}
b_{ij}=\sqrt{r_{ij}^{\prime 2}-(\vec{v}_{ij}^{\prime} \cdot \vec{p}_{ij}^{\prime}/|\vec{p}_{ij}^{\prime}|)^2}, \qquad
\vec{v}_{ij}^\prime=\vec{v}_i^{\prime}-\vec{v}_j^{\prime}=\frac{\vec{p}_i^{\prime}}{E_i^{\prime}}-\frac{\vec{p}_j^{\prime}}{E_j^{\prime}}
\end{equation}

\begin{equation}
\vec{p}_i^{\prime}=((\gamma-1)\vec{p}_i\cdot\frac{\vec{\beta}}{\beta^2}-\gamma E_i)\vec{\beta}+\vec{p}_i,\qquad
\vec{r}_{ij}^{\prime}=(\gamma-1)\vec{r}_{ij}\cdot\frac{\vec{\beta}}{\beta^2}\vec{\beta}+\vec{r}_{ij},
\end{equation}

\begin{equation}
\vec{\beta}=\frac{\vec{p}_i+\vec{p}_j}{E_i+E_j}, \qquad \gamma=\frac{1}{\sqrt{1-\beta^2}},
\end{equation}

After generating a random number $\xi$, the collision for elastic channel is determined by $\xi <\sigma_{el}^*/\sigma_{total}^*$. 
The momentum direction of outgoing nucleons is determined by their differential cross section. The nucleon cross section and their
differential cross section in free space are taken from Ref.~\cite{Cugnon:1996kh}.

\subsubsection{Pauli blocking}
The outgoing nucleons of an attempted collision are checked for Pauli blocking, which consists of the evaluation of two criteria.
The prejudgment is $\frac{4 \pi}{3} r_{i^{\prime}k}^3 \frac{4\pi}{3}p_{i^{\prime}k}^3\ge h^3/8$, where $i'$ 
is the outgoing nucleon, $k$ represents other surrounding nucleons. It means that the outgoing nucleon should not be too close 
to others in phase-space. If this relation is satisfied, the Pauli blocking with the probability, $1-(1-P'_{i} )(1-P'_j)$ is then checked, where $P'_i$ is calculated as $P'_i=4\sum_{k,k\ne i'}\exp [-(\vec{r}_{i'}-\vec{r}_k)^2/(2\sigma_r^2 )]\exp 
[-(\vec{p}_{i'}-\vec{p}_k)^2/(2\sigma_p^2 )]$, and $P'_i=1$ if $P'_i\ge 1$.

\subsubsection{Phase-space constraints}
Versions IQ2 and IQ3:
To describe the Fermionic nature of the N-body system and to improve the stability of an individual nucleus, the phase-space
occupation constraint method \cite{Papa01} is adopted. The phase-space occupation constraint is an effective approach to improve
the momentum distribution in the nuclear system. In this approach, the phase-space occupation number of each particle is checked
at each time step. If the phase-space occupation number is larger than 1 for particle $i$, that is, $\bar{f}_i>1$, the momentum
of particle $i$ is randomly changed between $i$ and its partner $j$ under the condition of total momentum and total kinetic energy conservation in the process. In the ImQMD model, the new sample for the
momenta of the particles is constrained by the Pauli blocking condition (either $\bar{f}_i>1$ or $\bar{f}_j>1$) coming from
the Fermionic nature of nucleon. Actually, the momenta of two particles obtained in this way will not only influence the motion of particles
at this time step but also influence the motions in subsequent steps. It is not known whether the system will follow in the most
suitable motion path. In this method~\cite{Wang01,Wang02}, we perform one further step; that is, we calculate the total energy of the system at step
$t$ and the total energy $E(t+\Delta t)$ at the next time step $t+\Delta t$ simultaneously~\cite{Wang_2016}. If the value of $E(t+\Delta t)$
is larger than that of $E(t)$, the momentum of nucleon $i$ and $j$ are rearranged. The number of times to re-execute the
procedure is small (zero to four) at each time step for fusion reactions. This additional constraint can further improve the
stability of an individual nucleus (reducing the spurious emission of nucleons) and is helpful for the study of the formation
process of compound nuclei, which lasts several thousand fm/$c$ or longer. We have checked that the total energy of the system
is well conserved for thousands of fm/$c$ with this new procedure.

Versions 05 and Sky:
Since these are mainly used for the beam energy ranging from 20-400A MeV, the method of phase-space constraint is similar to that of Ref.~\cite{Wang01,Wang02}. 

\subsection{The isospin-dependent quantum molecular dynamics at BNU (IQMD-BNU) code}
\vskip 0.1in
J. Su, F. S. Zhang 
\vskip 0.2in

In the following, the main features of the IQMD-BNU model are presented. The main references are
~\cite{jsu2011,jsu2013,jsu2014}.
\subsubsection{Code history}

\begin{itemize}

\item 1989:\\
The Isospin-QMD (IQMD) code \cite{firstIQMD}  was developed from the VUU code.
\item 1997-2000:\\
The IQMD code, which includes isospin-dependent Coulomb potential, symmetry potential, NN cross-sections
and Pauli blocking, was used by L.W. Chen and
F.S. Zhang to investigate the isospin effects on nuclear collective flow as well as the
multifragmentation phenomenon in heavy-ion collisions at intermediate energies~\cite{Chen98, Chen99, Zhang99}.
\item 2009-2014:\\
The original version of IQMD code was written in FORTRAN 77.
The version of IQMD code in FORTRAN 95 was written by J. Su and F.S. Zhang at Beijing Normal University,
and was called IQMD-BNU hereafter. In this version of code, the statistical-decay model
GEMINI \cite{Charity88} and the interface program were included.
It is used to study the multifragmentation and nuclear temperature~\cite{jsu2012,jsu2013}.

\item 2014-2020:\\
IQMD-BNU is further improved to study the symmetry energy, effective mass splitting~\cite{jsu2016}, and yields in spallation and fragmentation reactions at Sun Yat-sen University~\cite{jsu2018,jsu2019}.

\end{itemize}

\subsubsection{Initialization}

In coordinate space, the initial nuclei are initialized by randomly distributing the nucleons in a sphere of radius $r = 1.12 A ^{1/3}$ fm, while a minimum distance between two nucleons, $r_{ij}^{min}\ge 1.5$ fm, is required.
For the initialization of the momentum, the local potential $U(r_{i})$ of the i-th nucleon generated by all the other nucleons is evaluated, and the local Fermi momentum is then determined by the relation $p_{F}^2(r_{i}) = 2mU(r_{i})$.
The momentum of the i-th nucleon is chosen randomly between zero and the local Fermi momentum $p_{F}(r_{i})$.
The initialization also requires two nucleons to satisfy the phase-space relation $(\vec{r}_{i}-\vec{r}_{j})^{2}(\vec{p}_{i}-\vec{p}_{j})^{2} \ge d_{min}$, where $d_{min} = 0.038^2 (\text{fm}^2\text{MeV}^2/\text{c}^2)$.
Finally, the momenta of all nucleons are scaled in the same proportion by fitting the experimental binding energy.

The stability of the initialized nuclei is checked by performing the dynamical evolution. Generally, the initialized nuclei are stable for 1000 fm/$c$.

\subsubsection{Forces}
The forces are calculated from the Hamiltonian $H$, which is expressed as

  \begin{equation}
    H = T + U_{Coul} + \int V_{nucl} (\rho (\mathbf{r})) d\textbf{r}.
    \label{Heff}
  \end{equation}
Here, the first term $T$ is the kinetic energy, the second term $U_{Coul}$ is the Coulomb potential energy, and the third term is the nuclear potential energy. Each term of the nuclear potential energy-density functional $V_{nucl}$ reads as

  \begin{equation}
  \begin{aligned}
      V_{nucl}= &V_{Sky}+V_{sur}+V_{sym}+V_{mdi}.
    \label{Vloc}, \\
      V_{Sky} = &\frac{\alpha}{2} \frac{\rho^2}{\rho_0} + \frac{\beta}{\gamma+1} \frac{\rho^{\gamma+1}}{\rho_0^{\gamma}}, \quad
      V_{sur} = \frac{g_{sur}}{2} \frac{(\nabla\rho)^2}{\rho_0}, \quad
      V_{sym} = \frac{C}{2} \frac{(\rho_n -\rho_p)^2}{\rho_0}, \quad
      V_{mdi} = g_{\tau}\frac{\rho^{8/3}}{\rho_0^{5/3}}.
  \end{aligned}    
  \end{equation}
Here, $\rho$ is the density and $\rho _{0}$ is the saturation density.
$V_{Sky}$ contains the two- and three-body Skyrme interaction terms.
$V_{sur}$ is the surface term to describe the surface property of finite nuclei.
$V_{sym}$ is the symmetry term, which is crucial for reproducing isospin-dependent effects in the dynamics.
$V_{mdi}$ is the momentum-dependent interaction term.

Besides the default nuclear potential energy-density functional, another option is provided to study the symmetry energy and the effective k-mass splitting. In Refs. \cite{jsu2016, jsu2017}, the nuclear potential energy density of the asymmetric nuclear matter with density $\rho$ and asymmetry $\delta$ reads, 

\begin{equation}
  \begin{aligned}
    V(\rho, \delta) = & \frac{\alpha}{2} \frac{\rho^2}{\rho_0} + \frac{\beta}{\gamma+1} \frac{\rho^{\gamma+1}}{\rho_0^{\gamma}} + \frac{C_{sp}}{2}(\frac{\rho}{\rho_{0}})^{\gamma_{i}} \rho \delta ^{2} \\
    &+\sum_{\tau} (1+x) \iint v(\mathbf{p},\mathbf{p}') f_{\tau}(\mathbf{r},\mathbf{p}) f_{\tau}(\mathbf{r},\mathbf{p}')d\mathbf{p} d\mathbf{p}'  \\
    &+\sum_{\tau} (1-x) \iint v(\mathbf{p},\mathbf{p}') f_{\tau}(\mathbf{r},\mathbf{p}) f_{-\tau}(\mathbf{r},\mathbf{p}')d\mathbf{p} d\mathbf{p}'
    , \\
    v(\mathbf{p},\mathbf{p}') = & \frac{C_{m}/\rho_{0}}{1+(\mathbf{p}-\mathbf{p}')^{2}/\Lambda^{2}},
    \label{V}
  \end{aligned}
\end{equation}
where $\mathbf{p}$ and $\mathbf{p}'$ are the momenta of the nucleon, $f_{\tau}(\mathbf{r},\mathbf{p})$ is the phase-space density, with $\tau$ = 1/2 for neutrons and $\tau$ = -1/2 for protons.
For infinite nuclear matter at zero temperature, the phase-space density can be approximated as a step function, $f_{\tau}(\mathbf{r},\mathbf{p}) = \frac{3\rho_{0}}{4\pi p_{F\tau}^{3}} \Theta (p_{F\tau}-p)$.
The parameters $\alpha$, $\beta$, $\gamma$, $C_{sp}$, $\gamma_{i}$, $x$, $C_{m}$ and $\Lambda$ are temperature-independent.
In Eq. (\ref{V}), the first and second terms refer to the local two-body and three-body interactions, which are the same as in the default case.
The form of the local symmetric potential (the third term) is the extension of the default form, for which $\gamma_{i}$ = 1.
The fourth and fifth terms refer to the momentum-dependent interactions.
Values of the parameters for this nuclear potential energy density functional can be found in Refs.~\cite{jsu2016, jsu2017} .

\subsubsection{Collision term}
The binary collisions in the IQMD-BNU code are performed according to the following differential cross sections,

\begin{equation}
\left( \frac{d \sigma}{d \Omega} \right)_{el(inel)} = \sigma _{el(inel)} ^{free}\, f_{el(inel)}^{angl}\, f _{el(inel)}^{med},
\end{equation}
where $\sigma^{free}$ is the cross section of NN collisions in free space, $f^{angl}$ gives the angular distribution.
The isospin-dependent parametrizations of $\sigma ^{free}$ and $f^{angl}$ adopted in this work are taken from Ref. \cite{Cugnon:1996kh}.
$f^{med}$ gives the in-medium corrections to the NN cross section.
Five types of in-medium factor $f^{med}$ \cite{jsu2016a} can be chosen in the IQMD-BNU code.
The first in-medium factor (1F for short) is written as \cite{fmed2}

  \begin{equation}
    f _{el}^{med}=1 -0.2 \frac{\rho}{\rho _{0}}.
    \label{fmed1}
  \end{equation}
This in-medium factor is density-dependent and energy-independent.

The second one (2F for short) has been used by Wang et. al. \cite{Wang:2013wca01}

\begin{equation}
 f _{el}^{med} = \left\{
 \begin{aligned}
   &1  \qquad  &p_{NN} > 1 GeV/c, \\
    &1+\frac{1/6+5/6\,\exp(-3\rho/\rho _{0})-1}{1+(p_{NN}/0.3)^{8}} \qquad  &p_{NN} \leqslant 1 GeV/c,
    \label{fmed2}
 \end{aligned}
 \right.
 \end{equation}
where p$_{NN}$ in GeV/c denotes the relative momentum of two colliding nucleons.
This in-medium factor is density-dependent and energy-dependent.

The third one (3F for short) is proposed by Cai et. al. \cite{fmed3}

  \begin{equation}
  \begin{aligned}
    &f _{pp}^{med} =f _{nn}^{med}
    =\frac{1.0+7.772E_{lab}^{0.06}\rho^{1.48}}{1.0+18.01\rho^{1.46}}, \quad
    f _{np}^{med}
    =\frac{1.0+20.88E_{lab}^{0.04}\rho^{2.02}}{1.0+35.86\rho^{1.90}},
  \end{aligned}
  \label{fmed3}
  \end{equation}
where E$_{lab}$ in MeV is the incident energy in the laboratory frame of the two colliding nucleons.
These in-medium factors depend on density, energy and isospin.

The fourth in-medium factor (4F for short) is \cite{fmed4}

  \begin{equation}
  \begin{aligned}
    &f _{el}^{med} = \sigma_{0}/\sigma ^{free} \tanh (\sigma ^{free}/\sigma_{0}),\quad
    \sigma_{0} = 0.85\rho ^{-2/3}.
  \end{aligned}
  \label{fmed4}
  \end{equation}
Since the cross section in free space $\sigma ^{free}$ depends on the energy and isospin, the fourth in-medium factor is also energy-dependent and isospin-dependent.

For the fifth one, the effective k-masses of the nucleons are extracted from the momentum-dependent interaction, and then the in-medium factor of scattering between $i^{th}$ and $j^{th}$ nucleons is calculated by \cite{persram2002,Li05,fe12,padh1992}

  \begin{equation}
  \begin{aligned}
    &f_{ij}^{med}                 =\left[\frac{m_{i}^{*}m_{j}^{*}/(m_{i}^{*}+m_{j}^{*})}
                             {m_{i}m_{j}/(m_{i}+m_{j})}\right]^{2},\quad
    m_{i}^{*} =\left[\frac{1}{m_{i}}+\frac{\partial U}
                      {p_{i}\partial p_{i}}\right]^{-1}.
   \label{fmed5}
  \end{aligned}
  \end{equation}
Here, U is the single-particle potential, m$^{*}$ is the effective k-mass.
Considering Eq. (\ref{V}), the in-medium factor is density-, energy- and isospin-dependent.
Moreover, the isospin-dependence can be adjusted by changing the value of the parameter x.

\subsubsection{Pauli-blocking}

In the IQMD-BNU code, two Pauli-blocking methods have been implemented.
In the first Pauli-blocking method (named PAULI1 in the code), the phase-space density $f'_{i}$ in the final state is calculated and interpreted as a blocking probability
  \begin{equation}
     f_{i} = \sum_{n\neq i} \exp\bigg[-\frac{(\mathbf{r}_{n}-\mathbf{r}_{i})^{2}}{2L}
     -\frac{(\mathbf{p}_{n}-\mathbf{p}_{i})^{2}L}{\hbar^{2}}\biggr].
  \end{equation}
Thus, the collision is only allowed with a probability of $(1-f'_{1})(1-f'_{2})$.

The second Pauli blocking method (named PAULI2 in the code) is related to the phase-space density constraint \cite{Papa01}.
In fact the phase-space occupation $\overline{f}_{i}$ is calculated by performing the integration on an hypercube of volume $h^{3}$ in the phase space centered around the final state, i.e.,
  \begin{eqnarray}
     \overline{f}_{i}&=&\sum_{n} \delta _{\tau_{n},\tau_{i}} \delta _{s_{n},s_{i}}
     \int _{h^{3}} d^{3}rd^{3}p\, \frac{1}{\pi ^{3} \hbar ^{3}}
     \,\exp\bigg[-\frac{(\mathbf{r}_{n}-\mathbf{r}_{i})^{2}}{2L}
     -\frac{(\mathbf{p}_{n}-\mathbf{p}_{i})^{2}L}{\hbar ^{2}}\biggr] .
     \label{fi}
  \end{eqnarray}
The scattering is accepted only if the fraction of the final phase-space of both particles is less than 1.

\subsection{The isospin-dependent quantum molecular dynamics at SINAP (IQMD-SINAP) code}
\vskip 0.1in
G. Q. Zhang 
\vskip 0.2in

In this section, the relevant features of the IQMD-SINAP code are described.
Some applications to the analysis of experimental data are given in Refs.~\cite{Cao:2010def,Zhang:2011yv}.

\subsubsection{Code history and parentage}
The Object Oriented C++ version of IQMD-SINAP code derives from several QMD codes.

\begin{itemize}

\item SINAP, F77 IDQMD:\\
The IDQMD (Isospin dependent QMD) code was written in F77 language. The basic IQMD framework was introduced,
including initialization, mean field, collision and Pauli-blocking. This code is used at SINAP and mainly originates
from Ref.~\cite{Aic91,Hartnack:1997ez}

\item Wei Guo, C++ IDQMD:\\
The IDQMD (Isospin dependent QMD) code has been developed in C++ language by Wei Guo~\cite{guo_c++_2007}, at SINAP,
since 2007. The basic C++ framework was introduced, including particle class, nucleus class, nuclear matter class,
meanfield class and collision class.

\item Koji Niita, C++ G4QMD:\\
The G4QMD code in C++ language, deriving from its corresponding Fortran version JQMD~\cite{JQMD}, was
developed by Koji Niita as a part of the model library in Geant4. Its aim was to provide neutron spectra relevant for
application purposes.

\item Christoph Hartnack, IQMD :\\
The IQMD code was written in F77 language~\cite{firstIQMD,Aic91,Hartnack:1997ez}, where the interaction
density, the force and the collision were introduced.

\item Toshiki Maruyama, EQMD~\cite{maruyama_extension_1996}: \\
Implementation of the frictional cooling method for the ground-state nuclei for the initialization at event-by-event level.

\item ROOT library: \\
The ROOT library~\cite{brun_root_1997} was also introduced to IQMD-SINAP to enhance the efficiency of the code, especially
the TLorentzVector class and the TRandom3 class. The output of the code and the analysis are also performed by making use of
the ROOT framework.

\end{itemize}

\subsubsection{Initialization}

In IQMD-SINAP, the tentative distribution of nucleons in a nucleus can be obtained by using the Fermi function or other distribution
functions in coordinate space and a local density-dependent sampling distribution in momentum space. However, the following frictional
cooling procedure smears the tentative distribution in coordinate space and momentum space and leads the nucleus to a minimum-energy
state, which is adopted for the ground-state. The initial distributions of nucleons in coordinate space and momentum space depend on
the Hamiltonian of the nuclear system. Different tentative distributions affect only the time evolution of the system of nucleons
to reach the initial distribution. The nucleons in a nucleus obey the following equations of motion during the cooling process,
\begin{eqnarray}
\frac{d\vec{r}_i}{dt}&=&               \frac{\partial \langle H \rangle}{\partial{\vec{p}}_i}
               +\mu_{\vec{r}}\frac{\partial \langle H \rangle}{\partial{\vec{r}}_i} \ , \\
\frac{d\vec{p}_i}{dt}&=&              -\frac{\partial \langle H \rangle}{\partial{\vec{r}}_i}
         +\mu_{\vec{p}}\frac{\partial \langle H \rangle}{\partial{\vec{p}}_i}\ . \nonumber
      \label{eqDamp} 
\end{eqnarray}
Here ${\vec{r}}_i$ and ${\vec{p}}_i$ are the centers of position and momentum of the $i$-th wave packet;
$\langle H \rangle$ are the Hamiltonian of the nuclear system; and $\mu_{\vec{r}}$ and $\mu_{\vec{p}}$ are damping coefficients.
With negative values for these coefficients, the system goes to its local minimum energy state,
\begin{eqnarray}
\frac{d\langle H \rangle}{dt}&=&\sum_i\left[
\frac{\partial \langle H \rangle}{\partial{\vec{r}}_i}\dot{\vec{r}}_i+
{\frac{\partial \langle H \rangle}{\partial{\vec{p}}_i}}\dot{\vec{p}}_i\right]\\
&=&\sum_i\left[
\mu_{_{\vec{r}}}\left({\frac{\partial \langle H \rangle}{\partial{\vec{r}}_i}}\right)^2+
\mu_{_{\vec{p}}}\left({\frac{\partial \langle H \rangle}{\partial{\vec{p}}_i}}\right)^2\right]\leq 0. \nonumber
\label{eqmin}
\end{eqnarray}
Other factors, such as the minimum distance between nucleons, the single-nucleon energy, phase-space constraint and local
density-dependent momentum sampling, do not affect the initial distribution of nucleons in the nucleus, but help the tentative
distribution get closer to the initial distribution and save CPU time to prepare the initial distribution.

An initialization option using the tentative distribution as the initial distribution, but without the friction process, has been
implemented in order to perform the calculations proposed as part of the Code Comparison Project. Various transport codes share the same tentative
distribution for the following collision process to understand the difference among the codes. However, without the friction process,
the stability of the nucleus is reduced, leading to virtual emission of nucleons during the collision processes. It should also be
noted that the Hamiltonian of the nuclear system determines the initial distribution. A simple Hamiltonian may result in an unreasonable
initial distribution. The Skyrme-type interaction with soft compressibility and an initialization without an initial stability test,
as proposed for the comparison after the Shanghai Transport2014 meeting, result in an artificial state with an arbitrary binding 
energy rather than the minimum energy. 

\subsubsection{Forces}
The interaction potential adopted in the IQMD-SINAP is given by a local Skyrme-type interaction, the surface interaction, a
Coulomb interaction, an optional momentum-dependent interaction, an optional Pauli potential, and an isospin asymmetry potential.
The two-body potential interaction can be written as,
\begin{eqnarray}
V^{ij} = V^{ij}_{\rm Skyrme} + V^{ij}_{\rm Sur} + V^{ij}_{\rm mdi} +
           V^{ij}_{\rm Coul} + V^{ij}_{sym}.
\end{eqnarray}

In QMD codes each nucleon wave function is represented as a Gaussian form, which is defined as:
\begin{equation}
\label{eq:gauspack}
  \phi_i(\vec{r},t) = \frac{1}{{(2\pi L)}^{3/4}}
\exp\big[-\frac{{(\vec{r}- \vec{r_i}(t))}^2}{4L}\bigr] \, \exp\big[-\frac{i\vec{r}
\cdot \vec{p_i}(t)}{\hbar}\bigr],
\end{equation}
where $L$ is the width parameter for the Gaussian wave packet.
By applying the convolution of the interaction, the potential energy part then reads as:
\begin{eqnarray}
\label{meanfield}
 \langle V \rangle&=&\frac{1}{2} \sum_{i} \sum_{j \neq i}\int d\vec{r}\, d\vec{r}\,'
 \phi_i(\vec{r},t) \phi^{*}_i(\vec{r},t)\,V^{ij}  \phi_j(\vec{r}\,'\,t)\phi^{*}_j(\vec{r}\,',t)\,.
\end{eqnarray}
The equations of motion for the center of wave packets of nucleons then read as:
\begin{eqnarray}
\dot{\vec{p}}_i &=& - \frac{\partial \langle H \rangle}{\partial \vec{r}_i} = - \frac{\partial \langle V \rangle}{\partial \vec{r}_i}
\quad {\rm and} \nonumber\\
\dot{\vec{r}}_i &=& \frac{\partial \langle H \rangle}{\partial \vec{p}_i} = \frac{\vec{p}_i}{\sqrt{m_i^2+p_i^2}} + \frac{\partial \langle V \rangle}{\partial \vec{p}_i} ,
\end{eqnarray}

Due to the undetermined nuclear force, the interactions adopted in IQMD-SINAP are also kept in a flexible and extensible way. 
Benefited from the Object Oriented C++ framework, it is reasonably straightforward to add or modify the interaction, 
for example, to add the tensor force or the spin-orbit interaction term.

\subsubsection{Collision term}
The collision term in the IQMD-SINAP code originates from IQMD~\cite{Hartnack:1997ez}, where the
experimental cross-sections for elastic and inelastic collision of a pair of nucleons are parametrized.
At present, only the pion and $\Delta$ production inelastic channels are considered, while it is also possible to extend to
the strangeness degree of freedom. Two particles can collide if the minimum relative distance is less than
$\sqrt{\sigma_{\rm tot}/\pi}$, within the time step during the evolution of the system.

Besides the cross-section, there are some other conditions for the colliding pair of nucleons.
\begin{itemize}

\item Conservation laws:\\
Before and after the collision, each colliding pair of nucleons respect the law of the energy and momentum
conservation in their c.m.~reference. After the collision, the direction of momentum is generated at random with the constraint
of describing, on average, the experimental differential cross-sections. However, this results in breaking the angular momentum
conservation law at the same time, due to the classical description of the collision.

\item First collisions:\\
During the reaction, the first colliding partner of each nucleon in the projectile (target) is constrained to originate from the 
target (projectile), which reduces the chance of the virtual emission of nucleons.

\item Too soft collisions:\\
Collisions between nucleons that are too soft are blocked to save the CPU time and remove fake collisions  between nucleons
inside a forming fragment. Usually, the threshold kinetic energy between the colliding pair of nucleons is about 20 MeV, 
well below the Fermi energy.

\item Pauli blocking:\\
A Pauli blocking procedure is applied to decide whether a collision could happen.
\end{itemize}

\subsubsection{Pauli-blocking}
The overlap between two nucleons is
\begin{equation}
\langle \phi_i,\phi_j \rangle= \rm{exp}[  -\frac{(\vec{r}_{i} - \vec{r}_{j} )^2}{2L}-  \frac{2L(\vec{p}_{i} - \vec{p}_{j})^2}{\hbar^2} ].
\end{equation}
The Pauli-blocking factor for $i$ nucleons at the phase point $(\vec{r}_{i},\vec{p}_{i})$  is defined as
\begin{equation}
f(\vec{r}_{i},\vec{p}_{i})\equiv\sum_{j(i\neq j)}\delta(S_i,S_j)\,\delta(T_i,T_j)\,\langle \phi_i\phi_j \rangle,
\end{equation}
where $S$ and $T$ are the spin and isospin.
If the blocking factor is larger than a generated random number, the collision is blocked.
\subsection{Jet AA Microscopic transport (JAM) code\label{sec:JAM}}
\vskip 0.1in
A. Ono, N. Ikeno, Y. Nara, A. Ohnishi
\vskip 0.2in

\subsubsection{Code history}

Jet AA Microscopic transport model (JAM) is a transport model which is developed by Nara \textit{et al.}~\cite{JAM}.
This model has been successfully applied to high-energy collisions up to more than one hundred GeV/nucleon.
In JAM, hadrons and their excited states are explicitly propagated in space-time by the cascade method.
In order to describe nuclear collisions consistently from low to high energy, elementary hadron-hadron collisions
are modeled by resonance production at low energies, string excitations in the soft region and hard parton-parton
scattering at collider energies according to the important physics in each energy range.


The main features included in JAM  are as follows.
 (1) At low energy, inelastic hadron-hadron collisions are modeled by
     the resonance productions based on the idea from RQMD \cite{rqmd1a,rqmd1b,rqmd1c,rqmd2} and UrQMD \cite{Bass:1998ca,UrQMD}.
 (2) There are options to include the nuclear mean field.
  The nuclear mean field is simulated based on the BUU theory or
     Quantum Molecular Dynamics (RQMD/S) \cite{JAMmf,Maruyama:1996rn,Mancusi:RQMDS}.
 (3) Above the resonance region, soft-string excitation is implemented
     along the lines of the HIJING model~\cite{hijing01a,hijing01b,hijing01c}.
 (4) Multiple mini-jet production is also included in the same way
     as in the HIJING model in which jet cross sections and the number
     of jets are calculated using an eikonal formalism for
     perturbative QCD (pQCD), and hard parton-parton scatterings
     with initial- and final-state radiation are simulated
     using the PYTHIA~\cite{pythia} program.



\subsubsection{Initialization}
In usual calculations for heavy-ion collisions, the nucleons in the nucleus are randomly distributed
according to the appropriate Woods-Saxon density distribution $\rho(r)$.  The nucleons are distributed
in the order from protons to neutrons, and a nucleon is not allowed to be too close ($d_{\text{min}}=0.8$ fm) to any of the
already distributed nucleons. However, recent calculations choose $d_{\text{min}}=0$ to sample nucleon coordinates independently. The Fermi motion of each nucleon is assigned according to the local Fermi
momentum $p_{\text{f}} = \hbar (\frac{3}{2}\pi^2 \rho(r))^{1/3}$. The initialized phase space is then Lorentz-boosted.

\subsubsection{Potentials/Forces}
In JAM, there is an option for incorporating the nuclear mean field. However, in our comparison calculations, no mean field
is included. Therefore, the trajectories of hadrons are straight lines in between two-body collisions or decays.


\subsubsection{Collision prescription}
In the energy domain relevant for our work, only hadron-hadron collisions play a role.  Detailed explanations
of collisions are given in Ref.~\cite{JAM}. In JAM without mean field, the time step $\Delta t$ can be as
large as the whole calculation time of each event, though we may take $\Delta t= 10$ or 20 fm/$c$ for
output purposes. The particles are propagated along classical trajectories until they interact (two-body
scattering, absorption, or decay). The interaction possibilities are determined by the method of the so-called
``closest distance approach''. If the minimum relative distance $b_{\text{rel}}$ for any pair of particles
becomes less than the interaction range specified by $\sqrt{\sigma_{\text{tot}}(s)/\pi}$, then the particles
are assumed to collide.  However, there are some ambiguities in a relativistic treatment because a
collision occurs when the two particles are located at different space-time points. Here we adopt a similar
procedure to that in Refs.~\cite{PRC_JAM9_01a,PRC_JAM9_01b,PRC_JAM9_01c,PRC_JAM37} to mimic the reference-frame dependence.

The minimum relative distance $b_{\text{rel}}$ is defined as follows. Let us denote the coordinates and momenta of two particles in the computational frame
by $(t_1, \bm{x}_1)$, $(t_2, \bm{x}_2)$ and $(E_1, \bm{p}_1)$, $(E_2, \bm{p}_2)$, respectively, where
$t_i$ ($i=1, 2$) denote the production time (namely the time of the last collision) whose initial values
are set to 0.  In the c.m.\ frame of two particles, we have the trajectories of particles $i=1$ and 2, 
\begin{equation}
\bm{x}_{i}^{*}(t^{*}) = \bm{x}_{i}^{*} + \bm{v}_{i}^{*}(t^{*} - t^{*}_{i})
\end{equation}
where asterisks represent quantities in the two-body c.m.\ frame, and $\bm{v}_i^{*} = \bm{p}_i^{*}/E_{i}^*$
are velocities there.
One can minimize $|\bm{x}_1^*(t^*)-\bm{x}_2^*(t^*)|^2$ with respect to $t^*$ as in Ref.~\cite{PRC_JAM37} to
obtain the collision time and the minimum distance
\begin{equation}
 t_{\text{min}}^* = - \frac{\bm{x}_{12}^{*} \cdot \bm{v}_{12}^*
  }{\bm{v}_{12}^*}, \quad
b_{\text{rel}}^2 = \bm{x}_{12}^{*2} - \frac{(\bm{x}_{12}^* \cdot \bm{v}_{12}^*)^2}{\bm{v}_{12}^{*2}}
\end{equation}
with $\bm{x}_{12}^{*} = \bm{x}_1^*(0) -\bm{x}_2^{*}(0)$ and $\bm{v}_{12}^* = \bm{v}_1^{*} - \bm{v}_2^{*}$.
We note that the definition of $b_{\text{rel}}$ is the same as in Refs.~\cite{Bass:1998ca,UrQMD, Kodama:1983yk, WOLF1990615}.
By transforming to the computational frame, the collision times for these particles are given by
\begin{equation}
 t_i^{\text{coll}} = \gamma t_{\text{min}}^* + \gamma\bm{\beta} \cdot \bm{x}_i^*(t_{\text{min}}^*),
\end{equation}
where $\bm{\beta} = (\bm{p}_1 + \bm{p}_2)/(E_1 + E_2)$ and $\gamma$ is the corresponding
Lorentz $\gamma$ factor. We assume that the sequence of collisions are ordered
by the average time~\cite{PRC_JAM9_01a,PRC_JAM9_01b,PRC_JAM9_01c,PRC_JAM32} 
\begin{equation}
 t_{\text{c}} = \tfrac{1}{2}(t_1^{\text{coll}} + t_2^{\text{coll}}).
\end{equation}
The particles cannot collide if $\pi b_{\text{rel}}^2>\sigma_{\text{tot}}(s)$. Furthermore, the collision can
occur in the current time step only if all the following time conditions are met; (a) $t_1^{\text{coll}} > t_1$
and $t_2^{\text{coll}} > t_2$, i.e., the collision cannot occur before the previous interaction (collision or
production) for each particle, (b) $t_{\text{c}}\ge \max(t_1, t_{\text{start}})$ and
$t_{\text{c}}\ge \max(t_2, t_{\text{start}})$,  where $t_{\text{start}}$ is the beginning of the current
time step, (c) $t_{\text{c}}\le t_{\text{end}}$ where $t_{\text{end}}$ is the end of the current time step,
and (d) $t_{\text{c}}\le t_1^{\text{decay}}$ and $t_{\text{c}}\le t_2^{\text{decay}}$ where $t_i^{\text{decay}}$
is the time of decay, which was randomly decided when the particle $i$ was produced, in case it is unstable.
[There is an option to replace the condition (b) with  (b$^\prime$) $t_{\text{c}} > t_{\text{start}}$.
The condition (d) may also be replaced by (d$^\prime$) $t_1^{\text{coll}}\le t_1^{\text{decay}}$ and
$t_2^{\text{coll}}\le t_2^{\text{decay}}$.]

The predicted collisions are processed according to the order of $t_{\text{c}}$.  For a collision,
the space-time coordinates of the two particles $(t_1,\bm{x}_1)$ and $(t_2,\bm{x}_2)$ are first
propagated to $t=t_{\text{c}}$, namely $t_i:=t_{\text{c}}$ and $\bm{x}_i:=\bm{x}_i+(\bm{p}_i/E_i)(t_{\text{c}}-t_i)$.
[There is another option to propagate them to $t=t_1^{\text{coll}}$ and $t_2^{\text{coll}}$ as
$t_i:=t_i^{\text{coll}}$ and  $\bm{x}_i:=\bm{x}_i+(\bm{p}_i/E_i)(t_i^{\text{coll}}-t_i)$.]
We then generate elastic scattering according to the probability $P_{\text{el}} = \sigma_{\text{el}} /\sigma_{\text{tot}}$.
Otherwise, we select an inelastic channel. After updating the momenta of the particles in the final state, the
possibilities of future collisions associated with these particles have to be recalculated.

Each particle carries a variable to store the ID of the last collision or decay that the particle has
experienced most recently. Particles having the same collision/decay ID are not allowed to collide.

\subsubsection{Cross sections}
In low energy region, hadron-hadron reactions are treated by the cross sections based on experimental
data and the detailed balance. Here we describe the treatments of the $\Delta$ resonances and pions that
we adopted in our recent publication \cite{Ikeno}. The cross section for the $N + N \rightarrow N +\Delta$
reaction is written as,
\begin{eqnarray}
\frac{d \sigma [NN\rightarrow N\Delta(m)]}{dm}= \frac{C_I}{p_N(s)s} 
\frac{|\mathcal{M}|^2}{16\pi} F(m) p_\Delta(m,s),
\end{eqnarray}
where $m$ is mass of $\Delta$ resonances sampled in the region $m_N + m_\pi < m < \sqrt{s} - m_N$, and $p_N(s)$ and $p_\Delta(m,s)$ are the initial and final momenta, respectively, in the c.m.\ frame.  The matrix element $|\mathcal{M}|$ is assumed to be
\begin{equation}
|\mathcal{M}|^2 =A \frac{s\Gamma_{\Delta}^2}{(s-m_{\Delta}^2)^2 + s\Gamma_{\Delta}^2} ,
\end{equation}
with $A/16 \pi= 64400$ mb GeV$^2$.  The mass distribution function $F(m)$ is written as
\begin{equation}
 F(m) = \frac{2}{\pi} \frac{m m_{\Delta} \Gamma(m)} {(m^2-m_{\Delta}^2
  )^2 + m^2_{\Delta} \Gamma(m)^2},
\label{Fm}
\end{equation}
where $ \Gamma(m)$ is
\begin{equation}
 \Gamma(m) = \Gamma_{\Delta} \frac{m_{\Delta}}{m} 
\left( \frac{p_\pi(m)}{p_{\pi}(m_{\Delta})} \right)^3 
\frac{1.2}{1+0.2(p_\pi(m)/p_\pi(m_{\Delta}))^2}
\label{eq:Gamma}
\end{equation}
with $\Gamma_\Delta=0.118$ GeV, $m_\Delta=1.232$ GeV and $p_\pi(m)$ being the pion momentum when the $\Delta$
resonance with mass $m$ decays in its rest frame.  This is a similar parametrization to UrQMD in Ref.~\cite{Bass:1998ca,UrQMD}.
The Clebsch-Gordan factor is 
$C_I =\frac{1}{4}$ for the channels $nn \rightarrow n\Delta^0$, $np \rightarrow p\Delta^0$, $np \rightarrow n\Delta^+$,
and $pp \rightarrow p\Delta^+$, and $C_I = \frac{3}{4}$ for the channels $nn \rightarrow p\Delta^-$ and
$pp \rightarrow n\Delta^{++}$.



The cross section for the $N + \Delta \rightarrow N + N$ reaction
can be obtained by the spin averaged matrix elements
$|\mathcal{M}|^2=|\mathcal{M}_{NN\rightarrow N\Delta}|^2$ as
\begin{equation}
 g_1 |\mathcal{M}_{NN\to N\Delta}|^2 = g_2|\mathcal{M}_{N\Delta\to NN}|^2
\end{equation}
where $g_1$ and $g_2$ are the spin degeneracy factors
$g_1=(2s_N+1)(2s_N+1)=4$ and $g_2=(2s_N+1)(2s_\Delta+1)=8$.
Thus the cross section is
\begin{equation}
  \sigma [N\Delta(m) \rightarrow NN' ]  = \frac{g_1}{g_2}
\frac{C_I}{1+\delta_{NN'}}
\frac{1}{p_{\Delta}(m,s) s} \frac{|\mathcal{M}|^2}{16\pi} p_N(s).
\end{equation}
where the factor $(1+\delta_{NN'})$ takes into account the limitation
in angle integral for a final state with identical particles.

The decay width for $ \Delta(m) \rightarrow N + \pi$ is given by  $\Gamma(m)$ with Eq.~(\ref{eq:Gamma}).
The partial decay widths for isospin channels are $\Gamma[ \Delta^-\rightarrow n \pi^- ] = 
\Gamma[ \Delta^{++} \rightarrow p \pi^{+} ] = \Gamma(m) $, $\Gamma[ \Delta^{0} \rightarrow p \pi^- ] = 
\Gamma[ \Delta^{+} \rightarrow n \pi^{+} ] = \frac{1}{3}\Gamma(m) $, and $\Gamma[ \Delta^{0} \rightarrow n \pi^{0} ] = 
\Gamma[ \Delta^{+} \rightarrow p \pi^{0} ] = \frac{2}{3}\Gamma(m) $.

The cross section for the $ N + \pi \rightarrow \Delta(m)$ reaction
with $m=\sqrt{s}$ is written as
\begin{equation}
\sigma[N\pi \rightarrow \Delta(m)]
= \frac{4}{2\cdot 1}
\frac{\pi}{p_\pi(m)^2}\frac{\Gamma(m)\Gamma[\Delta(m)\rightarrow N\pi]}
{(m-m_\Delta)^2 + \frac{1}{4}\Gamma(m)^2 }.
\end{equation}
The first factor is the ratio of the spin degeneracies in this reaction.

\subsubsection{Pauli Blocking}
Pauli blocking  is considered for the nucleon(s) in the final state.
The blocking factor is calculated by the following expression using Gaussian wave packets as
\begin{equation}
 f_{i} =\frac{1}{2}  \sum_{k \in\tau_{i} (k \neq i)} 
2^3\, \exp\big[- { \bm{r}_{ik}^2/2L} - {2L \bm{p}_{ik}^2}/ \hbar^2\bigr],
\end{equation}
where $\tau_i$ is the species (proton or neutron) of the particle $i$, and the factor $\frac{1}{2}$ is for the spin averaging.  For each term in the right hand side, the distances $\bm{r}_{ik}$ and $\bm{p}_{ik}$ in the phase space are evaluated in the c.m.\ frame of the particles $i$ and $k$.  The parameter is fixed to be $L=2.0$ fm$^2$.




%

\subsection{JQMD 2.0 code} \label{sec:JQMD2.0}
\vskip 0.1in
T. Ogawa, K. Niita, S. Hashimoto, T. Sato
\vskip 0.2in
\subsubsection{Code history}
JQMD is one of the QMD models developed by Niita {\it et al.} in 1995 \cite{JQMD}. It served
for a lot of studies such as fundamental research \cite{qmd_fundam1, qmd_fundam2, qmd_fundam3}, accelerator
shielding design \cite{tanaka2011radiation, kusaka2015beam} and cosmic-ray dosimetry \cite{sihver2008dose} for
approximately 20 years without any modifications until JQMD Ver. 2.0 was developed \cite{JQMD2.0}. 
 JQMD was intended to describe various aspects of nucleus-nucleus
collisions by a simple approach (i.e., JQMD was designed to reproduce as many observables as possible 
using minimum number of parameters). Owing to its simple coding, one can easily run JQMD by conventional computers. 
JQMD also featured accurate reproductions of secondary particle production cross sections
(\cite{Satoh2007507}, see Figs. 7-15 of \cite{JQMD}), which was attributed to a reasonable configuration of initial states
and the parametrization of elastic and inelastic cross sections. 

JQMD was incorporated into general-purpose radiation transport codes such as PHITS \cite{phits2013} and
Geant4 \cite{GEANT4, GEANT-QMD}. JQMD Ver.2 is an improved version of JQMD Ver.1.
The description of the mechanisms of peripheral collisions was refined, and the reproduction of fragment yields was improved by this update. JQMD Ver.2 was incorporated into
PHITS Ver.2.76 and the later versions as the event generator for nucleus-nucleus collisions. 


The initialization, potential, collision and Pauli blocking algorithms of JQMD Ver.2 combined with the reactions above 3 GeV/n, such as production of resonances heavier than 1500 MeV, string formation and string decay, was published as JAMQMD Ver.2 \cite{JAMQMD}. By switching from JQMD Ver.2 and JAMQMD Ver.2 at 3 GeV/n, nucleus-nucleus collisions from 10 MeV/n to 1 TeV/n can be simulated. 

\subsubsection{Initialization}

The initial state of the nucleus is configured by a random packing method associated with frictional cooling/heating,
which adjusts the excitation energy of the nucleus. The excitation energy is adjusted until the binding energy
agrees with that of the nuclear ground state taken from the statistical decay model GEM \cite{GEM}. Random packing
is carried out by the scheme described below. 

The (x,y,z) coordinates of the nucleons are randomly sampled based on a Wood-Saxon distribution defined by the diffuseness equal to  0.2 fm and radius $r= 1.124 \times A^{1/3}$, 
%
%
where $A$ is the mass number of the nucleus. The minimum distances of nucleon pairs which the same and different isospin are 1.5 fm and 1.0 fm, respectively. If the sampled coordinate does not
satisfy this condition, the coordinate is rotated (i.e., angular coordinates were randomly changed without changing
the distance from the center) at random until the nucleon reaches unoccupied space. The momentum of nucleons is
randomly sampled below the local Fermi momentum.
%
%
The configured ground state nuclei is transferred to the center-of-mass frame of target and projectile by a
Lorentz-transform. 
%
%
%
%
%
The nucleus-nucleus impact parameter is sampled below the maximum obtained by 
%
$b_\mathrm{AAmax} = 1.2 \times (A_\mathrm{t}^{1/3}+ A_\mathrm{p}^{1/3})$. 
%
At the beginning of QMD calculation, the impact parameter $b$ is corrected considering a Coulomb trajectory,
which depends on the charge and mass of projectile and target, and incident energy. 

\subsubsection{Potentials/Forces}

The strong repulsive forces between baryons close to each other are treated as stochastic two-body collisions,
whereas the long-range interaction between particles is treated as the potential term of the Hamiltonian.
The single particle potential on particle $i$ is written as follows,
\begin{eqnarray} 
\label{potential}
 V_i &=& \frac{1}{2     }       \frac{A}{\rho_\mathrm{s}}      \langle \rho_i \rangle
 + \frac{1}{1+\tau}       \frac{B}{\rho_\mathrm{s}^\tau} \langle \rho_i \rangle^\tau + \frac{1}{2} \sum_j \frac{c_i c_j \: e^2}{|\mathbf{R_i} - \mathbf{R_j}|} \mathrm{erf}(\frac{|\mathbf{R_i} - \mathbf{R_j}|} {\sqrt{4L}})
 + \frac{C_\mathrm{s}}{2\rho_\mathrm{s}} \sum_j (1 - 2|c_i - c_j|) \rho_{ij}, 
\end{eqnarray} 
where  $A, B,$ and $C$ are a force parameters with values of 219.4, 165.3, and 25 MeV, respectively,
$\rho_\mathrm{s}$ is the saturation density (=0.168 fm$^{-3}$), $\langle \rho_i \rangle$ is the overlap integral
of wave packets between the $i$-th particle and all the other particles, $\tau$ is 4/3, $c_i$ is particle charge number, $\mathbf{R_i}$ denotes
the position of $i$-th particle, $L$ is the width of wave packet representing nucleons (= 2 fm$^2$), and $\rho_{ij}$ is the overlap integral of wave functions of
the $i$-th and $j$-th nucleons. The first, second, third, and fourth terms are two Skyrme-type terms, the Coulomb term,
and the symmetry term, respectively. All the forces act on baryons, whereas pions are affected only by the electric force. Forces explicitly dependent on momentum were not included. 

The equation of motion for particles is written as follows:
\begin{eqnarray} 
 \label{equation-of-motion-R} 
 & \mathbf{\dot{r_i}} = \frac{\mathbf{p_i}}{2p_i^0} + \sum_j^N \frac{m_j}{p^0_j} \frac{\partial \langle \hat{V_j} \rangle }{\partial \mathbf{p_i}}; ~~~~\mathbf{\dot{p_i}} = - \sum_j^N \frac{m_j}{p^0_j}  \frac{\partial \langle \hat{V_j} \rangle }{\partial \mathbf{r_i}}; 
 ~~~~ p_i^0 = \sqrt{ \mathbf{p_i}^2 + m_j^2 + 2 m_j \langle \hat{V_i} \rangle }, 
 \end{eqnarray} 
where $\langle \hat{V_j} \rangle$ is the potential of $j$-th particle, and $N$ is the number of particles in the system. 

%
%

\subsubsection{Collision term}
Two-body collisions are treated as the stochastic two-body collision process with the inclusion of the Pauli blocking. The prescription of two-body collision term adopted in JQMD is similar to that used in the BUU calculation done by Wolf {\it et al.} \cite{WOLF1990615,extended_detailed_balance}, and it was modified to extend the energy range up to 3 GeV/n. The extension above 3 GeV/n is discussed elsewhere \cite{JAMQMD}. 
In the calculation routine of JQMD, collision pairs are searched in sequence. The baryon with id number
1 and that with id number 2 are selected and their impact parameter is calculated by,
%
%
\begin{equation}
 b_\mathrm{col} = \sqrt{ \vec{r}^2 + (\frac{ \vec{p_1} \cdot \vec{r} }{m_1})^2 - \frac{( \frac{\vec{p_1} \cdot \vec{r} }{m_1} - \frac{\vec{p_2} \cdot \vec{r} m_1}{ p_1 \cdot p_2 }   )^2}{1 - (\frac{m_1 m_2}{p_1 \cdot p_2 } )^2} }, 
\label{covariant-impact parameter} 
\end{equation}
where $\vec{r}$ is the spatial distance between the baryons, $\vec{p_i}$ is the momentum of $i$-th baryon, $m_i$
is the rest mass of $i$-th baryons, and $p_i$ is the four-dimensional momentum of $i$-th baryon. All these
quantities are those defined in the nucleus-nucleus c.m. frame.  A collision is attempted if the
impact parameter is smaller than $b_\mathrm{max}$ calculated by, 
%
\begin{eqnarray} 
\label{temporal-bmax} 
 && b_\mathrm{max} = \sqrt{ \frac{\sigma}{\pi} } \quad (\sqrt{s'} \leq 428.6 \mathrm{MeV} \land b_\mathrm{AA} \leq 0.6 b_\mathrm{AAmax}) \nonumber\\ 
 && b_\mathrm{max} = \sqrt{ \frac{7.5 \sigma}{\pi} } \quad (\sqrt{s'} \leq 428.6 \mathrm{MeV} \land b_{AA} \geq 0.6 b_\mathrm{AAmax}) \nonumber \\
 && \quad \qquad = 1.32 \mathrm{fm} \quad  (\sqrt{s'} \geq 428.6 \mathrm{MeV}), \nonumber 
\end{eqnarray}
where $\sqrt{s'} = \sqrt{s} - m_1 - m_2$, $\sigma$ is the total reaction cross section, $b_\mathrm{AA}$ is the impact parameter between the
colliding nuclei. 
The collision is skipped if it did not occur
in the time step of 1 fm/c. The timing condition is defined by, 
\begin{eqnarray} 
\label{collision_timing}
 && |t_{12} + t_{21} | \leq 1 \mathrm{fm/c},\qquad
  t_{ij} = \left( \frac{ \vec{p_i} \cdot \vec{r}}{m_i} - \frac{ \frac{ \vec{p_i} \cdot \vec{r}}{m_i} 
 - \frac{ \vec{p_j} \cdot \vec{r} m_i}{p_i \cdot p_j}   }{1 - ({\frac{m_i m_j}{p_i \cdot p_j}})^2} \right)
 \gamma_i   , 
\end{eqnarray}
where $\gamma_i$ is the Lorentz gamma factor of the $i$-th baryon in the nucleus-nucleus c.m. frame. 
In the collision calculation routine, a reaction channel is assigned to each particle pair with the respective channel cross section. If $ b_\mathrm{max} \geq \sqrt{\sigma/\pi}$, no reaction channel is assigned and the collision is skipped. 

In order to treat reactions at high bombarding energies, nucleons N, deltas $\Delta$(1232), N*(1440)s,
and pions with their isospin degree of freedom are considered. Therefore the applicable energy range of JQMD is
3 AGeV. The creation and absorption of these particles are treated as collisions. The reaction channels considered in JQMD are 
baryon  elastic scattering $B + B \rightarrow B + B 	$, and the reactions
  $N + N		\leftrightarrow N + \Delta$ 	,
  $N + N 		\leftrightarrow N + N^*  $	,
  $N + \pi 		\rightarrow \Delta $ 	,
  $N + \pi		\rightarrow N^*	  $	, 
  $\Delta + \pi	\rightarrow N^* $. 
Elastic reaction cross sections have been parametrized by,
\begin{eqnarray} 
\label{elastic-X-sec}
 && \sigma = \frac{C_1}{1+100 \sqrt{s'}}+C_2 \qquad (\sqrt{s''} \leq 0.4286), \nonumber\\
 && \sigma = C_3 \left[ 1 - \frac{2}{\pi} \arctan(1.5 \sqrt{s'} - 0.8) \right] +C_4 \qquad (\sqrt{s''} \geq 0.4286), \\
 &&  \nonumber 
\end{eqnarray}%
where $\sqrt{s''} = \max [0,\sqrt{s}-m_1-m_2-C_0]$, $C_0$ is a reaction cut-off energy (0.02 GeV for nucleon-nucleon and 0 for the others), $C_{1,2,3,4}$ are
fitting parameters listed elsewhere \cite{JQMD}, and $m_{1,2}$ are the masses of the colliding
particles. Below 0.4286 GeV, the elastic reaction cross section is described by the Cugnon's parametrization
\cite{Cugnon:1980rb, Cugnon2} in central collisions, whereas in peripheral collisions
($b_\mathrm{AA} \geq 0.6 b_\mathrm{AAmax}$) the cross section in free space is adopted. 
%
%
%

The total baryonic resonance production cross sections ($N + N \rightarrow N + \Delta$ and $N + N \rightarrow N + N^*  $) are parametrized based on the method proposed by VerWest and Arndt \cite{VerWest}. Because their
parametrization was optimized to reproduce pion yields up to 1.5 AGeV of laboratory incident energy
whereas JQMD is applied up to 3 AGeV, the fitting parameters have been slightly modified. 

The cross sections can reproduce the experimental data on total, elastic, and inelastic
nucleon-nucleon cross sections (see Fig.~1 of \cite{JQMD}). In addition, calculated single pion
production cross sections and two-pion production cross sections are in good agreement with the
experimental data \cite{total_X-section} up to 3 AGeV of laboratory incident energy (see Figs.2
and 3 of \cite{JQMD}). Pionic fusion and s-wave pion production is not explicitly included, but
fitting parameters have been determined to include these channels effectively. 
 

The masses of the resonances are determined by considering them to have a statistical distribution. Specifically, the mass is
randomly sampled according to the Breit-Wigner distribution with width defined by, 
\begin{eqnarray} 
\label{resonance_mass_distribution} 
 & \Gamma(M) = (\frac{q}{q_r})^3 (\frac{M_r}{M}) (v(q)/v(q_r) )^2 \Gamma_r, \qquad 
 v(q) = \frac{\beta_r^2}{\beta_r^2 +q^2} , 
\end{eqnarray}
where $q$ denotes the c.m. momentum in the $\pi N$ channel. The parameters in Eq. \eqref{resonance_mass_distribution}
are given elsewhere \cite{JQMD}.
%
%
 
Cross sections of inverse reactions (Resonance + Baryon $\rightarrow$ Baryon + Baryon) are determined by the extended
detailed balance formula \cite{extended_detailed_balance} and the corresponding resonance production cross sections. 
%
\begin{eqnarray} 
 && \frac{d\sigma^{N\Delta \rightarrow NN}}{d\Omega} = \frac{1}{N_f} \frac{p_f^2}{p_i^2} \frac{d\sigma^{NN \rightarrow N\Delta}}{d\Omega}
 \frac{1}{\int^{(\sqrt{s}-m_\mathrm{N})^2}_{(m_\mathrm{N}+m_\mathrm{\pi})^2} F(M^2)dM^2} 	, \\ \nonumber 
 && F(M^2) = \frac{1}{\pi} \frac{m'_2\Gamma(M)}{(M^2-m'^2_2)^2 + m'^2_2\Gamma(M) }		, \nonumber 
\label{inverse reaction} 
\end{eqnarray}
where $\sigma^{N\Delta \rightarrow NN}$ is inverse reaction cross section, $N_f$ is isospin factor (2 for
reactions whose final state is n-p, 4 for reactions whose final state is n-n or p-p), $p_f$ is the momentum
at the final state, $p_i$ is the momentum at the initial state, $\sigma^{NN \rightarrow \Delta N}$ is resonance
production cross section, $m_\mathrm{N}$ is nucleon mass, $m_\mathrm{\pi}$ is pion mass, $m'$ is the mass at the
final state, and $\Gamma$ is the resonance mass width given elsewhere \cite{JQMD}. 

Pions produced by the decay of resonances can be absorbed. 
The reactions described so far are simulated by checking baryon-baryon pairs sequentially. After that, pion-baryon
pairs are sequentially examined to simulate pion absorption. Pion-baryon pairs closer than
$\sqrt{(\sigma_\Delta + \sigma_{N*})/\pi}$ are sent to the absorption check routine. $\sigma_\Delta$ and $\sigma_{N*}$
are listed in Table \ref{Pion_absorption_cross_section}.
\begin{table}
\caption{Pion absorption cross section maximum (mb). $\sigma_\Delta$ : $\Delta$ production cross section,
$\sigma_{N*}$ : $N*$ production cross section. }
\begin{center}
\begin{tabular}{ccccc} \hline
		& n+$\pi^-$	& n+$\pi^+$ 	& n+$\pi^0$	& $\Delta^- + \pi^+$, $\Delta^0 + \pi^0$, $\Delta^0 + \pi^+$	\\ 
		& p+$\pi^+$	& p+$\pi^-$ 	& p+$\pi^0$	& $\Delta^+ + \pi^-$, $\Delta^+ + \pi^0$, $\Delta^++ + \pi^-$	\\ \hline
$\sigma_\Delta$	& 200		& 70		& 135		& 0								\\ 
$\sigma_{N*}$	& 0		& 30		& 30		& 30								\\ \hline
\end{tabular}
\end{center}
\label{Pion_absorption_cross_section}
\end{table}
For a pair sent to the absorption check routine, if $1/(1+[4(\sqrt{s} - m_\mathrm{prod})/\Gamma]^2)$,
where $m_\mathrm{prod}$ is the product mass, and $\Gamma$ is the width given elsewhere \cite{JQMD}, is greater than a random number [0:1], the pion is absorbed by the baryon. 
 
 
When resonances ($\Delta$ or N$^*$) are present, their decay is calculated before collisions and pion absorption. The decay probability
is determined by the exponential decay law with the decay width calculated by Eq. \eqref{resonance_mass_distribution}.
%
%
 
\subsubsection{Pauli Blocking}
 
Every time when a collision or decay is attempted and the final state contains one or more nucleons, the Pauli-blocking
probability is calculated. After determining the momentum and the isospin at the final state of the collision,
the occupation probability is calculated. 
%
%
The one-body distribution function is obtained by the Wigner transform of the wave function. 
%
%
The Pauli-blocking probability of a reaction is obtained from
\begin{equation}
P_\mathrm{pauli} = (1 - \frac{h^3}{2}f(\vec{R_j},\vec{P_j}))(1 - \frac{h^3}{2}f(\vec{R_k},\vec{P_k})), 
\label{block-probability}
\end{equation}
where $j$ and $k$ denote the id of reacting nucleons. When a collision is Pauli-blocked, this reaction attempt is canceled and the next collision with a different partner is attempted.  
\subsection{The Lanzhou isospin-dependent quantum molecular dynamics code (LQMD)}
\vskip 0.1in
Z. Q. Feng, H. G. Cheng 
\vskip 0.2in

In this short write-up, we provide information on the LQMD code. The main references are~\cite{Chen98,fe08a1,fe08a2,fe11a1,fe11a2,fe11a3,fe14}.

\subsubsection{Code history}

\begin{itemize}
\item Lanzhou, 1995-2000: \\
Implementation of the isospin degree of freedom in the quantum molecular dynamics model including initialization,
mean-field potential, nucleon-nucleon collisions, and Pauli-blocking by Lie-Wen Chen and Feng-Shou Zhang~\cite{Chen98,Chen99,Zhang99},
named the isospin-dependent quantum molecular dynamics model (IQMD).
\item Lanzhou, 2002-2007:\\
Implementation of Skyrme-energy density functional and inclusion of shell effect in the IQMD model for describing
low-energy heavy-ion collisions and the fusion dynamics in the formation of superheavy nuclei by
Zhao-Qing Feng {\it et al.}~\cite{fe08a1,fe08a2}.
\item Lanzhou, 2008-2012:\\
Inclusion of all possible inelastic and elastic hadron-hadron collisions and improvements of the density-dependent symmetry energy and an isospin- and momentum-dependent single-particle potential \cite{fe11a1,fe11a2,fe11a3}.
Improvements of in-medium and isospin effects on strange particle production \cite{fe13a1,fe13a2}.
\item Lanzhou, 2011-2017:\\
Further improvements to treat antiproton induced reactions at FAIR energies \cite{fe14}.
\item Guangzhou, 2018-2020:\\
Treatment of hypernucleus formation in heavy-ion collisions and in hadron (proton, antiproton, pion, kaon, etc.) induced reactions \cite{fe20a}.
\end{itemize}

\subsubsection{Initialization}
The initial distribution of neutrons and protons in coordinate space is sampled by using a Fermi function
\begin{equation}
F(r_{i}) =\frac{1}{1+\exp [(r_{i}-R)/a]} .
\end{equation}
The radius $R=1.28A^{1/3}-0.76+0.8A^{-1/3}$ fm and diffuseness parameter $a=0.5-0.65$ fm are used for
 the nuclei of interest. The sampled radial distributions of the nuclei are constrained by the root-mean-square
radii of protons and neutrons from Skyrme-Hartree-Fock or relativistic mean-field models.

Similar to the RBUU code, the particle momenta are randomly distributed in an isotropic Fermi sphere,
$f_{n,p} ({\bf r}, {\bf k}) =\Theta \left( k_{{\rm F},n,p}({\bf r}) -|{\bf k}| \right)$, whose radius is determined
by the local Thomas-Fermi approximation, $k_{{\rm F},n,p}({\bf r}) = \left(3\pi^2 \rho_{n,p} ({\bf r}) \right)^{1/3}$.
Finally, the binding energy of selected nucleus is checked against experimental data.

\subsubsection{Forces}
In the LQMD code, an isospin- and momentum-dependent Skyrme-type force is used. The single-particle potential in nuclear
matter is expressed as follows:
\begin{eqnarray}
U_{\tau}(\rho,\delta,\textbf{p})&=&\alpha\frac{\rho}{\rho_{0}}+\beta\frac{\rho^{\gamma}}{\rho_{0}^{\gamma}}
+\frac{\partial E_{sym}^{loc}(\rho)}{\partial\rho}\rho\delta^{2}+ 
E_{sym}^{loc}(\rho)\rho\frac{\partial\delta^{2}}{\partial\rho_{\tau}} \\
&&+\frac{1}{\rho_{0}}C_{\tau,\tau} \int d\textbf{p}' f_{\tau}(\textbf{r},\textbf{p})[\ln(\epsilon(\textbf{p}-\textbf{p}')^{2}+1)]^{2}
+\frac{1}{\rho_{0}}C_{\tau,\tau'} \int d\textbf{p}' f_{\tau'}(\textbf{r},\textbf{p})[\ln(\epsilon(\textbf{p}-\textbf{p}')^{2}+1)]^{2}. \nonumber
\end{eqnarray}
Here $\tau\neq\tau'$, $\partial\delta^{2}/\partial\rho_{n}=4\delta\rho_{p}/\rho^{2}$ and $\partial\delta^{2}/\partial\rho_{p}=-4\delta\rho_{n}/\rho^{2}$.
The local potential energy is given by $U_{loc}=\int V_{loc}(\rho(\mathbf{r}))d\mathbf{r}$. The energy-density functional reads as
\begin{eqnarray}
V_{loc}(\rho)= \frac{\alpha}{2}\frac{\rho^{2}}{\rho_{0}} +
\frac{\beta}{1+\gamma}\frac{\rho^{1+\gamma}}{\rho_{0}^{\gamma}} + E_{sym}^{loc}(\rho)\,\rho\,\delta^{2} 
+\frac{g_{sur}}{2\rho_{0}}(\nabla\rho)^{2}+ \frac{g_{sur}^{iso}}{2\rho_{0}}[\nabla(\rho_{n}-\rho_{p})]^{2},
\end{eqnarray}
where the $\rho_{n}$, $\rho_{p}$ and $\rho=\rho_{n}+\rho_{p}$ are the neutron, proton and total densities, respectively, and the $\delta=(\rho_{n}-\rho_{p})/(\rho_{n}+\rho_{p})$ being the isospin asymmetry. The coefficients $\alpha$, $\beta$, $\gamma$, $g_{sur}$, $g_{sur}^{iso}$ and $\rho_{0}$ are set to have the values of -215.7 MeV, 142.4 MeV, 1.322, 23 MeV fm$^{2}$, -2.7 MeV fm$^{2}$ and 0.16 fm$^{-3}$, respectively. A compression modulus of K=230 MeV for isospin symmetric nuclear matter is produced with these parameters. A Skyrme-type momentum-dependent potential is used in the LQMD model via
\begin{eqnarray}
U_{mom}=\frac{1}{2\rho_{0}}\sum_{i,j,j\neq i}\sum_{\tau,\tau'}C_{\tau,\tau'}\delta_{\tau,\tau_{i}}\delta_{\tau',\tau_{j}}\int\int\int d \textbf{p}d\textbf{p}'d\textbf{r}\,
f_{i}(\textbf{r},\textbf{p},t)\, [\ln(\epsilon(\textbf{p}-\textbf{p}')^{2}+1)]^{2}\, f_{j}(\textbf{r},\textbf{p}',t).
\end{eqnarray}
Here $C_{\tau,\tau}=C_{mom}(1+x)$, $C_{\tau,\tau'}=C_{mom}(1-x)$ ($\tau\neq\tau'$) and the isospin symbols $\tau$($\tau'$) represent proton or neutron. The parameters $C_{mom}$ and $\epsilon$ were determined by fitting the real part of optical potential as a function of incident energy from the proton-nucleus elastic scattering data. In the calculation, we take the values of 1.76 MeV, 500 c$^{2}$/GeV$^{2}$ for the $C_{mom}$ and $\epsilon$, respectively, which result in the effective mass $m^{\ast}/m$=0.75 in nuclear medium at saturation density for symmetric nuclear matter. The parameter $x$ as the strength of the isospin splitting is set to be -0.65 and 0.65 for the cases of $m^{\ast}_{n}>m^{\ast}_{p}$ and $m^{\ast}_{n}<m^{\ast}_{p}$, respectively. In low-energy heavy-ion collisions near the Coulomb barrier, the shell effect is also considered within a phenomenological approach \cite{fe08a1,fe08a2}.

\subsubsection{Collision term}
In the LQMD code, a hard core scattering in two-particle collisions is assumed by using Monte Carlo procedures, in which the scattering
of two particles is determined by a geometrical minimum distance criterion weighted by the Pauli blocking of the final states. The total
elastic and inelastic nucleon-nucleon cross-sections are parametrized in accordance with the available
experimental data \cite{fe09}. One can also make a choice of the in-medium elastic cross sections evaluated through the
reduced masses of the colliding nucleon pairs \cite{fe12}. The resonances ($\Delta$(1232), $N^{\ast}$(1440), 
$N^{\ast}$(1535), etc) are produced by inelastic NN collisions. We have included the reaction channels as follows:
\begin{eqnarray}
&& NN \leftrightarrow N\triangle, \quad  NN \leftrightarrow NN^{\ast}, \quad  NN
\leftrightarrow \triangle\triangle,  \nonumber \\
&& \Delta \leftrightarrow N\pi,  N^{\ast} \leftrightarrow N\pi,  NN \leftrightarrow NN\pi (s-state),  \nonumber \\
&& N^{\ast}(1535) \rightarrow N\eta.
\end{eqnarray}
Energy and momentum-dependent decay widths are used in the model for the $\Delta$(1232) 
and $N^{\ast}$(1440) resonances~\cite{fe09}. We have taken a constant width of $\Gamma$=150 MeV for the $N^{\ast}$(1535) decay.
The reaction channels of strange particles are included as
\begin{eqnarray}
&& BB \rightarrow BYK,  BB \rightarrow BBK\overline{K}, \nonumber\\ && B\pi \rightarrow YK,  B\pi \rightarrow NK\overline{K}, \\
&& Y\pi \rightarrow B\overline{K}, \quad  B\overline{K} \rightarrow Y\pi, \quad YN \rightarrow \overline{K}NN.\nonumber
\end{eqnarray}
Here $B$ stands for (N, $\triangle$, N$^{\ast}$) and similarly  Y for ($\Lambda$, $\Sigma$), K for (K$^{0}$, K$^{+}$), and 
$\overline{K}$ for ($\overline{K^{0}}$, K$^{-}$). The elastic scattering between strange hadrons and baryons  
includes the channels of $KB \rightarrow KB$, $YB \rightarrow YB$ and $\overline{K}B \rightarrow \overline{K}B$.
The charge-exchange reactions between the $KN \rightarrow KN$ and $YN \rightarrow YN$ channels are included by using the
same cross sections as for the elastic scattering, e.g., $K^{0}p\rightarrow K^{+}n$, $K^{+}n\rightarrow K^{0}p$ etc.
Effects of the effective masses of kaons in the nuclear medium on the elementary cross section are considered through changes of the 
threshold energy, which result in the reduction of kaon and the enhancement of anti-kaon yields in heavy-ion collisions.
The annihilation channels, charge-exchange reaction, elastic and inelastic scattering in antinucleon-nucleon collisions
have also been included in the LQMD model : $\overline{N}N \rightarrow \overline{N}N$, $\overline{N}N \rightarrow \overline{B}B$,
$\overline{N}N \rightarrow \overline{Y}Y$ and $\overline{N}N \rightarrow \texttt{annihilation}(\pi,\eta,\rho,\omega,K,\overline{K},
K^{\ast},\overline{K}^{\ast},\phi)$.
The overline on $B$ (Y) denotes its antiparticle.

\subsubsection{Pauli-blocking}
Two options for checking the Pauli-blocking of the final state of a two-body collision have been implemented in the LQMD model. 
One approach is to evaluate the phase-space
volume $V$ of the scattered nucleons occupied by the other nucleons of the same isospin after the NN scattering~\cite{Chen98}. If the factor $2V/h^{3}$ is larger than a generated random number, the collision is blocked.
Otherwise the collision is allowed. The other one is to compute the blocking factor 
$(1-\overline{f}_{i})(1-\overline{f}_{j})$ with the $\overline{f}_{i}$ and $\overline{f}_{j}$ being the phase-space
occupation number of the two colliding nucleons after the collision~\cite{Guo_2018}. If the factor is larger than a generated random number,
the collision is blocked. Once the collision is blocked, the momenta of the associated nucleons are set to their original values.

\subsection{The TuQMD/dcQMD code} \label{tuqmd}
\vskip 0.1in
D. Cozma
\vskip 0.2in

\subsubsection{Code history}
\label{introduction}

The T\"ubingen QMD (TuQMD) was developed in the 90's and first half of the following decade in T\"ubingen, Germany~\cite{Khoa:1992zz,Fuchs:1996pa,Wang:1997my,Shekhter:2003xd} to address heavy-ion related physics problems in the 0.5-2.0 AGeV energy regime. It has common features with several contemporary transport models: initialization of nuclei resembles that of BQMD~\cite{Hartnack:1997ez}, Pauli blocking algorithm is similar to the one implemented in one of the earliest versions of QMD~\cite{Aic91} while propagation and computation of forces mirror IQMD~\cite{Hartnack:1997ez}. In addition to the nucleonic degrees of freedom, all well established baryonic resonances with isospin 1/2 and 3/2 and a pole mass below 2 GeV have also been included; a comprehensive list can be found in Ref.~\cite{Shekhter:2003xd}. In this context, the detailed balance algorithm described in Ref.~\cite{Bass:1995pj} has been implemented. Several mesonic degrees of freedom have also been included: $\pi$(138), $\eta$(547), $\rho$(770), $\omega$(782), $\Phi$(1020) and $K$(494). The large variety of particle species and elementary reactions has allowed the study of several heavy-ion related topics: the equation of state of symmetric nuclear matter~\cite{Fuchs:2000kp}, the impact of in-medium effects on heavy-ion collision dynamics~\cite{Khoa:1992zz,Fuchs:1996pa}, emission of dilepton pairs and in-medium change of vector meson properties as precursors of chiral symmetry restoration and/or deconfinement~\cite{Shekhter:2003xd,Faessler:2000md,Santini:2008pk}.

The final version of TuQMD has been adopted by the code correspondent as the transport model of choice for the study of the isospin asymmetric part of nuclear matter at supra-normal densities in heavy-ion reactions in the few hundred MeV/nucleon regime and has been upgraded for that purpose~\cite{Cozma:2011nr,Cozma:2013sja}. This upgraded version will be referred as dcQMD. In the following, the latest available version of dcQMD will be described and the upgrades with respect to the last version of TuQMD, where they exist, will be pointed out.

\subsubsection{Initialization}
The coordinate space initialization of a nucleus proceeds as follows: nucleons are generated randomly
with their distance from the center of the nucleus satisfying the following distribution

\begin{displaymath}
P(r) = \left\{ \begin{array}{ll}
cr^2, & \textrm{$r<r_0$}\\
a-(r-b)^2,& \textrm{$r \geq r_0$,}
\end{array} \right.
\end{displaymath}
which is modified close to the surface in order to better reproduce the experimental measured
diffuseness of nuclei. In addition, the distance between neighboring nucleons is required to be larger than 1.55 fm. In dcQMD the $r$-space initialization enforces that the center-of-mass of neutron and proton distributions lie closer to each other than a predefined distance of 0.05 fm. A neutron skin with thickness $R_{skin}$=0.9$\beta$-0.03 fm and thickness uncertainty $\delta R_{skin}$=0.15$\beta$+0.02 fm depending on isospin asymmetry $\beta$,, consistent with theoretical predictions, can also be required.

Momentum initialization in TuQMD takes into account the potential energy of the nucleon in question and generates a random momentum with the constraint that the nucleon is bound (4 MeV $\leq E_{bind}\leq $12 MeV) and there is a certain minimal distance in phase-space relative to other nucleons ($\Delta r_{ij} \Delta p_{ij}>0.28$ fm$\,$GeV). The potential used in this context is an isospin independent one. The Coulomb interaction is scaled down by a factor of $(Z/A)^2$ to counterbalance its implemented charge independence such that the binding energy is in agreement with the empirical mass formula. The process of momentum initialization is iterative and it is terminated after a predefined (large) number of iterations and can lead to as many as 15$\%$ unbound nucleons for momentum-dependent interactions. This procedure leads to nuclei with values of the average binding energy per nucleon in the range 7.5 MeV $\leq$ $E_{bind}/N  \leq $ 8.5 MeV.

The dcQMD upgrade implements isospin-dependent potentials already at the momentum initialization stage. The momenta of nuclei are initialized according to the local density (ignoring isospin dependence of the local Fermi momentum) but the requirement that initialized nucleons are bound is still enforced by lowering the value of the Fermi momentum if needed. With this procedure the number of final unbound nucleons can be lowered to less than 3$\%$. The interaction considered in this version of the momentum initialization routines is identical with the one used in the mean-field propagation routines. This variant allows the option to select initialized nuclei that exhibit predefined good stability of the rms and neutron skin thickness in nucleus's rest frame over a time period comparable to the duration of a heavy-ion collision. For momentum-independent interactions, rms variations of less than 5$\%$ can be achieved over a time frame of 150 fm/c. Momentum-dependent interactions raise that fraction to 25-30$\%$.


\subsubsection{Forces}
The QMD equations of motion are given by the well known Hamilton equations. The original TuQMD model considered only an isospin-independent interaction which was later supplemented by a power-law parametrization of the symmetry energy. The parametrization used for the Hamiltonian reads

\begin{eqnarray}
 \langle H \rangle &=&\sum_i\,\sqrt{\vec{p}_i^{\;2}+m_i^2}+\sum_i\,\frac{\alpha}{2}\,u_i+\sum_i\,\frac{\beta}{\gamma+1}\,{u_i}^\gamma 
+\sum_{i,j,i<j}\,\delta\,\mathrm{ln^2}[\,\varepsilon\,(\vec{p}_i-\vec{p}_j)^2+1\,]\,u_{ij}\\ &+&S_0\,\sum_i\,\tilde\tau_i\,u_i^{\gamma-1}\,\sum_j\,\tilde\tau_j\,u_{ij}\nonumber+\sum_{i,j,i<j}\,U_{ij}^{Coul}, \nonumber
\end{eqnarray}
with the reduced interacting density $u_i$ and Coulomb term given respectively by
\begin{eqnarray}
u_i&=&\sum_{i,j}\,u_{ij},\quad u_{ij}=\frac{1}{(2\pi L)^{3/2}\rho_0}\,\mathrm{exp}\,\bigg\{ -\frac{(\vec{r}_i-\vec{r}_j)^2}{L^2}\,\bigg\},\\
U_{ij}^{Coul}&=&\bigg( \frac{Z}{A}\bigg)^2\,\frac{e^2}{|\vec{r}_i-\vec{r}_j|}\,\mathrm{Erf}\bigg(\frac{|\vec{r}_i-\vec{r}_j|}{\sqrt{L^2}}\bigg).
\end{eqnarray}

The values of the parameters of the potential can be found in~\cite{Aic91}, while for the newer version of the optical potential of Ref.~\cite{Hartnack:1994zz}, the corresponding values used are mentioned in Ref.~\cite{Cozma:2013sja}. When a symmetry energy term is included, the $(Z/A)^2$ factor in the expression of the Coulomb interaction is set equal to 1. The wave function width $L$ takes values in the range of 1-1.4 fm.

The dcQMD version uses a different parametrization for the momentum-dependent part of the potential (Gogny-like) and a charge-dependent Coulomb, with the above suppression factor $(Z/A)^2$ removed. The effective Hamiltonian reads:
\begin{eqnarray}
 \langle H \rangle&=&\sum_i \sqrt{\vec{p}_i^{\;2}+m_i^2}+\sum_{i,j,j>i}\,\bigg[\frac{A_u+A_l}{2}+\tilde\tau_i\,\tilde\tau_j\,\frac{A_l-A_u}{2}\bigg]\,u_{ij} \\
 &+&\sum_{i,j,j>i}\bigg[(C_l+C_u)+\tilde\tau_i\,\tilde\tau_j\,(C_l-C_u)\bigg]\frac{u_{ij}}{1+(\vec{p}_i-\vec{p}_j)^2/\Lambda^2}\nonumber\\
 &+&\sum_i\,\frac{B}{\sigma +1}\,[1-x\tilde\tau_i\,\beta_i\,]\,u_i^\sigma+\frac{D}{3}\,[1-y\tilde\tau_i\,\beta_i]\,u_i^2 +\sum_{i,j,j>i} U_{ij}^{Coul} . \nonumber
\end{eqnarray}

The notation for the Gogny parametrization of the potential adheres to that of Ref.~\cite{Das:2002fr}. An additional term, proportional to a parameter denoted $D$ has been introduced in order to allow independent variations of the compressibility/skewness or slope/curvature parameters of the equations of state of symmetric and asymmetric nuclear matter. The parameters denoted by $x$ and $y$ allow the adjustment of the stiffness of the symmetry energy. The variable $\tilde\tau_i$ is defined as $\tilde\tau_i=-\tau_i/T_i$, with $T_i$ and $\tau_i$ the isospin and its projection for particle $i$.  
This ansatz thus introduces a isospin dependent potential for baryonic resonances, most notably $\Delta$(1232), the particular values for the $\tilde\tau$ variable being motivated by the branching ratios of each charge state into the possible nucleon-pion pairs. An isospin-dependent $\Delta(1232)$ potential has a large impact on observables such as the $\pi^-/\pi^+$multiplicity ratio close to threshold and consequently on constraints extracted for the density dependence of symmetry energy from heavy-ion reactions~\cite{Cozma:2014yna,Cozma:2016qej}. The newest version of the dcQMD model~\cite{Cozma:2021tfu} allows the parameters of the above interaction to depend on particle's specie, removing the requirement of the nucleon and $\Delta(1232)$ being identical.

In the original TuQMD model, meson degrees of freedom are propagated both taking into account only a point-like Coulomb interaction. In dcQMD model, mesons are treated on the same footing as baryons, by assigning them Gaussian wave functions. For pions, the wave-function width is set to $L^2_\pi=0.5\,L^2$ such as to approximately describe the experimental pion-to-proton charge rms ratio. The strong pion-nucleon interaction is accounted for by propagating pions under the influence of S and P wave optical potentials. The Ericson-Ericson parametrization is used,
\begin{eqnarray}
V_{opt}(r)&=&\frac{2\pi}{\mu}\Big[\,-q(r)+\,\vec\nabla\,\frac{\alpha(r)}{1+\frac{4}{3}\pi\lambda\alpha(r)}\,\vec\nabla\,\Big]\,,
\end{eqnarray}
where
\begin{eqnarray}
 q(r)&=&\epsilon_1(\bar{b}_0\rho+\bar{b}_1\beta\rho)+\epsilon_2B_0\rho^2, \nonumber\\
\alpha(r)&=&\epsilon_1^{-1}(c_0\rho+c_1\beta\rho)+\epsilon_2^{-1}(C_0\rho^2+C_1\beta\rho^2). \nonumber
\end{eqnarray}
Here, $\rho$ and $\beta$ stand for the pion-nucleon interaction density and isospin asymmetry. The wave-function width required to evaluate these quantities ($\rho$ and $\beta$) is approximately equal to $L^2_{\pi N}=0.5(L^2_\pi+L^2)$. Details regarding the set of parameters used can be found in Ref.~\cite{Cozma:2016qej}.

The integration algorithm of choice for mean-field propagation is Runge-Kutta of order 4, but higher order routines (6$^{th}$ order Fehlberg and 8$^{th}$ order Dormand-Prince) have also been implemented.

\subsubsection{Implementation of Collisions}

Collision acceptance or rejection is based on a minimum distance criterion, $\pi\,d_{min}^2\leq \sigma$, with $\sigma$ being the total scattering cross-section. During each time step, a given particle is allowed to undergo at most one collision. To avoid a preferential reaction outcome, which depends on how the initialization of baryon arrays was performed, the labels of all baryons are randomly reassigned each time the collision routines are called. Such a procedure is important for example in the case of scattering of identical nuclei for which the directed flow should vanish: $v_1$=0.

Originally, TuQMD included only the (vacuum) Cugnon parametrizations~\cite{Cugnon:1980rb} for elastic nucleon-nucleon cross-sections ($\sigma_{nn}=\sigma_{pp}$ and $\sigma_{np}$). The upgraded version includes the Li-Machleidt parametrizations as well~\cite{Li:1993ef,Li:1993rwa} and their density-dependent version below the pion production threshold. The isospin asymmetry dependence is implemented by a scaling factor equal to the product of in-medium masses of the final and initial states relative to the density dependent ones. Above the pion production threshold, the in-medium modification scaling factor retains both the density and asymmetry dependence of in-medium nucleon masses. The latest version of dcQMD includes also an empirical isospin-independent modification factor of elastic nucleon-nucleon cross-sections~\cite{li11} whose parameters have been determined by requiring a good description of stopping, directed and elliptic flows of protons and light fragments in intermediate energy HICs~\cite{Cozma:2021tfu}.

Excitation and absorption of single and double resonances are also included in the 2-baryon scattering routines. All well established resonances with isospin 1/2 and 3/2 and pole masses below 2.0 GeV have been included. Inelastic cross-sections for $\Delta$(1232) and $N$(1440) resonances are determined using a parametrization of the one-boson-exchange model of Huber $et~al.$~\cite{Huber:1994ee}. The transition matrix elements needed for production cross-sections of other resonances are assumed to depend only on masses of final state particles~\cite{Shekhter:2003xd}. A medium modification of inelastic cross-sections has been implemented only in dcQMD~\cite{Cozma:2021tfu}. It consists of an approximation of the general expression for two-body in-medium cross-section formula~\cite{Larionov:2003av}. Cross-sections of inelastic processes in which resonances are absorbed are determined via the method of detailed balance~\cite{Bass:1995pj}. Several processes of resonance decay, meson absorption, elastic and inelastic scattering of mesons off baryons are also included.

An upgrade particular only to dcQMD is the consideration of total energy (kinetic+potential) balance during collision, decay or absorption processes due to in-medium potentials. This has the effect of shifting the production thresholds upwards or downwards, enhancing or suppressing a reaction based on the difference of the initial and final potential energies. Two scenarios have been implemented: a) only the potential energies of the scattering particles are considered b) the potential energy of the whole system is taken into account. Scenario b) enforces conservation of the total energy of the system for the whole duration of the reaction~\cite{Cozma:2014yna}.

\subsubsection{Pauli Blocking}
Whenever a collision occurs, the occupancy in phase-space in the vicinity of the scattering nucleons is checked. The collision is blocked with a probability
\begin{eqnarray}
 P_{blocked}=1-(1-P_1)(1-P_2),
\end{eqnarray}
where $P_1$ and $P_2$ are the occupation fractions of phase-space around nucleons 1 and 2, respectively. In the vicinity of nucleus's surface a correction is performed, such that only the classically allowed portion of the phase-space is taken into account. Baryonic resonances are considered to be unaffected by Pauli blocking. In TuQMD, the phase-space occupancy is determined by considering nucleons as hard spheres in both position and momentum space, the algorithm being the same as in the first version of QMD~\cite{Aic91}. The newest versions of dcQMD includes several additional approaches of computing the occupancy by making use of the Gaussian wave-function associated to nucleons. One of them is a straightforward extension of the hard spheres method to a different nucleon shape~\cite{Cozma:2017bre}, while a second makes use of the requirement that the integrated occupancy over a sphere of volume $h^3$ in phase-space has to be smaller than the degeneracy factor for fermionic systems. 

\subsection{The ultra-relativistic quantum molecular dynamics (UrQMD) code}
\vskip 0.1in
Y. J. Wang, Q. F. Li, Y. X. Zhang
\vskip 0.2in

\subsubsection{History}
\begin{itemize}
\item The development of UrQMD started in the middle of 1990s at Frankfurt. The detailed model description can be found in Refs \cite{Bass:1998ca,Bleicher:1999xi}. The first version already contains 50 different baryon species and 25 different meson species, as well as their corresponding antiparticles and all isospin-projected states. It can be used to study collisions of pp, pA, and AA over the vast energy range from GSI-SIS to CERN-SPS, i.e., the entire available range of energies at 1990s. Since then, the UrQMD model has been widely applied for simulating heavy-ion collisions at relativistic energies. \\
\item Since 2005, with the implementation of several ingredients into UrQMD, it has been used to study symmetry energy effects in heavy-ion collisions at intermediate energies~\cite{Li:2005zza,LI:2005zi,Li:2005kqa}. 
Since 2014, a mean-field potential derived from the Skyrme potential energy density functional (DF) has been implemented into the code at Huzhou University~\cite{Wang:2013wca01,Wang:2013wca02,Liu:2020jbg}. This version is different from the codes maintained at Frankfurt by the same name used at ultra-relativistic energies. To avoid misunderstandings, this version will be named ‘UrQMD-DF’ in the future.
\\
\item In 2008, to employ the UrQMD transport approach at higher energies, a second version (UrQMD-2.3) was published, in which Pythia was implemented to perform the initial hard collisions \cite{urqmd2.3}. \\

\item In 2009, UrQMD-3.3 version was published. In this version, running UrQMD with a hydrodynamic evolution for the hot and dense stage of heavy-ion collision is possible.  \\

\end{itemize}

In this section, several important ingredients of UrQMD model to study heavy-ion collisions at intermediate energies will be addressed.

\subsubsection{Initialization}
In the UrQMD model, two different choices for the initialization have been implemented, called Woods-Saxon and hard-sphere ones. The hard-sphere mode is usually employed for describing heavy-ion collisions (HICs)
at intermediate energies when the potentials are considered. First the radius of projectile and target nuclei are determined using
the following expression~\cite{Bass:1998ca},
\begin{equation}
 R = \left( \frac{3}{4\pi\rho_{0}} \right)^{1/3} \left\{\frac{1}{2}[A+(A^{1/3}-1)^3]\right\}^{1/3}
 \label{eq1urqmd},
\end{equation}
where $A$ denotes the mass number of projectile and target nuclei, while $\rho_{0}$ is the normal nuclear density.
Eq.~(\ref{eq1urqmd}) considers the fact that the finite width of Gaussians (nucleons) results in a large surface, therefore
the radius calculated with this formula is smaller than the one usually used $R=1.12A^{1/3}$ fm. The centroids of the
Gaussians are randomly distributed within a sphere of the radius $R$, but those for which the minimum distance
between two nucleons inside a nucleus is smaller than 1.6 fm are rejected. The initial momenta of the protons and neutrons are randomly
chosen between 0 and the respective local Fermi momenta $\hbar(3\pi^2\rho_{p,n})^{1/3}$. The next step is to
calculate the binding energy per nucleon $B/A$ of the initialized nucleus. If the difference between the calculated
and experimental $B/A$ is larger than 1 MeV, the momenta will be re-sampled.

\subsubsection{Mean-field potential}
\begin{itemize}
\item Baryons\\
In Ref.~\cite{Li:2005zza}, the total Hamiltonian includes the two-body and three-body Skyrme-, Yukawa-, Coulomb-,
Pauli-, and symmetry terms, as well as the momentum-dependent one. Polynomial interpolation is used to determine derivatives
of potentials so as to allow a much faster numerical simulation of the reaction. The accuracy of this method has been checked,
and it is found that the error introduced by the interpolation is small enough to be neglected. In Ref.~\cite{Li:2007yd}, potentials for ``pre-formed” particles (string fragments) from color fluxtube fragmentation as well as for confined particles were considered for studying HICs at energies from AGS, SPS, to RHIC. It was found that the inclusion of potential interactions provides stronger pressure at the early stage and describes observables better than the default cascade mode. In Refs.~\cite{Wang:2013wca01,Wang:2013wca02},
the main potential is updated by considering the Skyrme potential energy density functional, while the momentum-dependent term used in Ref.~\cite{Li:2005zza} is still taken into account. In Ref. \cite{Tong:2020dku}, the neutron-proton effective mass splitting effect was studied by incorporating an isospin-dependent form of the momentum-dependent potential. In Ref. \cite{Liu:2020jbg}, mean-field potentials with different parameter sets (e.g., different stiffness of the nuclear equation of state and different nucleon effective mass) were introduced, and potentials for Delta resonances were described in detail. More details about mean-field potentials for nucleons can be found in review papers \cite{Wang:2020vwb,Zhang:2020dvn}.

\item Mesons\\
In Ref. \cite{Liu:2018xvd}, the pion potentials obtained from the in-medium dispersion relation of the $\Delta$-hole model were implemented. It was found that the relatively weak pion potential from the $\Delta$-hole model can provide a good description for the FOPI data of both collective flows as functions of both centrality and rapidity. In Ref. \cite{Du:2018ruo}, the kaon-nucleon (KN) potential was introduced, where both the scalar and the vector (also dubbed Lorentz-like) parts were considered. The influence of the KN potential on the collective flow of $K^{+}$ meson produced in Au+Au collisions at $E_{lab}$=1.5 GeV/nucleon was revisited, and it was found that the corresponding KaoS data of both directed and elliptic flows of $K^{+}$ meson can be simultaneously reproduced well. The Coulomb potential for all charged particles is considered in the same way as for protons.

\end{itemize}

\subsubsection{Collision term}
The UrQMD collision term is based on analogous principles of the RQMD model~\cite{rqmd1a,rqmd1b,rqmd1c}. More details are presented in Refs.~\cite{Bass:1998ca,Bleicher:1999xi}.
Two particles may collide if their distance $d$ fulfills the relation:
\begin{equation}
 d \leq d_0 = \sqrt{\frac{\sigma_{tot}}{\pi}}, \qquad \sigma_{tot}=\sigma(\sqrt{s},type)
 \label{eq2urqmd}.
\end{equation}
The total cross section $\sigma_{tot}$ depends on the cm energy $\sqrt{s}$ and the type of the incoming particles.
For the theoretical studies of the symmetry energy with beam energies, such as relevant to the ALADIN and FOPI collaborations at SIS,
the total nucleon-nucleon cross section is modified by the nuclear medium according to
\begin{equation}
\sigma_{tot}^{*} = \sigma_{inelastic}+\sigma_{elastic}^{*} = \sigma_{inelastic}^{free}+F\cdot\sigma_{elastic}^{free}
 \label{eq3urqmd}.
\end{equation}
The inelastic and elastic nucleon-nucleon cross sections in free space are taken from experimental data. The in-medium
nucleon-nucleon elastic cross section is treated as the product of a medium correction factor $F$ and the free cross sections.
The factor $F$ depends on the relative momentum of the two colliding nucleons, the density and isospin asymmetry at the location
of the collision. The relevant details were presented in Ref.~\cite{li02}. The UrQMD approach uses an analytical expression
for the differential cross-section of in-medium nucleon-nucleon elastic scattering derived from the collision term of the
self-consistent relativistic Boltzmann-Uehling-Uhlenbeck (RBUU) equation~\cite{mao01a,mao01b} to determine the scattering angles between
the outgoing partners of all elementary hadron-hadron collisions. However, this analytical expression is not used for the corresponding total cross sections.

For studying pion production at intermediate energies, the cross sections of 
the process $NN\rightarrow N\Delta$ and its reversed channel are very important. In Ref. \cite{Liu:2020jbg}, the cross section of $ N\Delta\rightarrow NN $ is modified based on the one-boson exchange model (OBEM) \cite{Cui:2018dex}, which is:
\begin{equation}
\sigma _{N\Delta(m)\rightarrow NN}=\frac{1}{64 \pi^{2}s}\frac{p_{NN}}{p_{N\Delta ( m )}}\frac{I_{i}^{2}\emph{M}}{(2S_{N}+1)(2S_{\Delta }+1)}\frac{1}{1+\delta_{N_{1}N_{2}} }
\end{equation}
with
\begin{equation}
\emph{M}=\frac{1}{I_{i}^{2}}\int \sum_{s_{1}s_{2}s_{3}s_{4}} |M_{N\Delta\to NN}|^{2}d\Omega,
\end{equation}
where the $| M_{N\Delta(m)\to NN} |^{2}$ is determined from $|M_{NN\to N\Delta(m)} |^{2}$ based on the time reversal invariance. The form of $|M_{NN\to N\Delta(m)} |^{2}$ is obtained by fitting the experimental data of $\sigma_{NN\to N\Delta}$ in OBEM model. In this case, the dependence on the Delta mass $m$ of the momentum $p_{NN}$ and the matrix element are taken into account. $I_i$ is the isospin factor.

The decay rate of $\Delta\to N\pi$ is:
\begin{equation}
\label{eq:gammam}
\Gamma_{  N\pi }(m)=\Gamma _{\Delta }^{N\pi }\frac{m _{\Delta }}{m} \Bigl( \frac{p_{  N\pi }(m)}{p_{  N\pi }(m_{\Delta })}  \Bigl)^{2l+1}\frac{1.2}{1.0+0.2 ( \frac{p_{  N\pi }(m)}{p_{  N\pi }(m_{\Delta })}  )^{2l} .}
\end{equation}
$ m _{\Delta } $ is the pole mass of $\Delta $ resonance, $m _{\Delta } $=1.232 GeV and $m$ is the mass of $\Delta$. $\Gamma _{\Delta }^{N\pi }$ is the partial decay width into the channel N and $\pi$, $\Gamma _{\Delta }^{N\pi }$=0.115 GeV, and $l$ is the decay angular momentum of the exit channel ($l$=1).

The pion-nucleon cross section $\sigma _{N\pi\to \Delta}$ is calculated through the following equation:
\begin{equation}
\sigma _{N\pi\to \Delta}=<j_{N}m_{N}j_{\pi }m_{\pi}||J_{\Delta }M_{\Delta }> \frac{(2S_{\Delta }+1)}{(2S_{N}+1)(2S_{\pi }+1)}\,
\frac{\pi}{p_{cm}^{2}}\,\frac{\Gamma _{\Delta\rightarrow N\pi}\Gamma _{tot}}{ ( m_{\Delta }-\sqrt{s} )^{2}+\Gamma _{tot}^{2}/4}.\nonumber
\end{equation}

\subsubsection{The procedure for binary collision}
At the beginning of each time step (for the potential update), besides the distance $d$, the collision time $t_{coll}$
(for the time of closest approach) between particles $i$ and $k$ is calculated with the following formula~\cite{Bass:1998ca}
\begin{equation}
t_{coll} = -\frac{(\textbf{r}_i-\textbf{r}_k)\cdot(\textbf{p}_i/E_i-\textbf{p}_k/E_k)}{(\textbf{p}_i/E_i-\textbf{p}_k/E_k)^2}
 \label{eq4urqmd}.
\end{equation}
Here $\textbf{r}_i$, $\textbf{p}_i$, and $E_i= \sqrt{m_i^2 + \mathbf{p_i}^2}$ are the coordinate, momentum, and total energy of the particle $i$ in
the reference frame of the nucleus-nucleus collision. After scanning all possible collisions, the first collision ($t_{coll}$) that can happen in the time step is determined. The next step is to check whether this collision is allowed or not according to the cross section and Pauli blocking. If it is allowed, the collision happens.
The same processes will be re-done until the $t_{coll}$ is larger than the time step. Thus, collisions happen
one by one, each with its own time. This is called the time-step-free method.

\subsubsection{Pauli blocking treatment}
For each collision, the phase space densities $f_i$ and $f_j$ of the two outgoing particles are first determined,
in order to assure that they are in agreement with the Pauli principle ($f<1$), by using the expression
\begin{equation}
f_{i} = \frac{1}{(2s+1)}\sum_k  \frac{1}{(\pi\hbar)^3}\,\exp\big[-(\textbf{r}_i-\textbf{r}_k)^2/(2\sigma^2_r)\bigr] \, \exp\big[-(\textbf{p}_i-\textbf{p}_k)^2\cdot2\sigma^2_ r/\hbar^2\bigr]
 \label{eq5urqmd}.
\end{equation}
Here, $k$ represents particles with the same type around the outgoing particle $i$ (or $j$). The degeneracy factor (2$s$+1) is the number of the spin states of each particle, for nucleons $s=1/2$. The following two criteria are
then considered at the same time~\cite{li11},
\begin{equation}
\frac{4\pi}{3}r_{ik}^3\frac{4\pi}{3}p_{ik}^3 \geq (2s+1)\left(\frac{h}{2}\right)^3
 \label{eq6urqmd};
\end{equation}

\begin{equation}
P_{block}=1-(1-f_i)(1-f_j) <\xi
 \label{eq7urqmd}.
\end{equation}
Here, $r_{ik}$ and $p_{ik}$ denote the relative distance and momentum between particles $i$ and $k$ (particles of the same species as $i$). The above conditions are also considered for particle $j$. The symbol $\xi$
denotes a random number uniformly sampled between 0 and 1. If one of the two criteria is not fulfilled the collision is
not allowed and the two particles remain with their original momenta.



\newpage
\section{Conclusion}

The Transport Model Evaluation Project (TMEP) was launched to improve the consistency and predictive power of transport descriptions of intermediate-energy heavy-ion collisions, i.e., of beam energies from Fermi to relativistic energies of a few GeV. For the viability of the field, it is important that different simulations of transport models reach similar conclusions from the same data when using similar physical models. However, this was not always the case in recent studies. 
In this project, we choose not to attempt to develop a universal code, but rather to compare different implementations under controlled conditions. This was done on one hand by calculations in a box with periodic boundary conditions, which are able to test the various ingredients of a simulation separately. These calculations thus can be considered as benchmark calculations, since in most cases exact analytical or numerical results are available. On the other hand, we performed calculations of full heavy-ion reactions to see how the experience reached in nuclear matter box calculations is reflected in real collisions. 

In the present paper, we have first given a brief review of the different types and aspects of the transport approach. We then have reviewed in a compact form the results of the different studies within the TMEP project. Through these studies, we have gained much insight into the workings and problems of transport simulations. Yet, the project is not at its end. The box calculations have been systematically analyzed, ineffective or incorrect treatments have been improved and corrected, and the remaining differences
among the codes are well understood.
However, when the codes were then applied to a prediction of pion production allowing different physical models, the results still differed too much to allow a robust conclusion of the density dependence of the symmetry energy.
We think that these differences are due to effects that have not yet been tested in detail, such as the momentum dependence of the mean fields, which lead to threshold shifts in particle production and to questions of strict energy conservation. 
In an ongoing study we will test the codes again in a heavy-ion collision with controlled physical models to demonstrate the progress reached in this project. Also future developments of transport models are suggested, such as the dynamical treatment of spectral functions (off-shell effects), short-range correlations, clustering with controlled fluctuations, and the use of microscopic interactions and cross sections.
A more detailed summary of the findings of our study and a discussion and outlook on further development has been given in
Secs.~\ref{sec:summary} and \ref{sec:discussion}.


The number of transport codes in use by the community has risen substantially in the past decades. We have made an effort to enlist the participation from all major transport codes in this project. Altogether 14 codes of BUU type and 12 codes of QMD type participated in some part of our project.  A compact description of these codes presented here provides for an easy and compact reference for scientists who use these codes or want to assess their results. 
We recommend for the future to give version numbers to the codes to indicate, which features of a code are used in a particular calculation. It would also be desirable in the long run that more codes are openly available, e.g., in a repository.
We have also briefly highlighted the importance  of reliable transport models for heavy-on collisions in fundamental research and in applications.

Through the published and future comparison studies, we have established benchmark calculations that could assist the development of transport codes and ensure a more uniform quality. Our evaluation studies often suggest more effective strategies in implementing different algorithms for solving the transport simulations and thus provide guidance to new code developers (see, e.g., ~\cite{Lin:2018wjj}). We encourage existing codes that have not participated in our studies to independently verify our benchmark calculations before being used to compare with data.
Differences in the results of transport models using the same physical model can be regarded as a systematical theoretical error. In this project we aim to quantify and reduce this error. We believe that this interim report shows, that we have already made essential progress in this direction. The ongoing further comparisons will move closer to realistic heavy-ion collisions and will thus show the extent to which we can achieve this goal.

\section*{Acknowledgments}
{

The results and the publications of the TMEP project have been discussed at many video meetings especially in the last years with many participants, not all of whom are authors of this paper. In particular, we would like to thank William Lynch, Joe Natowitz, Christian Drischler, and Arnau Rios for very valuable contributions to these discussions. We also like to thank the hosts of meetings and conferences, who provided financial support to allow the writing group of the paper to meet before or after the main meetings. In particular, we thank the host institutions of the Symposium on Nuclear Symmetry Energy (NuSYM): the National Superconducting Laboratory (NSCL) at Michigan State University, USA (2013), IFJ-PAN and Jagellonian University, Krakow, Poland (2015), Tsinghua University, Beijing, China, (2016), the laboratory of GANIL, Caen, France (2017); the University of Busan, South Korea, (2018); the University of Hanoi, Vietnam (2019); and of the transport workshops at Shanghai Jiao Tong University, Shanghai, China (2014); Shanghai Institute of Applied Physics, SINAP, Shanghai, China  (2015); Beijing National University, Beijing, China (2016);  NSCL at Michigan State University, East Lansing, USA (2017); the China Institute of Atomic Energy CIAE, Beijing, China (2017); the ECT* Center at Trento, Italy (2019); and Sun Yat Sen University, Zhuhai, China (2019); and we thank the Laboratori Nazionali del Sud (LNS), INFN, Catania, Italy, for support during the IWM-EC meeting 2018 and with computational resources.

L. W. Chen acknowledges the support by the National Natural Science
Foundation of China under grant no.~11625521 and the National SKA Program of China grant no.~2020SKA0120300.
P.~Danielewicz acknowledges support from the U.S. Department of Energy
under grant no.~DE-SC0019209.
S. Jeon is supported in part by the Natural Sciences and Engineering
Research Council of Canada.
M. Kim and C.-H. Lee were supported by grants of the National Research Foundation of
Korea (NRF) funded  by the Korean government (Ministry of Science
and ICT and Ministry of Education) (nos.~2016R1A5A1013277 and 2018R1D1A1B07048599).
K. Kim and Y. Kim were supported by the Rare Isotope Science Project of Institute
for
Basic Science funded by Ministry of Science, ICT and Future Planning,
and National Research Foundation of Korea (2013M7A1A1075764).
C. M. Ko acknowledges the support by the U.S. Department of Energy under award no.~DE-SC0015266 and the Welch Foundation under grant no.~A-1358.
R. Kumar and Betty Tsang acknowledge the support by the U.S. Department of Energy (Office of Science) under grant no.~DE-SC0014530.
B.A. Li acknowledges the U.S. Department of Energy under award number DE-SC0013702. P. Danielewicz, C.M. Ko, B.A. Li and Betty Tsang also acknowledge the CUSTIPEN
(China-U.S. Theory Institute for Physics with Exotic Nuclei) under the
U.S. Department of Energy grant no.~DE-SC0009971. 
A.~Ono acknowledges support from Japan Society for the Promotion of
Science KAKENHI, grant nos.\ 24105008, 17K05432, and 21K03528. 
A. Sorensen and D. Oliinychenko received support through the U.S.
Department of Energy, Office of Science, Office of Nuclear Physics,
under contract number DE-AC02-05CH11231 and received support within the
framework of the Beam Energy Scan Theory (BEST) Topical Collaboration.
M.~B.~Tsang acknowledges the support by the U.S. National Science
Foundation
grant no. PHY-1565546 and the U.S. Department of Energy under grant nos.~DE-SC0021235, DE-SC0014530, and DE-NA0003908.
H.~Wolter acknowledges support by the Deutsche Forschungsgemeinschaft
(DFG, German Research Foundation) under Germany's Excellence Strategy EXC-2094-390783311, ORIGINS.
Y. J. Wang and Q.F. Li acknowledge support in part by the National Natural
Science Foundation of China grant nos.~U2032145, 11875125 and 12147219, and partly by
the National Key Research and Development Program of China under grant
no.~2020YFE0202002.
J.~Xu acknowledges
the support by the National Natural Science Foundation of China under
grant no.~11922514.
F. S. Zhang acknowledges National Natural Science Foundation of China
under grant nos.~12135004, 11635003, 11025524, and 11161130520.
Y. X. Zhang acknowledges support in part by National Natural Science
Foundation of China grant nos. 11875323, 11961141003, 11475262, and 11365004,
by the National Key Basic Research Development Program of China under grant no.~2018YFA0404404, and by the Continuous Basic Scientific Research Project
(nos.~WDJC-2019-13 and BJ20002501).
Z. Zhang acknowledges the support by the National Natural Science
Foundation of China under grant no.~11905302. Z. Q. Feng acknowledges the support by the National Natural Science Foundation of China under grant nos.~12175072 and 11722546.
}


\bibliographystyle{elsarticle-num}
\bibliography{9-reference}   

\end{document}